\documentclass[useAMS,usenatbib]{mn2e}
\usepackage{subfigure}
\usepackage{graphicx}

\title[The Optical morphologies of the 2Jy sample]{The optical morphologies of the 2Jy sample of radio galaxies: evidence for galaxy interactions}
\author[C. Ramos Almeida et al.]
{\parbox{\textwidth}{C. Ramos Almeida$^{1}$\thanks{E-mail:C.Ramos@sheffield.ac.uk},
C.~N.~Tadhunter$^{1}$,
K.~J.~Inskip$^{2}$,
R.~Morganti$^{3,4}$,
J.~Holt$^{5}$, \&
D. Dicken$^{6}$}\vspace{0.4cm}\\
\parbox{\textwidth}{$^{1}$Department of Physics and Astronomy, University of Sheffield, Sheffield, S3 7RH, UK\\
$^{2}$Max-Planck-Institut f\"ur Astronomie, K\"oningstuhl 17, D-69117 Heidelberg, Germany\\
$^{3}$Netherlands Institute for Radio Astronomy, Postbus 2, 7990 AA Dwingeloo, the Netherlands\\
$^{4}$Kapteyn Astronomical Institute, University of Groningen, Postbus 800, 9700 AV Groningen, the Netherlands\\
$^{5}$Leiden Observatory, Leiden University, PO Box 9513, 2300 RA Leiden, the Netherlands\\
$^{6}$Department of Physics and Astronomy, Rochester Institute of Technology, 84 Lomb Memorial Drive, Rochester NY 14623, USA}}
\begin{document}

\date{}

\pagerange{\pageref{firstpage}--\pageref{lastpage}} \pubyear{2010}

\maketitle

\label{firstpage}

\begin{abstract}
We present deep GMOS-S/Gemini optical broad-band images for a complete sample of 46 southern 2Jy radio galaxies at 
intermediate redshifts (0.05$<$z$<$0.7). 
Based on them, we discuss the role of galaxy interactions in the triggering of powerful radio galaxies (PRGs). 
The high-quality observations presented here show for the first time that the overall majority of PRGs 
at intermediate redshifts (78-85\%) show peculiarities in their optical morphologies at relatively high levels of 
surface brightness ($\tilde{\mu}_V=23.6~mag~arcsec^{-2}$; $\Delta\mu_V\simeq[21,26]~mag~arcsec^{-2}$). 
The observed morphological peculiarities include tails, fans, bridges, shells, dust lanes, irregular features, 
amorphous haloes, and multiple nuclei. 
While the results for many of the galaxies are consistent with them being observed at, or after, the time of
coalescence of the nuclei in a galaxy merger, we find that more than one-third of the sample are observed in 
a pre-coalescence phase of the merger, or following a close encounter between galaxies that will not 
necessarily lead to a merger.
By dividing the sample into Weak-Line Radio Galaxies (WLRGs; 11 objects) and 
Strong-Line Radio Galaxies (SLRGs; 35 objects) 
we find that only 27\% of the former show clear evidence for interactions in their optical morphologies, in contrast to the SLRGs, of which 
at least 94\% appear interacting. 
This is consistent with the idea that many WLRGs are fuelled/triggered by Bondi accretion of hot gas.
However, the evidence for interactions and dust features in a fraction of them indicates that the accretion of cold gas cannot always be 
ruled out. 
Of the 28\% of the sample that display evidence for significant starburst activity, we find that 92\% present 
disturbed morphologies, following the same general trend as the total and SLRG samples.
By comparing our PRGs with various samples of quiescent ellipticals from the literature, we conclude that the 
percentage of morphological disturbance that we find here exceeds that found for quiescent ellipticals
when similar surface brightnesses are considered. 
Overall, our study indicates that galaxy interactions are likely to play a key role in the triggering of AGN/jet activity.
\end{abstract}

\begin{keywords}
galaxies: active -- galaxies: nuclei -- galaxies: interactions -- galaxies: photometry.
\end{keywords}

\section{Introduction}
\label{intro}

Over recent years, substantial evidence has accumulated for links between Active Galactic Nuclei (AGN) activity 
and the evolution of galaxies (e.g., \citealt{Cattaneo09}). Not only are the masses of the central black holes (BHs) 
which power the AGN tightly correlated with the properties of the bulges of the host galaxies 
\citep{Gebhardt00,Ferrarese00,Greene06}, but the variation in the co-moving number density of PRGs 
\citep{Dunlop90} bears a striking similarity to the evolution of the global star formation rate 
in the Universe \citep{Madau96}. 
Simulations of hierarchical galaxy evolution predict that the periods of BH growth and AGN activity are intimately 
tied to the growth of the host galaxy, and that the triggering of the main phase of AGN activity in gas-rich 
mergers will always be accompanied by a major galaxy-wide starburst \citep{Kauffmann00,diMatteo05,Springel05,Hopkins08a,
Hopkins08b,Somerville08}. 
However, the timescales involved in both the triggering of the merger-induced starburst and the 
AGN activity remain uncertain. 

In order to understand the symbiosis between galaxy evolution and AGN activity, it is important 
to test the models for the triggering of nuclear activity and starbursts (starbursts) using observations of well-defined samples of nearby AGN. 
Radio galaxies are particularly useful in this context because they are invariably associated with early-type hosts, 
allowing cleaner searches for signs of morphological disturbance and recent star formation activity. 

Although morphological evidence for mergers/interactions has been found in some nearby radio galaxies (Heckman et al.~1986; 
Smith \& Heckman 1989, hereafter \citealt{Heckman86},\citealt{Smith89}), 
this mechanism for triggering the AGN activity is not without controversy \citep{Dunlop03,Grogin05}. 
On the one hand, using ground-based deep imaging observations 
\citealt{Smith89} found that over 50\% of their sample of PRGs display morphological deviations from elliptical symmetry at high 
levels of surface brightness, and that about half of the PRGs with strong optical emission lines exhibit peculiar 
optical morphologies. These peculiarities include tails, fans, bridges, shells and dust lanes. 
More recently, \citet{Roche00} found that at least 13 out of 15 3CR radio galaxies with ground-based optical imaging 
appear to be interacting with neighbours, based on the projected separations of the galaxy pairs. 
On the other hand, the high spatial resolution Hubble Space Telescope (HST) images of a sample of nearby
radio galaxies, radio-quiet quasars (QSOs) and radio-loud QSOs presented by \citet{Dunlop03} indicated that they are
hosted by relatively undisturbed giant elliptical galaxies. Indeed, the main conclusion of the latter work is that 
the hosts of radio-loud and radio-quiet AGN are indistinguishable from quiescent 
ellipticals of similar mass. 

At least part of the discrepancy between the previous imaging results may be
due to the differences in the depth of the ground- and space-based observations. 
Despite the higher spatial resolution of the HST images, 
the short exposure time data obtained by \citet{Dunlop03} using the Wide Field Planetary Camera 2 (WFPC2)/HST 
are relatively insensitive to large-scale diffuse structures such as tidal tails, fans or shells.
Indeed, \citet{Canalizo07} and \citet{Bennert08} recently presented deeper Advanced Camera for Surveys (ACS)/HST 
optical images (5 orbits) of five nearby QSOs classified by \citet{Dunlop03} as undisturbed elliptical galaxies. 
These deeper images reveal shells and tidal tails in four of the five galaxies
that were not apparent in the earlier HST observations. 

The morphological features found by \citealt{Heckman86}, \citealt{Smith89}, and \citet{Roche00} can be the result of 
either mergers between gas-rich galaxies (i.e., late-type galaxies), gas-poor galaxies (those involving two
eary-type galaxies: the so-called ``dry-mergers'') or  galaxy encounters that will not lead to a merger. 
Dry-mergers have been shown to explain the observed
properties of high-luminosity (M$_V\la$-21) boxy, velocity dispersion-suported early-type galaxies \citep{Naab06}, 
whereas simulations of merging disks easily reproduce the remnants of disky, rotationally-suported 
low-luminosity elliptical galaxies \citep{Toomre72,Bendo00,Naab99,Naab03}. 
The morphological signatures of gas-poor interactions are normally very weak: e.g., broad and low surface 
brightness tidal tails, fans, asymmetries, and double nuclei \citep{Borne84,Borne85,Combes95,Naab06,Bell06}. 
Given the small amount of cool gas involved in the interaction between ellipical galaxies, the gas disruption in such dry-mergers 
may result in only a small amount of star formation. 
In contrast, gas-rich interactions usually result in sharper, narrower and brighter tails, 
shells or fans. In this case, there is plenty of cold gas to trigger the star formation.   
In general, close pairs and gas-rich mergers are expected to be visible for a timescale between 0.5 and 1.5 Gyr 
\citep{Fevre00,Patton02,Conselice03,Kawata06}, whereas dry-mergers are visible for only $\sim$150 Myr \citep{Bell06}.   
Considering that the AGN/jet activity in PRGs is estimated to last no longer than 100 Myr \citep{Leahy89}, 
gas-rich merger signatures would be detectable once the radio/AGN activity faded, whereas the visible signs of a dry-merger
would be roughly coincident with the PRG lifetime.

Although evidence for mergers and interactions is found in many PRGs, a subset 
presents optical morphologies and emission line kinematics that do not support the 
idea of the triggering of the radio activity via mergers. These include some central cluster
galaxies surrounded by massive haloes of hot gas \citep{Tadhunter89,Baum92}. In such cases, the infall of cold
gas condensing from the X-ray haloes in cooling flows has been suggested as a triggering mechanism (e.g., 
\citealt{Tadhunter89,Baum92,Bremer97}).
Also, it has been shown that the direct accretion of hot gas from the X-ray haloes of galaxies 
is a plausible mechanism for fuelling radio galaxies that lack strong emission lines, namely the Weak-Line Radio Galaxies 
(WLRGs; \citealt{Allen06,Best06,Hardcastle07,Balmaverde08}).


In order to test the idea that the powerful radio galaxies are triggered in galaxy interactions, 
and also to examine the diversity of triggering mechanisms, it is crucial to use deep imaging observations to determine 
the interaction status of the galaxies. In this, the first of two papers, we present deep imaging data for a complete 
sample of 46 intermediate redshift (0.05$<$z$<$0.7) 2Jy radio galaxies \citep{Tadhunter93,Tadhunter98}, making
a first qualitative study of the morphologies of the host galaxies.  
We search for signs of nearby companions, secondary nuclei, shells and extended low surface brightness 
features such as tidal tails and bridges. 
This represents the first systematic imaging study of a major sample of radio galaxies using an 8m telescope. 
In a forthcoming paper we will make a full quantitative assessment of the interaction status of the radio galaxies, 
and also quantify their environments, relating these properties to those of the AGN and stellar populations.

In Section 2, we present details of our sample of PRGs. Section 3 describes the observations and data reduction. A 
description of the image enhancement techniques employed in this work can be found in Section 4, and in Section 5
we present the observational results. Discussion and conclusions are presented in Sections 6 and 7, respectively. 
Finally, notes on the individual galaxies are reported in Appendix \ref{individual}, and processed images for all the 
galaxies in the sample are shown in Appendix \ref{images}. 
Throughhout this paper we assume a cosmology with H$_0$=73 km~s$^{-1}$~Mpc$^{-1}$; $\Omega_m$=0.27, and $\Omega_{\Lambda}$=0.73.

\section{Sample Selection}
\label{selection}

The objects studied here comprise all powerful radio galaxies (PRGs) and quasars from the \citet{Tadhunter93} sample 
of 2Jy radio galaxies with S$_{2.7GHz}\ge$ 2.0 Jy, steep radio spectra $\alpha_{2.7}^{4.8} > 0.5~(F_{\nu}\propto\nu^{-\alpha})$, 
declinations $\delta<+10\degr$ and redshifts 0.05$<$z$<$0.7 (see Table \ref{data}). 
It is itself a subset of the \citet{Wall85} complete sample of 2Jy radio sources.
We also included the source PKS 0347+05 \citep{diserego94}, 
that was not in the original \citet{Tadhunter93} sample, but subsequently proved to fulfil the same selection criteria.
This sample is unique in terms of the depth, completeness, and quality of its supporting optical \citep{Tadhunter93,Tadhunter98,Tadhunter02,Holt07}, 
near-infrared (NIR; for the z $<$ 0.5 subset; \citealt{Inskip10}), mid- to far-infrared \citep{Dicken08,Dicken09,Dicken10}, 
and radio data \citep{Morganti93,Morganti97,Morganti99}, which provide an accurate picture, not only of the radio source and AGN properties, 
but also of the level of star formation in each host galaxy. Much of the existing information on this sample is summarized on
the 2Jy website\footnote{http://2jy.extragalactic.info}.
For further details on sample selection and completeness see \citet{Tadhunter93,Tadhunter98} and \citet{Dicken08}. 
The z $>$ 0.05 limit ensures that the radio galaxies are genuinely powerful sources, while the z $<$ 0.7 limit ensures
that sources are sufficiently nearby for detailed morphological studies.
In all the Tables and Appendices presented in this work the PRGs are ordered in redshift, from the most nearby 
sources to the most distant ones.

\begin{table*}
\centering
\begin{tabular}{lcllccccc}
\hline
\hline
PKS ID   & Other   &  Optical&   Radio     &        z  & Scale   &  Distance &   Morphology & Group    \\
\hline
0620-52	 &         &   WLRG* &    FRI      &	 0.051 &  965	 &  220  &  \dots 	& 5 \\
0625-53	 &         &   WLRG  &    FRI      &	 0.054 &  1014   &  232  &  B		& 1 \\  
0915-11	 & Hydra A &   WLRG* &    FRI      &	 0.054 &  1044   &  240  &  D		& 4 \\         
0625-35	 & OH-342  &   WLRG  &    FRI      &	 0.055 &  1027   &  236  &  J  	        & 5 \\  
2221-02	 & 3C445   &   BLRG  &    FRII     &	 0.057 &  1027   &  236  &  F,S 	& 1 \\ 
1949+02	 & 3C403   &   NLRG  &    FRII     &	 0.059 &  1083   &  250  &  S,D	        & 2 \\
1954-55	 &         &   WLRG  &    FRI      &	 0.060 &  1075   &  248  &  \dots 	& 5 \\ 
1814-63	 &         &   NLRG* &    CSS      &	 0.063 &  1160   &  270  &  2I,D      	& 2 \\
0349-27	 &         &   NLRG  &    FRII     &	 0.066 &  1215   &  285  &  2B,[S]  	& 1 \\
0034-01  & 3C15    &   WLRG  &    FRII     &	 0.073 &  1316   &  312  &  J		& 5 \\      
0945+07  & 3C227   &   BLRG  &    FRII     &	 0.086 &  1565   &  381  &  S		& 2 \\
0404+03  & 3C105   &   NLRG  &    FRII     &	 0.089 &  1593   &  398  &  [S]    	& 5 \\
2356-61  &         &   NLRG  &    FRII     &	 0.096 &  1708   &  423  &  2S,F,I	& 2 \\       
1733-56  &         &   BLRG* &    FRII     &	 0.098 &  1743   &  433  &  2T,2I,2S,[D]& 2 \\
1559+02  & 3C327   &   NLRG  &    FRII     &	 0.104 &  1854   &  467  &  2S,D,[2N]	& 2 \\
0806-10  & 3C195   &   NLRG  &    FRII     &	 0.110 &  1943   &  494  &  F,2S	& 2 \\
1839-48  &         &   WLRG  &    FRI      &	 0.112 &  1945   &  495  &  2N,S,[T]	& 2,3 \\
0043-42  &         &   WLRG  &    FRII     &	 0.116 &  2010   &  516  &  [2N],[B]  	& 5 \\
0213-13  & 3C62    &   NLRG  &    FRII     &	 0.147 &  2464   &  668  &  2S,[T]  	& 2 \\
0442-28  &         &   NLRG  &    FRII     &	 0.147 &  2474   &  671  &  S		& 2 \\
2211-17  & 3C444   &   WLRG  &    FRII     &	 0.153 &  2544   &  696  &  D,[F]	& 4 \\  								     
1648+05  & Herc A  &   WLRG  &    FRII?    &	 0.154 &  2574   &  707  &  D		& 4 \\  			   
1934-63  &         &   NLRG* &    GPS      &	 0.183 &  2961   &  854  &  2N,2T	& 1,3 \\
0038+09  & 3C18    &   BLRG  &    FRII     &	 0.188 &  3013   &  875  &  T		& 2 \\
2135-14  &         &  QSO    &    FRII     &	 0.200 &  3171   &  941  &  T,S,A,[B]	& 2 \\
0035-02  & 3C17    &  BLRG   &    FRII     &	 0.220 &  3408   & 1044  &  B,F,[S]	& 1 \\
2314+03  & 3C459   &  NLRG*  &    FRII     &	 0.220 &  3409   & 1044  &  2F,[T]  	& 2 \\   
1932-46  &         &  BLRG*  &    FRII     &	 0.231 &  3551   & 1109  &  2F,A,I	& 2 \\
1151-34  &         &  QSO*   &    CSS      &	 0.258 &  3873   & 1267  &  F,[S]	& 1 \\
0859-25  &         &  NLRG   &    FRII     &	 0.305 &  4362   & 1535  &  2N		& 3 \\
2250-41  &         &  NLRG*  &    FRII     &	 0.310 &  4393   & 1553  &  2B,[T],[F]	& 1 \\
1355-41  &         &  QSO    &    FRII     &	 0.313 &  4438   & 1580  &  S,T 	& 2 \\
0023-26  &         &  NLRG*  &    CSS      &	 0.322 &  4508   & 1623  &  A,[D]	& 2 \\
0347+05  &         &  WLRG*  &    FRII     &	 0.339 &  4673   & 1727  &  B,3T,D	& 1 \\
0039-44  &         &  NLRG   &    FRII     &	 0.346 &  4734   & 1767  &  2N,3S,[T],[D]& 2,3 \\  	    
0105-16  & 3C32    &  NLRG   &    FRII     &	 0.400 &  5192   & 2096  &  B		& 1 \\
1938-15  &         &  BLRG   &    FRII     &	 0.452 &  5589   & 2428  &  F		& 2 \\
1602+01  & 3C327.1 &  BLRG   &    FRII     &	 0.462 &  5667   & 2500  &  F,S,[J]	& 2 \\
1306-09  &         &  NLRG   &    CSS      &	 0.464 &  5686   & 2518  &  2N,S	& 2,3 \\
1547-79  &         &  BLRG   &    FRII     &	 0.483 &  5810   & 2636  &  2N,T	& 2,3 \\
1136-13  &         &  QSO    &    FRII     &	 0.554 &  6276   & 3147  &  T,J		& 2 \\
0117-15  & 3C38    &  NLRG   &    FRII     &	 0.565 &  6303   & 3180  &  3N,S,I,[D]  & 2,3 \\
0252-71  &         &  NLRG   &    CSS      &	 0.566 &  6324   & 3207  &  [A] 	& 5 \\
0235-19  & OD-159  &  BLRG   &    FRII     &	 0.620 &  6592   & 3565  &  2T,[B]	& 2 \\        
2135-20  & OX-258  &  BLRG*  &    CSS      &	 0.635 &  6663   & 3669  &  F		& 2 \\
0409-75  &         &  NLRG*  &    FRII     &	 0.693 &  6925   & 4095  &  2N  	& 3 \\
\hline		     			      		 
\end{tabular}						 
\caption{Full classification of the sample objects ordered by redshift. Columns 1 and 2 give the PKS and alternative names (if any) for the sample objects.  
Columns 3 and 4 give the spectroscopic class (* indicates evidence of starburst signatures) and radio morphology. Columns 5, 6, and 7 list the 
spectroscopic redshift as reported in the NASA/IPAC Extragalactic Database (NED), the scale in pc~arcsec$^{-1}$ 
(calculated using H$_0=73~km~s^{-1}~Mpc^{-1}$, $\Omega_{matter}$=0.27, 
$\Omega_{vacuum}$=0.73), and the luminosity distance (Mpc). 
Columns 8 and 9 correspond to our morphological classification (T: Tail; F: Fan; B: Bridge; S: Shell; D: Dust feature;  
2N: Double Nucleus; 3N: Triple Nucleus; A: Amorphous Halo; I: Irregular feature; and J: Jet; brackets indicate uncertain
identification of the feature), and sample division in 1) galaxy pair or group in tidal interaction; 2) galaxies 
showing T,F,S,D,A,I; 3) multiple nuclei (inside a 9.6 kpc); 4) galaxies with dust as the only detected feature, 5) isolated galaxies with no sign
of interaction.}
\label{data}
\end{table*}

In terms of the optical classification, based on both previous optical spectra \citep{Tadhunter98}
and on optical appearance \citep{Wall85}, the sample comprises 24\% WLRGs (sources with [O III]$\lambda$5007 
emission line equivalent widths below 10 \AA), 43\% Narrow-Line Radio Galaxies (NLRGs), and 33\%
Broad-Line Radio Galaxies and quasars (BLRGs and QSOs). The QSOs were classified by \citet{Wall85}
based on their stellar appearance on optical images, and later confirmed from optical spectroscopy
\citep{Tadhunter98}. BLRGs and NLRGs are defined on the basis of whether or not their optical spectra show 
broad line components of the permitted emission lines. 

Considering the radio morphologies, Fanaroff-Riley II (FRII) sources constitute
the majority of the sample (72\%), 13\% are Fanaroff-Riley I (FRI), and the remaining 15\% correspond to compact, 
steep-spectrum (CSS) or Gigahertz-peaked spectrum (GPS) sources (Table \ref{data}). 
All FRIs in our sample are WLRGs according to their optical spectra. In contrast, the majority
of FRII and CSS/GPS sources are classified as SLRGs, with a minority showing WLRG spectra. 
Only thirteen out of the 46 radio galaxies in the sample (28\%) present any 
evidence of recent starburst activity, based on their optical spectra \citep{Tadhunter93}, 
far-infrared (FIR) excess, and/or the detection of polycyclic aromatic hydrocarbons (PAHs) 
in their mid-infrared (MIR) spectra \citep{Dicken08,Dicken09,Dicken10}.

\section{Observations and Data Reduction}
\label{observations}

Deep optical imaging data were obtained for the full sample using the Gemini Multi-Object Spectrograph South (GMOS-S) 
on the 8.1 m Gemini South telescope at Cerro Pach\' on, Chile. The observations were carried out in queue mode between 
July 2008 and April 2009 under good seeing conditions (median seeing FWHM = 0.8\arcsec, ranging from 0.4\arcsec~to 1.15\arcsec), 
as required to allow the best chance to detect
subtle morphological features. The seeing values were measured individually for each of the 46 GMOS-S images 
using foreground stars. 
Details of the observations are reported in Table \ref{mag}.
The GMOS-S detector \citep{Hook04} comprises three adjacent 2048x4096 pixel CCDs separated by two gaps of $\sim$2.8\arcsec, 
giving a field-of-view (FOV) of 5.5x5.5 arcmin$^2$, with a pixel size of 0.146\arcsec~pixel$^{-1}$.  

\begin{table*}
\centering
\begin{tabular}{lccccccc}
\hline
\hline
PKS ID & Exptime (s) & Obs. date & Seeing & Filter & A$_{\lambda}$ (mag) & Mag (AB) & Mag (corr)  \\
\hline
0620-52& 100x8  &   2008-10-24   & 0.70\arcsec  & r' & 0.181 & 13.75    &	13.51         \\
0625-53& 100x8  &   2008-11-01   & 1.00\arcsec  & r' & 0.250 & 13.81    &	13.49         \\  
0915-11&  67x16 &   2009-02-20   & 1.00\arcsec  & r' & 0.109 & 14.33    &	14.15         \\	 
0625-35&  90x8  &   2008-11-01   & 1.00\arcsec  & r' & 0.176 & 14.22    &	13.98         \\  
2221-02&  16x16 &   2008-09-24   & 0.60\arcsec  & r' & 0.220 & 15.25    &	14.96         \\ 
1949+02& 100x8  &   2008-08-05   & 0.70\arcsec  & r' & 0.505 & 14.98    &	14.40         \\
1954-55&  62x16 &   2008-08-08   & 0.85\arcsec  & r' & 0.153 & 14.88*   &	14.65*         \\ 
1814-63&  67x16/15x4 &   2009-04-21/23   & 0.70\arcsec/0.65\arcsec  & r' & 0.228 & 15.84*   &	15.53*         \\
0349-27& 100x8  &   2008-08-30   & 0.85\arcsec  & r' & 0.024 & 15.83    &	15.72         \\
0034-01& 100x8  &   2008-10-06   & 0.85\arcsec  & r' & 0.057 & 15.23    &	15.08         \\      
0945+07&  67x16 &   2009-02-21   & 0.85\arcsec  & r' & 0.069 & 16.02    &	15.85         \\
0404+03& 250x4  &   2008-10-24   & 0.70\arcsec  & r' & 1.264 & 17.67    &	16.30         \\
2356-61& 100x8  &   2008-08-26   & 1.00\arcsec  & r' & 0.036 & 15.82    &	15.67         \\       
1733-56& 125x8  &   2009-04-03   & 0.55\arcsec  & r' & 0.264 & 16.00    &	15.62         \\
1559+02&  67x16 &   2009-03-25   & 0.95\arcsec  & r' & 0.237 & 15.84    &	15.48         \\
0806-10& 125x8  &   2009-02-19   & 0.70\arcsec  & r' & 0.225 & 16.06    &	15.70         \\
1839-48& 125x8  &   2009-04-03   & 0.75\arcsec  & r' & 0.178 & 15.67    &	15.36         \\
0043-42& 250x4  &   2008-08-10   & 0.80\arcsec  & r' & 0.031 & 16.20    &	16.03         \\
0213-13& 250x4  &   2008-10-28   & 0.85\arcsec  & r' & 0.054 & 16.69    &	16.46         \\
0442-28& 250x4  &   2008-10-23   & 0.55\arcsec  & r' & 0.087 & 16.80    &	16.54         \\
2211-17& 250x4  &   2008-08-26   & 0.85\arcsec  & r' & 0.067 & 17.01    &	16.76         \\								      
1648+05& 250x8  &   2009-03-25   & 0.85\arcsec  & r' & 0.251 & 17.79    &	17.35         \\			    
1934-63& 250x4  &   2008-07-25   & 0.65\arcsec  & r' & 0.226 & 17.75    &	17.30         \\
0038+09& 250x4  &   2008-10-24   & 1.00\arcsec  & r' & 0.422 & 18.34    &	17.69         \\
2135-14&  35x12 &   2008-09-24   & 0.55\arcsec  & r' & 0.139 & 16.41    &	16.03         \\
0035-02& 250x4  &   2008-08-09   & 0.60\arcsec  & r' & 0.062 & 18.16    &	17.83         \\
2314+03& 250x4  &   2008-10-06   & 1.10\arcsec  & r' & 0.173 & 17.48    &	17.04         \\   
1932-46& 300x4  &   2008-07-25   & 0.80\arcsec  & r' & 0.144 & 18.68    &	18.25         \\
1151-34& 250x4  &   2009-02-22   & 0.55\arcsec  & r' & 0.219 & 17.75    &	17.21         \\
0859-25& 300x4  &   2009-02-21   & 0.85\arcsec  & r' & 0.546 & 18.80    &	17.85         \\
2250-41& 167x6  &   2008-08-04   & 1.00\arcsec  & i' & 0.025 & 16.58    &	16.32         \\
1355-41&  67x16 &   2009-03-19   & 0.40\arcsec  & r' & 0.232 & \dots    &	\dots 	      \\
0023-26& 300x5  &   2008-08-04   & 0.90\arcsec  & r' & 0.042 & 19.08    &	18.60 	      \\
0347+05& 300x4  &   2008-10-07   & 0.50\arcsec  & r' & 0.745 & 19.68    &	18.46 	      \\
0039-44& 300x4  &   2008-08-09   & 0.55\arcsec  & r' & 0.021 & 18.82    &	18.31 	      \\		 
0105-16& 300x4  &   2008-10-26   & 0.85\arcsec  & i' & 0.041 & 19.02    &	18.66 	      \\
1938-15& 250x6  &   2008-07-25   & 0.80\arcsec  & i' & 0.465 & 19.30    &	18.47 	      \\
1602+01& 168x6  &   2009-03-02   & 0.70\arcsec  & i' & 0.250 & 18.93    &	18.31 	      \\
1306-09& 167x6  &   2009-02-24   & 0.55\arcsec  & i' & 0.087 & 18.79    &	18.33 	      \\
1547-79& 167x6  &   2009-02-25   & 0.80\arcsec  & i' & 0.402 & 18.36    &	17.56	      \\
1136-13&  83x12 &   2009-02-05   & 0.80\arcsec  & i' & 0.074 & 16.50    &	15.94	      \\
0117-15& 250x6  &   2008-10-22   & 0.50\arcsec  & i' & 0.035 & 19.19    &	18.66 	      \\
0252-71& 250x6  &   2008-10-28   & 1.00\arcsec  & i' & 0.058 & 20.46    &	19.90	      \\
0235-19& 250x6  &   2008-10-28   & 0.70\arcsec  & i' & 0.062 & 19.11    &	18.44	      \\         
2135-20& 250x6  &   2008-08-05   & 0.95\arcsec  & i' & 0.064 & 19.09    &	18.38	      \\
0409-75& 250x6  &   2008-10-01   & 1.15\arcsec  & i' & 0.148 & 20.54    &	19.57	      \\
\hline		     			      		 
\end{tabular}						 
\caption{Summary of observations. Columns 1, 2, and 3 give the PKS name, the exposure times, and the date of observation.
Columns 4 and 5 list the measured seeing and the filter employed in each observation. Column 6 gives either 
the A$_{r'}$ or A$_{i'}$ (see column 5) values determined using the E(B-V) values from the NASA/IPAC IRSA and the Galactic
extinction law of \citet{Cardelli89}. Two last columns (7 and 8) correspond to the apparent magnitudes of the galaxies in a 
30 kpc diameter metric aperture after removing the flux from any 
contaminating object and the K-corrected magnitudes in both $r'$ and $i'$ filters using \citet{Fukugita95} K-corrections. 
All the magnitudes are also corrected for Galactic extinction. * The determination of these magnitudes was 
particularly difficult due to the presence of a bright foreground star in front of the galaxies. We do not report any  
magnitude values for PKS 1355-41 because the galaxy nucleus is saturated.}
\label{mag}
\end{table*}

With the exception of the source PKS 2250-41\footnote{Observed with the $i'$-band filter in order to avoid contamination 
by off-nuclear [O III]$\lambda$5007 \AA~line emission, 
which we know is very prominent from optical spectra \citep{Tadhunter02}.}, all the 
galaxies with z $\la$ 0.4 were observed in the $r'$-band filter ($r_{-}G0326$, $\lambda_{eff}$=6300 \AA, $\Delta\lambda$=1360 \AA), 
while those with z $>$ 0.4 were observed in the $i'$-band ($i_{-}G0327$, $\lambda_{eff}$=7800 \AA, $\Delta\lambda$=1440 \AA), 
in order to cover the typical rest frame wavelength range 4500-6000 \AA. 
The GMOS-S $r'$ and $i'$ filters are very similar to the $r$ and $i$ used
by the Sloan Digital Sky Survey (SDSS; see \citealt{Fukugita96} for details of the latter photometric system).  

Depending on source brightness, and in order to avoid saturation of sources with either bright nuclei or close foreground 
stars, 
we took from four to sixteen images per filter, arranged in either 2x2 square or 
3x2 rectangular dither pattern with a step size of 10\arcsec. The purpose of the dithering is to eliminate the gaps between 
the three CCDs, remove other image blemishes, and improve flat fielding.
For details of the 
dither patterns and exposure times for each source see Table \ref{mag}. The total integration times 
(from 256s to 1500s depending on redshift) were designed 
to make the observations deep enough to detect features at relatively low levels of surface brightness.
In addition to longer exposure images, the radio source PKS 1814-63 also has a short exposure time observation of 4x15s,
which was obtained to avoid saturation of a foreground star in front of the galaxy (see Table \ref{mag}).
As well as the main science target fields, offset fields ($\sim$20\arcmin~offset) were observed after each radio 
galaxy observation, in order to better quantify the background galaxy population of the host galaxies for 
future environmental studies. The latter data will be analyzed in a forthcoming paper.

The data were reduced using the GMOS dedicated software packages within the IRAF\footnote{IRAF is distributed by the National 
Optical Astronomy Observatory, which
is operated by the Association of Universities for the Research in Astronomy, Inc., under cooperative agreement 
with the National Science Foundation
(http://iraf.noao.edu/).} environment. The reduction process included bias subtraction and flat fielding, 
merging of the three separate CCD images comprising each exposure into a single combined frame, and finally, 
production of a single image by combining the 
mosaicked frames. For the $i'$-band images, an additional step was necessary to remove the fringing that 
is significant for this filter. The dithered images were median-combined without aligning them after the objects were masked, 
resulting in an image of the fringe pattern only. This master fringe frame was later subtracted from the individual 
frames before coadding them.  
 
In order to calibrate the data in terms of a magnitude scale, observations of several Landolt photometric 
standard stars were taken on the same nights as the target radio galaxies. Using the latter to determine the photometric
zero points, we estimate that
the photometric accuracy of the observations is $\pm$0.04 mag in both $r'$ and $i'$ bands.
Column 7 of Table \ref{mag} lists the aperture magnitudes in the AB system 
using a fixed metric aperture of 30 kpc for each source. We have corrected these magnitudes for any 
contaminating objects (e.g., foreground stars and small companion galaxies)
that are close enough to the science target to be contained in our fixed aperture. In some cases, the presence of saturated 
or near-saturated stars very close to the radio galaxy complicates the determination of the aperture magnitudes 
(e.g., PKS 1954-55 and PKS 1814-63). 

Considering the relatively large range of redshift spanned by the galaxies in our sample, it is necessary
to apply K-corrections to the aperture magnitudes in order to make them comparable. In order to do this, 
we made use of the values reported in \citet{Frei94} and \citet{Fukugita95} for elliptical galaxies. 
Column 8 of Table \ref{mag} presents 
the aperture magnitudes from column 7, K-corrected and Galactic extinction-corrected. 
For the galactic extinction correction, we made use of the E(B-V) values from \citet{Schlegel98}
and the Galactic extinction law of \citet{Cardelli89} to derive the corresponding 
A$_{r'}$ and A$_{i'}$ values (see Table \ref{mag}).

\section{Data Analysis}
\label{analysis}

The main aim of this work is to take advantage of our deep GMOS-S images to detect fine structure in the optical morphologies
of powerful radio galaxies, such as tidal features, multiple nuclei, and dust (see description 
in Section \ref{morphology}).  
Our primary morphological assessement of each galaxy was based on visual inspection of the fully-reduced GMOS-S images. 
However, in order to improve our sensitivity to faint structure, we have also made use of three image enhancement techniques, 
which have been shown to greatly improve the search for faint features in elliptical galaxies:
1) image filtering; 2) unsharp-masking; and 3) smoothed galaxy subtraction. 
All these techniques have been applied to the data using the Interactive Data Lenguage (IDL). 

\begin{itemize}

\item {\it Image filtering}. 
This enhancement technique computes the median of the pixels contained in a moving square box of a given width, 
replacing the central pixel with the computed median (i.e., the image is smoothed; \citealt{McGaugh90}). It also processes the 
pixels at the edges and corners of the images, 
and it is possible to apply iterative smoothing. For the GMOS-S images, we used a box width of 5 pixels, with 
the exception of two galaxies with large-scale tidal features (PKS 0038+09 and PKS 0349-27) for which we employed
larger box width values (10 and 20 pixels, respectively). By median-filtering the images, large and/or broad 
features like bridges or tails, and broad fans or shells are generally well enhanced.

\item {\it Unsharp-masking}. This technique is appropiate for enhancing sharp features such as shells, ripples and tails 
in the outer parts of elliptical galaxies. When the features are faint, the radial intensity gradient of the galaxy 
makes them difficult to detect. This procedure consists of smoothing the image until the small-scale features disappear, 
and then subtracting this blurred image from the original to create a high-pass filtered image. 
As a result, the finer details of the morphology are enhanced. 
This method is not suitable for galaxy centers, but rather produces good results in the outher halos. Indeed, it is
the digital equivalent of the technique applied by \citet{Malin83} to photographic plates in order to reveal faint shells 
and filaments in nearby ellipticals. More recently, \citet{Colbert01} applied this technique to optical and NIR images of 
RC3 early-type galaxies, and \citet{Peng02} did the same for optical data of Centaurus A. 
For our galaxies we smoothed the images using a box width between 5 and 20 
pixels, depending on the size of the features and redshifts of the galaxies. 

\item {\it Smoothed galaxy subtraction}.
This technique is similar to the unsharp-masking. The difference is that, in this case, we subtract a highly 
smoothed image (using a filter of 10-20 pixels) from an slightly smoothed one (2 pixels). 
This generally produces a better result than the unsharp-masking for detecting dust or jet/tails at large scales. 
For example, \citet{Fabian06} applied this technique to Chandra X-ray observations of Perseus A. 

\end{itemize}

\begin{figure*}
\centering
\subfigure[]{
\includegraphics[width=8cm]{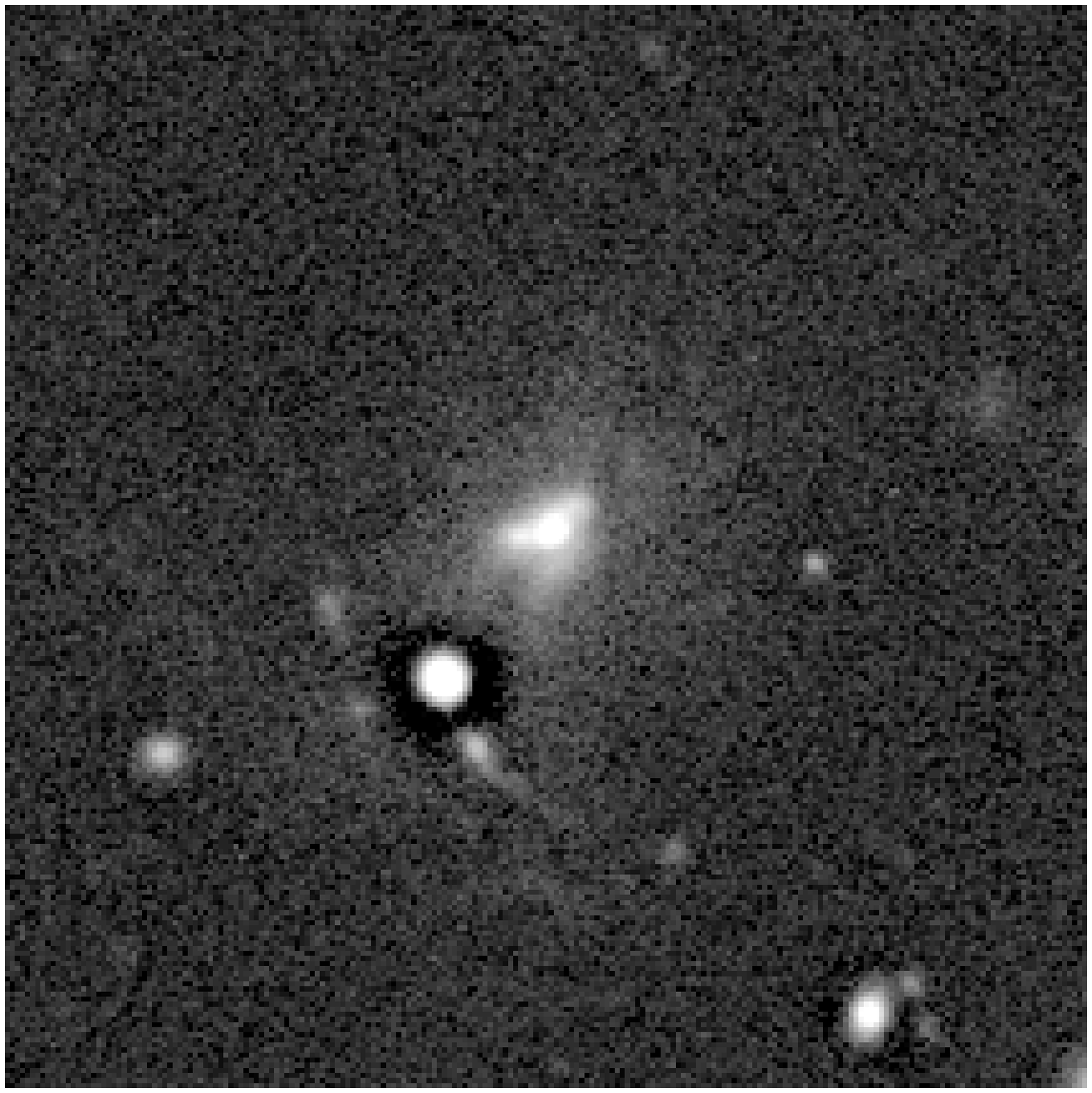}
\label{pks0347a}}
\subfigure[]{
\includegraphics[width=8cm]{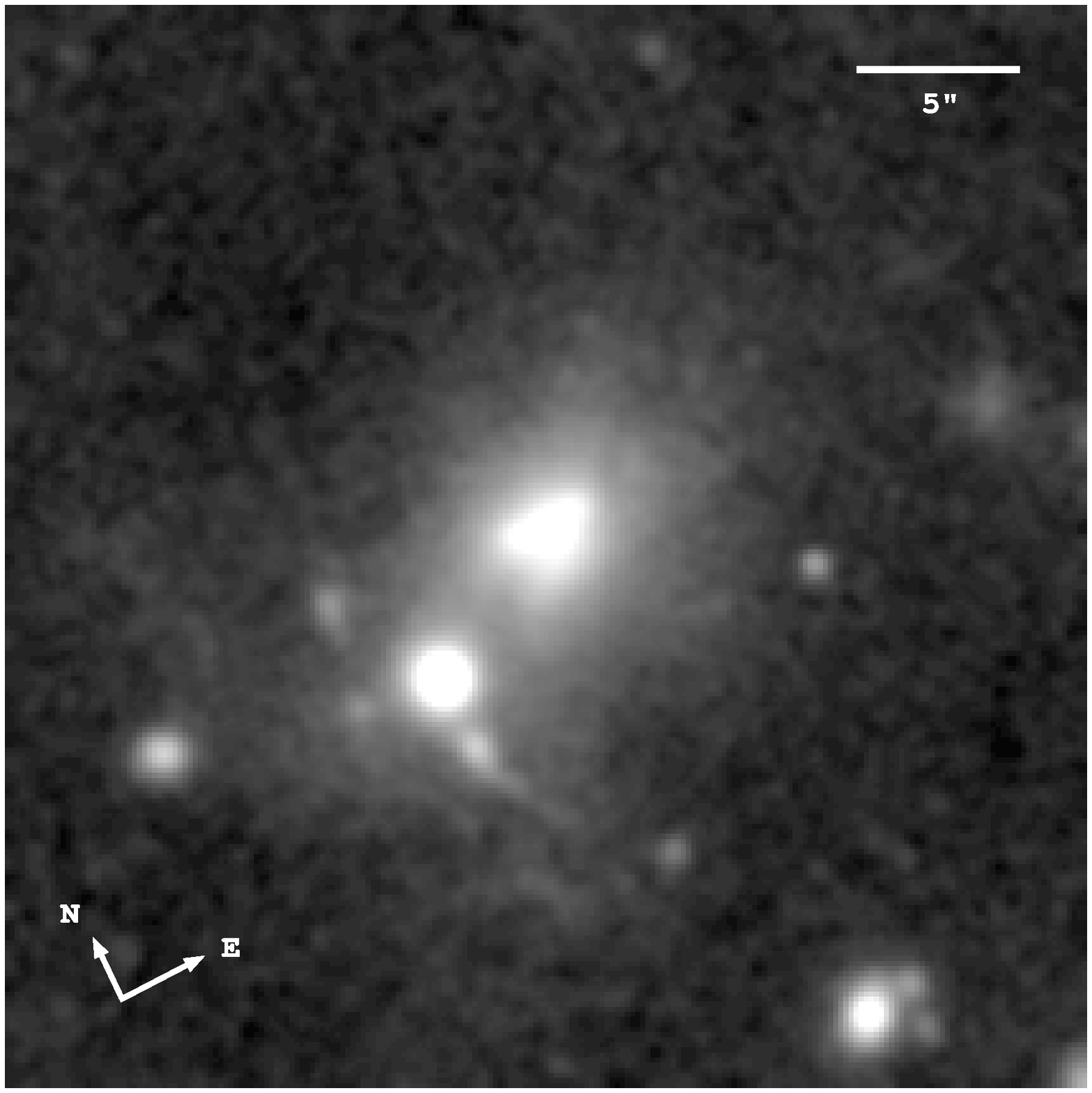}
\label{pks0347b}}
\caption{Example of the detection of tidal tails, a bridge, and dust in the interacting 
system between the WLRG PKS 0347+05 (center) and a QSO (SW of the radio galaxy). 
(a) Unsharp-masked image using a 10 pixel radius for the Gaussian smoothing filter. 
(b) Median filtered image using a 5 pixel box width. At least three tails, a bridge 
between the QSO and the radio galaxy, and dust are detected. (The images of all the galaxies in the sample 
are available in the electronic edition of the journal; Appendix \ref{images}. We strongly encourage the reader to look at 
the electronic versions of the figures in order to clearly distinguish the detail).}
\label{pks0347}
\end{figure*}

\begin{figure*}
\centering
\subfigure[]{\includegraphics[width=8cm]{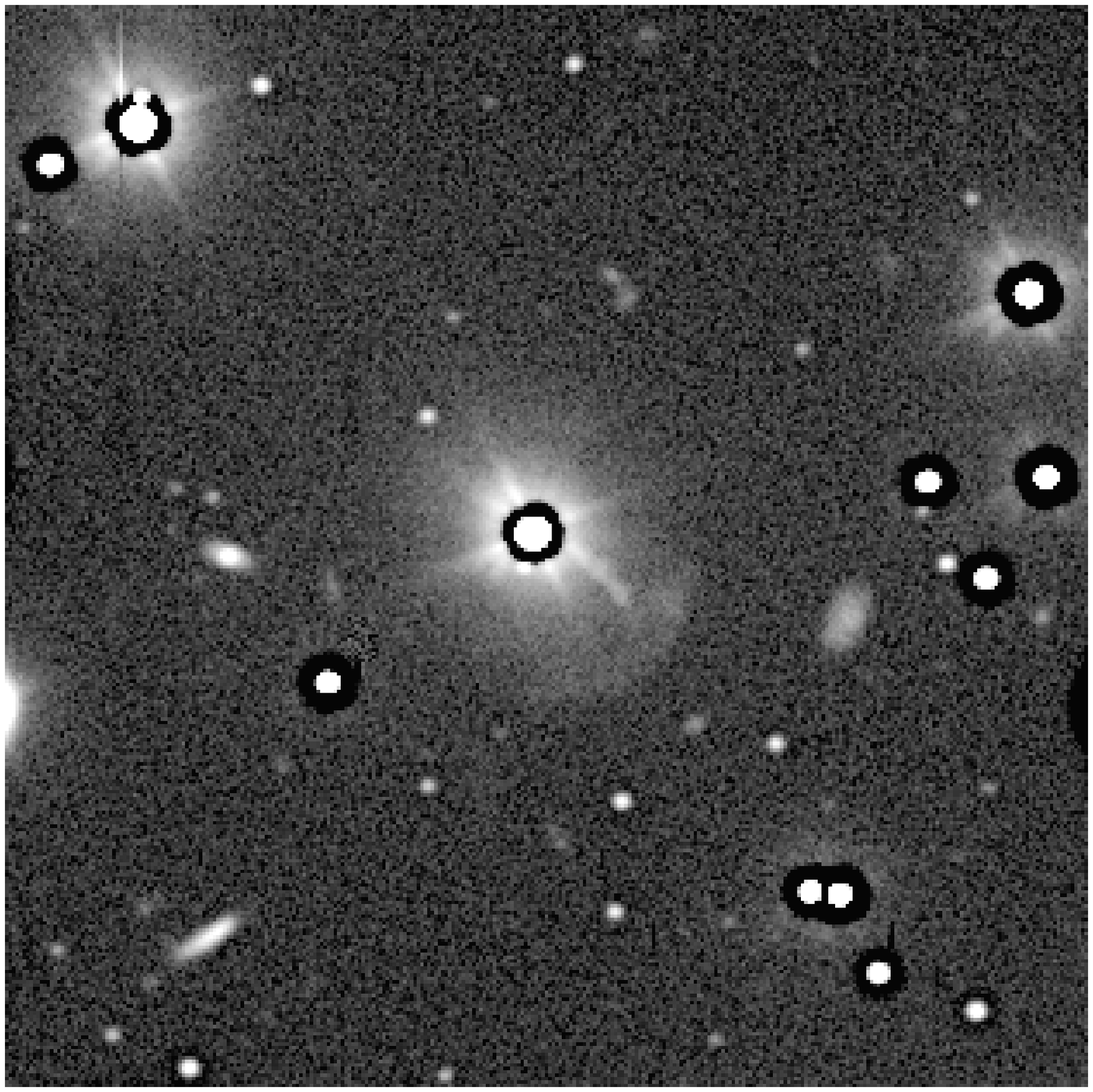}
\label{pks1355a}}
\subfigure[]{\includegraphics[width=8cm]{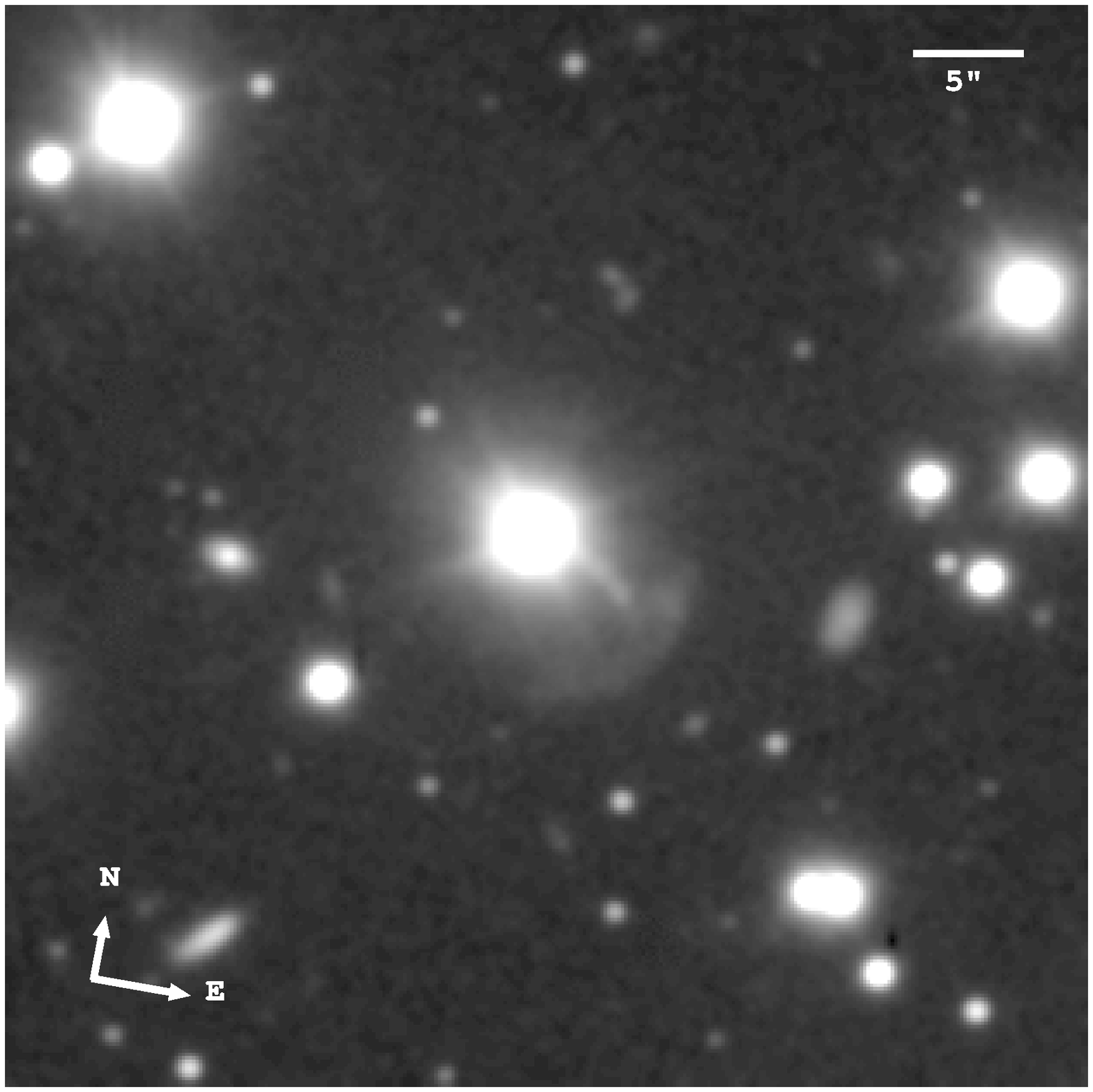}
\label{pks1355b}}
\caption{Illustration of the detection of a shell and a tail in the galaxy PKS 1355-41. 
(a) Unsharp-masked image using a moving box of 5 pixels width. 
(b) Median filtered image using a 5 pixel box width. A spectacular shell and a bright tail to SE 
of the radio galaxy nucleus are detected.}
\label{pks1355}
\end{figure*}

\begin{figure*}
\centering
\subfigure[]{\includegraphics[width=8cm]{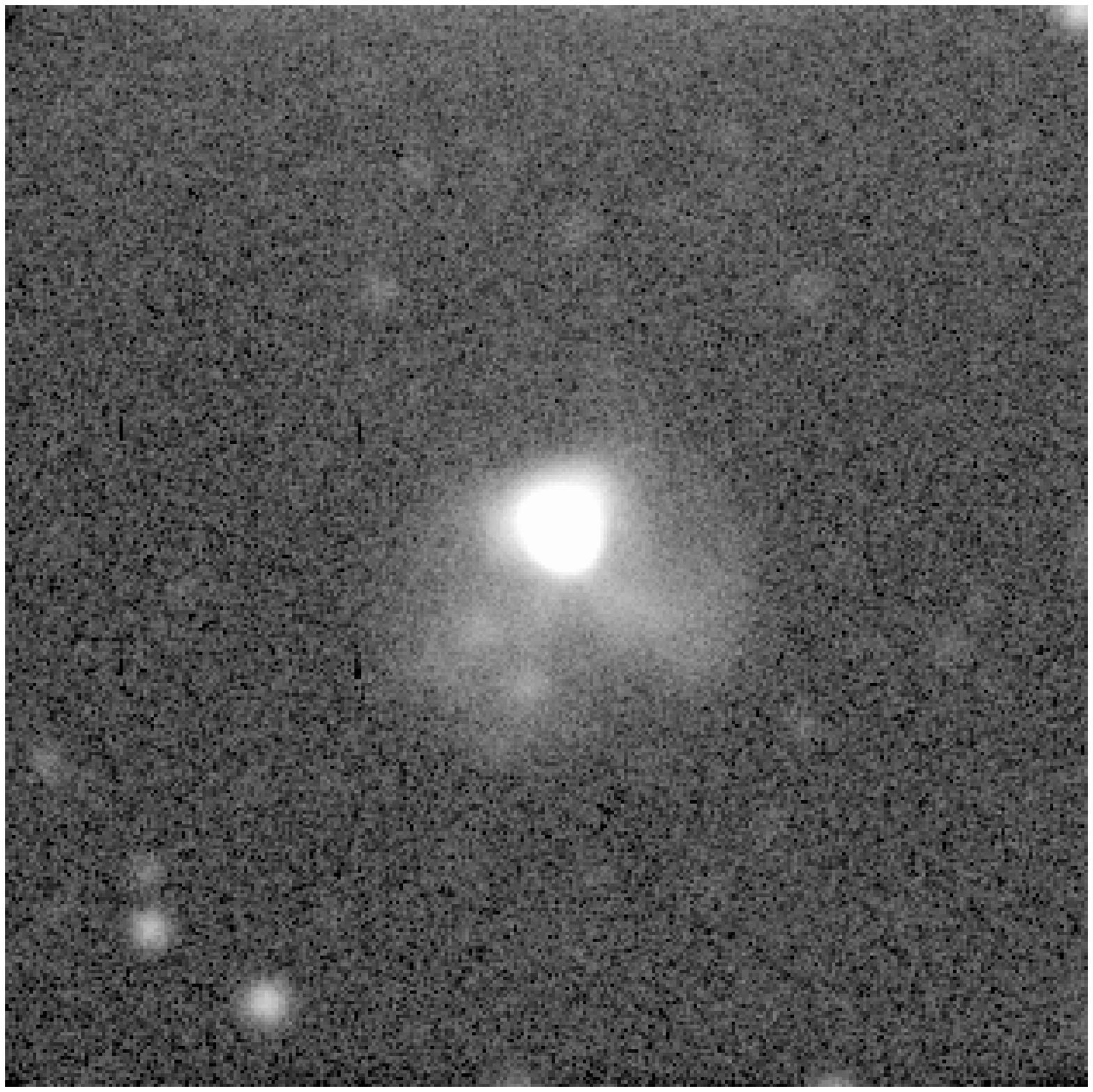}
\label{pks2314a}}
\subfigure[]{\includegraphics[width=8cm]{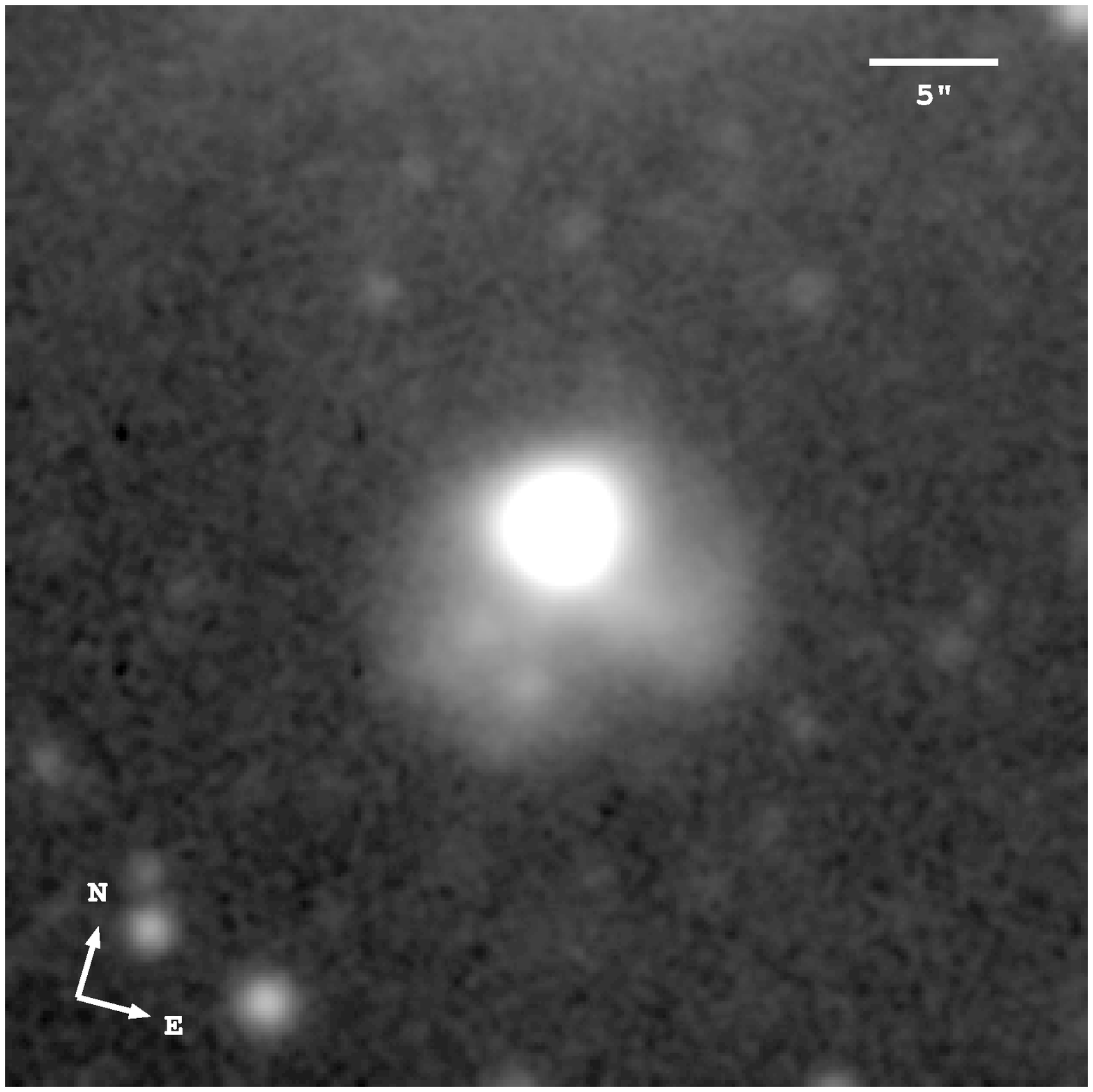}
\label{pks2314b}}
\caption{Example of the detection of fans and a tail in the galaxy PKS 2314+03.
(a) Unsharp-masked image using a moving box width of 10 pixels. 
(b) Median filtered image using a 5 pixel box width. Two bright fans are clearly 
detected in both images, together with a faint tail pointing towards the North.}
\label{pks2314}
\end{figure*}

\begin{figure*}
\centering
\subfigure[]{\includegraphics[width=8cm]{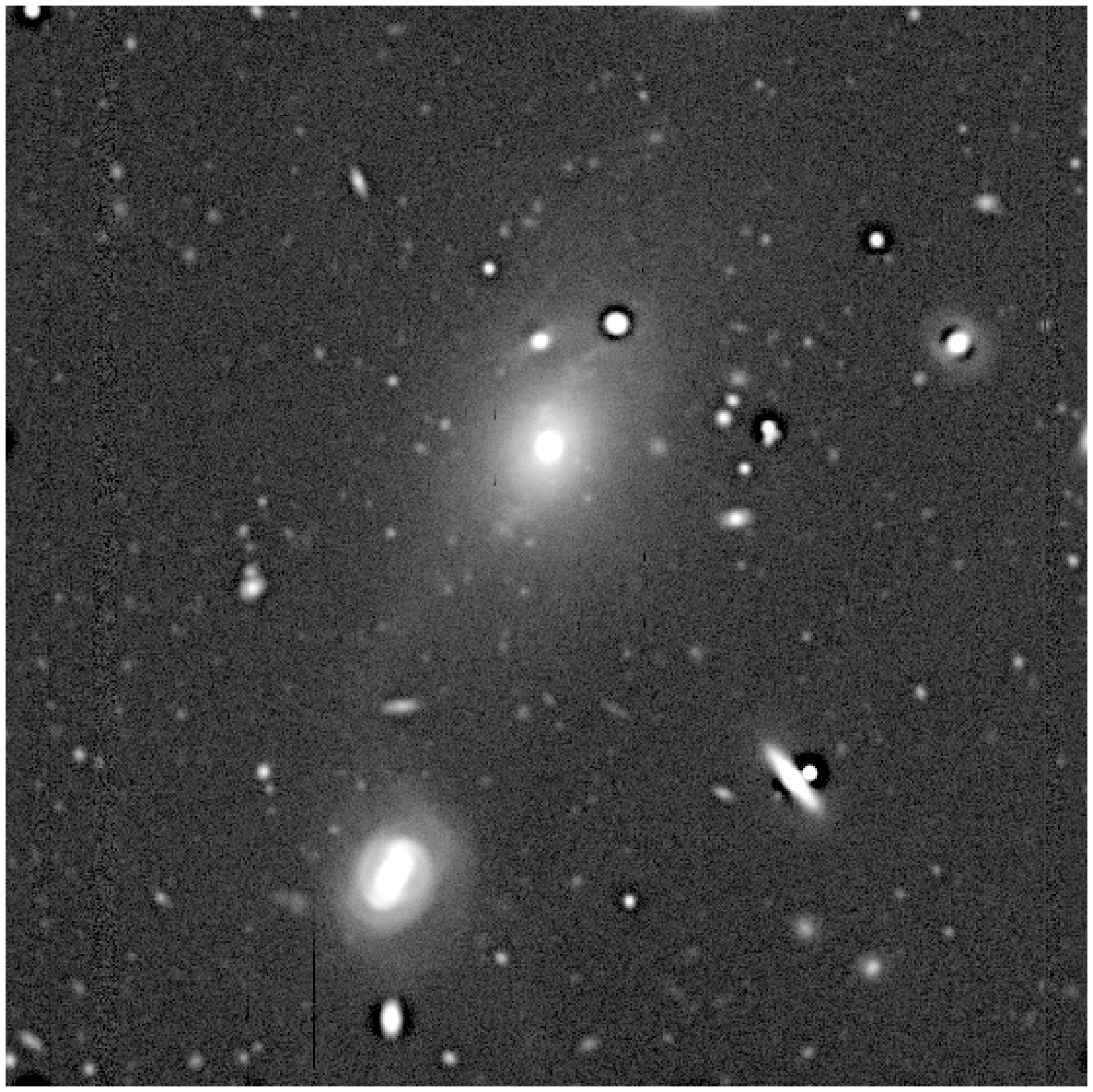}
\label{pks0349a}}
\subfigure[]{\includegraphics[width=8cm]{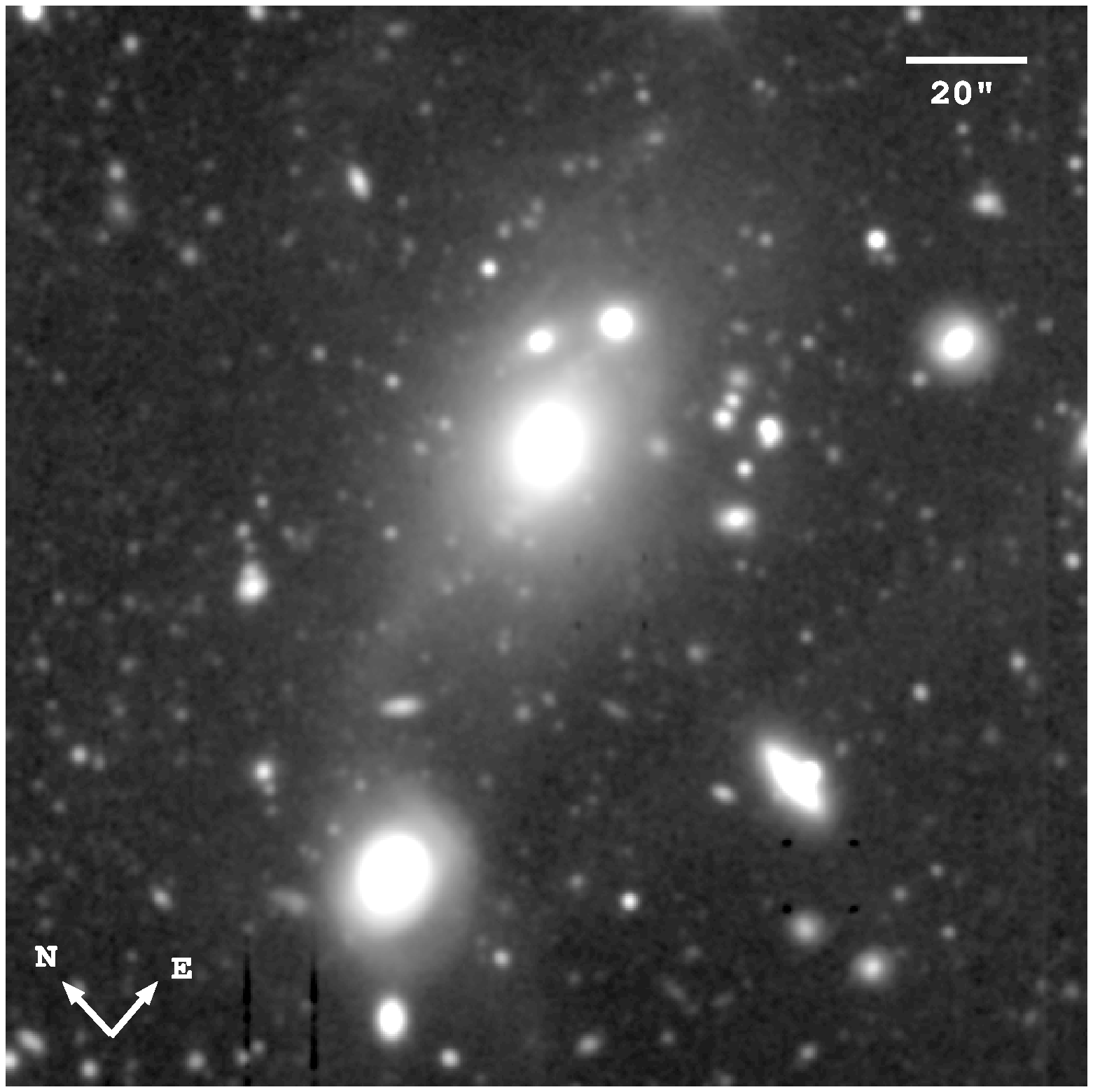}
\label{pks0349b}}
\caption{Illustration of the detection of bridges in the galaxy PKS 0349-27.
(a) Unsharp-masked image using a moving box width of 10 pixels. 
(b) Median filtered image using a 5 pixel box width. Two bridges linking the radio 
galaxy with the large western galaxy and the small galaxy towards the East of the radio source are clearly detected. 
Also noticeable is the high level of distortion of the large companion galaxy.}
\label{pks0349}
\end{figure*}

\begin{figure*}
\centering
\subfigure[]{\includegraphics[width=8cm]{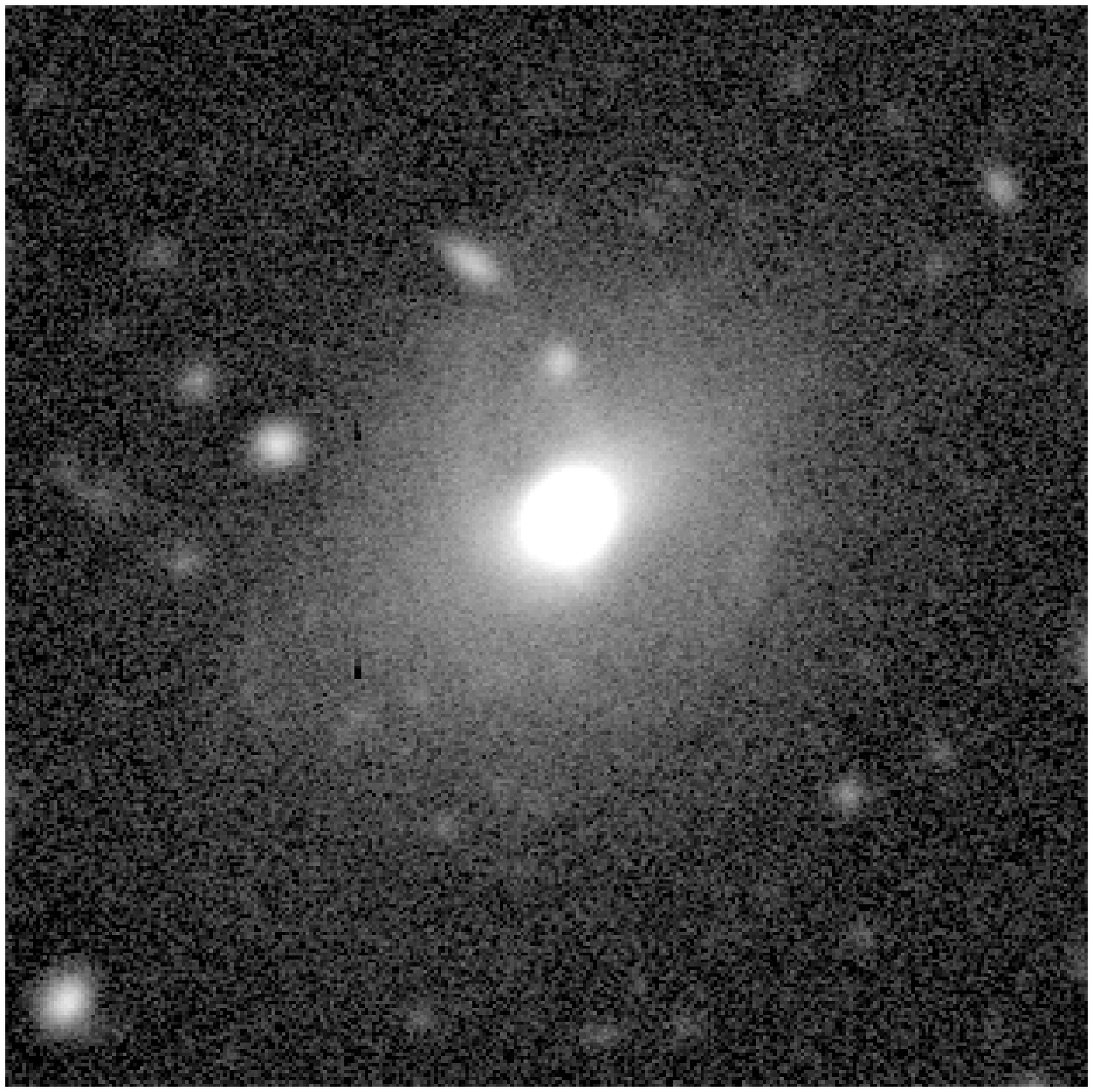}
\label{pks0213a}}
\subfigure[]{\includegraphics[width=8cm]{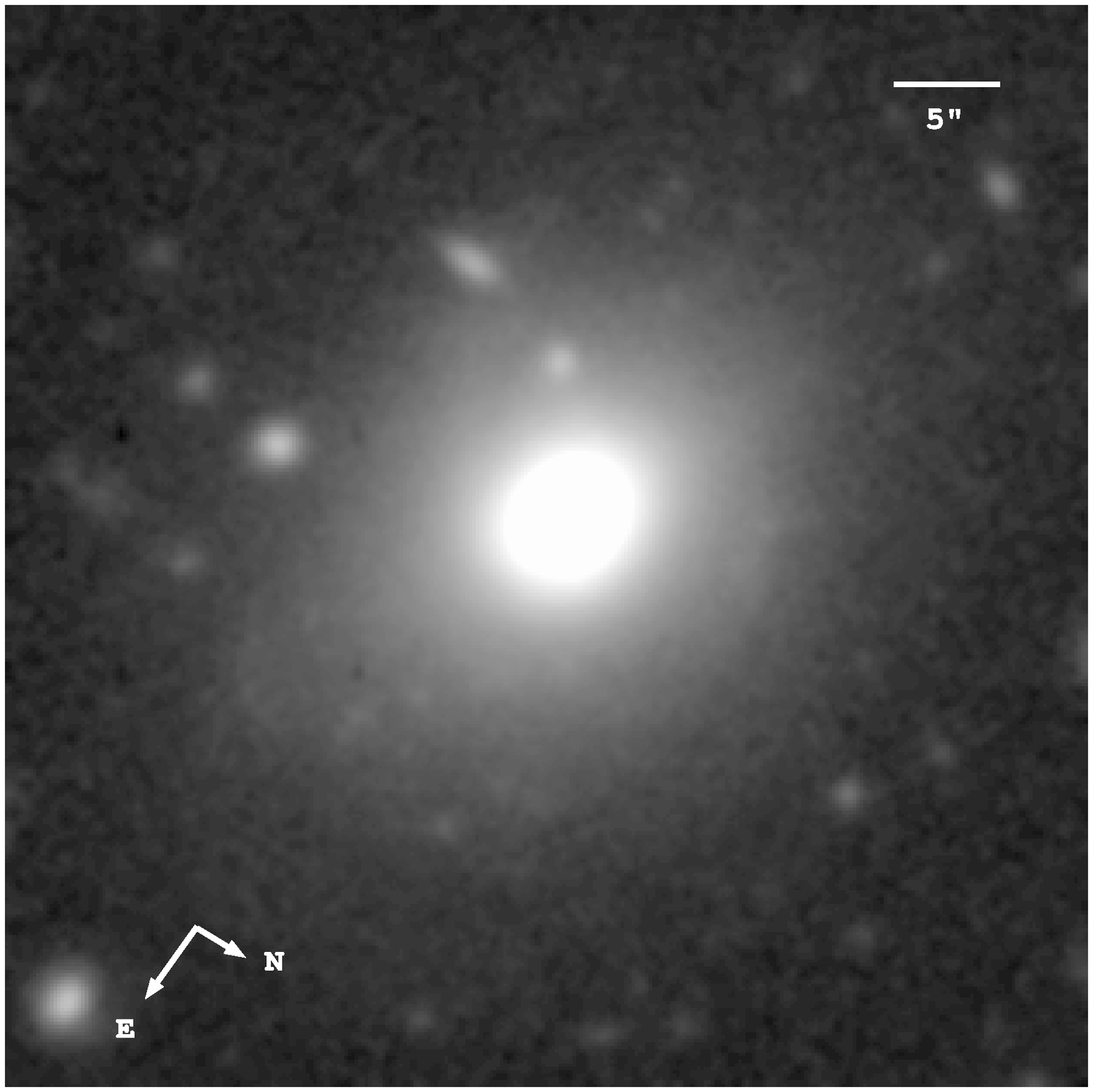}
\label{pks0213b}}
\caption{Example of the detection of shells and a tail in the galaxy PKS 0213-13.
(a) Unsharp-masked image using a 20 pixel radius for the Gaussian smoothing filter. 
(b) Median filtered image using a 5 pixel box width. Two shells are detected
towards East and North from the radio galaxy nucleus, respectively. A faint tail towards the 
SW is also revealed in the processed images.}
\label{pks0213}
\end{figure*}

\begin{figure*}
\centering
\subfigure[]{\includegraphics[width=8.0cm]{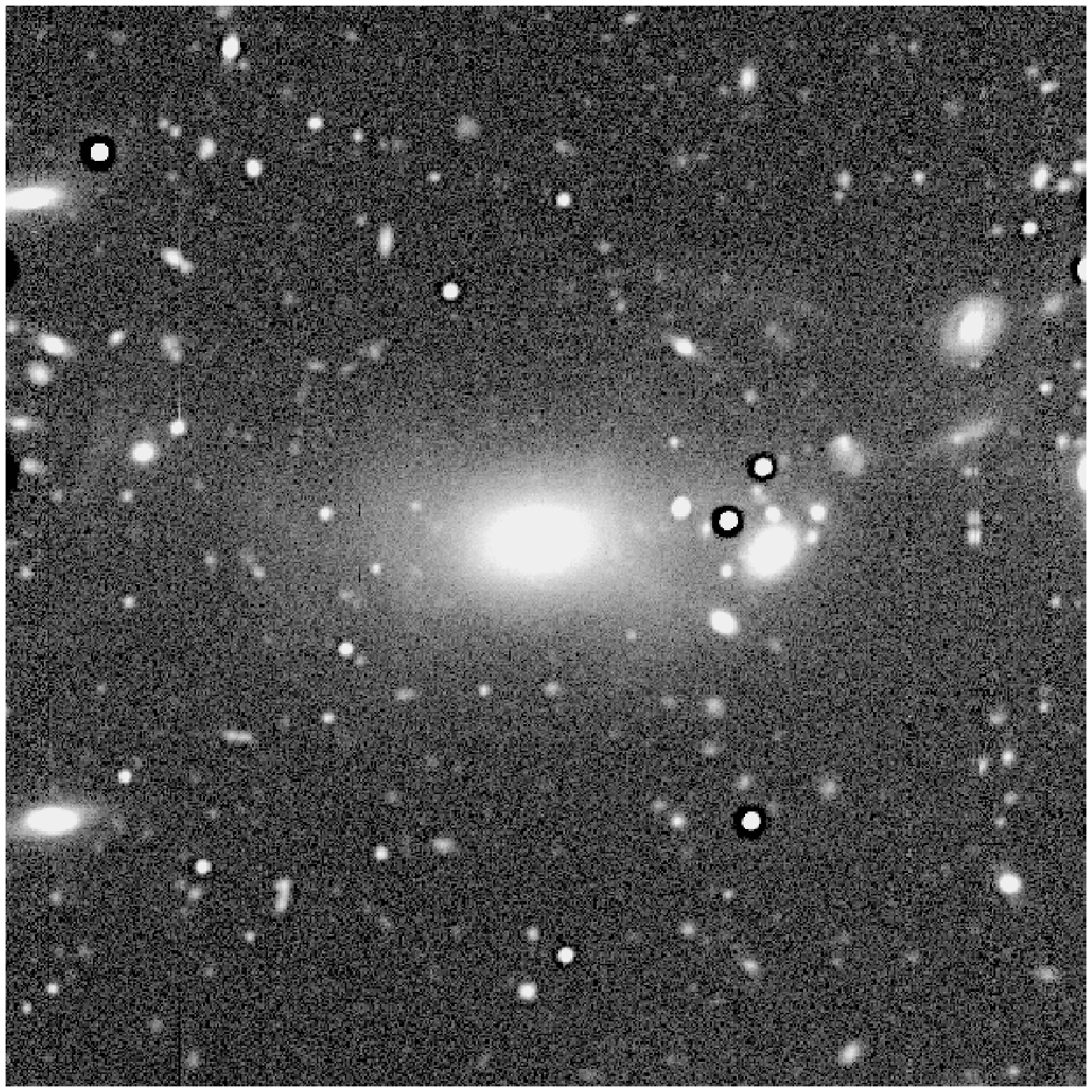}
\label{pks2356a}}
\subfigure[]{\includegraphics[width=8.0cm]{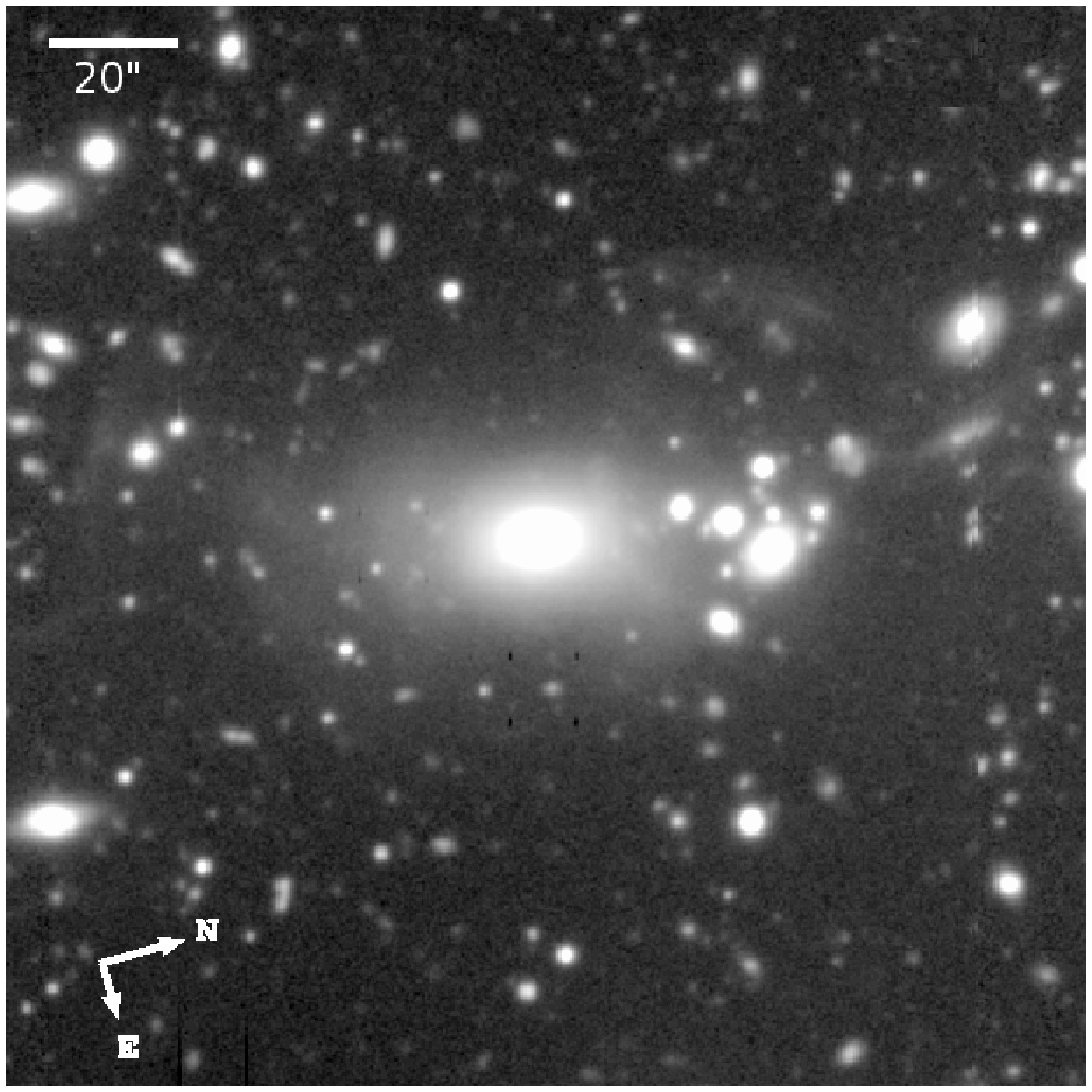}
\label{pks2356b}}
\caption{Illustration of the detection of shells, a fan, and irregular arc-like features in the galaxy PKS 2356-61.
(a) Unsharp-masked image using a moving box width of 10 pixels. 
(b) Median filtered image using a 5 pixel box width. Two shells are detected close to the galaxy center, 
one broad fan towards the South, and at least three irregular arc-like features.}
\label{pks2356}
\end{figure*}

\begin{figure*}
\centering
\subfigure[]{\includegraphics[width=8cm,angle=-90]{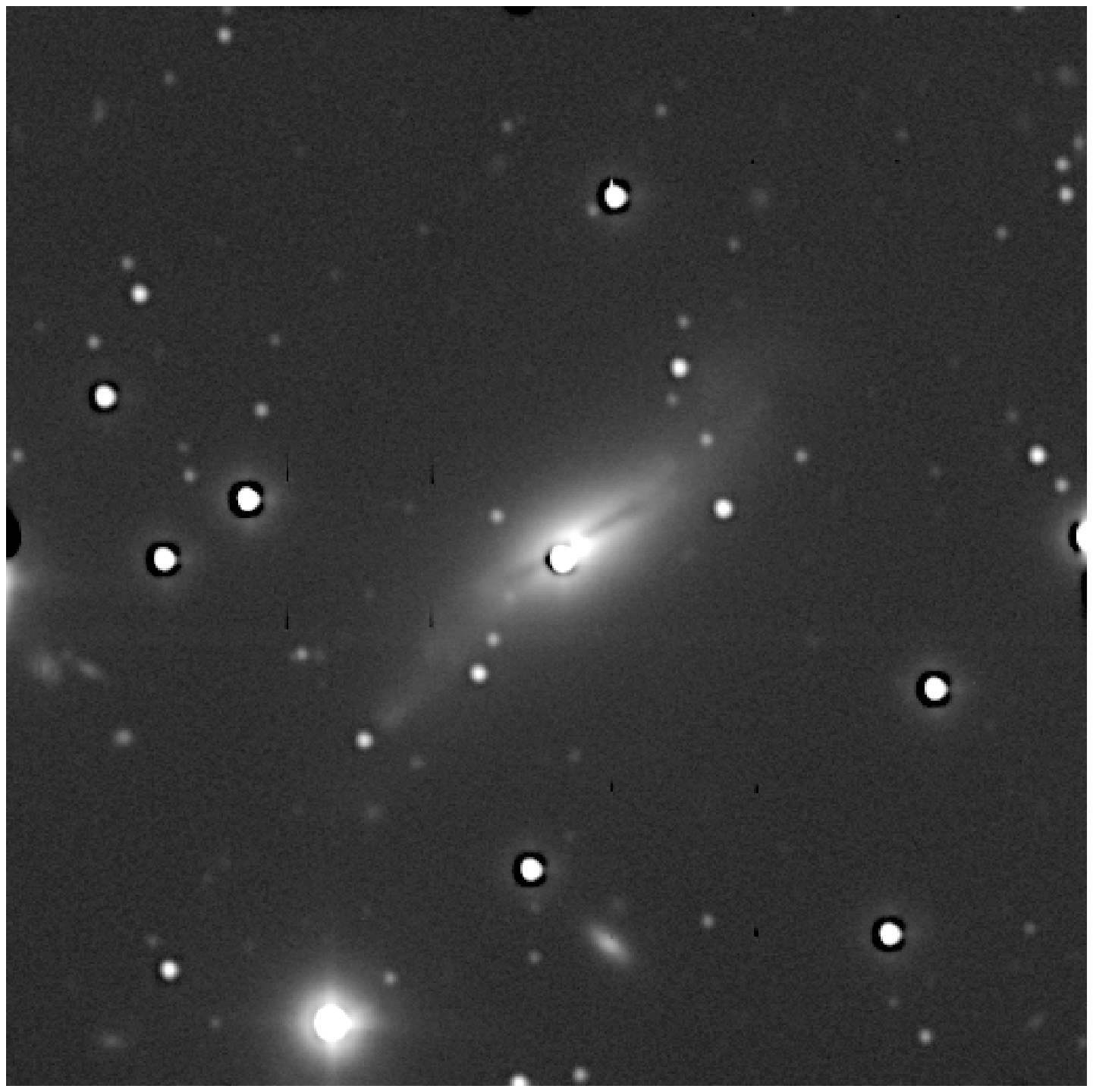}
\label{pks1814a}}
\subfigure[]{\includegraphics[width=8cm,angle=-90]{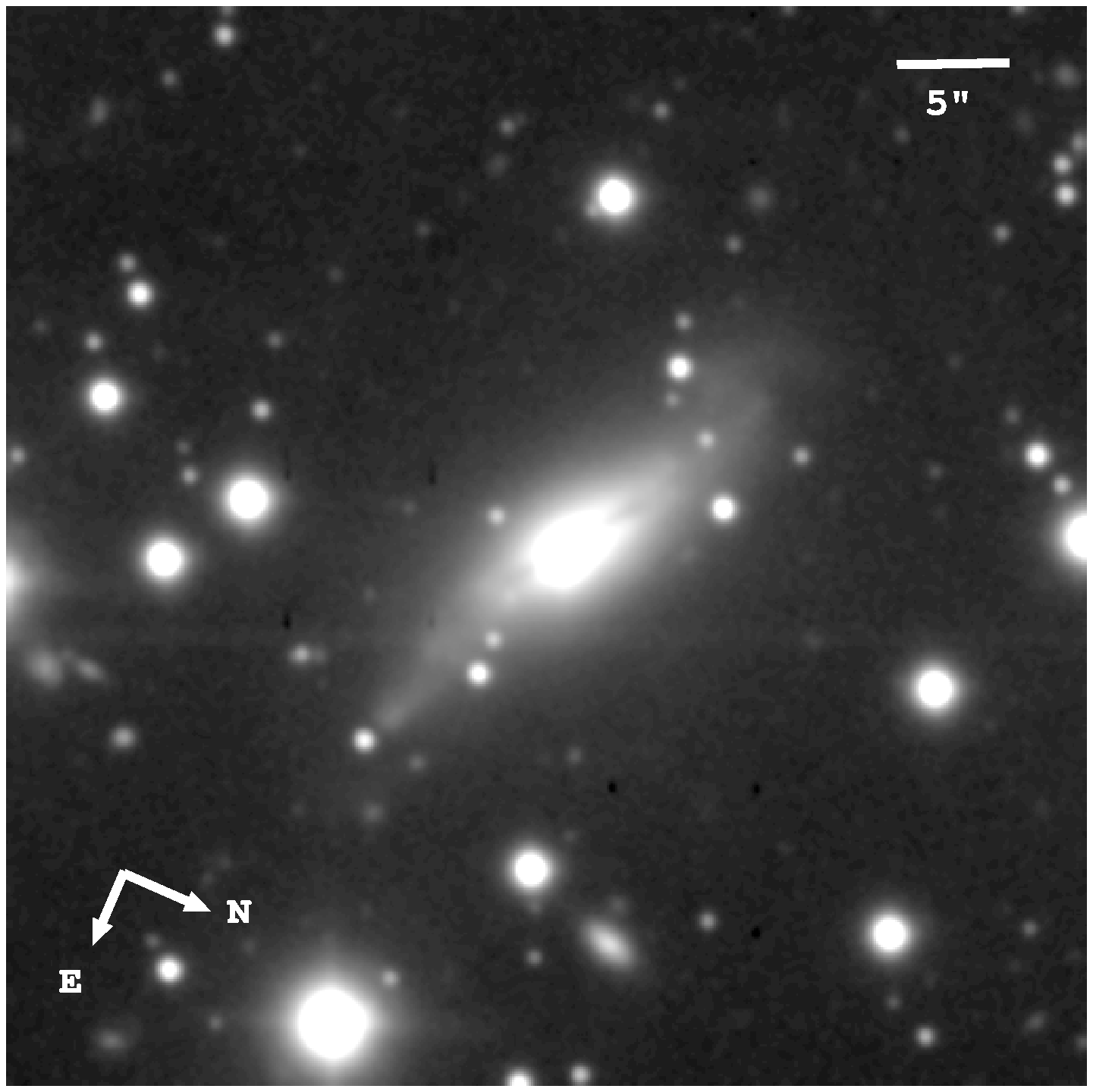}
\label{pks1814b}}
\caption{Illustration of the detection of dust and irregular features in the galaxy PKS 1814-63. 
(a) Unsharp-masked image using a moving box of 5 pixels width. 
(b) Median filtered image using a 5 pixel box width. A prominent dust lane crosses the 
galaxy center. Note also the high level of distortion in the outer disk of the galaxy, showing two irregular features in the edges.}
\label{pks1814}
\end{figure*}

\begin{figure*}
\centering
\subfigure[]{\includegraphics[width=8cm]{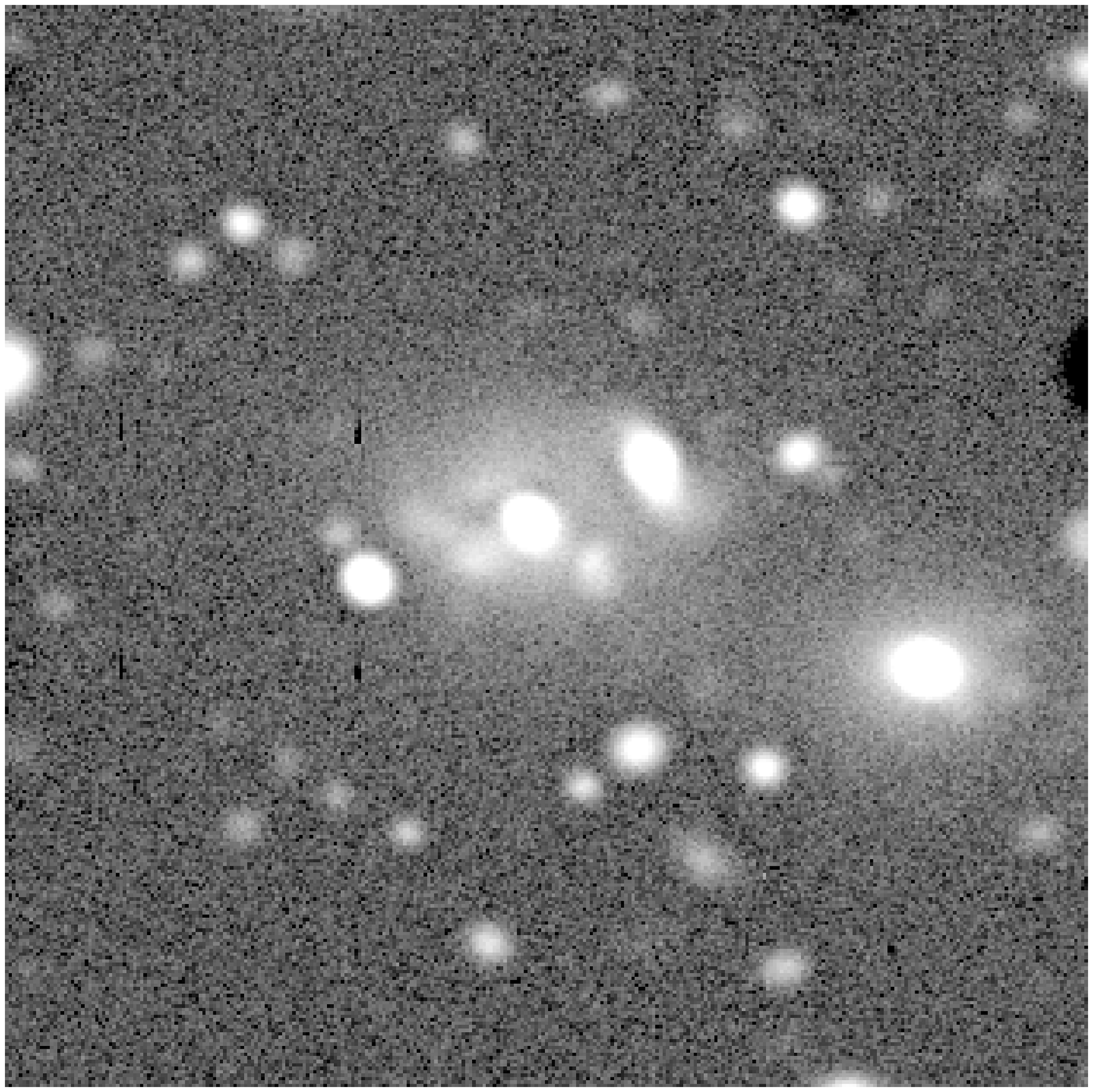}
\label{pks0023a}}
\subfigure[]{\includegraphics[width=8cm]{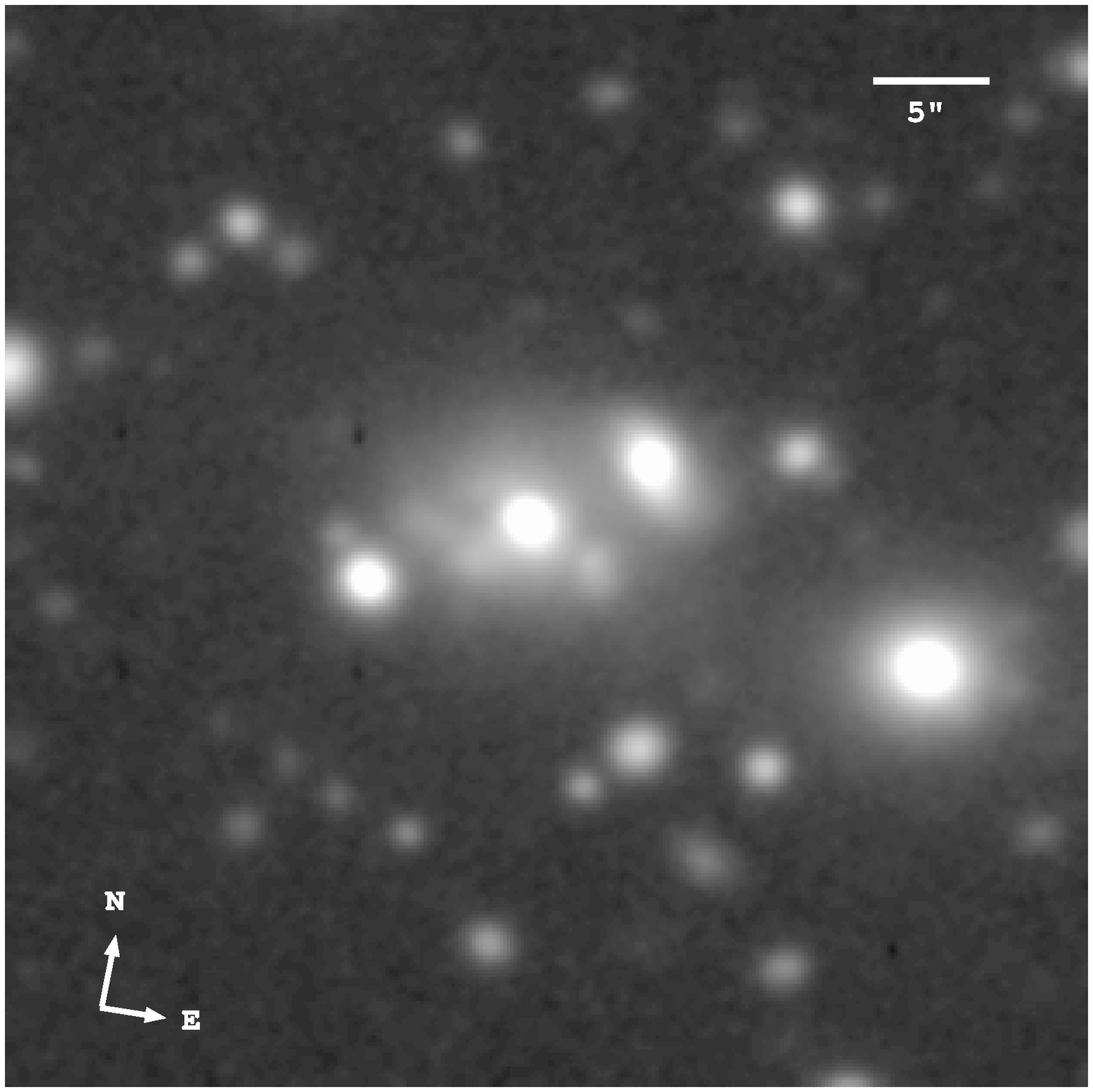}
\label{pks0023b}}
\caption{Example of the detection of an amorphous halo and dust in the galaxy PKS 0023-26.
(a) Unsharp-masked image using a moving box of 10 pixels width. 
(b) Median filtered image using a 5 pixel box width. Note the amorphous halo 
that lies between the galaxy and its companions. Dust also seems to be present.}
\label{pks0023}
\end{figure*}

\begin{figure*}
\centering
\subfigure[]{\includegraphics[width=8.0cm]{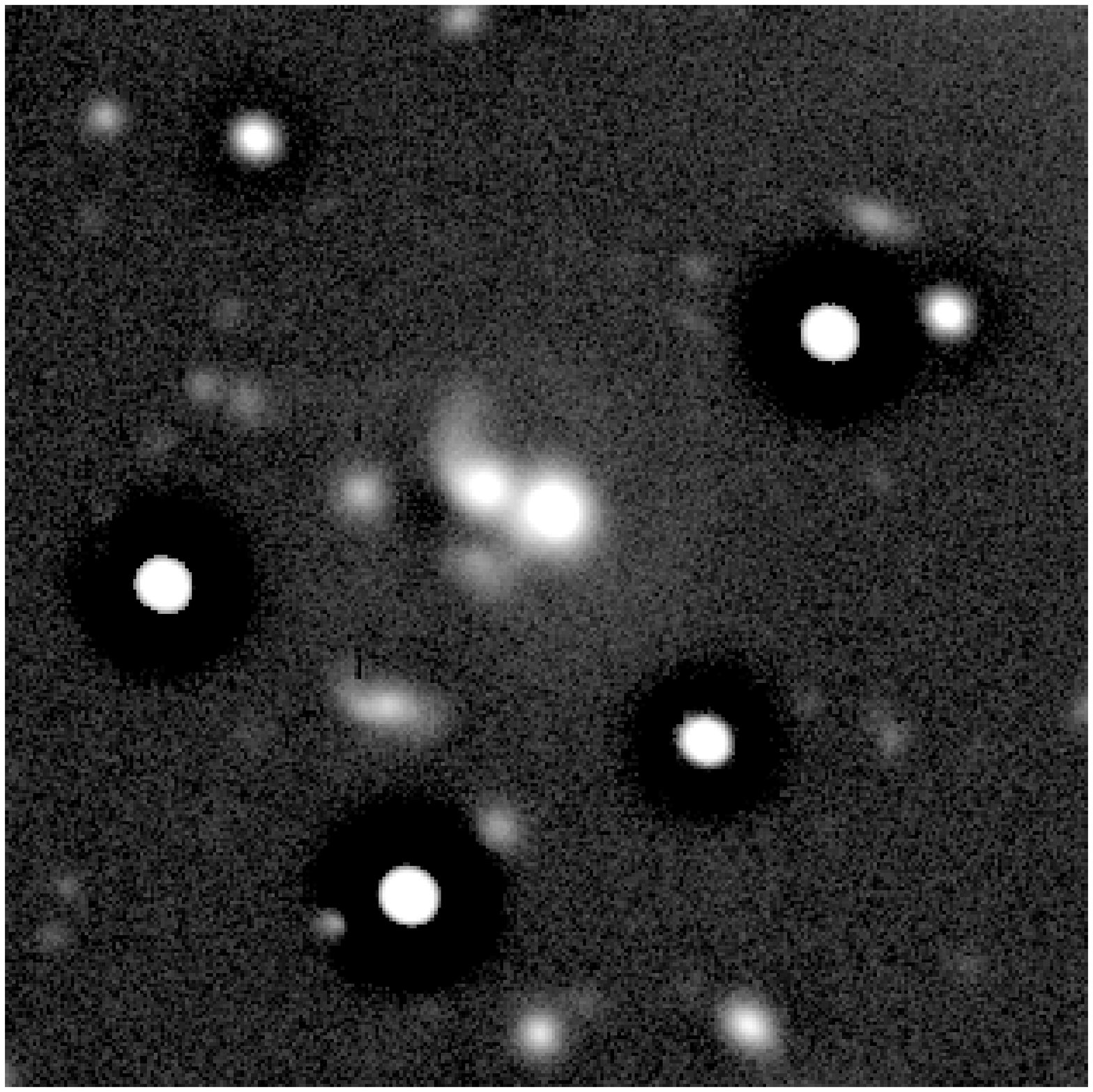}
\label{pks1934a}}
\subfigure[]{\includegraphics[width=8.0cm]{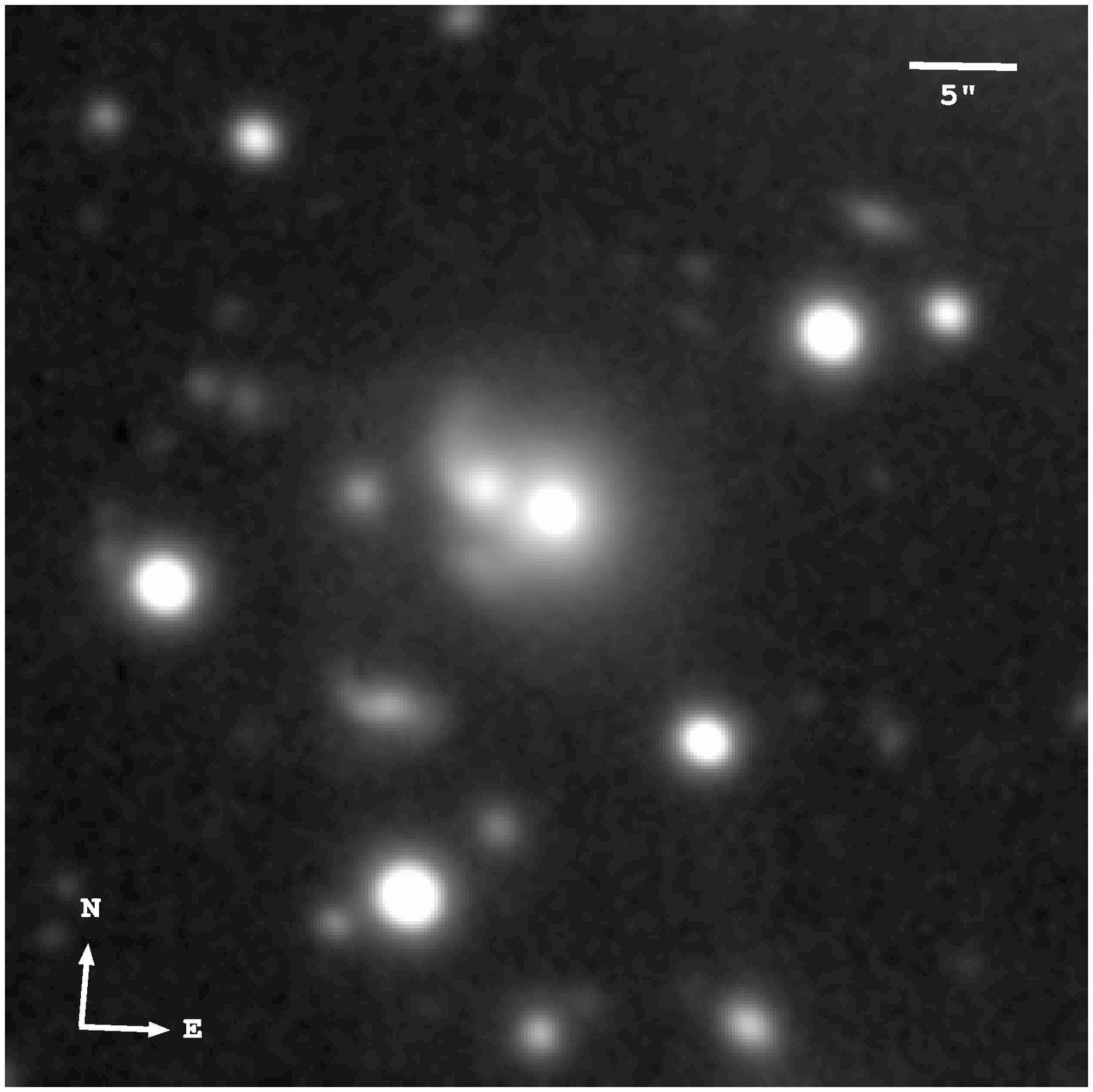}
\label{pks1934b}}
\caption{Example of the detection of a double nucleus and tidal tails in the galaxy PKS 1934-63.
(a) Unsharp-masked image using a moving box width of 20 pixels. 
(b) Median filtered image using a 5 pixel box width. Two nuclei, separated by 8.9 kpc, are
detected together with two tidal tails.}
\label{pks1934}
\end{figure*}

We tried the three techniques described above with all the galaxies in our sample, and present here the results for the two techniques
which gave the best result in each case, i.e., that in which the features appeared most clearly.
The purpose of all these image enhancement techniques is to allow us to display the faint features that we detect as clearly as possible.
Representative examples of the different features we detected are shown in Figures \ref{pks0347} 
to \ref{pks1934}. The processed images for the whole sample are shown in the electronic edition of the journal (Appendix \ref{images}), and
descriptions of the galaxy morphologies are given in Appendix \ref{individual}. 

\section{Results}

\subsection{Optical Morphologies}
\label{morphology}


\subsubsection{Morphological Features}
\label{morphological_features}

The morphological classification of the galaxies was done blind, with no information about any previous 
work on the sources, by CRA visually inspecting the GMOS-S images. The first categorization was made using the  
raw reduced images, and then confirmed by inspecting the enhanced images (using the three methods described in Section \ref{analysis}).
The classified features are listed in Tables \ref{data} and \ref{features}; note that these were all first detected 
in the original, rather than the enhanced images.

The classification of the various features detected in the GMOS-S images 
is based on that first used by \citealt{Heckman86}. In this context, a $tail$ corresponds to a narrow curvilinear 
feature with roughly radial orientation (see Figures \ref{pks0347} and \ref{pks1355});
a $fan$ is similar to a tail, but shorter and broader (see Figure \ref{pks2314}); by $bridge$ we mean a feature that 
links the radio galaxy with a companion (Figure \ref{pks0349}); and a $shell$ is a curving filamentary structure 
with a roughly tangential orientation to the main body of the galaxy (Figures \ref{pks1355}, \ref{pks0213}, and \ref{pks2356}). 
$Dust$ $lanes$ are also found in some of the galaxies (e.g., Figures \ref{pks0347}, \ref{pks1814} and \ref{pks0023}),
and some have $amorphous$ $haloes$ (Figure \ref{pks0023}). 
By $irregular$ we refer to any feature that cannot be classified as any of the previous (Figures \ref{pks2356} and \ref{pks1814}). 
Finally, we considered $multiple$ $nuclei$ when there are two or more brightness peaks inside 9.6 kpc, following the 
definition employed by \citealt{Smith89}, based on statistical studies of cluster galaxies \citep{Hoessel80}\footnote{\citet{Hoessel80} 
claimed that typical cluster members are expected to experience a close encounter or merger within this radius every 10$^9$ years. } 
and N-body simulations of interacting binary galaxies \citep{Borne84}. 
An example of a double nucleus system is shown in Figure \ref{pks1934}.
All of these features, apart, perhaps, from the dust, are very likely the result of galaxy mergers or close encounters.
In Tables \ref{data} and \ref{features} we also reported the detection of $jets$ emanating from the nuclei of a few of the galaxies. 
However, we do not consider such jets as genuine morphological disturbances related to galaxy interactions, since
they generally coincide with synchrotron-emitting jets already detected at radio wavelengths.

Simulations have shown how spiral-spiral (S-S), elliptical-spiral (E-S) and elliptical-elliptical (E-E)
interactions can produce all of the features that form the basis of our classification \citep{Quinn84,Hernquist92,Cattaneo05,Lotz08,Feldmann08}. 
To produce long and narrow morphological features such as $tails$ in elliptical systems, a dynamically cold gas supply from 
the companion galaxy is required \citep{Feldmann08}. However, tidal tails can also be produced as the result of a dry-merger 
according to some simulations \citep{Combes95}. The latter type of tails
are broader and shorter-lived (and consequently, more diffuse) than the narrow tails produced in a gas-rich merger \citep{Naab06,Bell06}. 
Broad $fans$ are also quite common in interacting systems involving two elliptical galaxies \citep{McIntosh08} and have been 
reproduced in dry-merger simulations with low surface brightness \citep{Borne84,Borne85}. 
$Shells$ are quite common in large ellipticals \citep{Schweizer80,Malin83,vanDokkum05,Sikkema07,Tal09}, 
and are generally considered to be the result of the accretion or capture of a small disk
(minor merger; \citealt{Quinn84}), although they can also be produced in major mergers (e.g., \citealt{Hernquist92}). Simple 
numerical simulations are able to reproduce the observed properties of these shells \citep{Dupraz86,Canalizo07}. 
\citet{Canalizo07} successfully reproduced the system of shells that they found for a QSO hosted by an elliptical galaxy
by simulating a minor merger as well as a major merger of two ellipticals. 


Note that, in order to distinguish between the dry/wet nature of the interactions that produce features presented here,
a more quantitative and finer-grained analysis of the optical morphologies is necessary. This will be addressed in forthcoming papers. 
Moreover, in the future it may also be possible to tackle the dry versus wet issue by directly quantifying the cool
dust and gas contents of the galaxies via their FIR emission (e.g., using combined Spitzer and Herschel observations).

\subsubsection{Emission-line contamination}
\label{contamination}

A potentially important issue is emission-line contamination which may affect the detected features.
For the low-redshift galaxies (z$<$0.065) the H$\alpha$ emission is included in the $r'$-band 
filter, and the same happens with the [O III]$\lambda$5007 \AA~for objects with z$>$0.14. 
To confirm whether or not some of the detected
features are emission-line gas, we made use of our own published and unpublished optical long-slit spectroscopic observations 
(e.g., \citealt{Tadhunter93,Tadhunter98,Tadhunter02,Holt07}), published optical spectra and narrow-band images obtained 
by other groups, as well as of the 
K-band images presented in \citet{Inskip10} for the objects in common with our sample.
By comparing these data sets with our optical images we find that, if we consider
the galaxies in the sample showing morphological features apart from dust and jets (36 objects; 78\% of the total sample),
for 20/36 (56\%) we can confirm either from their optical spectra or their NIR images (or both) that at least
some of the peculiar features are continuum rather than emission-line features. Only for two galaxies in the sample, namely
PKS 0349-27 and PKS 0235-19, do the long-slit spectra show that emission line contamination is likely to be 
a major issue for all the peculiar features. For the remaining 14 galaxies, there are neither optical spectra nor infrared data suitable 
for assessing the degree of emission line contamination. Thus, for those objects in the sample showing 
peculiar features for which we have sufficient data to either rule out or confirm emission line contamination (22 objects), 
20 galaxies (91\%) have at least some features which are continuum, rather than emission-line gas. By extrapolating this
result to the whole sample, we can conclude that for the overall majority of the detected features  
emission line contamination is not a serious issue.
See Appendix \ref{individual} for details on each individual object.

\subsubsection{Classification}
\label{classification}

In column 5 of Table \ref{features} we report the apparent surface brightnesses ($\mu_{AB}$) 
for all the secure detections of $tails$, $fans$, $shells$, $bridges$, $amorphous$ $haloes$ and 
$irregular features$ detected in our images. These surface brightnesses have been corrected for 
the sky and the diffuse host galaxy background. Using the same methodology that we used for the 
aperture magnitudes of the galaxies, in column 6 of Table \ref{features} we present the 
$\mu_{AB}^{corr}$ values, after applying Galactic extinction and K-corrections (see Section \ref{observations}). 
When considering surface brightness, it is also necessary 
to include the effect of the (1+z)$^4$ cosmological dimming. Thus, in addition to the previous corrections, we also 
subtract from our $\mu_{AB}$ values the surface brightness dimming values reported in the NASA/IPAC Extragalactic Database 
(NED) for each galaxy, in order to estimate the depth the features would appear to have at zero redshift (see Table \ref{features}). 
Assuming typical colors for elliptical galaxies \citep{Fukugita95}, we converted our $\mu_{AB}^{corr}$ measurements in 
both the $r'$ and $i'$ filters into $\mu_{V}$ values, in order 
to allow better comparison with the results of published studies. Thus, the median depth and range of surface brightness for the
detected features, once corrected, are $\tilde{\mu}_V=23.6~mag~arcsec^{-2}$ and $\Delta\mu_{V}=[21.3,~26.2]~mag~arcsec^{-2}$, 
respectively.

\begin{table*}
\centering
\begin{tabular}{lcccll}
\hline
\hline
PKS ID  & Filter & Dimming & Morphology & $\mu_{AB}$  (mag~arcsec$^{-2}$)  &  $\mu_{AB}^{corr}$  (mag~arcsec$^{-2}$)	  \\
\hline
0620-52    & r'  &   0.218     &  \dots        &    \dots			   &   \dots				 \\
0625-53    & r'  &   0.229     &  B	       &    22.81			   &	22.26				 \\  
0915-11    & r'  &   0.236     &  D	       &    \dots			   &   \dots				 \\	    
0625-35    & r'  &   0.232     &  J	       &    \dots			   &   \dots				 \\  
2221-02    & r'  &   0.232     &  F,S	       &    25.23, 24.93		   &	24.71, 24.41			 \\ 
1949+02    & r'  &   0.246     &  S,D	       &    23.03			   &	22.21				 \\
1954-55    & r'  &   0.244     &  \dots        &    \dots			   &   \dots				 \\ 
1814-63    & r'  &   0.264     &  2I,D         &    24.29, 24.52		   &	23.72, 23.95			 \\
0349-27    & r'  &   0.275     &  2B,[S]       &    26.19, 25.13		   &	25.81, 24.75			 \\
0034-01    & r'  &   0.303     &  J	       &    \dots			   &   \dots				 \\	 
0945+07    & r'  &   0.362     &  S	       &    24.03			   &	23.49				 \\
0404+03    & r'  &   0.369     &  [S]	       &    \dots			   &   \dots				 \\
2356-61    & r'  &   0.397     &  2S,F,I       &    25.30, 25.75, 26.03, 25.13     &	24.75, 25.20, 25.48, 24.58	 \\	  
1733-56    & r'  &   0.408     &  2T,2I,2S,[D] &    24.07, 24.00, 24.40, 24.02, 24.45, 24.86     &	23.28, 23.21, 23.61, 23.23, 23.66, 24.07	 \\
1559+02    & r'  &   0.434     &  2S,D,[2N]    &    23.61,24.07 		   &	22.81, 23.27			 \\
0806-10    & r'  &   0.453     &  F,2S         &    23.88, 23.78, 25.07 	   &	23.07, 22.97, 24.26		 \\
1839-48    & r'  &   0.456     &  2N,S,[T]     &    22.52			   &	21.75				 \\
0043-42    & r'  &   0.474     &  [2N],[B]     &    \dots			   &   \dots				 \\
0213-13    & r'  &   0.593     &  2S,[T]       &    25.56, 25.18		   &	24.74, 24.36			 \\
0442-28    & r'  &   0.595     &  S	       &    26.23			   &	25.37				 \\
2211-17    & r'  &   0.614     &  D,[F]        &    \dots			   &   \dots				 \\									 
1648+05    & r'  &   0.622     &  D	       &    \dots			   &   \dots				 \\			       
1934-63    & r'  &   0.729     &  2N,2T        &    24.21, 23.90		   &	23.03, 22.72			 \\
0038+09    & r'  &   0.744     &  T	       &    27.27			   &	25.88				 \\
2135-14    & r'  &   0.789     &  T,S,A,[B]    &    25.05, 24.35, 23.41 	   &	23.88, 23.18, 22.24		 \\
0035-02    & r'  &   0.858     &  B,F,[S]      &    25.78, 25.96		   &	24.59, 24.77			 \\
2314+03    & r'  &   0.859     &  2F,[T]       &    24.50, 24.22		   &	23.20, 22.92			 \\   
1932-46    & r'  &   0.900     &  2F,A,I       &    24.52, 24.57, 24.26, 25.19     &	23.19, 23.24, 22.93, 23.86	 \\
1151-34    & r'  &   1.001     &  F,[S]        &    25.95			   &	24.40				 \\
0859-25    & r'  &   1.159     &  2N	       &    \dots			   &	\dots				 \\
2250-41    & i'  &   1.170     &  2B,[T],[F]   &    25.62, 25.53		   &	24.19, 24.10			 \\
1355-41    & r'  &   1.185     &  S,T	       &    24.32, 23.61		   &	22.48, 21.77			 \\
0023-26    & r'  &   1.209     &  A,[D]        &    24.35			   &	22.66				 \\
0347+05    & r'  &   1.266     &  B,3T,D       &    25.46, 24.30, 25.92, 26.39     &	22.98, 21.82, 23.44, 23.91	 \\
0039-44    & r'  &   1.288     &  2N,3S,[T],[D]&    24.72, 24.77, 25.53 	   &	22.93, 22.98, 23.74		 \\	      
0105-16    & i'  &   1.458     &  B	       &    25.64			   &	23.82				 \\
1938-15    & i'  &   1.618     &  F	       &    24.85			   &	22.40				 \\
1602+01    & i'  &   1.651     &  F,S,[J]      &    22.82, 24.75		   &	20.55, 22.48			 \\
1306-09    & i'  &   1.659     &  2N,S         &    25.83			   &	23.71				 \\
1547-79    & i'  &   1.712     &  2N,T         &    26.03			   &	23.52				 \\
1136-13    & i'  &   1.929     &  T,J	       &    24.26			   &	21.78				 \\
0117-15    & i'  &   1.942     &  3N,S,I,[D]   &    22.99, 26.83		   &	20.51, 24.35			 \\
0252-71    & i'  &   1.953     &  [A]	       &    \dots			   &   \dots				 \\
0235-19    & i'  &   2.093     &  2T,[B]       &    24.38, 24.22		   &	21.62, 21.46			 \\	 
2135-20    & i'  &   2.133     &  F	       &    25.52			   &	22.68				 \\
0409-75    & i'  &   2.287     &  2N	       &    \dots			   &   \dots				 \\
\hline		     	 	        				        					  
\end{tabular}						 
\caption{Surface brightness measurements. Columns 1, 2, and 3 give the PKS ID, GMOS-S filter,
and surface brightness dimming from the NED (magnitude per unit area). Column 4 lists our morphological classification on the basis of the detected features 
(same as in Table \ref{data}). Apparent and corrected (including galactic extinction, 
K-correction and dimming) surface brightness for secure identifications of T, F, S, B, A and I are given in columns 5 and 6, respectively. Brackets in column 4 
indicate either uncertain identification of the feature, or detection in the processed images only. We do not report surface brightness measurements for the latter features.}
\label{features}
\end{table*}

The previous transformations of surface brightnesses into V-band measurements were done by assuming K-corrections and colors 
of elliptical galaxies \citep{Frei94,Fukugita95}. However, some of the features (e.g., shells) may be produced  
in mergers involving small disk galaxies. In this case, the color of the shell would likely be
similar to that of a spiral galaxy. 
In order to assess the importance of this effect we re-calculated the $\mu_{V}$ values for the detected features using K-corrections  
and colors of Sbc-type spiral galaxies from \citet{Frei94} and \citet{Fukugita95}, and found that they
do not change significantly ($\tilde{\mu}_V=23.7~mag~arcsec^{-2}$ and $\Delta\mu_{V}=[21.3,~26.3]~mag~arcsec^{-2}$ using Sbc colors and K-corrections).  


Considering only the secure identifications of morphological features\footnote{For some of the galaxies, 
we also report in Table \ref{features} the existence of features whose identification is not 
completely secure, or which are only detected in the processed images.}, the sample can be divided into five groups: 

\begin{enumerate}
\renewcommand{\theenumi}{(\arabic{enumi})}

\item {\it Galaxy pair or group in tidal interaction.} Galaxy pairs showing bridges (e.g., PKS 0349-27), or co-aligned distorted 
structures (e.g., PKS 2221-02). 20\% of the 2Jy PRGs are including in this category.

\item {\it Galaxies presenting any sign of morphological disruption.} Galaxies showing shells, fans, tails, 
amorphous haloes, and irregular features. 54\% of the sample show such features.

\item {\it Multiple nuclei.} Galaxies with a companion lying inside a 9.6 kpc radius (17\% of the sample), according to 
the theoretical definition employed by \citet{Hoessel80} and \citealt{Smith89}. 

\item {\it Dust features.} Galaxies presenting dust features as the only sign of disturbance (7\%). 

\item {\it Isolated galaxies with no sign of interaction}. 
Objects in which we cannot confidently identify morphological peculiarities (15\%).

\end{enumerate}

Note that these categories are not exclusive because some galaxies show more than one of the morphological 
features described above (see Table \ref{data}).
Initially we considered objects in groups 1, 2, 3, and 4 as showing disturbed morphologies consistent with them 
having been involved in a galaxy interaction/merger, whilst galaxies classified in 
the fifth group were classified as undisturbed. Based on this classification, 85\% of the sample are very likely interacting 
objects or the result of a past merger event. 
Three of the galaxies included in this 85\%, namely PKS 0915-11, PKS 2211-17 and PKS 1648+05, present dust features as the only 
detected sign of disturbance (see Table \ref{data}). It is worth noticing that all the three of these sources are WLRGs 
(see Section \ref{lines}). However, $dust$ features by themshelves may not necessarily be a sign of galaxy interactions. 
If we do not consider dust as a sign of morphological disturbance related to mergers and interactions, 
then the percentage of galaxies in the full sample presenting 
evidence for interactions/mergers is 78\%. Note that, while small-scale dust is often taken as an observational signature for
recent mergers (e.g., \citealt{vandokkum95}), it may also be associated with cooling flows in cluster galaxies 
(e.g., \citealt{Fabian94,Hansen95,Edge99,Edge10}). 

On the other hand, we also note that the two SLRGs classified in group 5, 
namely PKS 0404+03 and PKS 0252-71, are very likely undergoing an interaction/merger, based on their morphologies.
In the case of PKS 0404+03 our processed images reveal hints of a shell towards NE, 
but the presence of a bright star close to the radio galaxy, and 
high dust extinction (A$_V>1$ mag) that affects this area of the sky, prevent us from making any secure classification. 
The galaxy PKS 0252-71 appears to be surrounded by a distorted halo which points towards a faint companion galaxy 
at $\sim$33 kpc towards the South, indicating a possible interaction. Unfortunately, the relatively poor 
seeing for the image of this object (FWHM=1\arcsec) prevents us from confidently classifying it as disturbed. Thus, it is likely 
that all the SLRGs in our sample are interacting objects presenting morphological peculiarities.

\subsubsection{Sample Morphologies}
\label{coalescence}

The main result of this work is that 85\% (78\% if we do not consider galaxies with dust features only) 
of our sample of PRGs show peculiar optical morphologies at 
relatively high levels of surface brightness ($\tilde{\mu}_V=23.6~mag~arcsec^{-2}$ and $\Delta\mu_{V}=[21.3,~26.2]~mag~arcsec^{-2}$). 
This fraction of distorted morphologies is much higher than that found for radio 
quiet ellipticals at the same brightness level, as we will discuss in Section \ref{comparison}. 
It is also greater than the fraction presented in the literature for nearby PRGs (\citealt{Heckman86},
\citealt{Smith89}, \citealt{Dunlop03}).

In order to track any changes with redshift/radio power, and also to avoid any problems caused by the
variation in effective spatial resolution and limiting surface brightness with redshift, the sample can 
be divided into three redshift ranges for the analysis: 0.05$<$z$<$0.11 (15 objects); 0.11$\le$z$<$0.31 
(15 objects); and 0.31$\le$z$<$0.7 (16 objects). Across each of these ranges the effective spatial scale (kpc~arcsec$^{-1}$) and limiting 
monochromatic luminosity surface brightness vary by less than a factor of two.
Considering these three ranges of redshift ($\Delta$z$_1$, $\Delta$z$_2$, and $\Delta$z$_3$ from lower to higher z), the percentages of 
morphological disturbance for each subset are 67\% for the galaxies within $\Delta$z$_1$, 93\% for those in $\Delta$z$_2$, and 94\% 
considering sources in $\Delta$z$_3$ (60\%, 80\%, and 94\% respectively if we do not consider dust as a sign of disturbance). 
Indeed, there are only two galaxies in the higher redshift range ($\Delta$z$_2$+$\Delta$z$_3$) 
showing no clear sign of interaction (PKS 0043-42 and PKS 0252-71), indicating a higher proportion of interacting galaxies 
at higher redshifts/radio powers. 

However, it is important to note that four of the five undisturbed galaxies within $\Delta$z$_1$ are WLRGs, which
have been proposed to be powered by hot gas accretion instead of the typical AGN cold gas accretion 
(see Section \ref{lines} for a more detailed explanation). 
Thus, the apparently lower percentage of interacting galaxies 
at low redshift could be due to the higher number of WLRGs in $\Delta$z$_1$ subsample. 
The percentages of morphological disturbance found for the total sample, the SLRGs and the WLRGs in each redshift 
bin are given in Table \ref{redshift}.

\begin{table*}
\centering
\begin{tabular}{lccccccc}
\hline
\hline
Redshift bin & Range & \multicolumn{2}{c}{Total Sample} & \multicolumn{2}{c}{SLRGs} & \multicolumn{2}{c}{WLRGs} \\
 & & Galaxies & Disturbed & Galaxies & Disturbed  & Galaxies & Disturbed  \\
\hline
$\Delta$z$_1$ & 0.05$<$z$<$0.11	   & 15  & 67\% (60\%)	&  9   & 89\%  (89\%)  & 6  &  33\%  (17\%)  \\
$\Delta$z$_2$ & 0.11$\le$z$<$0.31  & 15  & 93\%	(80\%)  &  11  & 100\% (100\%) & 4  &  75\%  (25\%)  \\
$\Delta$z$_3$ & 0.31$\le$z$<$0.7   & 16  & 94\%	(94\%)  &  15  & 93\%  (93\%)  & 1  &  100\% (100\%) \\
\hline		     			      		 
\end{tabular}						 
\caption{Division of all the galaxies in the sample, SLRGs, and WLRGs in the three redshift ranges considered. 
The number of galaxies in a given redshift bin and the fraction of distorted morphologies found (i.e., objects classified 
in groups 1, 2, 3, and 4) are given for each group. Percentages given within parenthesis exclude objects with dust as the only
identified feature (group 4).}
\label{redshift}
\end{table*}

The morphological features detected in our sample of PRGs are very likely the result of galaxy mergers/interactions. 
In terms of the merger scenario, it is interesting that we appear to be observing many of the systems {\it before} the final
coalescence of the nuclei of the merging galaxies. The galaxies classified in 
groups 1 ({\it galaxy pair or group in tidal interaction}) and 3 ({\it multiple nuclei}) would correspond to
systems observed before the nuclei have coalesced, whereas those in group 2 ({\it galaxies presenting any sign of 
disturbance}), and possibly in 4 ({\it dust features}), would correspond to more evolved systems (coalescence or post-coalescence). 
By considering only galaxies in groups 1 and 3, we find that 35\% of the sample are clearly pre-coalescence systems, 
presenting bridges with close companions (e.g., PKS 0625-53, PKS 0349-27, PKS 0035-02, and PKS 2250-41), co-aligned distorted structures 
(e.g., PKS 2221-02, PKS 1934-63, and PKS 1151-34) and multiple nuclei (PKS 1306-09, PKS 1547-79, and PKS 0409-75).
This percentage increases up to 41\% if we consider the galaxies 
with tentative detections of bridges linking the radio source host with companion galaxies (PKS 0043-42, PKS 2135-14, and PKS 0235-19). 
In the case of PKS 0043-42, the bridge also appears to be detected in the 
K-band image presented in \citet{Inskip10}, and for the other two, the bridges are relatively clear in the median-filtered images, 
but not in the original reduced frames. 

Thus, the results for more than one-third of the sample are consistent with the systems being observed after the first peri-center
passage but before the final coalescence of the merging nuclei. It is clear that, if radio galaxies are indeed triggered
in galaxy mergers, they are not triggered at a unique phase of the merger (e.g., as the nuclei coalesce).
Moreover, since we do not know the relative velocities of the galaxies, it is not possible to rule out the idea 
that the activity in some systems has been triggered in galaxy encounters that will not eventually lead to a merger. 

\subsection{Comparison between Weak- and Strong-Line Radio Galaxies.}
\label{lines}

\subsubsection{Optical Morphologies}

As we described in Section \ref{selection}, according to the spectroscopic classification 24\% of the sample of PRGs presented here
are WLRGs, 43\% are NLRGs, and 33\% are BLRGs or QSOs (see Table \ref{data}). In order to compare them in terms of their
morphologies, we define the Strong-Line Radio Galaxies (SLRGs) to include NLRGs, BLRGs, and QSOs. 
Thus, we have 76\% of SLRGs versus 24\% of WLRGs, also known in the literature as High- and Low-excitation radio Galaxies 
(HEGs and LEGs, respectively; see \citealt{Buttiglione10} and references therein). 

Considering the radio morphologies of our PRGs, only 13\% of the sample are FRI sources (six objects), and all of 
them are WLRGs according to their optical spectra (see Table \ref{data}). In contrast, while all the SLRGs in the sample
are associated with FRII radio morphologies (or alternatively CSS/GPS),
some FRII sources are, in fact, WLRGs (PKS 0034-01, PKS 0043-42, PKS 2211-17, and PKS 0347+05).

The optical spectra of WLRGs are dominated by the stellar continua of the host galaxies; 
they are defined by a small [O III]$\lambda$5007 \AA~equivalent width (EW$_{[O III]}<$10 \AA; \citealt{Tadhunter98}). These objects
also show a low ionization state, as indicated by their [O II]$\lambda$3727/[O III]$\lambda$5007 ratios
(see also \citealt{Buttiglione10}). 
By definition, strong emission lines are detected in the optical spectra of SLRGs, but not in WLRGs. 
Previous studies failed to explain the differences between WLRGs and SLRGs as due to different or time-varying 
accretion rates \citep{Ghisellini01}. An alternative interpretation (see \citealt{Buttiglione10} and references therein) 
is that SLRGs are powered by cold gas accretion (e.g., provided by a recent merger), with the cold gas flowing to 
the central region of the AGN, while WLRGs are fuelled by accretion of hot gas provided by the reservoir 
of their X-ray gaseous coronae \citep{Allen06,Best06,Hardcastle07,Balmaverde08}.
The high temperature of this hot gas would prevent the formation of the ``cold'' structures (e.g., the Broad-Line Region and the torus). 
This hypothesis would explain the non-detection of broad lines in the vast majority of WLRGs, and also their X-ray properties. 
Therefore, it is interesting to compare the optical morphologies of the two types of radio galaxies, in order to check whether 
or not the percentage of mergers/interactions is higher in SLRGs than in WLRGs.


Despite the fact that, in our sample of PRGs, there are only 11 WLRGs versus 35 SLRGs, we can investigate possible
differences in their optical morphologies. Out of the 11 WLRGs, 6 show morphological peculiarities in our optical images (55\%), 
in contrast to the SLRGs, for which all but two (94\%) show such peculiarities. 

\begin{figure*}
\centering
\subfigure[]{\includegraphics[width=5.8cm]{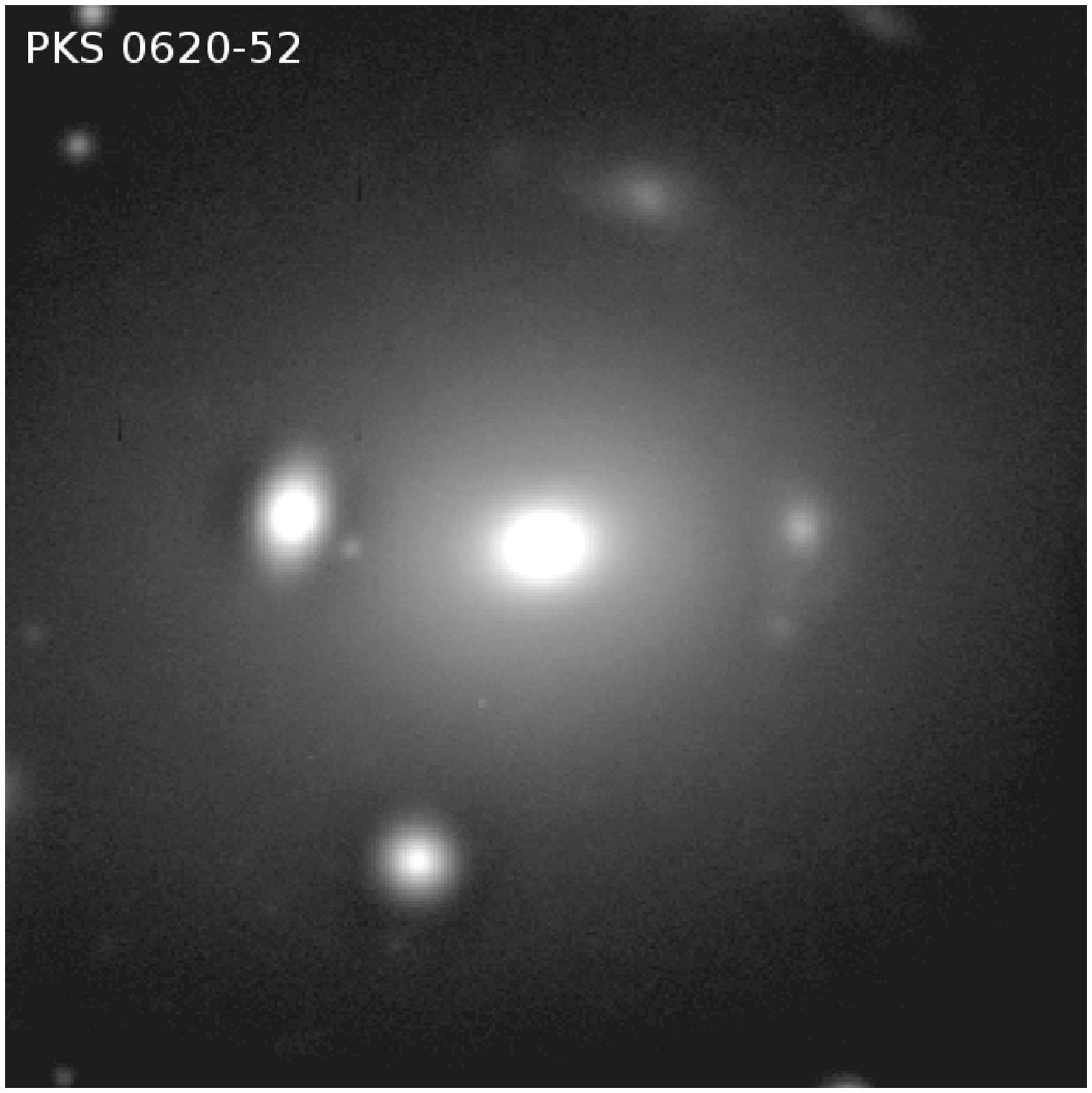}}
\subfigure[]{\includegraphics[width=5.8cm]{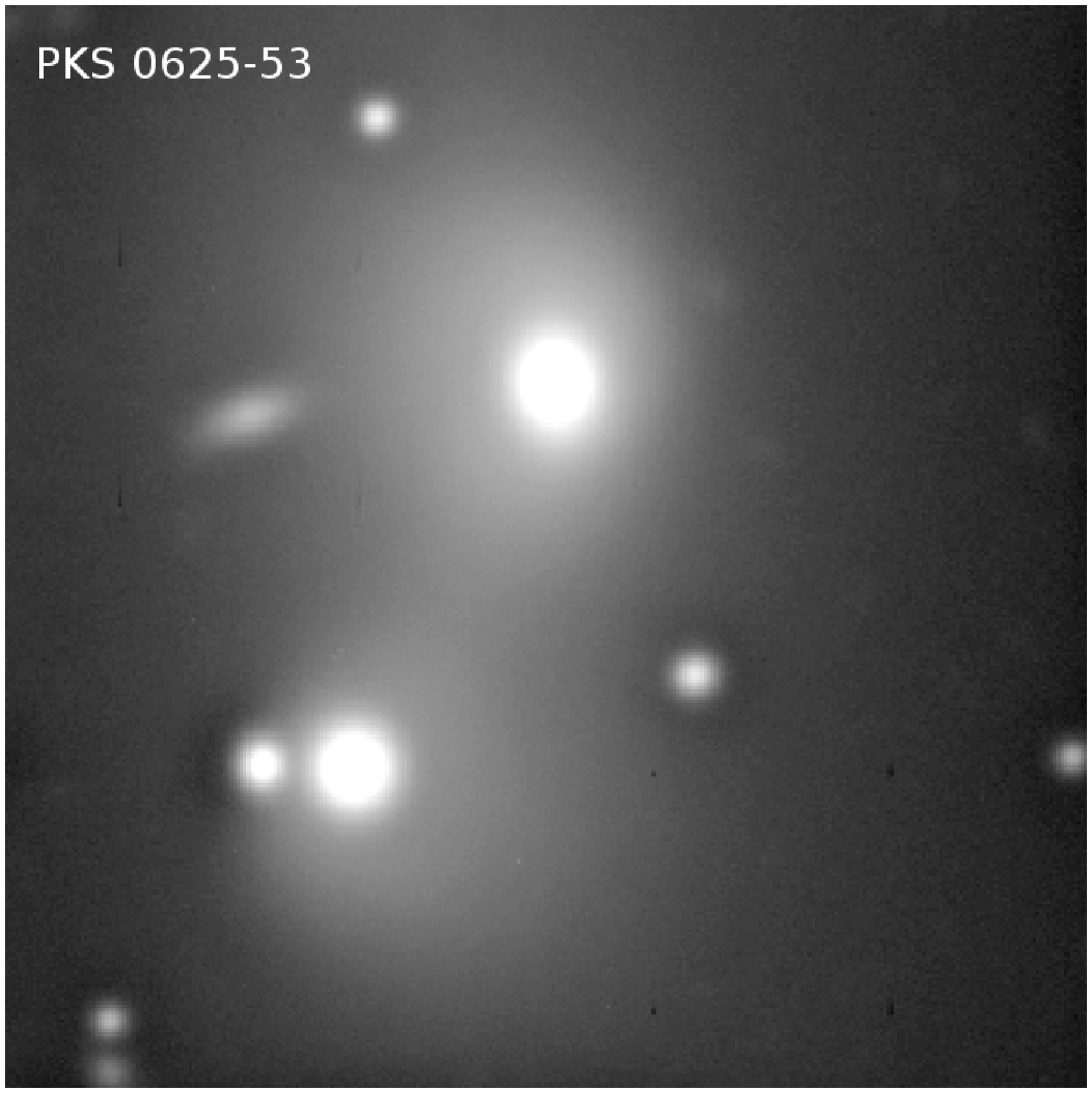}}
\subfigure[]{\includegraphics[width=5.8cm]{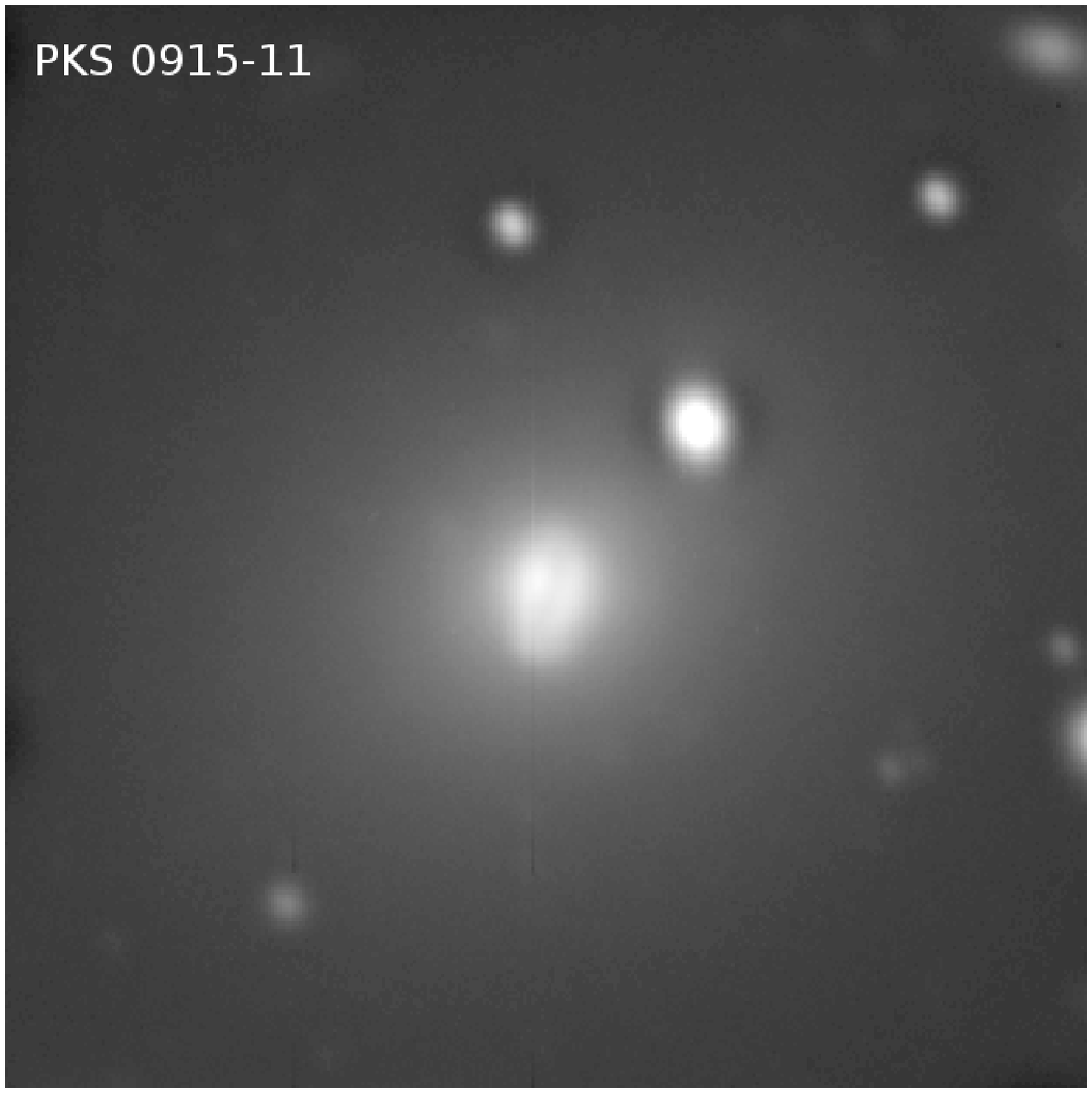}}
\subfigure[]{\includegraphics[width=5.8cm]{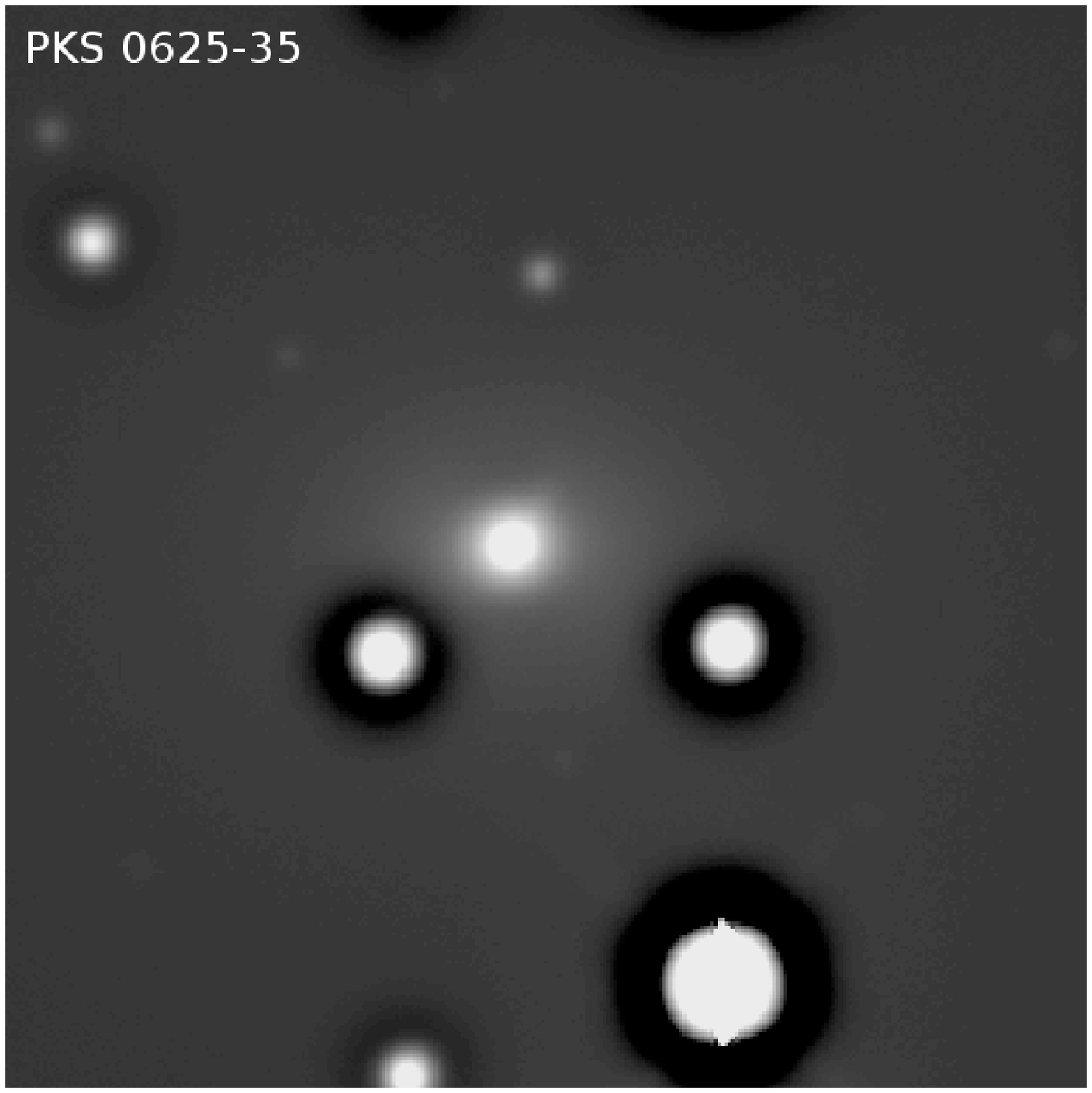}}
\subfigure[]{\includegraphics[width=5.8cm]{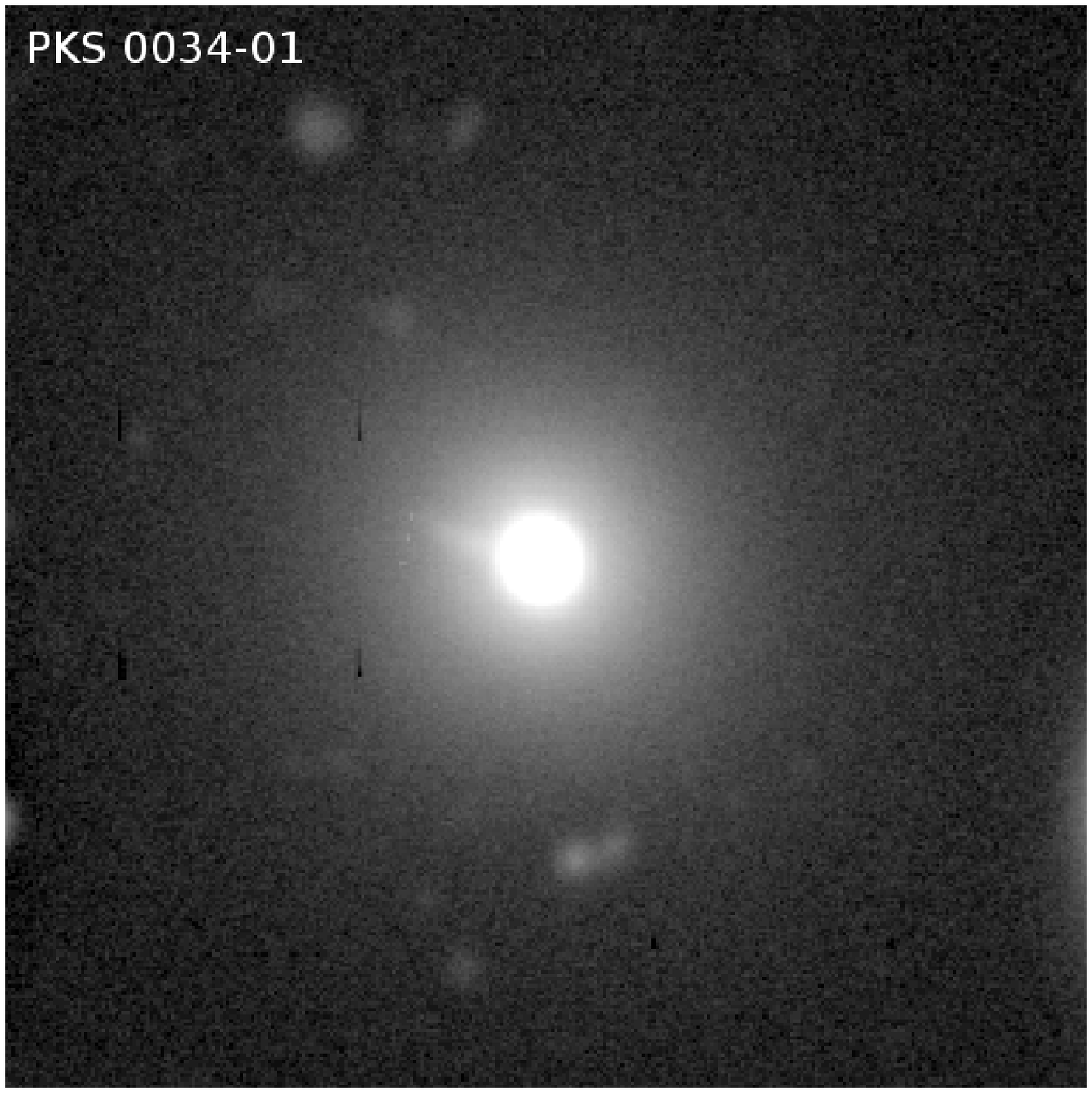}}
\subfigure[]{\includegraphics[width=5.8cm]{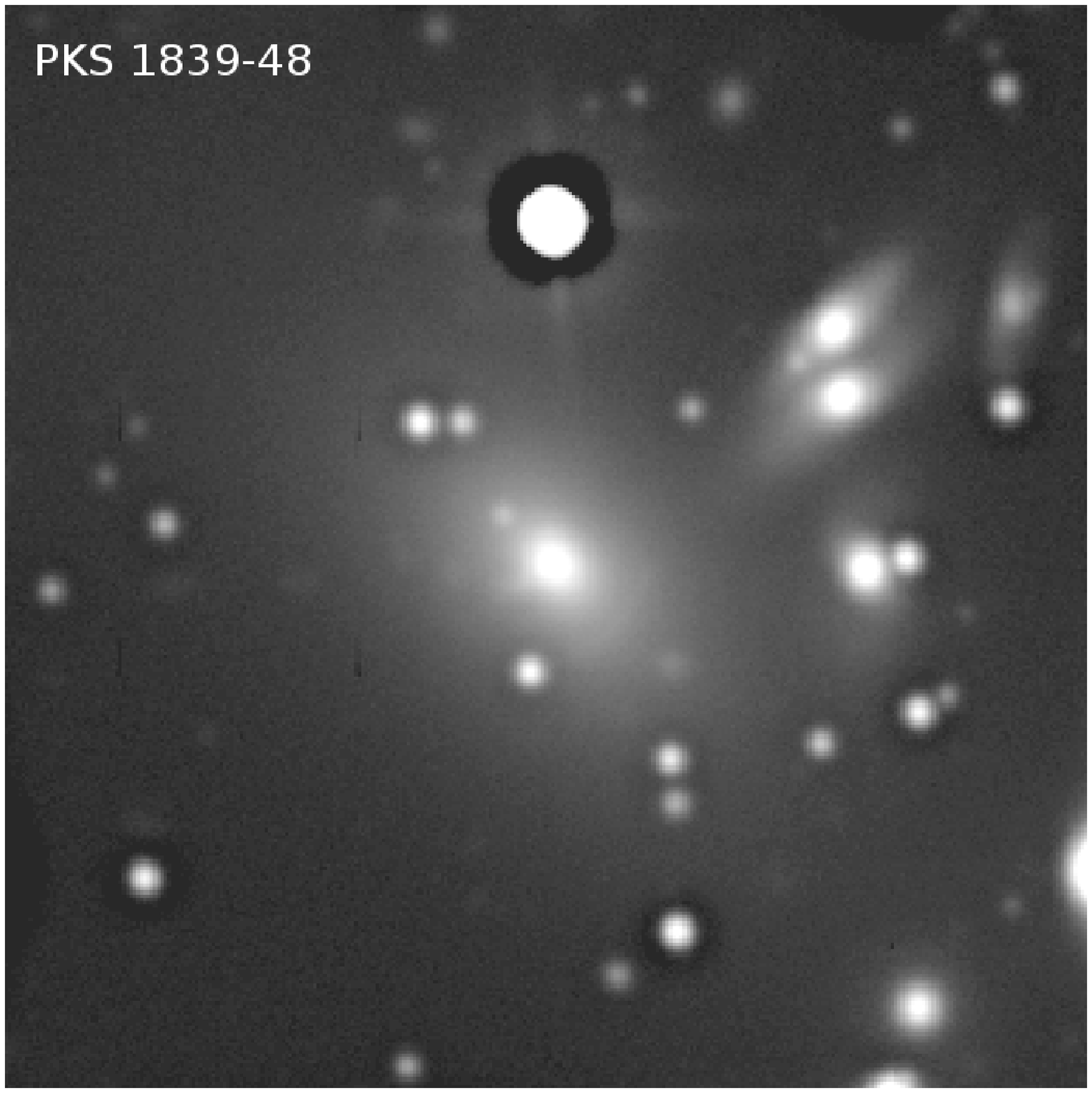}}
\subfigure[]{\includegraphics[width=5.8cm]{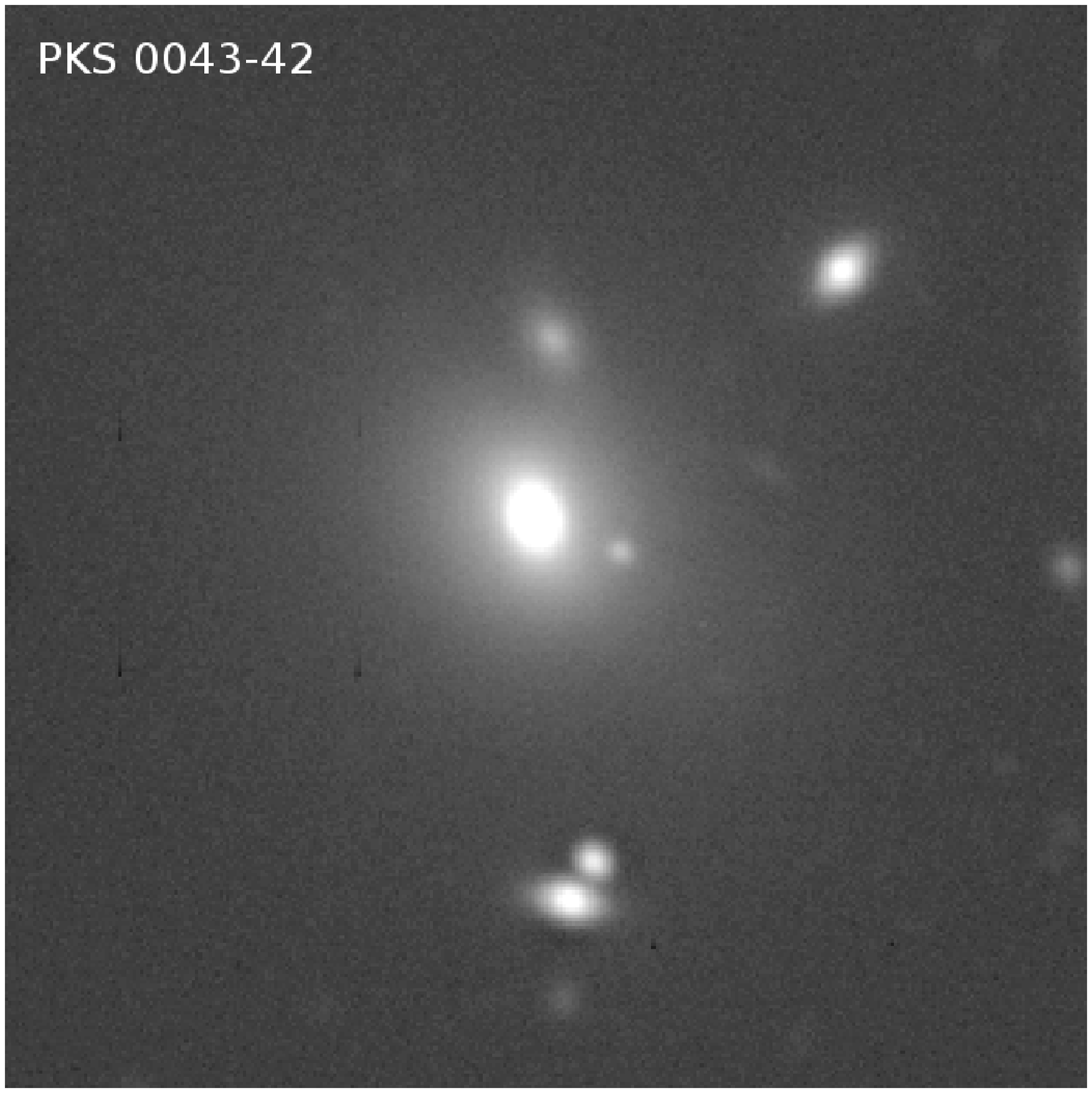}}
\subfigure[]{\includegraphics[width=5.8cm]{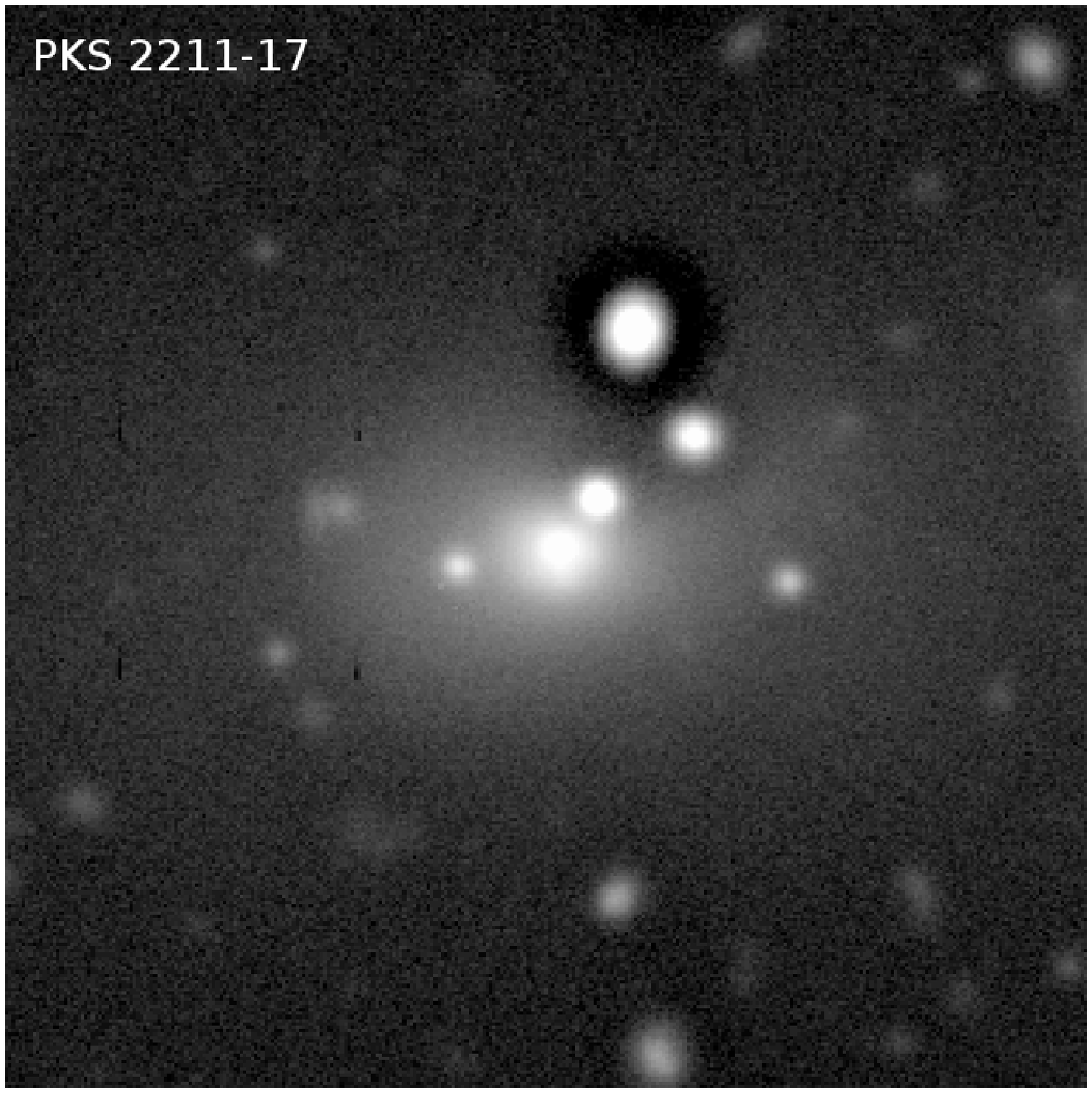}}
\subfigure[]{\includegraphics[width=5.8cm]{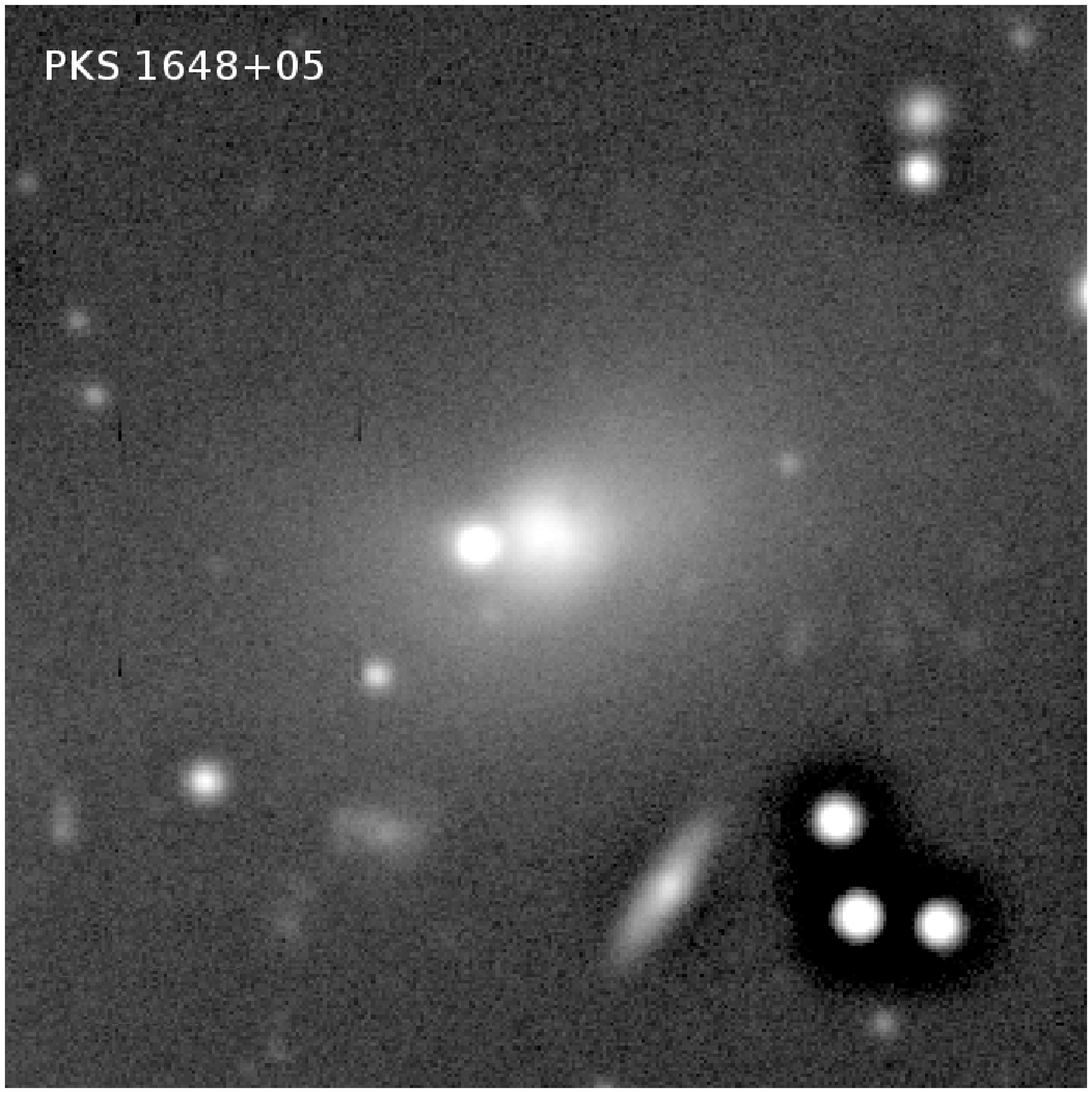}}
\caption{Processed images of 45x45 arcsec$^2$ size
of the nine WLRGs in the sample with redshift z $<$ 0.18 (we excluded PKS 1954-55 because of the bright foreground,
star in front of the galaxy).  
Only two out of the nine WLRGs shown here present signs of interaction, versus eleven out to the twelve SLRGs at z $<$ 0.18. Considering 
the sample as a whole, the percentages of interacting objects are 27\% for WLRGs and 94\% for SLRGs.}
\label{wlrg}
\end{figure*}

\begin{figure*}
\centering
\subfigure[]{\includegraphics[width=5.8cm]{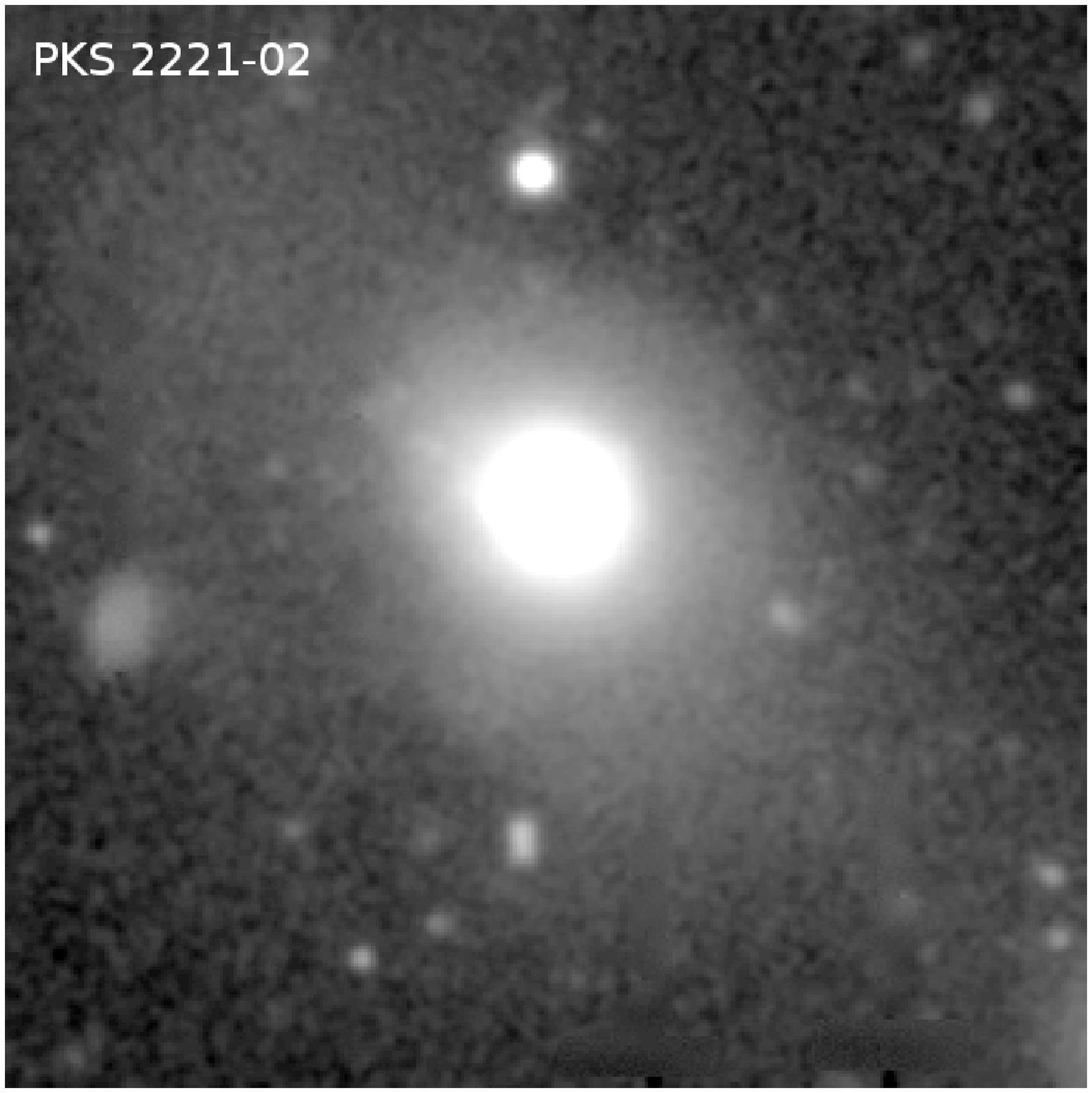}}
\subfigure[]{\includegraphics[width=5.8cm]{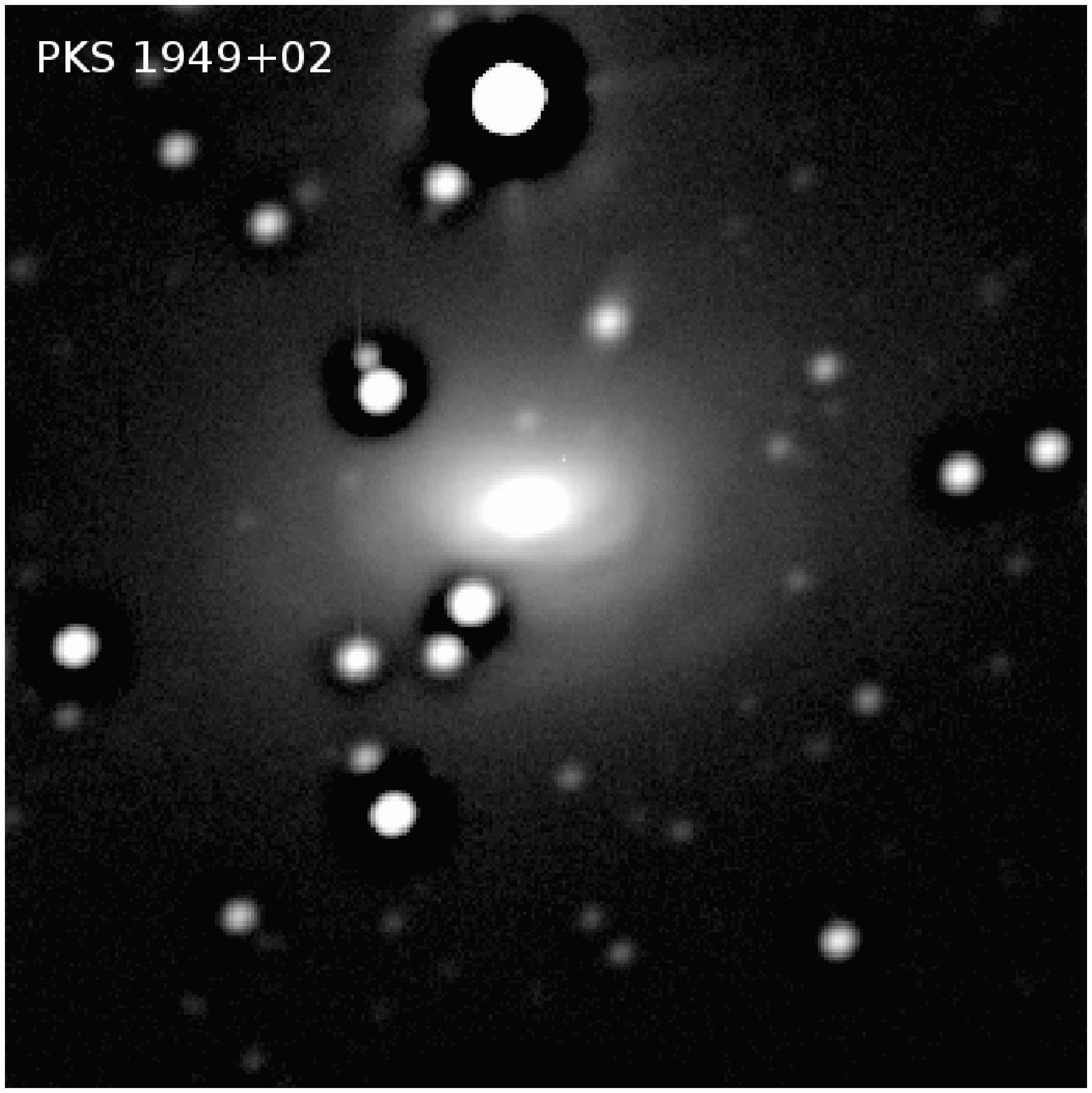}}
\subfigure[]{\includegraphics[width=5.8cm]{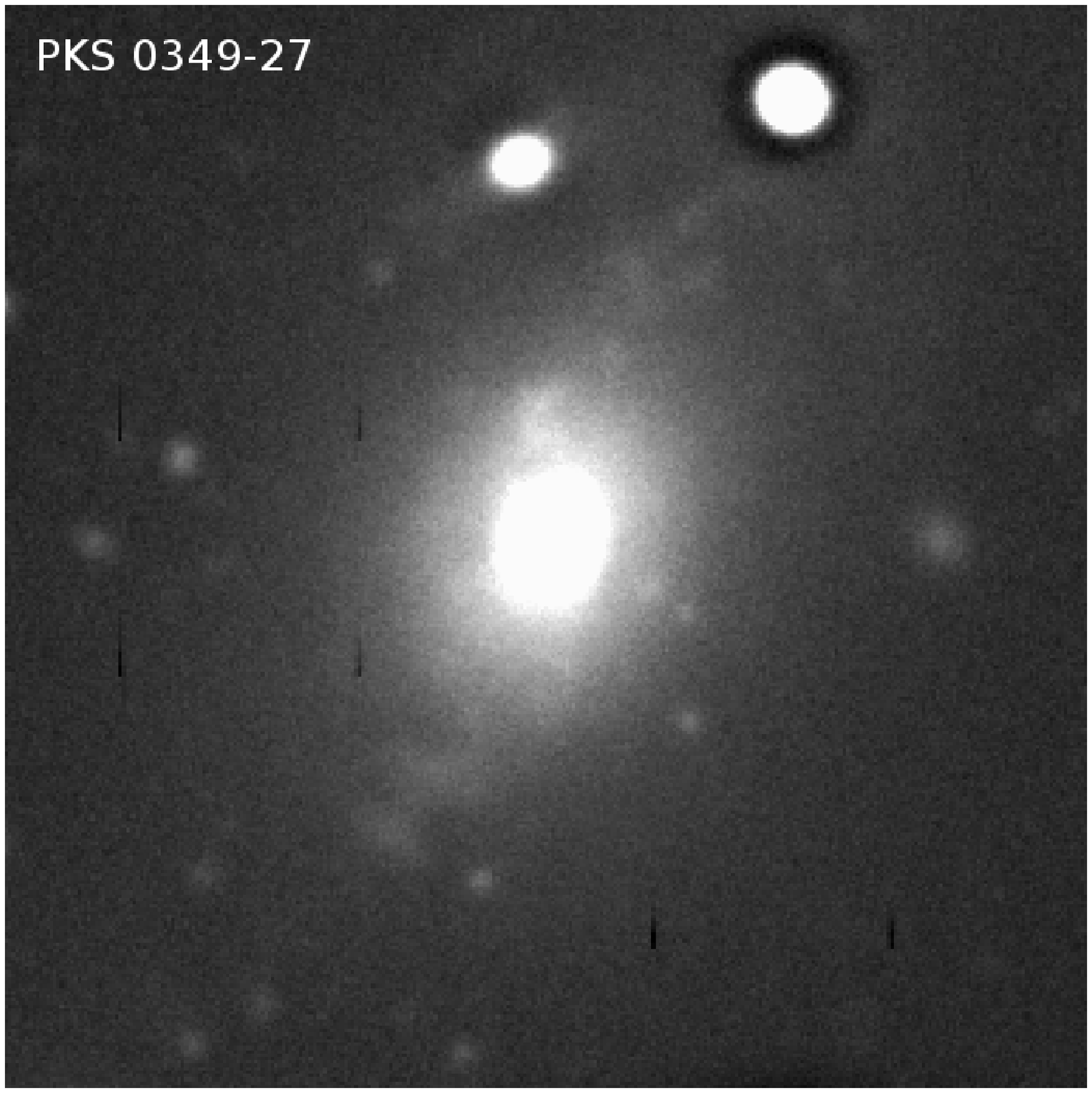}}
\subfigure[]{\includegraphics[width=5.8cm]{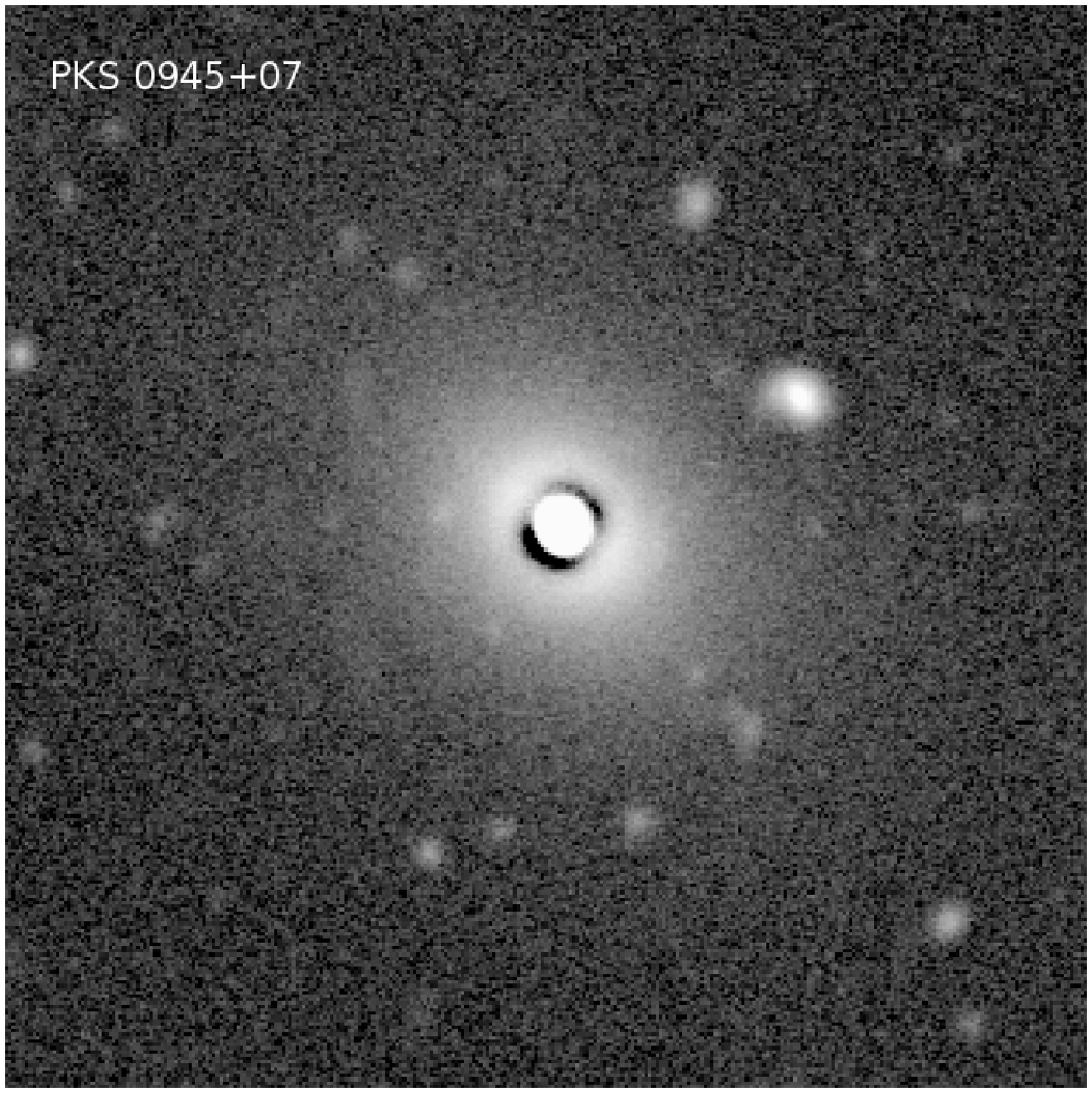}}
\subfigure[]{\includegraphics[width=5.8cm]{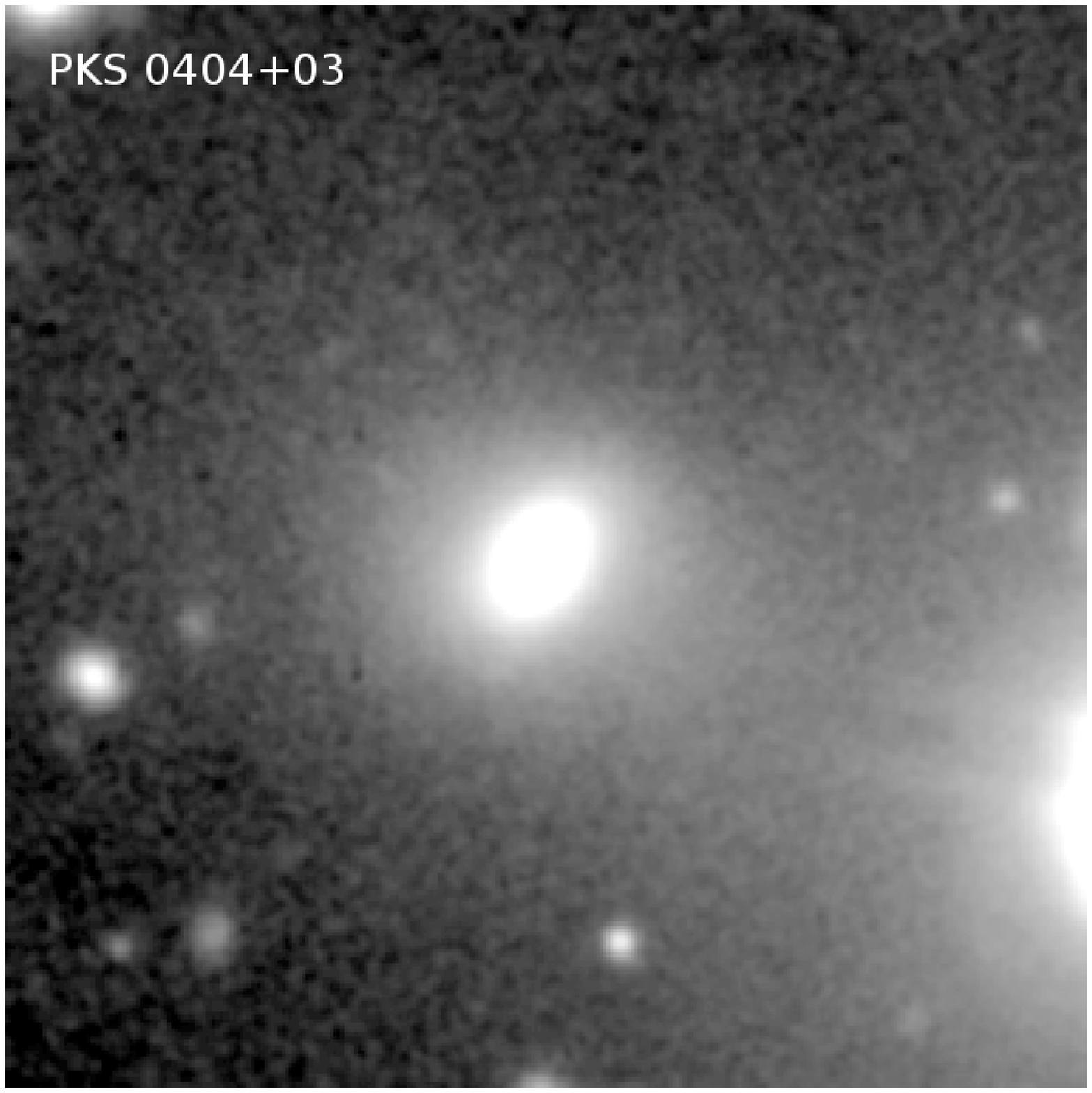}}
\subfigure[]{\includegraphics[width=5.8cm]{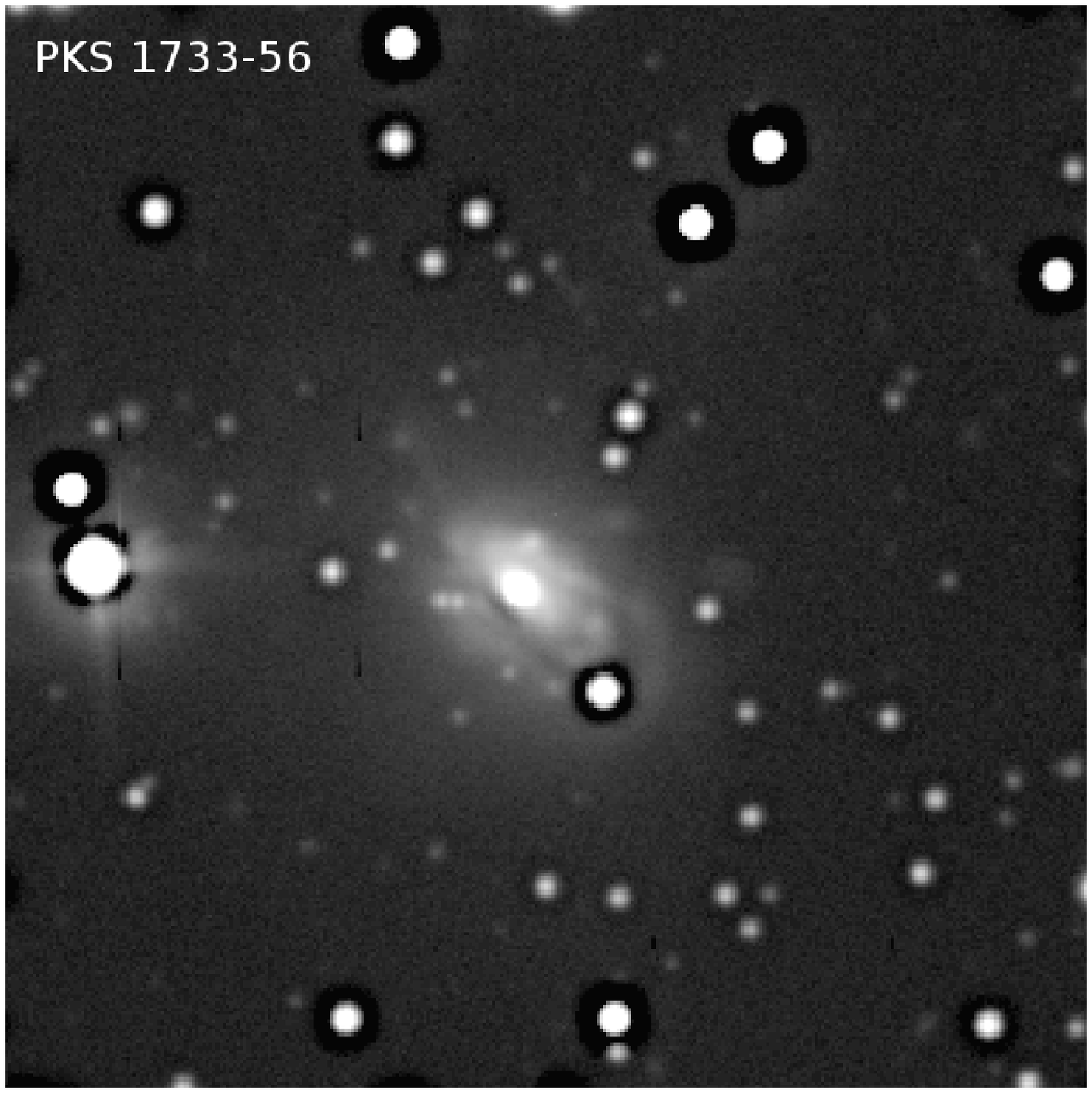}}
\subfigure[]{\includegraphics[width=5.8cm]{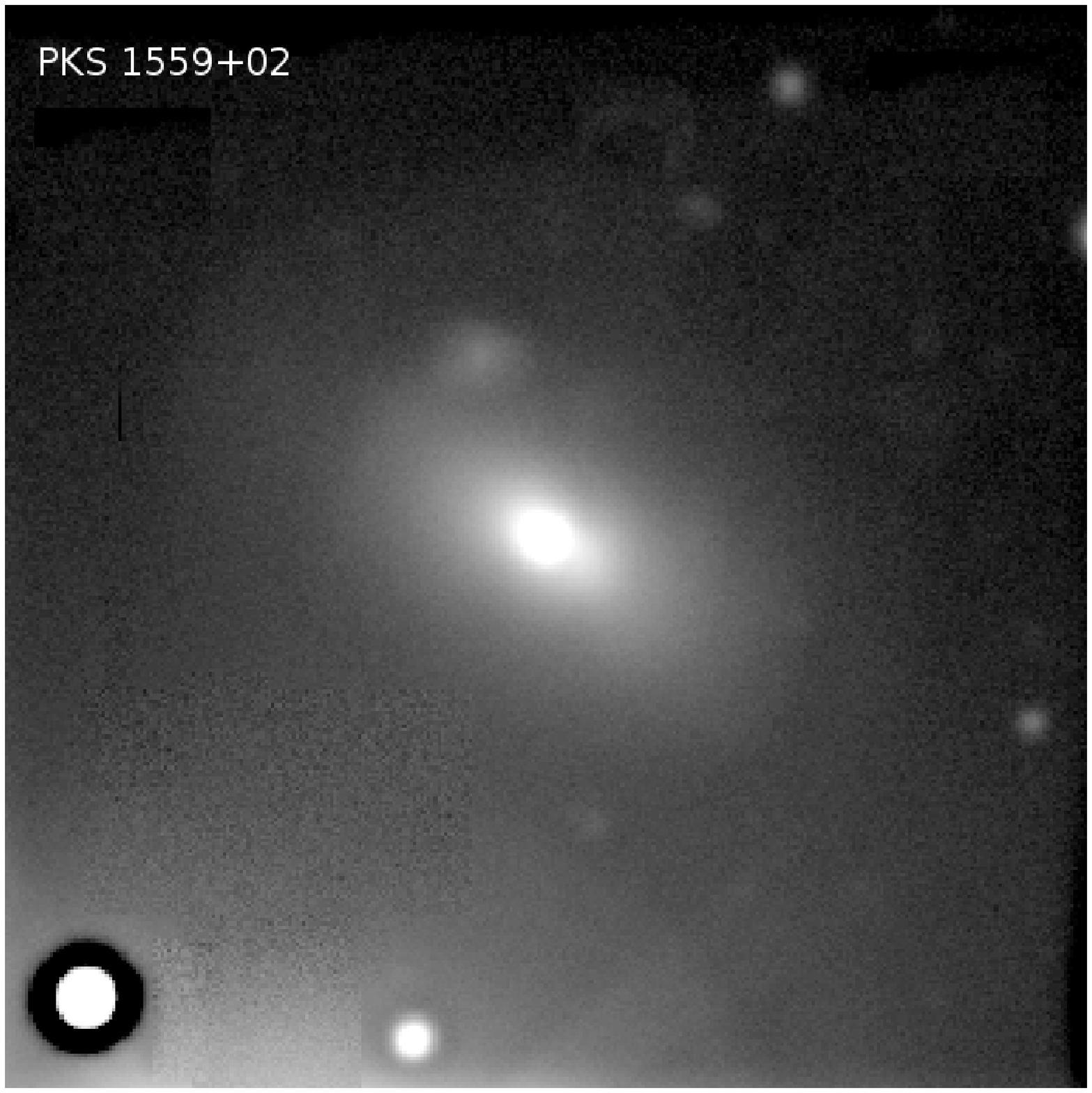}}
\subfigure[]{\includegraphics[width=5.8cm]{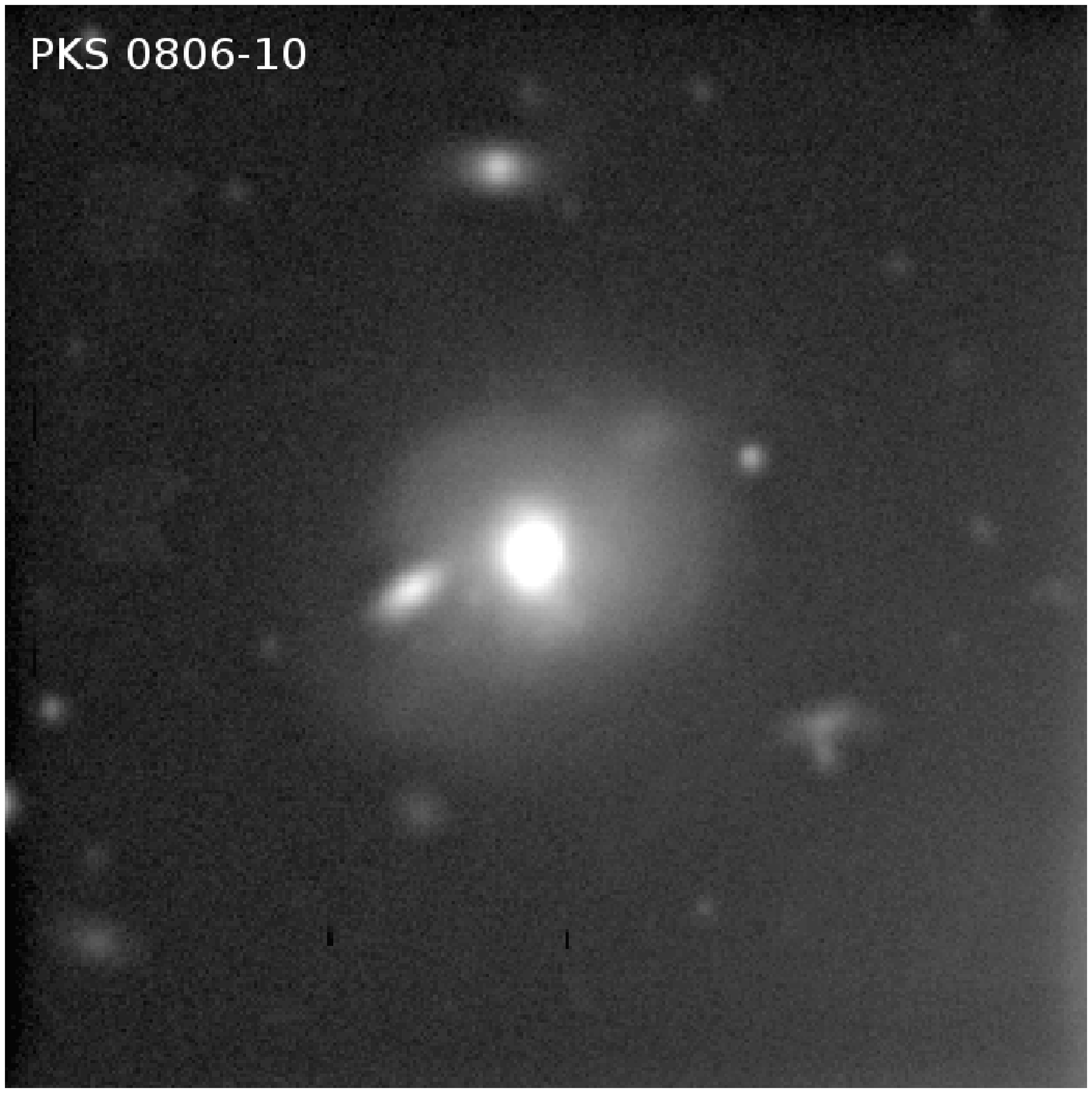}}
\subfigure[]{\includegraphics[width=5.8cm]{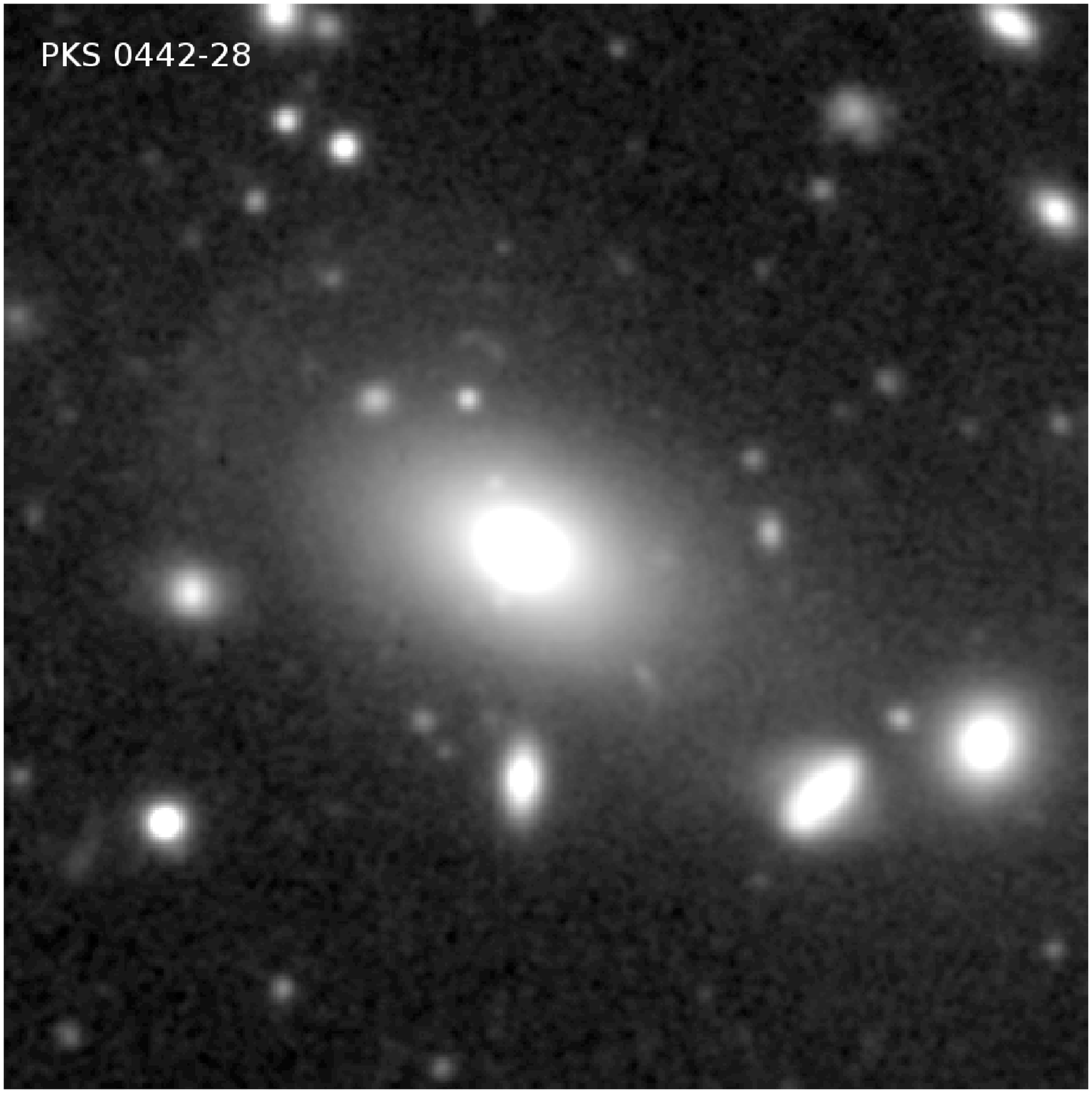}}
\caption{Same as Figure \ref{wlrg} but for nine of the SLRGs with z $<$ 0.18 (the box size for PKS 0442-28 is 60x60 arcsec$^2$). 
There are three more SLRGs with z $<$ 0.18 (PKS 1814-63, PKS 2356-61, and PKS 0213-13), which are shown in Figures  
\ref{pks0213}, \ref{pks2356}, and \ref{pks1814}.}
\label{slrg}
\end{figure*}

As we discussed at the end of Section \ref{classification},
three of the galaxies in the sample, namely
PKS 0915-11, PKS 2211-17 and PKS 1648+05, are found to present dust features as the only detected sign of disturbance, and 
all the three are WLRGs.  
If we do not consider dust as a morphological feature, then the percentage of WLRGs which 
show signs of morphological disturbance decreases to 27\% (3 galaxies), whilst for SLRGs it remains the same (94\%).
It is worth mentioning that all the WLRGs showing signs of interaction are pre-coalescence systems, whereas more 
than half of the SLRGs (57\%) are in the coalescence or post-coalescence phase of the interaction (see Table \ref{wlrg_slrg}).
In Figures \ref{wlrg} and \ref{slrg} we show the processed images 
of 9 WLRGs and 9 SLRGs with redshift z$<$0.18 (containing all-but two of the WLRGs) to visually illustrate 
the differences in the typical morphologies of the two radio galaxy types.

\begin{table*}
\centering
\begin{tabular}{lcccccc}
\hline
\hline
Morphology & Group & \multicolumn{1}{c}{Total Sample} & \multicolumn{1}{c}{SLRGs} & \multicolumn{1}{c}{WLRGs} & \multicolumn{1}{c}{Starbursts} & \multicolumn{1}{c}{Non-starbursts} \\
\hline
Pre-coalescence                  & 1,3  &  35\%  & 37\%  &  27\%   & 38\%  &  33\% \\
Coalescence or post-coalescence  & 2*   &  43\%  & 57\%  &     0   & 46\%  &  43\% \\
No sign of interaction           & 4,5  &  22\%  &  6\%  &  73\%   & 16\%  &  24\% \\
\hline		     			      		 
\end{tabular}						 
\caption{Classification of all the galaxies in the sample, SLRGs, WLRGs, starburst, and non-starburst galaxies based on the detected features. Sources
belonging to groups 1 and 3 are considered as pre-coalescence systems, those in group 2 are likely coalescence or post-coalescence 
scenarios, and finally, galaxies classified in groups 4 and 5 do not show signs of interaction. * Those galaxies classified as (2,3) in Table \ref{data} 
(the WLRG PKS 1839-48 and the SLRGs PKS 0039-44, PKS 1306-09, PKS 1547-79, and PKS 0117-15) are considered as pre-coalescence systems here, although 
they belong to group 2 as well.}
\label{wlrg_slrg}
\end{table*}

\subsubsection{Triggering scenarios for WLRGs}

Our result follows the similar trend of undisturbed morphologies
found by \citealt{Smith89} for a sample of 72 PRGs. They claim that 7\% and 50\% of the WLRGs and SLRGs in their sample, respectively, 
show morphological features such as tails, fans, bridges or shells. 
On the other hand, based on the analysis of K-band images of a different (but strongly overlapping) subset of the 2Jy sample, \citet{Inskip10} 
found that a higher proportion of disturbed WLRG hosts: 62\% compared with 40\% for the SLRGs.
Given the small number statistics, the rate of morphological disturbance for the WLRGs is consistent between the optical 
and NIR studies, and the major difference lies in the much lower rate of interaction detected in the NIR 
observations of the SLRGs than in our optical observations of the same objects. The latter is to be expected, 
since the optical observations are more sensitive to subtle, low-surface-brightness signs of galaxy interactions 
than the NIR observations. Moreover, the SLRGs are at higher redshifts on average than the WLRGs, decreasing the
likelihood of morphological disturbances being detected for the higher-redshift SLRGs in the K-band.  

For those WLRGs which do not display any of the tidal features reported in Table \ref{data}, we speculate on two 
possible scenarios which could explain the lack of any sign of past interactions. On the one hand, they could be
fuelled by accretion of hot gas from their X-ray coronae (i.e., Bondi accretion; see e.g. \citealt{Allen06}).
In this case, a hot rather than cold gas supply would trigger the AGN/radio activity, and consequently, 
no interaction signatures would necessarily be present. On the other hand, it is also possible that 
these PRGs were triggered in dry-mergers or gas-poor interactions. 
In this scenario, broad tails or fans can be produced as result of the interaction (e.g., \citealt{Naab06,Bell06}), 
but without cold and young gas component, these are expected to be more diffuse and shorter-lived ($\sim$150 Myr; \citealt{Bell06})
than those seen in mergers involving at least one late-type (gas-rich) galaxy (see Section \ref{morphological_features}).  
Indeed, by analysing a large sample of early-type galaxy pairs with SDSS spectra, \citet{Rogers09} claimed that 
even in the absence of visual signs of disturbance, it is possible to detect the effects of a close
pair interaction from the activity of the AGN as shown by the spectra of elliptical galaxies. Thus, it is possible to trigger the activity 
in interactions involving early-type galaxies that involve relatively small quantities of warm/cold gas. 
This is clearly the case for the WLRG PKS 0625-53, which in our 
GMOS-S image appears to be interacting with an eary-type galaxy companion at 18.6 kpc.
Both the accretion of hot gas and the gas-poor interaction scenarios could explain the non-detection of tidal features in 
the WLRGs in our sample for which they are absent (73\%), whereas for the 6\% of SLRGs showing no morphological peculiarities, 
the second scenario would be the more probable, since accretion of hot gas cannot explain the SLRG spectra.


\subsection{Starburst vs non-starburst radio galaxies.}
\label{starburst} 
 
In terms of the merger hypothesis for the triggering of AGN activity, hydrodynamic simulations predict 
that, if the triggering mergers are both major and gas-rich,
merger-induced star formation activity will occur in two main phases: around the the first 
peri-centre passage of the merging nuclei, and 
close to the final coalescence of nuclei (e.g., \citealt{Mihos96,Barnes96,Springel05,Cox06,diMatteo07}).
The relative importance of these two phases, and the time lag between them, depends on the details of the models.

Strong observational support for the general idea that gas-rich mergers trigger
substantial star formation activity is provided the fact that the overwhelming
majority of ultra luminous infrared galaxies (ULIRGs) -- representing
the most luminous star forming galaxies in the local Universe --
show signs of severe morphological disturbance consistent with them having undergone major mergers 
\citep{Sanders96,Veilleux02}. Moreover, analysis of the
stellar populations of nearby ULIRGs provides strong evidence that we are observing most of them
within $\sim$100~Myr the final coalescence of the merging nuclei in major, gas-rich mergers \citep{Rodriguez09,Rodriguez10}.

In this context it is interesting that we find evidence for significant recent star formation activity -- based both on optical
and MIR/FIR data -- in  only 28\% of the 2Jy sample of radio galaxies considered in this paper, and only two objects in the sample 
(4\%) are sufficiently luminous at infrared wavelengths to be 
classified as ULIRGs. These results immediately imply that
most radio galaxies are not triggered in the final coalesence phase of major galaxy mergers that are also gas-rich. 
Note, however, that other phases of such mergers, or different types of galaxy interactions (e.g. more minor and/or less gas-rich mergers) 
are not ruled out as triggers for the AGN activity, based on the evidence of the stellar populations alone 
(see \citealt{Tadhunter10} and references therein). Indeed, as we saw in the previous 
sections, a large fraction of 2Jy radio galaxies in fact show 
morphological evidence that they are/have been involved in recent galaxy interactions. The question then arises: is there any evidence 
for morphological differences between the starburst and non-starburst radio galaxies which might suggest that the two groups have been 
triggered in different types of galaxy interactions, or at different phases of the same type of interaction?

First, if we simply
consider the rates of morphological disturbance in the two sub-samples, we find no significant differences: 92\% of the starburst 
radio galaxies and 82\% of the
non-starburst radio galaxies appear morphologically disturbed. Excluding objects in which dust features are the only sign of 
morphological disturbance, or excluding WLRGs, makes little difference to this comparison. 
Second, looking at the surface brightnesses of the detected features, we again
fail to find any significant differences: the median and range of surface brightness of the morphological features 
in the starburst radio galaxies ($\tilde{\mu}_V=23.6~mag~arcsec^{-2}$ and $\Delta\mu_{V}=[22.2, 25.0]~mag~arcsec^{-2}$ respectively) 
are both similar to those of the non-starburst galaxies. In contrast, if starburst radio galaxies were triggered in more major 
or more gas-rich mergers than non-starburst radio galaxies, we might expect them to show higher surface brightness features, 
because of the larger stellar masses in the infalling galaxies, and the potential for merger-induced star formation.

Looking at the classification of the galaxies in more detail we find no significant differences between the rate of 
disturbance measured for starburst and non-starburst radio galaxies, at least at the crude level of the five major morphological 
classes considered here (two last columns of Table \ref{wlrg_slrg}).

However, on a more detailed level we find that, whereas shells are the most commonly detected
feature in the 2Jy sample as a whole, and are detected in 15 out of 33 (45\%) of the non-starburst systems, 
they are found in only 1 out of 13 (8\%) of the starburst systems. 
If shells are predominately associated with minor mergers (e.g., \citealt{Quinn84,Dupraz86}), then these results suggest 
that the non-starburst radio galaxies 
are more likely to be associated with minor mergers than their starburst counterparts. 
This in turn would be consistent with the lack of evidence
for recent star formation activity in the former group.

We emphasise that the lack of greater evidence for differences between the morphologies of 
starburst and non-starburst radio galaxies may be partly a 
consequence of our relatively crude classification scheme and small sample size; more 
quantitative morphological analysis with a larger sample
may in the future uncover larger differences.

Overall, the diversity of morphologies observed in the starburst radio galaxies 
(and indeed the 2Jy sample as a whole) suggests that they are not solely triggered 
at a single stage of a particular type of galaxy interaction. This ties in with recent 
analysis of the ages of the young stellar populations detected at optical 
wavelengths in starburst radio galaxies \citep{Tadhunter05,Wills08,Tadhunter10}, 
which suggests that, while some ULIRG-like systems (e.g., PKS 2135-20 and PKS 2314+03) 
are almost certainly triggered close ($<$100~Myr) to the peaks of starburst activity 
in major, gas-rich mergers, others are triggered in a substantial 
period ($>$200~Myr) {\it after} the main merger-induced starbursts.

\section{Discussion}

\subsection{Comparison with other samples}
\label{comparison}

In this Section we discuss the results found for our complete sample of PRGs in the context of published
morphological studies of both radio galaxies and quiescent ellipticals. All of these comparisons are summarized 
in Table \ref{literature}.

\begin{table*}
\centering
\begin{tabular}{lccccc}
\hline
\hline
Work & Objects & Sample & Redshift & $\mu_V$ \scriptsize{(mag~arcsec$^{-2}$)} &  \% features \\
\hline
\citet{Malin83}     & QE           & 137 & $<$0.01  & $\la$25.5-26        & $\sim$10 \\
\citet{Tal09}       & QE           &  55 & $<$0.01  & $\la$27.7           & 53-73    \\
\citet{vanDokkum05} & QE           &  86 & 0.1      & $\la$28.7*          & 71       \\
\hline
\citealt{Smith89}   & PRGs         & 72  & $<$0.3   & 21.0-25.0           & $\sim$50 \\
\citet{Dunlop03}    & RG, RQQ, RLQ & 33  & 0.2      & $\la$24.6*          & **    \\            
This work           & PRGs         & 46  & 0.05-0.7 & 23.6,(21.3-26.2)*   & 85      \\ 
\hline		     			      		 
\end{tabular}						 
\caption{Comparison among different morphological studies of quiescent ellipticals (QE), radio galaxies (RGs), 
PRGs and radio quiet and radio loud QSOs (RQQ and RLQ, respectively). The number of objects in each
sample, redshift range or median redshift, median $\mu_V$ values or/and range and the percentage of 
galaxies with peculiar optical morphologies are given. 
* $\mu_V$ values obtained by assuming typical colors of elliptical galaxies
from \citet{Fukugita95} at given redshift. 
** Comparable to that of QE of similar redshift and masses analyzed in the same work.}
\label{literature}
\end{table*}

\subsubsection{Comparison with other samples of radio galaxies.}

Our measured percentage of morphological peculiarities (85\%) is higher than the $\sim$50\% found by \citealt{Smith89} for 
a similar sample of PRGs at z $<$ 0.3 (see Table \ref{literature}). 
The latter authors analyzed ground-based optical images taken using telescopes ranging from 1 to 4 m and detected features at 
$\mu_V$ $\la$ 25 $mag~arcsec^{-2}$. For our PRGs, we are detecting features at a median surface brightness depth of 
$\tilde{\mu}_V=23.6~mag~arcsec^{-2}$, ranging from 21.3 to 26.2 $mag~arcsec^{-2}$, once 
cosmological dimming, galacting extinction and K-corrections are considered (see Section \ref{classification}).

To investigate the reasons for the apparent discrepancy between the detection rates of morphological disturbance
reported in \citealt{Heckman86},\citealt{Smith89} ($\sim$50\%) and in this work (85\%), 
we have searched for the galaxies that we have in common, and find 14 in total. 
We detect tidal features in all of them, whereas \citealt{Heckman86} reported signs of morphological peculiarities in the form 
of fans, shells, tidal tails, etc. in only six 
(PKS 2221-02, PKS 0349-27, PKS 0945+07, PKS 1559+02, PKS 1934-63, and PKS 2314+03).
By comparing their and our detected features for the latter galaxies, in general we find good agreement for the morphologies of 
these six galaxies, except for PKS 0945+07. We detect a shell in the case of this galaxy, whereas \citealt{Heckman86} reported
the existence of a tail and a fan, but associated with line-emitting gas (see Appendix \ref{individual}). 
For the remaining 8 objects, \citealt{Heckman86} and \citealt{Smith89} did not detect any features. Three of them 
are the three galaxies presenting dust as the only sign of disturbance in our images (PKS 0915-11, PKS 2211-17 and PKS 1648+05), and
for PKS 0035-02, PKS 0213-13, PKS 1949+02, and PKS 2135-14 we detect bridges, shells, 
tails, fans, and amorphous haloes. The higher percentage of features that we report here is likely due to the higher quality of our
GMOS-S images, taken using an 8m telescope and with good seeing conditions (median FWHM$\sim$0.8\arcsec, in contrast with 
the 1-2\arcsec~range of seeing of the \citealt{Heckman86},\citealt{Smith89} observations), our better pixel sampling 
(0.146\arcsec~pixel$^{-1}$ with GMOS-S, compared with the 0.29\arcsec-1.3\arcsec~pixel sizes employed by 
\citealt{Heckman86},\citealt{Smith89}), and the fact that our images go $\sim$1 mag deeper.

It is also necessary to consider the results of \citet{Dunlop03}, based on a sample of 33 AGN comprising radio galaxies, 
radio quiet QSOs (RQQs) and radio loud QSOs (RLQs) at a median redshit of z=0.2. 
They analyzed high spatial resolution HST/WFPC2 images, and showed that all the AGN in their study appear
to be hosted by relatively undisturbed giant elliptical galaxies. Indeed, by comparing with a control sample of 
quiescent ellipticals in clusters, they concluded that the hosts of radio-loud and radio-quiet AGN are indistinguishable from quiescent 
ellipticals of similar masses and redshifts. As discussed in Section \ref{intro}, we believe that the discrepancy between 
the results presented by \citet{Dunlop03} and ours is mainly based on the differences in the depth of the observations. 
As can be seen in Table \ref{literature}, the features measured by \citet{Dunlop03} reach a limiting faintness of 24.6 mag~arcsec$^{-2}$ in the 
V-band (assuming typical colors of elliptical galaxies), 
whereas in our case, we are detecting features as faint as $\mu_V$=26.2 mag~arcsec$^{-2}$, i.e., 1.6 mag fainter. 

Thus, as discussed in Section \ref{intro}, the discrepancy between the previous results and ours is very likely due to 
the differences in depth of the observations. 
The high-quality deep Gemini observations presented here 
show for the first time that the overall majority of PRGs at intermediate redshifts present disturbed morphologies, 
which are very like the result of galaxy mergers and/or interactions.

\subsubsection{Comparison with samples of quiescent elliptical galaxies.}

If galaxy interactions are the main triggering mechanism for radio-loud AGN activity in our sample, 
then we should expect to find the signs of 
morphological disturbance to be stronger in the radio source host galaxies than in the general population 
of quiescent (i.e. non-active) early-type galaxies. 
Fortunately, in the general context understanding the importance of galaxy mergers for the 
formation of such galaxies, a number of detailed morphological 
studies have been made of the quiescent elliptical galaxy population in the local Universe 
\citep{Schweizer80,Malin83,vanDokkum05,Sikkema07,Tal09}.

Perhaps the most relevant study -- both in terms of its surface brightness depth and its general approach 
to morphological classification -- is that 
by \citet{Malin83}. They used visual inspection of deep photographic images to detect shell and ripple features in $\sim$10\% of nearby 
elliptical galaxies in the RC2 catalogue with declinations $\delta < -17\degr$. Importantly, the limiting depth of their observations 
($\mu_V\la25.5-26~mag~arcsec^{-2}$) is similar to the effective depth reached by our Gemini observations of the 2Jy sample. 
Clearly, the rate of morphological disturbance found by \citet{Malin83} is significantly lower than the one we find for the 2Jy sample, 
even if we only make the comparison with the lowest redshift bin ($\Delta z_1$) of our PRGs, which contains a number of 
relatively undisturbed WLRG objects (see section \ref{lines}). Unfortunately, \citet{Malin83} give only the limiting surface 
brightness for their survey as a whole, so a 
more detailed comparison between the surface brightnesses of the detected features is not possible.

More recently, there have been two CDD-based studies of nearby elliptical galaxies that reach significantly fainter surface brightness limits.
First, \citet{vanDokkum05} reported that 71\% of a sample of 86 colour- and morphology-selected 
bulge-dominated early-type galaxies at a median redshift
of $z=0.1$ show morphological evidence of broad fans, tails of ripples at very faint levels of surface brightness: $\mu_V <$ 28.7~mag arcsec$^{-2}$ 
(assuming typical colours for elliptical galaxies at z=0.1 from
\citealt{Fukugita95}). A more recent study by \citet{Tal09} has presented an analysis of the optical morphologies of a complete sample of 55 luminous 
and nearby elliptical galaxies (z $<$ 0.1). Reaching a limiting depth of 
$\mu_V <$ 27.7~mag arcsec$^{-2}$, they detect signs of morphological disturbance in $\sim$53-73\% (depending on the criterion used) of their sample. 
Clearly, the rates of morphological disturbance detected in these recent, 
CDD-based studies approach those we find in the radio galaxies in the 2Jy sample. 
However, it is notable that their limiting surface brightnesses are much fainter than we 
achieve for the 2Jy radio galaxies: 2.5 and 1.5 magnitudes fainter 
in the cases of \citet{vanDokkum05} and \citet{Tal09} respectively. Therefore it is not clear that the comparison is reasonable.

Finally we note that there are some important caveats to bear in mind when making the comparison 
with the existing studies of quiescent elliptical galaxies. 
First, an ``ideal'' comparison sample would be matched in galaxy mass, environment, 
redshift and depth/resolution of observations; clearly no such sample exists at present. 
Second, emission line contamination might affect the comparison. For example,  
any neutral gas present in  quiescent elliptical galaxies that falls below the detection limits 
of existing HI 21cm surveys and is not visible at optical wavelengths (because it is not ionized), 
might be rendered visible by AGN illumination in the radio-loud AGN population. 

Although none of the three comparison samples considered above is completely ideal, all show a strong overlap 
with the 2Jy sample in terms of the absolute magnitudes and 
environments of the host galaxies. Any bias would likely be in the direction 
that the radio galaxy hosts are more luminous than the 
general quiescent elliptical galaxy population covered by the comparison 
samples\footnote{Whereas the radio galaxy hosts are generally hosted by giant
elliptical galaxies, at the upper end of the E-galaxy luminosity function (typically L $>$ 2L$_*$; \citealt{Dunlop03}), 
the RC2 catalogue that forms the basis
of \citet{Malin83} study samples down to much lower optical luminosities.}. 
Therefore, if radio galaxies were not triggered by galaxy interactions, we would require a strong 
positive correlation between luminosity 
and degree of morphological disturbance, in order to explain the fact that the 2Jy sample displays a higher rate of 
disturbance than the sample considered by \citet{Malin83}. As far as we are aware, no such correlation exists.

In terms of emission line contamination, even if we make the most pessimistic assumptions about 
the degree of emission line contamination (see section \ref{contamination}), 
the rate of morphological disturbance due to genuine continuum-emitting 
structures in the 2Jy sample still comfortably exceeds that of the \citet{Malin83} sample.

\section{Conclusions}

We have analyzed deep GMOS-S/Gemini optical broad-band imaging for a complete sample of southern PRGs at 
intermediate redshifts (0.05$<$z$<$0.7). We visually classify their morphologies, and divide the sample 
into five groups, depending on the detected features. In order to confirm the detections, we made use 
of three different data analysis techniques with the purpose of enhancing the morphological features. 
Our major results are as follows:

\begin{itemize}

\item The high-quality observations presented here 
show for the first time that 85\% of our sample of PRGs at intermediate redshifts present peculiar morphologies 
at relatively high levels of surface brightness ($\tilde{\mu}_{V}=23.6~mag~arcsec^{-2}$).
If we do not consider dust as a sign of morphological disturbance, then 78\% present peculiarities.
In any case, this fraction of distorted morphologies is much higher than for radio quiet ellipticals 
at the same sample brightness level and also for other samples of PRGs.

\item The morphological peculiarities of the galaxies include tails, fans, bridges, shells, dust lanes, irregular features, 
amorphous haloes, and multiple nuclei. We propose that these features are the result of the merger or close encounter 
of galaxies in pairs or groups.

\item The results for more than one-third of the sample are consistent with the galaxies being observed after the first 
peri-center passage but before the final coalescence of the merging nuclei. It is clear that, if radio galaxies are
indeed triggered in galaxy mergers, it does not happen at a unique phase of the merger. Moreover, since we do not
know the relative velocities of the galaxies, it is not possible to rule out the idea that the activity in some
galaxies has been triggered in galaxy encounters that will not eventually lead to a merger. 

\item By dividing the sample in WLRGs and SLRGs
we find that 55\% of the former show peculiarities in their optical morphologies, contrary to the SLRGs, of which 
at least 94\% present any sign of disturbance.  Indeed, the percentage of distorted WLRGs decreases down to 27\% if 
we do not consider dust as a sign of morphological disturbance. Based on these results, the majority of WLRGs 
would be fuelled/triggered by Bondi accretion of hot gas. However, the evidence for interactions and dust in a fraction of them 
indicates that the cold-gas accretion cannot be ruled out.

\item For those WLRGs which do not show any evidence of morphological disturbance, we propose two 
possible scenarios to explain the lack of any sign of past interactions. Either they are 
fuelled by Bondi accretion of hot gas from their X-ray coronae or they are the result of dry-mergers or gas-poor
interactions.

\item We find that 92\% of the starburst radio galaxies in the sample present 
peculiar morphologies, following the same trend as the total and SLRG samples.
There are no significant differences between the optical morphologies of starburst and non-starburst galaxies, neither 
in the rate of detected features, nor in their surface brightnesses. The only possible distinction between 
the two groups is the lower proportion of identified shells in the former. 


\item By comparing with different samples of quiescent ellipticals from the literature, we conclude that the 
percentage of morphological disturbance that we find here for PRGs greatly exceeds those found for quiescent ellipticals
when similar surface brightness limits are considered.

\end{itemize}

\appendix
\section{Notes on Individual Objects}
\label{individual}

Below, we comment individually on the optical morphologies of the galaxies in our sample, based
on our GMOS-S images. Details on the possible emission-line contamination in the detected features 
are also given, based on the comparison with the K-band data presented in \citep{Inskip10}, and  
with optical long-slit spectra and narrow-band images of the galaxies from various published and unpublished sources.
The quoted surface brightnesses have been corrected to the rest frame V-band, as described in 
Section \ref{classification} of the main body of the text. All the images referred to are presented in the electronic
edition of the journal (Figures \ref{pks0620_online} to \ref{pks0409_online}).  

\subsection{PKS 0620-52.}  

This WLRG is the closest object in our sample, and it has been found to show clear 
evidence for a YSP \citep{Wills04}. \citet{Tadhunter10} analyzed new spectra for 
this source, determining an upper limit of 0.9 Gyr for the age of the YSP.
Our GMOS-S image (Figure \ref{pks0620_online}) does not reveal any evidence for morphological disturbance, apart from
a  relatively high incidence of companion galaxies, as expected for a galaxy at the centre of a 
rich cluster. A cluster environment for this radio galaxy is further supported
by the existence of a moderately-luminous X-ray halo 
\citep{Siebert96,Trussoni99}. Statistically, it is more difficult to detect tidal features such as 
shells or broad fans in regions of high galaxy density, since the tidal effects rapidly disrupt
these features. 

\subsection{PKS 0625-53.}

This WLRG is hosted by the eastern component of a dumbbell system in a rich galaxy environment. Our Gemini 
image (Figure \ref{pks0625_53_online}) shows a 
broad bridge linking the two galaxies, which are separated by a distance of $\sim$19 kpc, have
highly distorted isophotes, and are clearly interacting. 
The same morphology is found in the NIR images analyzed in \citet{Inskip10}, who did not detect strong nuclear 
point source component in either object.
Also noticeable in our GMOS-S image is the degree of distortion in the two galaxy companions 
located to the NW of the dumbbell system in Figure \ref{pks0625_53_online}a. 

\subsection{PKS 0915-11 (3C218, Hydra A).}

Situated in the Hydra cluster of galaxies, this FRI is one of the most powerful radio sources in
the local universe. Its optical spectrum (WLRG) show weak emission lines, along with absorption lines and 
a continuum SED that provide evidence 
for a young stellar population  of age $\sim$0.05 Gyr \citep{Aretxaga01,Wills04,Holt07}.
Our GMOS-S $r'$-band image (Figure \ref{pks0915_online}) shows an extended diffuse halo similar to that of PKS 0620-52, which is typical 
of central cluster galaxies; it also reveals for the first time a dust lane of diameter 6~kpc crossing the galaxy 
nucleus. This dust lane appears to be aligned with the rotating disk structure detected by \citet{Melnick97}.

\subsection{PKS 0625-35 (OH-342).}

This low-redshift WLRG does not reveal any morphological peculiarity associated with a merger or interaction
in our Gemini image (Figure \ref{pks0625_35_online}). The only feature that we detected is a one-sided jet 
pointing to the SE, which is also found by \citet{Inskip10} in their residual NIR image. The detection of 
this jet supports the classification of this object as a BL-Lacertae (BL-Lac) by \citet{Wills04}. 

\subsection{PKS 2221-02 (3C445).}

This BLRG shows a very broad and asymmetric H$\alpha$ profile in the optical spectrum first reported by 
\citet{Osterbrock75}.
In the NIR, its morphology appears dominated by a nuclear point source, and the host
galaxy is well-reproduced by a de Vaucouleurs elliptical in both the NIR \citep{Inskip10} and the optical
\citep{Govoni00}. Our deep Gemini image (Figure \ref{pks2221_online}) reveals the spectacular morphology of this galaxy, which is 
clearly interacting with a close companion that appears obviously distorted in the direction of the
radio galaxy host. The separation of the radio source 
and the companion galaxy is $\sim$37 kpc. The radio galaxy shows a broad $fan$ of surface brightness $\mu_V$=25.6 mag~arcsec$^{-2}$ towards
the SW that is co-aligned with the axis of the double galaxy system, and a $shell$ towards the East with $\mu_V$=25.3 mag~arcsec$^{-2}$. 
The optical morphology of this
galaxy is described in \citealt{Heckman86}. They also identify the $fan$, as well as a bright tail extending out from the
galaxy center for $\sim$10 kpc to the SW, which they identify as a strong source of emission lines.  
In our median-filtered image this feature appears more like a series of blobs that curve to the SW of the nucleus. 
Long-slit spectra of this source \citep{Tadhunter86} demonstrate that the flux in the outer distorted structure is
not dominated by emission lines. 

\subsection{PKS 1949+02 (3C403).}

This NLRG/FRII galaxy with an X-shaped radio morphology  \citep{Black92} was observed in the 
optical with HST by \citet{Martel99} and \citet{deKoff00}, 
who reported the existence of dust lanes and a low-surface brightness halo with a sharp boundary surrounding the central 
region at a radius of $\sim$2.5\arcsec. Our Gemini image (Figure \ref{pks1949_online}) confirm the latter feature, and, for the first time, also reveal an extraordinary
series of concentric arc/shell features extending to a maximum radius of $\sim$12~kpc to the West, and $\sim$20~kpc
to the South. The brightest of these shell features has a corrected surface brightness of $\mu_V=22.6~mag~arcsec^{-2}$. 
The K-band model-subtracted
residual images presented in \citet{Inskip10} reveal hints of the shells, indicating their continuum-emitting nature. 
Our long-slit spectra confirm that the outer arc and shell features are not dominated by emission line radiation.
\citealt{Heckman86} also studied the optical morphology of the galaxy, but did not report any type of disturbance. 

\subsection{PKS 1954-55.} 

This WLRG/FRI galaxy is located in a rich environment. 
The presence of a bright star close (in projection) to the nucleus precludes detailed examination of the morphology of
the central regions of the galaxy, but our Gemini image (Figure \ref{pks1954_online}) shows no sign of morphological disturbance
in its halo, consistent with the earlier results of 
\citet{Fasano96}. The galaxy belongs then to the group of WLRGs in our sample which do not present evidence for mergers/interactions. 

\subsection{PKS 1814-63.}

This NLRG/GPS has been found to show starburst activity in the MIR/FIR, based on its FIR excess and strong PAH features \citep{Dicken09,Dicken10}.
Our GMOS-S optical image (Figure \ref{pks1814_online}), which shows a straight dust lane of diameter $\sim$20~kpc along with highly elongated, 
disky isophotes, is consistent with the morphological classification of this system as 
an S0 or Sa galaxy viewed close to edge-on. However, the outer disk of the galaxy is clearly
highly distorted. Due to the presence of a bright star 
close to the galaxy nucleus we took two different sets of images with short and long exposure time (see Table \ref{mag}). The first 
allows us to unveil details of the nuclear region, avoiding star
saturation, whereas the second longer exposure time set reveals more details of the irregular features in the outer parts of the
galaxy. Similar features (a faint extended disk feature and a possible dust lane) are observed in the model-subtracted K-band 
image analysed by \citet{Inskip10}. 
Long-slit spectra of this galaxy confirm that emission line contamination is not a serious issue for the outer
distorted features. This source will be the subject of a detailed study of its optical and radio properties \citep{Holt10}.

\subsection{PKS 0349-27.}

PKS 0349-27 shows one of the most spectacular morphologies of the sample in our GMOS-S image (Figure \ref{pks0349_online}). 
The galaxy was studied by \citealt{Heckman86}, who analysed narrow-band [O III]$\lambda$5007 and broad V-band images. They 
reported the existence of two close companion galaxies at 26 and 20 kpc towards the East and NE, respectively. In their 
images (both narrow- and broad-band) they detected a bridge linking the radio galaxy with the eastern companion located at 26 kpc
and a tail-like structure pointing to the opossite direction and extending up to 20 kpc. According to the study of the 
gas kinematics of this system presented by \citet{Danziger84}, these features are very likely the result of 
an interaction with the eastern galaxy companion, despite the fact that they are line-emitting gas features. 
Our deep GMOS-S image shows in much more detail the scenario of this interaction. We confirm the existence of the 
structures detected by \citealt{Heckman86}, and measure a surface brightness for the bridge linking the radio 
galaxy with the eastern companion of $\mu_V$=25.5 mag~arcsec$^{-2}$. In addition, we also detect a more extended, and fainter, 
bridge linking the radio galaxy with the clearly distorted galaxy at 83 kpc to the West 
($\mu_V$=26.5 mag~arcsec$^{-2}$). The latter bridge feature was also detected by \citet{Hansen87} in their deep 
H$\alpha$+[NII] image, at a surface brightness level which suggests that much of the emission from the bridge detected in our GMOS-S
image could be emission line, rather than continuum radiation. Whereas \citet{Hansen87} classified the distorted galaxy at the end of the 
larger bridge
as an elliptical, it is clearly revealed as a disk galaxy in our GMOS-S image. 
The median-filtered image also reveals the presence of a faint shell extending $\sim$10 kpc to the West of the 
radio galaxy nucleus.

\subsection{PKS 0034-01 (3C15).}

This WLRG has a radio morphology that is intermediate between FRI and FRII classes \citep{Morganti99}. The 
only feature that we detect in our GMOS-S image of this object (Figure \ref{pks0034_online}) is the optical synchrotron jet reported for the first time
in \citet{Martel98}. It extends to a projected distance of 7.5 kpc ($\sim$40\arcsec) in the NW direction. The jet is also detected
in the K-band images presented in \citet{Inskip10}.
WFPC2/HST images of this source reveal subarcsecond-scale features including a dust lane and three emission filaments or arms 
\citep{Martel98,Sparks00}. These latter features are not resolved in our GMOS-S image.

\subsection{PKS 0945+07 (3C227).}

This BLRG/FRII is studied in detail by \citet{Prieto93}, who found spectacular extended emission line structures. Also, 
the optical morphology of this system was studied by \citealt{Heckman86}, and they reported the detection of several knots and tails 
which very likely correspond to line-emitting gas. Our GMOS-S image (Figure \ref{pks0945_online}) does not reveal any of 
these features, but it shows a shell of $\mu_V$=24.4 mag~arcsec$^{-2}$ $\sim$7~kpc to the NW of the nucleus. In fact, emission line 
contamination of the shell is not a serious issue 
because strong emission lines do not fall in the $r'$ filter employed in our observations.
Indeed, this is confirmed by the presence of this shell in the K-band images of this galaxy \citep{Inskip10}.

\subsection{PKS 0404+03 (3C105).}

The presence of a very bright 
star close (in projection) to  this NLRG/FRII, along with the high Galactic dust extinction measured for this field (A$_V>$1), 
mean that our observations of this
source are less sensitive to low surface brightness features than for the
other objects in our sample. Nonetheless, our processed images (Figure \ref{pks0404_online}) reveal hints of a faint shell towards the NE. 

\subsection{PKS 2356-61.}

The optical morphology of this NLRG, as shown in our GMOS-S image, appears to be full of features which are very likely the 
result of a past merger/interaction (Figure \ref{pks2356_online}). At shorter scales, we detect one arc-like irregular feature of depth
24.9 mag~arcsec$^{-2}$ in the V-band at the northern side of the galaxy center. There is also a shell towards the south of the galaxy, 
and a fainter one to the NE ($\mu_V$=25.1 and 25.6 mag~arcsec$^{-2}$ respectively). Further from the brighter shell
there is also a broad fan/shell in the same direction, at a 25.8 mag~arcsec$^{-2}$ depth level. Finally, 
at $\sim$82 and 118 kpc distance from the galaxy center, there are two faint arcs towards the NW and SW, respectively. Note that
the shell features detected in this galaxy are more irregular in appearance than the sharp shells detected in other
objects in our sample (e.g., PKS 1559+02 and PKS 0442-28) and in the nearby ellipticals studied by \citet{Malin83}. 
Emission line contamination is not a serious issue in the case of this galaxy
because strong emission lines do not fall in the $r'$ filter employed in our observations.

\subsection{PKS 1733-56.}

This BLRG/FRII, with evidence for starburst activity provided by its FIR excess and strong PAH features in its MIR spectrum \citep{Dicken09,Dicken10},
lies in a relatively crowded field. Its optical morphology, as revealed by our GMOS-S image (Figure \ref{pks1733_online}), 
is highly disturbed, showing at least two
tidal tails: a shorter one pointing to the West and a longer one to the SE both having $\mu_V$=23.6 mag~arcsec$^{-2}$.
We also detect several arc-like irregular features, including a smooth, inner one (of $\mu_V$=23.6 mag~arcsec$^{-2}$) up to 
6.5 kpc in the SE direction and a larger one of $\mu_V$=24 mag~arcsec$^{-2}$ extended up to 11.6~kpc to the NW. 
It is likely that at least part of the complex
structure visible in the central regions of the galaxy is due to obscuration by a complex system of dust lanes.
The brightest inner arc-like irregular structure is detected in the K-band, model-subtracted images presented by \citet{Inskip10}, 
who also claim that the residuals from the subtraction  display an excess of emission aligned with the major axis of the galaxy. 
In the outermost part of the radio galaxy host we also detect an interlocking  series of fainter
shells that extent to a maximum radius of $\sim$25~kpc to the SE and $\sim$17~kpc to the South. We measure a surface brightness of 
$\mu_V$=24.4 and 24.0 mag~arcsec$^{-2}$ for these outer shells, respectively.
Emission line contamination is not a serious issue in our GMOS-S image, since
strong emission lines do not fall in the $r'$-band filter used for the observations.
Based on integral-field spectroscopic observations, \citet{Bryant02} found evidence for a disturbed morphology, 
disrupted gas rotation, and patchy starburst emission for this source. They claimed that the most
plausible explanation for all of these features is an interaction or merger with a SE companion from which the radio galaxy would
be accreting gas, but they did not find such a merging galaxy in the DSS-II and SuperCOSMOS images. Our deeper GMOS-S image 
does no show any merging candidate either. 

\subsection{PKS 1559+02 (3C327).}

The WFPC2/HST images of this NLRG/FRII source reveal a bifurcated dust lane structure extending $\sim$3~kpc to the NE of the 
galaxy nucleus, with the radio axis lying roughly 
perpendicular to the dust lanes. The nuclear structure, as revealed by the HST image, appears to be complex and possibly 
double \citep{deKoff96,deKoff00}. The presence of a bright star close to the galaxy complicates its morphological classification. 
However, by looking at our processed images ({\it unsharp-masked} and specially {\it smoothed galaxy-subtracted}; 
Figure \ref{pks1559_online}) the presence 
of dust lanes becomes clear. The galaxy nucleus appears double as well, but it is difficult to say whether is intrinsically
double or just apparently divided by the dust lane. On larger scales the GMOS-S image reveals a system of sharply-defined
shells at radii of $\sim$16 and 30~kpc to the SE, and $\sim$14 and 22~kpc to the NW; these sharply-defined shell features are
reminiscent of the systems of shells detected around some elliptical galaxies in the nearby Universe by \citet{Malin83}. 
We measured surface brightnesses of $\mu_V$=23.2 and 23.6 mag~arcsec$^{-2}$ for the inner and outer shells, respectively. 
The larger of the shells to the SE was also detected by \citealt{Heckman86}, while \citet{Inskip10} detected the inner shell to
the SE (16 kpc radial distance form the nucleus) in the K-band,
claiming that an elongated tidal feature appears to be connecting the galaxy nucleus with a satellite object to the South.
Emission line contamination can be ruled out, since strong emission lines do not fall in the r' filter used for the observations.

\subsection{PKS 0806-10 (3C195).}

The GMOS-S image of this NLRG/FRII (Figure \ref{pks0806_online}) shows a clearly disturbed morphology, including features 
such as a broad fan towards the
East ($\mu_V$=23.4 mag~arcsec$^{-2}$), a bright shell to NW, and a fainter shell towards the West ($\mu_V$=23.3 and 24.6 mag~arcsec$^{-2}$, 
respectively). The {\it unsharp-masked} image also show several bright knots of emission closer to the galaxy center 
(see Figure \ref{pks0806_online}a). Emission line contamination of the features is not a serious issue here
because strong emission lines do not fall in the $r'$-band  filter employed in our observations.
\citet{Inskip10} also report the detection of the NW shell (or arc-like feature) 
in their K-band images. They claim also that the galaxy is possibly interacting with the small disk galaxy companion 
at $\sim$10 kpc East. Finally, based in optical imaging of the galaxy using the Canada-France-Hawaii Telescope (CFHT), 
\citet{Hutchings88} reported that the radio galaxy appears asymmetrical, and probably undergoing an interaction, from the high
concentration of of faint objects located within a 20\arcsec~radius from the galaxy center.

\subsection{PKS 1839-48.}

This WLRG/FRI galaxy is the brightest member of a group or cluster, with its spectrum presenting several absorption lines 
typical of early-type galaxies, but no strong emission lines \citep{Tadhunter93}. 
Our GMOS-S image (Figure \ref{pks1839_online}) shows an apparent secondary nucleus within a 5.6 kpc radius to the SE, a bright arc-like 
shell $\sim$7~kpc to the 
NW, and a tentative narrow and sharp tail NE. 
It is also possible that the radio galaxy is interacting with the two galaxies 
aligned along the SW direction, since at least one them appears clearly disturbed.  
Emission line contamination can be ruled out, since strong emission lines do not fall in the filter used.
The bright and sharp shell-like feature described above appears different to the shells/arcs detected for other galaxies
in our sample. Indeed, considering that PKS 1839-48 is a massive galaxy immersed in a rich cluster, we cannot discard the 
possibility of this arc/shell being the result of gravitational lensing. 

\subsection{PKS 0043-42.}

This WLRG/FRII appears to be located in the center of a group or  cluster, surrounded by an extended diffuse halo (Figure \ref{pks0043_online}). 
There is faint companion located $\sim$8~kpc NE of the galaxy center. However, the lack of any other 
features and of galaxy distortion in this direction prevents us from classifying the radio galaxy as a double nucleus system,
based on its optical morphology.
However, from the K-band image presented in \citet{Inskip10}, the authors reported a central isophotal twist 
and an excess of emission along the NS direction. They claimed that both are likely to be associated either to the presence of a 
dust lane or to the apparent interaction with the companion object $\sim$15~kpc to the North. Indeed, 
the residuals in the model-subtracted image reported by \citet{Inskip10} show a bridge-like structure linking with the 
northern companion. Although we have classified the radio galaxy as undisturbed, by looking at our  {\it smoothed galaxy-subtracted}
optical image (see Figure \ref{pks0043_online}a), it is possible to unveil the existence of a bridge linking with the small galaxy at the north.

\subsection{PKS0213-13 (3C62).}

The GMOS-S image of this NLRG/FRII galaxy (Figure \ref{pks0213_online}) reveals one of the most clear systems of shells in our sample. 
We measured a surface brightness 
of $\mu_V$=24.7 mag~arcsec$^{-2}$ for the brighter shell ($\sim$20~kpc to the North) and 25.1 mag~arcsec$^{-2}$ for the eastern and outer 
shell ($\sim$27~kpc from the nucleus). Our {\it unsharp-masked} image also shows the presence of a narrow and sharp tidal tail to the SW
(see Figure \ref{pks0213_online}a).
The galaxy morphology of 3C62 was also studied by \citealt{Heckman86}, but they did not report the existence of any feature. 
Our long-slit spectrum of this galaxy shows that emission line contamination is not a serious issue 
for the shell structure to the NE.

\subsection{PKS 0442-28.}

This NLRG/FRII galaxy shows a faint, but sharply defined shell $\sim$50~kpc to SW of its nucleus in our GMOS-S image 
(Figure \ref{pks0442_online}). This
shell has a surface brightness $\mu_V$=25.7 mag~arcsec$^{-2}$. 
There are several galaxies within $\sim$20\arcsec, any of them possibly interacting with the radio galaxy. There are neither 
optical long-slit spectra nor infrared data suitable for assessing the degree of emission line contamination of this shell. 

\subsection{PKS2211-17 (3C444).}

This WLRG/FRII is in the center of a cluster, and its optical morphology was studied by \citealt{Heckman86} in the past, but their
optical images did not reveal any features. In our GMOS-S image (Figure \ref{pks2211_online}) the galaxy appears surrounded 
by the diffuse halo typical 
of central cluster galaxies, and also shows hints of a broad fan or shells towards the NE. By looking at our 
processed images, we also detect 
a knotty sub-structure in the near-nuclear regions of the galaxy, which may be due to patchy dust obscuration.
Evidence for a dust lane is also found in the model-subtracted NIR image of 
\citet{Inskip10}. The NIR data also reveal an excess of diffuse emission towards the East, coinciding with our tentative 
detection of a fan or a shell. 
This source is one of the three galaxies in our sample with dust as the only feature with secure detection. 

\subsection{PKS 1648+05 (3C348, Herc A).}

This WLRG/FRI source is at the center of a cooling flow cluster of galaxies at z=0.154. A recent analysis of Chandra X-ray
data showed that the cluster has cavities and a shock front associated with the radio source \citep{Nulsen05}.
\citealt{Heckman86} did not report the existence of any feature for this galaxy from their optical imaging. 
However, our Gemini data (Figure \ref{pks1648_online}) reveal patchy dust features crossing the galaxy nucleus 
in a roughly East-West direction, out to a radius of
$\sim$6.5~kpc to the East. Note that \citet{deKoff96} and \citet{Baum96} have used the HST to
detect fine dust features in the nuclear regions of this galaxy, and they claim that the dust is distributed as two interlocking
rings. Unfortunately our Gemini observations do not have sufficient spatial resolution to allow us to confirm the ring-like 
morphology of the near-nuclear dust.

\subsection{PKS 1934-63.}

This compact NLRG/GPS source is hosted in a double and highly disturbed system. Our optical image (Figure \ref{pks1934_online}) 
shows two bright nuclei separated
by $\sim$9 kpc, embedded in a common and irregular envelope. Two tidal tails extending towards the SW from the radio galaxy 
and to the North from its companion are detected with surface brightnesses of 23.1 and 23.4 mag~arcsec$^{-2}$, respectively. 
All the previous features were also detected by \citealt{Heckman86} with the same $\mu_V$ values. They observed the radio galaxy
and companion in both B and V bands, reporting an integrated color of B-V=1.25, which is $\sim$0.3 mag bluer  than the normal color of
elliptical galaxy at z=0.2 \citep{Fukugita95}. Although much of the blue light excess in this source is likely to be due
to AGN-related continuum components such as scattered light and nebular continuum (see object-specific discussion
in \citealt{Tadhunter94} and \citealt{Tadhunter02}), evidence for on-going star formation activity
in this source is provided by the detection of PAH emission features at MIR wavelengths by \citet{Dicken10}. 
Note that presence of the tidal features described here is also reported in the K-band imaging study of \citet{Inskip10}, indicating 
their continuum-emitting nature. Our long-slit spectra also confirm that the companion galaxy and its associated tidal tail
are dominated by continuum emission.

\subsection{PKS 0038+09 (3C18).}

This BLRG/FRII at z=0.19 is found to have about two thirds of its K-band flux in the nuclear unresolved component \citep{Inskip10}.
The galaxy appears to be in a dense enviroment, with several companions within $\sim$20\arcsec. Our GMOS-S image 
(Figure \ref{pks0038_online}) reveals a faint and 
long tidal tail of surface brightness $\mu_V$= 26.2 mag~arcsec$^{-2}$ extending up to 40~kpc to the NW --- this is the faintest tidal
feature detected in our imaging survey. The galaxy was observed
with a seeing of FWHM$\sim$1\arcsec, which makes it difficult to identify other possible signs of disturbance. 
In the K-band images shown in \citet{Inskip10}, the galaxy appears elongated in the direction of the tidal tail. However, due 
to the faintness of the latter feature, it is not clearly detected in the NIR. 
Our long-slit spectrum of the galaxy does not allow us to confirm/discard the emission-line nature of this tail.

\subsection{PKS 2135-14.}

The optical image of this QSO/FRII (Figure \ref{pks2135_14_online}) shows a disk galaxy companion at $\sim$18 kpc SE
and a close-in companion, or secondary nucleus, at a distance of 6 kpc. However, analysis of its spectrum  
showed that the latter object is actually a foreground star \citep{Canalizo97}. 
Our GMOS-S image reveals a very disturbed morphology, including a shell in the West side
of the quasar nucleus, which is embedded in an amorphous halo, and a faint tidal tail pointing to the SE.
This tidal tail, which in the original image extends up to $\sim$36 kpc and has a surface brightness of 
$\mu_V$=24.2 mag~arcsec$^{-2}$, is very likely the brightest part of a bridge linking the radio galaxy and 
the disturbed galaxy at $\sim$134 kpc to the SE. However, the bridge is only detected in our {\it median-filtered image} (see Figure
\ref{pks2135_14_online}b), 
and thus we do not consider it as a secure  morphological classification of a feature in the galaxy.
The galaxy morphology was also studied by \citealt{Heckman86}, but they did not report the detection of any feature. 
The results on the determination of the continuum or line-emitting nature of the detected features are inconclusive based on long-slit
spectra \citep{Tadhunter98}.

\subsection{PKS 0035-02 (3C17).}

This BLRG/FRII appears clearly disturbed in our optical image (Figure \ref{pks0035_online}). 
The radio galaxy shows a bridge linking with a companion galaxy 
$\sim$46~kpc to the South. We also detected a fan of $\mu_V$=25.1 mag~arcsec$^{-2}$ towards the NW. 
\citealt{Heckman86} did not report the detection of any feature for this galaxy. The residuals from the model subtraction 
of the K-band image reported in \citet{Inskip10} display an excess of emission extending linearly from the NE to the SW, 
which roughly coincides with the direction of the bridge linking with the companion galaxy. This confirms the
continuum-emitting nature of the bridge.

\subsection{PKS 2314+03 (3C459).}

This NLRG/FRII with strong evidence for a YSP at both optical \citep{Wills08} and MIR/FIR wavelengths \citep{Dicken09,Dicken10},
is classified as ULIRG based on its FIR luminosity, and appears clearly disturbed in 
our GMOS-S image (Figure \ref{pks2314_online}). We detect two broad and symmetrical fans of $\mu_V$=23.6 (South) 
and 23.3 (East) mag~arcsec$^{-2}$, extending
to a maximum distance of $\sim$33~kpc to the South of the nucleus, and giving an overall butterfly-like appearance.
The southern fan shows some bright knots. Pointing to the North, we detect a faint tail in our processed images. 
This galaxy is included in the \citealt{Heckman86} sample of PRGs, and they also detected the two fans in their optical
images. Our VLT spectra demonstrate that the fan features are dominated by continuum rather than emission line radiation.

\subsection{PKS 1932-46.}

This BLRG/FRII appears to be a member of a small interacting group of galaxies which includes a starburst galaxy at
a similar redshift at
$\sim$100 kpc to the NE \citep{Inskip07}. Optical spectroscopy of the radio galaxy reveals an extended population 
of very young ($<$10 Myr) stars in the galaxy halo \citep{Villar05,Holt07}. Our Gemini image (Figure \ref{pks1932_online}) shows an extremely 
disturbed morphology, including two symmetrical fans towards North and East ($\mu_V$=23.6 and 23.5 
mag~arcsec$^{-2}$, respectively), very similar to those detected in PKS 2314+03. The galaxy is embedded in an 
amorphous halo showing some bright knots. But the most spectacular feature is an extraordinary series of 
arc-like irregular features up to $\sim$70 kpc distance from the galaxy center, which almost connect
with the highly disturbed starburst galaxy mentioned above. Our long-slit spectra demonstrate that
the latter arcs emit a combination of continuum and line radiation, and are not solely due to emission line contamination.

\subsection{PKS 1151-34.}

The presence of a YSP in this QSO/CSS is indicated by the presence of PAH features in its Spitzer MIR spectrum \citep{Dicken10}.
Our GMOS-S image (Figure \ref{pks1151_online}) reveals a spectacular extended structure for this system, with a prominent tidal tail/arc feature
$\sim$34~kpc to the NW that connects to a compact secondary nucleus $\sim$27~kpc to the North of the quasar, as well as a broad
fan of emission extending $\sim$38~kpc to the East of the nucleus. The overall morphology of the extended structure to the N-NW
suggests that the quasar host galaxy is strongly interacting with a late-type spiral galaxy (possibly a barred spiral galaxy). Since
the spiral structure of the interacting galaxy is reasonably coherent (although somewhat asymmetric), it appears that we are
observing the system around the first peri-centre of the interacting galaxies, rather than in the final stages of a merger as the
nuclei coalesce. For this galaxy there are neither 
optical spectra nor infrared data suitable for assessing the degree of emission line contamination of the detected features.  

\subsection{PKS 0859-25.}

The source PKS 0859-25, which is inmersed in a relatively crowded field, appears in our GMOS-S image as a
double nucleus system including the radio galaxy nucleus and a faint component $\sim$6~kpc the SW (Figure \ref{pks0859_online}). 
This morphological classification is confirmed by the NIR observations and analysis presented by \citet{Inskip10}.

\subsection{PKS 2250-41.}

This NLRG is likely lying within a group, and possibly interacting with one or even more objects according the 
detailed analysis presented by \citet{Inskip08} using imaging and spectroscopy in the optical and NIR. 
The presence of a YSP is confirmed by the presence of PAH features in the Spitzer MIR spectrum of this source \citep{Dicken10}.
The disk galaxy at 47 kpc to the NE of the radio source is at an almost identical redshift to PKS 2250-41 \citep{Inskip08}, 
and appears clearly disturbed in the direction of the radio galaxy. Our $i'$-band GMOS-S image (Figure \ref{pks2250_online}) 
reveals a faint bridge 
linking with this NE companion, and a shorter one linking the radio galaxy with the smaller galaxy at 27 kpc towards the SW. 
The galaxy was observed in using the $i'$-band filter in order to avoid contamination with the extended [O III]$\lambda$5007 \AA~line emission, 
which we know is very prominent from optical spectra \citep{Tadhunter02}. Indeed, the long-slit spectrum of the galaxy 
shows that emission line contamination is not a serious issue for all the peculiar features detected. 
Finally, it is possible to distinguish a faint arc-like tail extending to the West from the small galaxy companion 
and a fan towards the East of the radio galaxy. 

\subsection{PKS 1355-41.}

This QSO at redshift z=0.31 has the best seeing among all the GMOS-S observations presented here (FWHM$\sim$0.4\arcsec). This 
fact is clearly reflected in the high-quality of the images and the level of detail of the identified features
(Figure \ref{pks1355_online}). There is 
a sharp shell feature $\sim$24~kpc to the SE of the quasar and covering an angle of $\sim$60\degr, along
with a bright and sharp tail extending up to 
$\sim$22 kpc to the SE along PA=121$\degr$ and pointing towards the arc. Although the latter feature is closely
aligned with the axis of the large-scale radio structure (PA=125$\degr$), the high frequency radio
map presented in \citet{Dicken08} shows no sign of any radio emission from the tail that might suggest that
it is non-thermal in nature. For this galaxy, there are neither optical long-slit spectra nor infrared data 
suitable for assessing the degree of emission line contamination of the detected features. 

\subsection{PKS 0023-26.}

This NLRG/CSS source appears to lie within a dense cluster environment. Indeed, the two galaxies to either side 
(at 25 kpc SW and 33 kpc NE) have similar redshifts to the radio galaxy host
\citep{Tadhunter10}. In our GMOS-S image (Figure \ref{pks0023_online}), all three galaxies appear embedded in a common envelope, 
which has amorphous appearance close to the radio galaxy, and is possibly crossed by dust lanes. 
The presence of dust here is consistent with the detection of a YSP in the optical spectrum
of the radio galaxy \citep{Holt07}, and the FIR excess revealed by Spitzer \citep{Dicken09}.
Our optical spectrum of the galaxy confirms that emission line contamination is not a serious issue for this amorphous halo. 

\subsection{PKS 0347+05.}

The WLRG/FRII PKS 0347+05 is part of a spectacular interacting system, which includes a QSO/Seyfert 1
nucleus $\sim$26~kpc to the SW of the radio galaxy host itself (Figure \ref{pks0347_online}). This object provides some of the best evidence
for  the triggering of radio/AGN activity as a result of a galaxy interaction in our sample. 
In addition, Spitzer MIR and FIR data for the WLRG reveal the presence of a YSP, based on the detection of a FIR excess \citep{Dicken09}
and strong PAH features 
\citep{Dicken10}. Our GMOS-S image shows a bridge linking the two interacting galaxies, and at least three tidal 
tails extending to the SW of the system. The closest tail to the QSO is the brightest, presenting a bright knot at 
the closest edge. The presence of dust features close to the radio galaxy is also clear, and consistent with a star-forming
region. Long-slit spectra of the galaxy show that emission line contamination is not a serious issue for most of the 
peculiar features detected in this system. This object will be the subject of a detailed study of its combined optical, infrared,
and radio properties (Ramos Almeida et al., in prep.).

\subsection{PKS 0039-44.}

The nucleus of this NLRG appears to be double in our GMOS-S optical image (separation $\sim$4 kpc; see inset in Figure \ref{pks0039_online}). 
Shell features are detected $\sim$16~kpc to the
NW ($\mu_V$=23.3 mag~arcsec$^{-2}$) and $\sim$16 and $\sim$27~kpc ($\mu_V$=23.3 and 24.1 mag~arcsec$^{-2}$ respectively) to the SE of the nucleus. 
Although we do not consider it for the morphological 
classification of PKS 0039-44, it is possible to unveil a faint tail pointing to the star-like companion at $\sim$70 kpc to the West. 
This tentative tail appears co-aligned with the system of shells.
Dust seems to be also present in the galaxy nucleus. 
Emission line contamination cannot be ruled out based on long-slit spectra  for the case of the shells and 
the tail. However, our spectra show that the apparent secondary nucleus is likely to 
be dominated by [O III]$\lambda$5007 \AA~emission line radiation.

\subsection{PKS 0105-16 (3C32).}

Our GMOS-S $i'$-band image of this NLRG/FRII (Figure \ref{pks0105_online}) reveals a bridge of $\mu_V$=24.6 mag~arcsec$^{-2}$ 
that appears to link the radio galaxy
host with an early-type galaxy of similar brightness $\sim$70 kpc NW. 
Emission line contamination of the detected feature cannot be ruled out based on the existing long-slit spectra. 

\subsection{PKS 1938-15.}

This BLRG/FRII lies in a very dense environment, and in our optical image appears to have several diffuse companions within a 
10\arcsec~radius (Figure \ref{pks1938_online}). We detect a tail/fan of $\mu_V$=23.2 mag~arcsec$^{-2}$ extending up to 20 kpc 
to the NE of the radio galaxy.
It also looks very probable that PKS 1938-15 is interacting with the companion lying at 18~kpc to the East. 
There are neither optical long-slit spectra nor infrared data suitable for assessing the 
degree of emission line contamination of the detected features in this galaxy.

\subsection{PKS 1602+01 (3C327.1).}

This BLRG/FRII at redshift z=0.46 shows several signs of interaction in our $i'$-band GMOS-S image (Figure \ref{pks1602_online}). 
Towards the North of the 
galaxy nucleus we detect a bright and smooth fan of $\mu_V$=21.34 mag~arcsec$^{-2}$. On the opposite side 
there is a shell of $\mu_V$=23.3 mag~arcsec$^{-2}$. Finally, and immersed in the shell envelope, we detect 
a sharp knot-like feature that extends up to 23 kpc and that is closely aligned (PA=148\degr) with the inner axis 
of the radio jet (PA=145\degr; see radio images
in \citealt{Morganti99}). At a distance of 18 kpc from the nucleus, the latter feature is close to, but does not exactly 
coincide with the third knot in the radio jet leading to the South of the nucleus. 
The WFPC2/HST images of this galaxy only revealed an elliptical nucleus with a faint halo of emission surrounding it
\citep{deKoff96}.
For this galaxy, there are neither optical long-slit spectra nor infrared data suitable for assessing the 
degree of emission line contamination of the detected features.

\subsection{PKS 1306-09.}

Our GMOS-S image (Figure \ref{pks1306_online}) reveals a secondary nucleus for this NLRG/CSS source, 
which is also detected in the residuals of the 
model-subtracted K-band image analyzed by \citet{Inskip10}. The galaxy lies in a group of galaxies. From
both the optical and NIR data, the radio galaxy is very likely undergoing interactions with 
other objects in the group. In addition to the double nucleus, we detect a sharply-defined shell $\sim$20~kpc to SE of the radio galaxy, of 
surface brightness $\mu_V$=24.5 mag~arcsec$^{-2}$.

\subsection{PKS 1547-79.}

We classify this BLRG/FRII as a double nucleus system, since the radio source has a fainter companion  
$\sim$9~kpc to the South of the nucleus, as revealed by our GMOS-S image (Figure \ref{pks1547_online}). 
In addition, our GMOS-S optical image reveals
a tidal tail extending  $\sim$27~kpc to the NE of the galaxy center of surface brightness $\mu_V$=24.3 mag~arcsec$^{-2}$.
In both the optical and the NIR, PKS 1547-79 appears as a clear interacting system, surrounded by several diffuse 
companions. Note that the companion galaxy $\sim$46~kpc to the West of the radio galaxy host has
weak emission lines that suggest a similar redshift to PKS1547-79. On the other hand, the much brighter
early-type galaxy $\sim$70~kpc to the East, which also shows some signs of morphological
disturbance, is at a much lower redshift. The continuum-emitting nature of the tidal tail of PKS1547-79 
cannot be either confirmed or discarded based on existing long-slit spectra 
\citep{Tadhunter98}, but the secondary nucleus is detected in the NIR \citep{Inskip10}.

\subsection{PKS 1136-13.}

Our GMOS-S image of this QSO/FRII at redshift z=0.55 (Figure \ref{pks1136_online}) reveals the optical counterpart of a radio jet 
extending up to 60 kpc in the 
NW from the radio galaxy \citep{Uchiyama07}. We also detect a shorter tidal tail pointing to the West of the quasar, 
for which we measure a surface brightness of $\mu_V$=22.6 mag~arcsec$^{-2}$. It is also possible to unveil another tail  
SE of the nucleus. The lack of either optical long-slit spectra or infrared data suitable for assessing the 
degree of emission line contamination of the detected features prevents us from confirming whether the extended structures are
dominated by continuum radiation.

\subsection{PKS 0117-15 (3C38).}

Our optical image of this NLRG/FRII, taken in excellent seeing conditions (FWHM=0.5\arcsec), shows a high-surface-brightness 
triple nucleus system of $\sim$9~kpc diameter co-aligned along PA=8\degr~(Figure \ref{pks0117_online}). 
In addition, we detect a shell on the western side of the radio galaxy extending up to 21 kpc, for which we measure a surface 
brightness of $\mu_V$=25.1 mag~arcsec$^{-2}$, as well as an 
arc-like structure (identified as "I" in Table \ref{data}) of $\mu_V$=21.3 mag~arcsec$^{-2}$ and a companion galaxy 
$\sim$9~kpc and $\sim$17~kpc to the NE of the central
nucleus respectively. Interestingly, the NE arc shows a gap close to its interception with the axis of the large 
scale radio structure (PA=35\degr). 
This suggests a possible interaction between the arc and the radio jet. However, although the central triple structure shows some
resemblance with the structures found closely aligned with the radio axes of some high redshift radio galaxies (e.g.
\citealt{McCarthy87,Best96})
it is, in fact, significantly misaligned from the radio axis by 27 degrees. Therefore, it is uncertain whether the latter 
structure is due to jet/cloud interactions.
Dust features between the triple system components 
are tentatively detected in our processed images. The results on the degree of emission line contamination
of the detected features are inconclusive, based on our existing long-slit spectra. Clearly, further optical 
spectroscopic and NIR imaging observations are required to determine whether the 
central triple structure is stellar in nature, or represents a manifestation of the ``alignment effect'' frequently observed
in high redshift radio galaxies.  
 
\subsection{PKS 0252-71.}

The NLRG/CSS galaxy PKS 0252-71 appears to be surrounded by an amorphous halo which points towards a fainter companion galaxy 
at $\sim$33 kpc to the South, thus indicating a possible interaction between the galaxies (Figure \ref{pks0252_online}). Note that
the bright object $\sim$3\arcsec~to the SE of the radio galaxy host is a star. Unfortunately, the relatively poor 
seeing of this image (FWHM=1\arcsec) prevents a clear detection of other possible signs of disturbance. 

\subsection{PKS 0235-19 (OD-159).}

Our GMOS-S image of this BLRG/FRII at redshift z=0.62 (Figure \ref{pks0235_online}) shows tidal tails of similar surface brightness to both the 
NE and SW of the nucleus along PA=52\degr, with a maximum extent of $\sim$15~kpc to the NE 
($\mu_V$=22.2 and 22.4 mag~arcsec$^{-2}$ for the NE and SW tails, respectively). 
Apart from the tidal tails, another interpretation of the observed
structures is that they represent a warped disk structure with the BLRG nucleus at its centre.
The tail/disk structure is significantly mis-aligned by 42 degrees from the axis of the large scale radio structure (PA=94\degr). 
Although we do not consider it for the morphological classification of the galaxy, 
there is tentative evidence for a bridge linking the radio galaxy with the smaller galaxy 165 kpc to the NE. The bridge appears
as a elongation of the NE tidal tail, but much fainter. Our long-slit spectra of this galaxy suggest that line
contamination may be a serious issue for the tail/disk structure. 

\subsection{PKS 2135-20 (OX-258).}

Our optical image of this BLRG/CSS (Figure \ref{pks2135_20_online}) shows a broad fan on the northern side of the radio galaxy, of 
surface brightness $\mu_V$=23.5 mag~arcsec$^{-2}$. This feature is very likely the result of a past interaction, consistent
with the detection of a YSP in the nuclear region of this galaxy from optical spectroscopy \citep{Holt07}, FIR
excess \citep{Dicken09}, and the detection of strong PAH features \citep{Dicken10}. The results regarding the continuum or 
line-emitting nature of this fan are inconclusive based on our long-slit spectra. Note that PKS2135-20 is one of the only two
galaxies (the other is PKS 2314+03) in our sample that would be classified as ULIRGs based on their MIR and FIR luminosities.

\subsection{PKS 0409-75.}

This NLRG/FRII with evidence for a YSP based on its optical spectrum \citep{Holt07}, FIR excess and infrared colours \citep{Dicken09}, 
is the most distant object in our sample. Our GMOS-S $i'$-band image reveals a secondary nucleus of similar brightness to
the radio galaxy  $\sim$8 kpc to the East along PA=84\degr~(Figure \ref{pks0409_online}). Note that the axis of the double nucleus
is misaligned by $\sim$40 degrees from the axis of the large-scale radio structure (PA=124\degr). The double nucleus 
system lies in a crowded field, and there are
several galaxies within a 20\arcsec~radius. Although the system looks disturbed, unfortunately the high value of 
the seeing measured for this image (FWHM=1.15\arcsec) prevents any more detailed morphological classification. 
The continuum or line-emitting nature of the secondary nucleus is uncertain based on long-slit
spectra of this source.

\section*{Acknowledgments}

CRA ackowledges financial support from STFC PDRA (ST/G001758/1). 
KJI is supported through the Emmy Noether
programme of the German Science Foundation (DFG).
This work is based on observations obtained at the Gemini Observatory, which is operated by the
Association of Universities for Research in Astronomy, Inc., under a cooperative agreement
with the NSF on behalf of the Gemini partnership: the National Science Foundation (United
States), the Science and Technology Facilities Council (United Kingdom), the
National Research Council (Canada), CONICYT (Chile), the Australian Research Council
(Australia), Minist\'{e}rio da Ci\^{e}ncia e Tecnologia (Brazil), and Ministerio de Ciencia, 
Tecnolog\'{i}a e Innovaci\'{o}n Productiva (Argentina). 
The Gemini programs under which the data were obtained are GS-2008B-Q-44 and GS-2009A-Q-54.
This research has made use of the NASA/IPAC Extragalactic Database (NED) which is 
operated by the Jet Propulsion Laboratory, California Institute of Technology, under 
contract with the National Aeronautics and Space Administration.
C.R.A. acknowledges Pieter van Dokkum, and Rub\' en D\' iaz for their valuable help.
We finally acknowledge useful comments from the anonymous referee.

\section{Online-only processed images}
\label{images}

\begin{figure*}
\centering
\subfigure[]{\includegraphics[width=8.0cm]{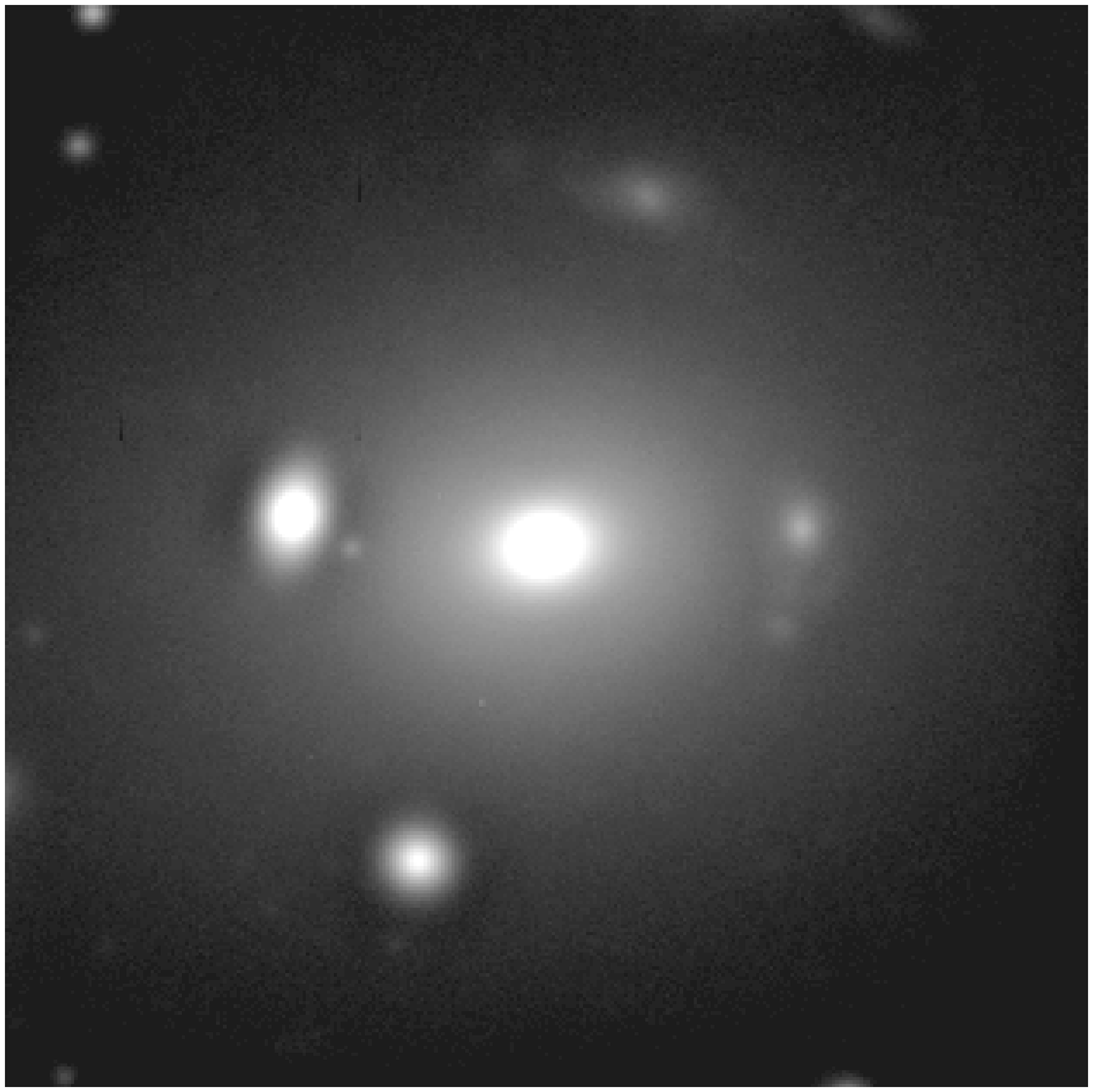}}
\subfigure[]{\includegraphics[width=8.0cm]{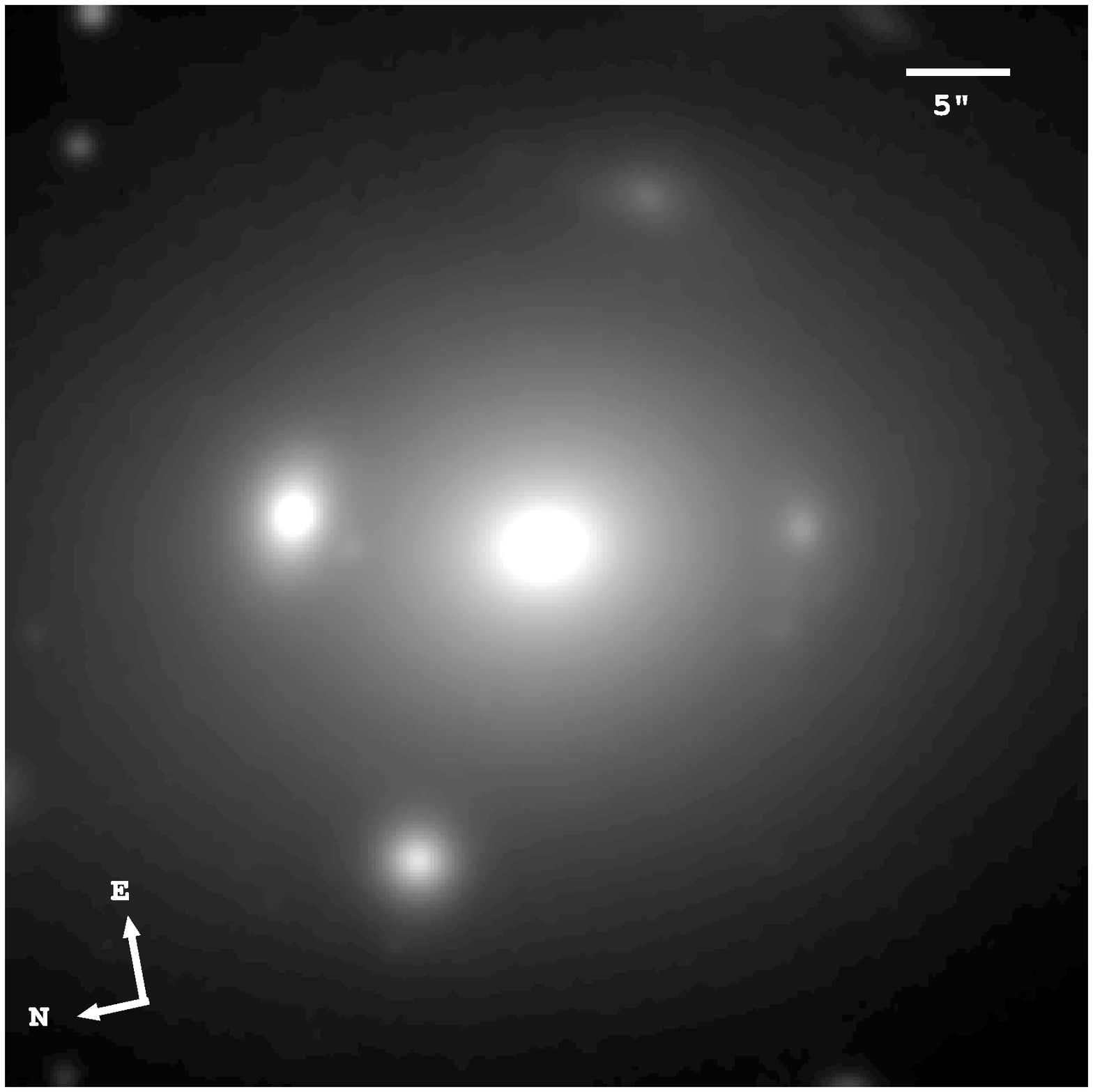}}
\caption{PKS 0620-52. (a) Unsharp-masked image. (b) Median filtered image.}
\label{pks0620_online} 
\end{figure*}

\begin{figure*}
\centering
\subfigure[]{\includegraphics[width=8.0cm,height=8.0cm]{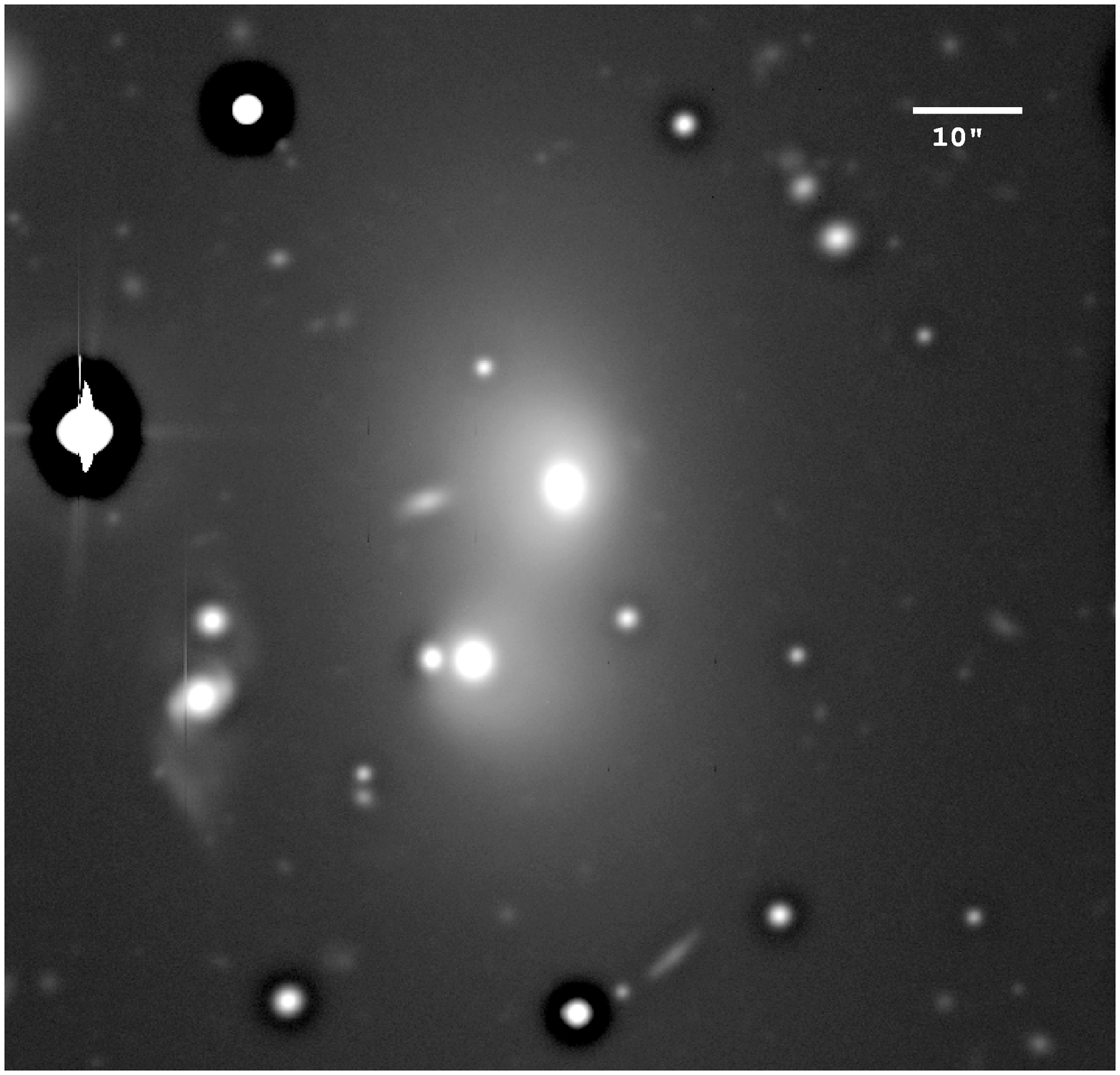}\label{pks0625_53_online_a}}
\subfigure[]{\includegraphics[width=8.0cm,height=8.0cm]{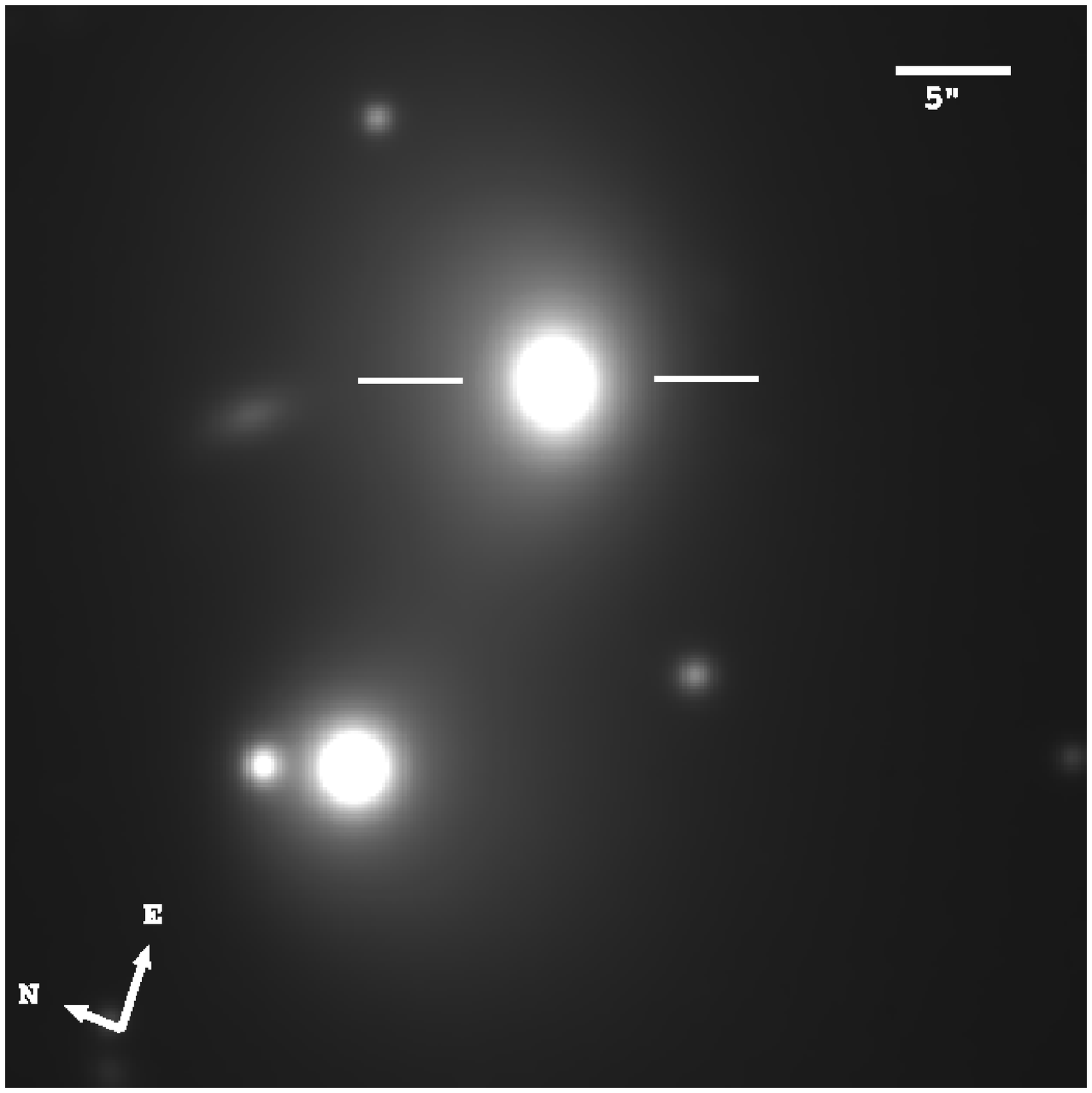}\label{pks0625_53_online_b}}
\caption{PKS 0625-53. (a) Unsharp-masked image. (b) Median filtered image.}
\label{pks0625_53_online} 
\end{figure*}

\begin{figure*}
\centering
\subfigure[]{\includegraphics[width=8.0cm]{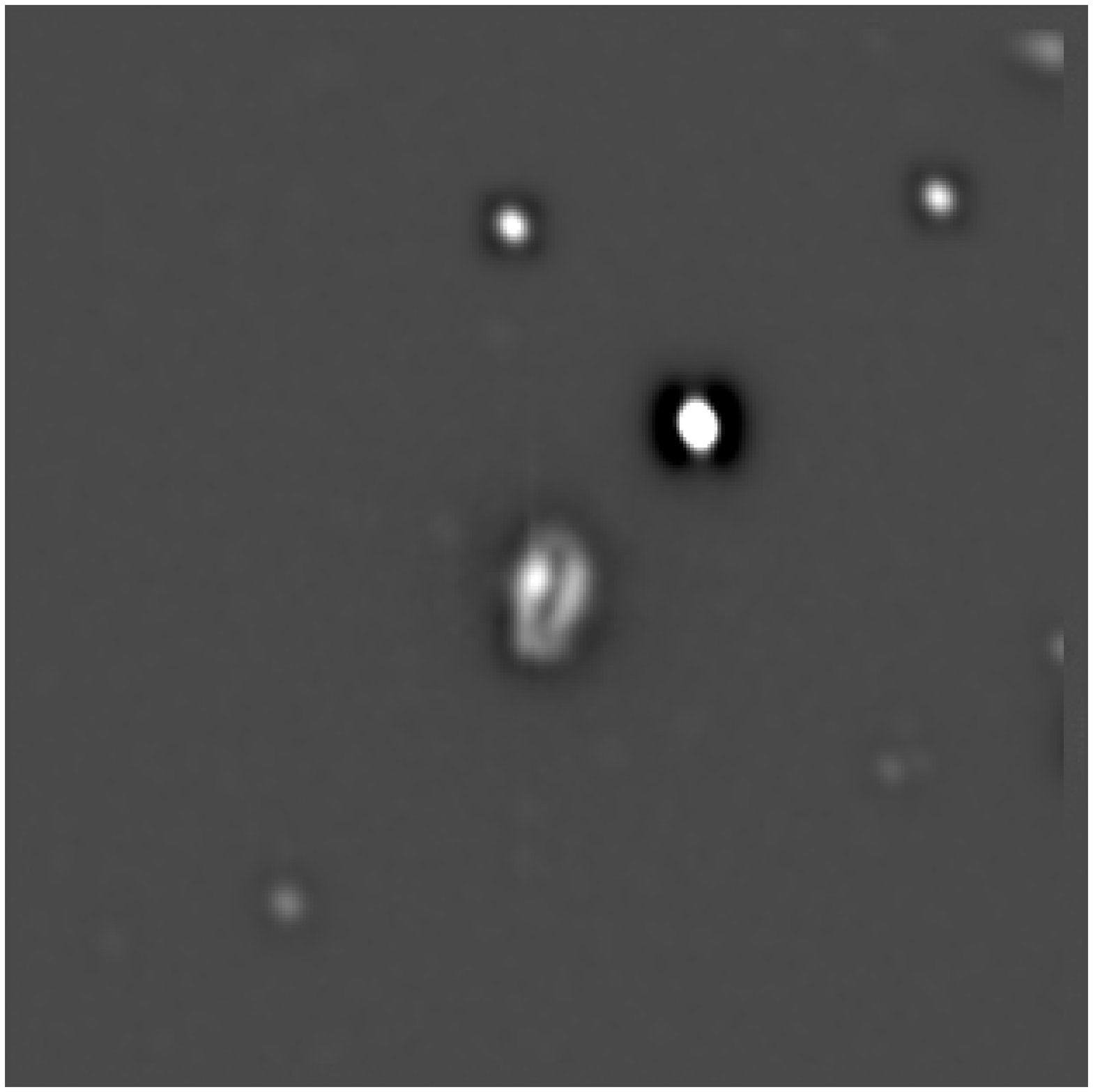}}
\subfigure[]{\includegraphics[width=8.0cm]{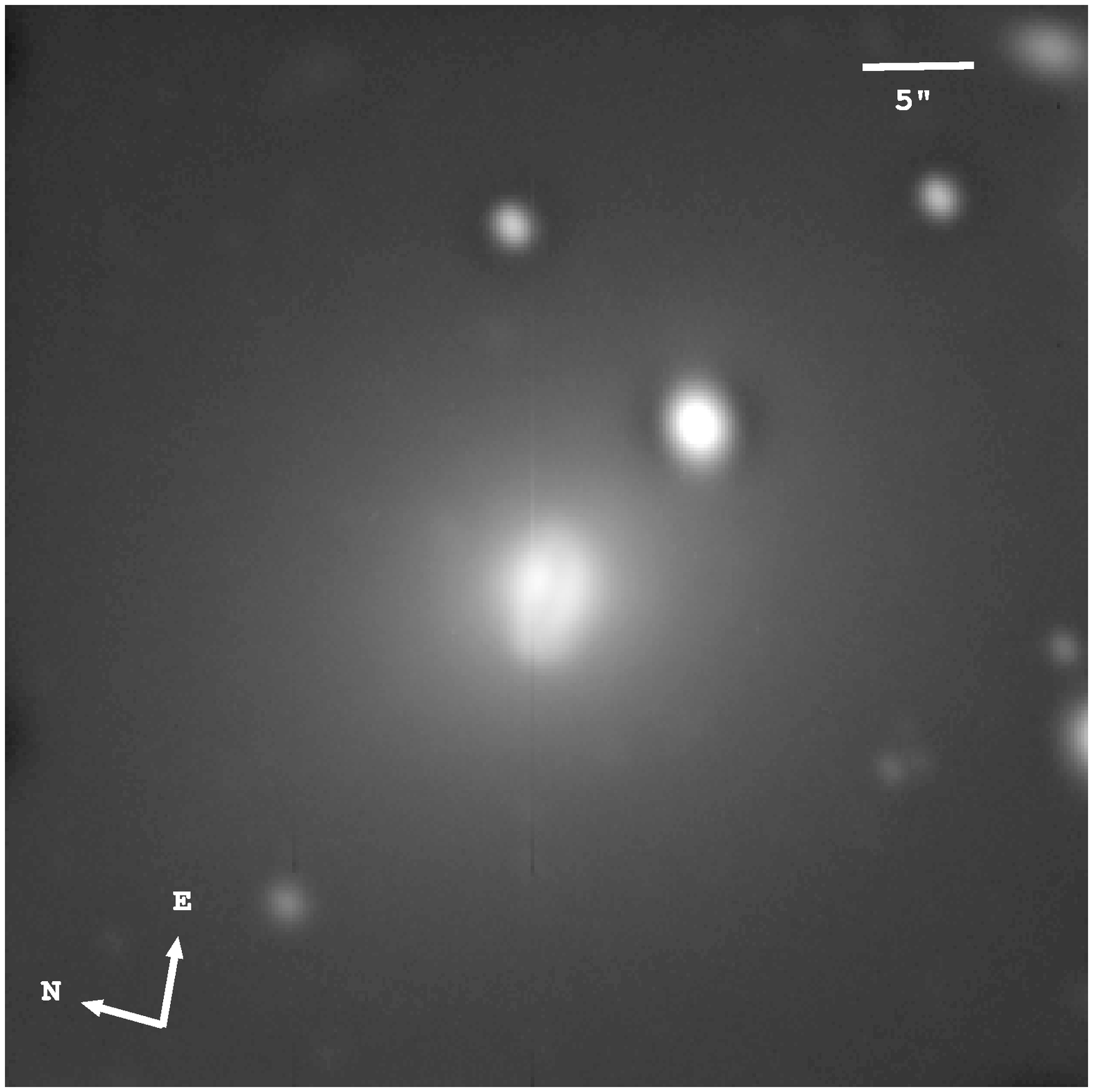}}
\caption{PKS 0915-11. (a) Smooth galaxy-subtracted image. (b) Unsharp-masked image.}
\label{pks0915_online} 
\end{figure*}

\begin{figure*}
\centering
\subfigure[]{\includegraphics[width=8.0cm]{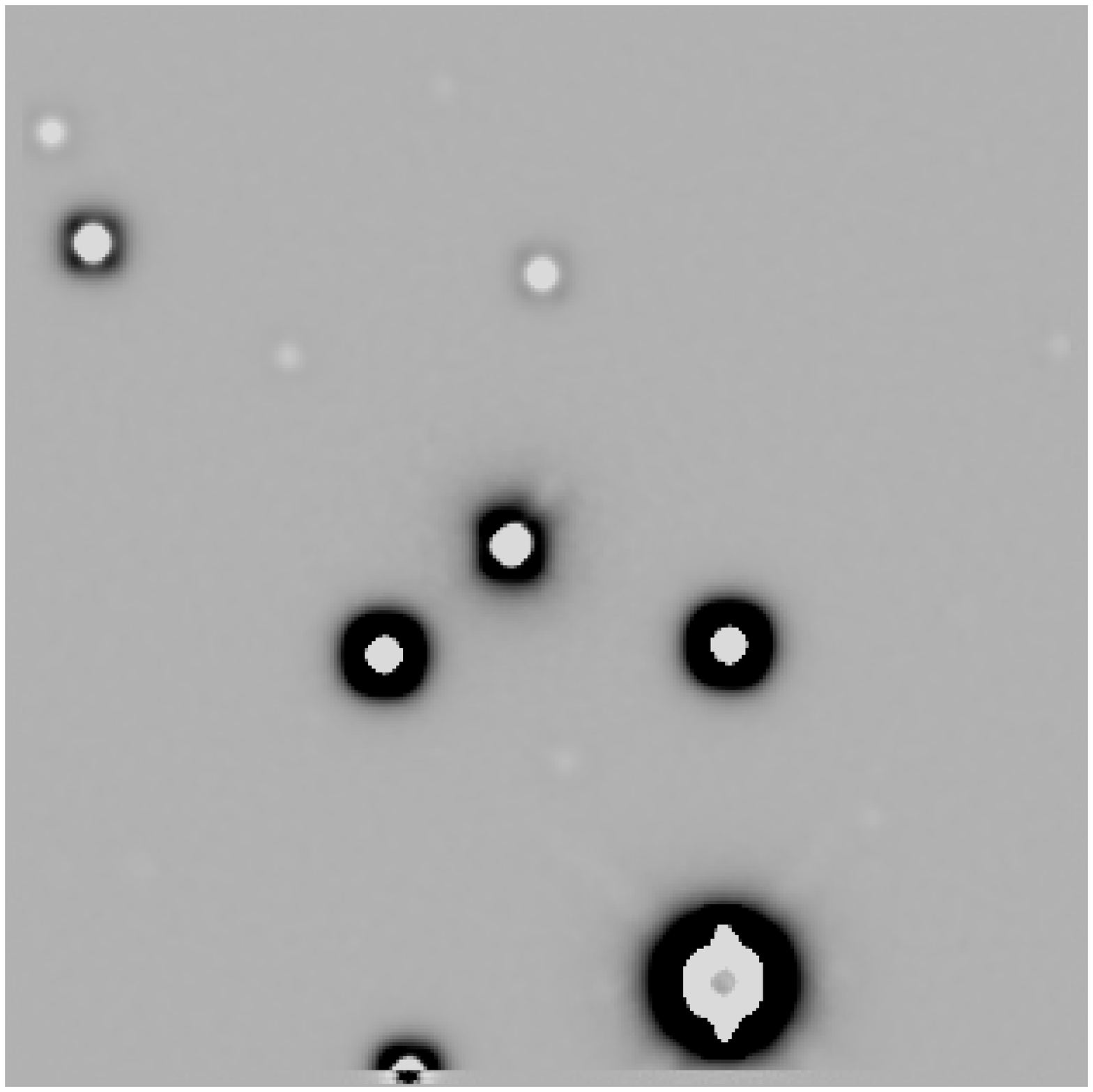}}
\subfigure[]{\includegraphics[width=8.0cm]{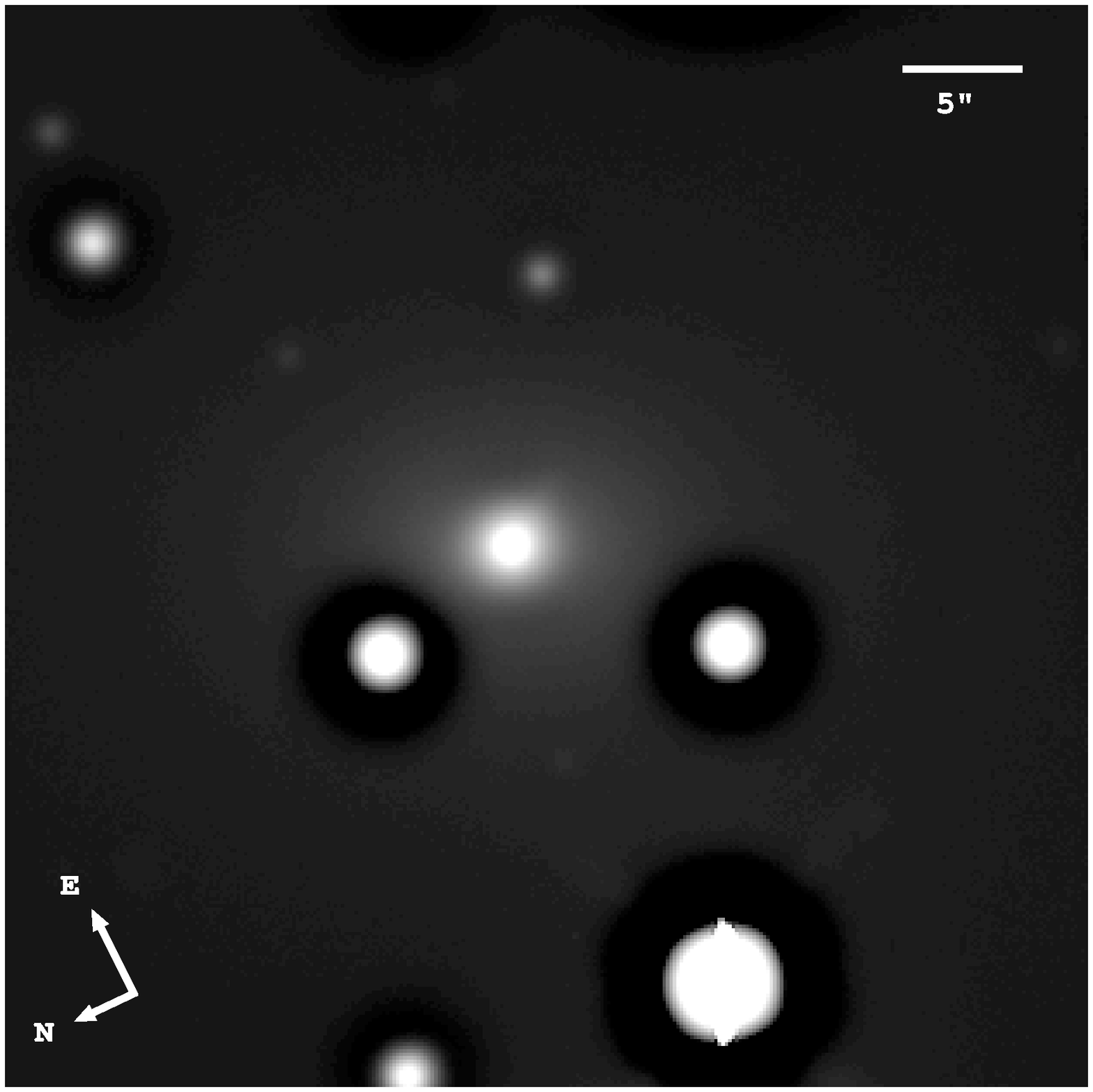}}
\caption{PKS 0625-35. (a) Smooth galaxy-subtracted image. (b) Unsharp-masked image.}
\label{pks0625_35_online} 
\end{figure*}

\begin{figure*}
\centering
\subfigure[]{\includegraphics[width=8.0cm]{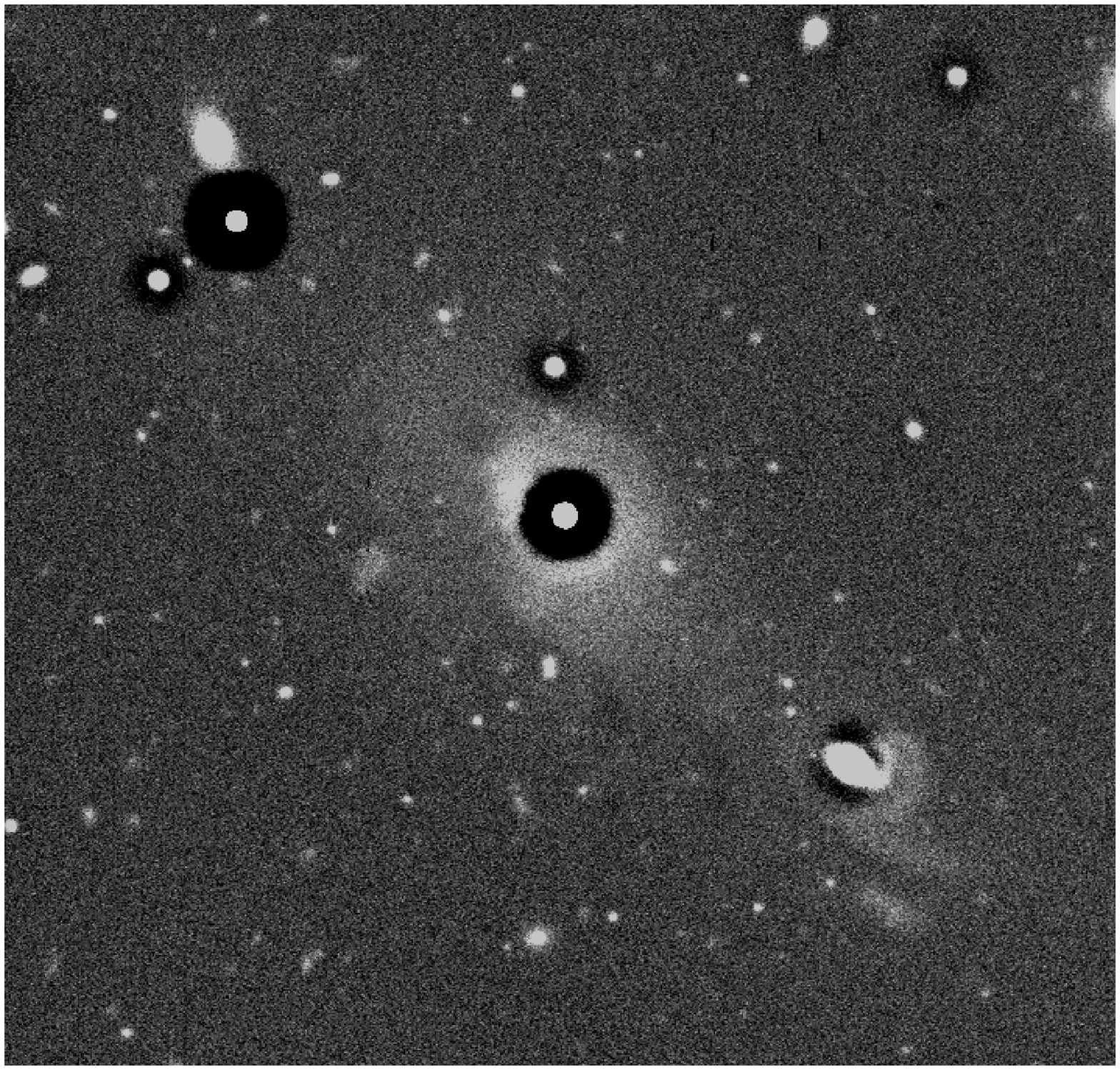}}
\subfigure[]{\includegraphics[width=8.0cm]{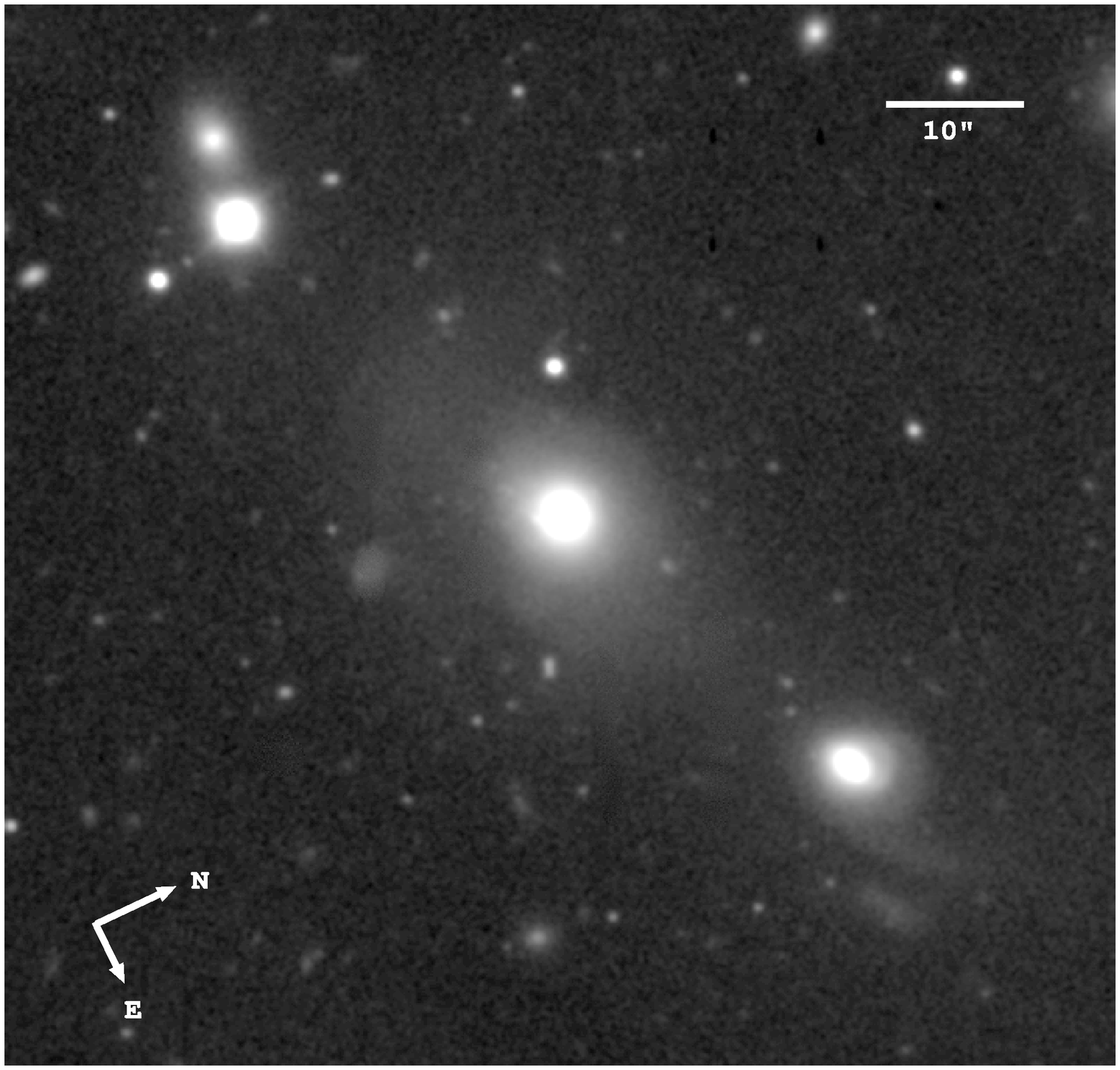}}
\caption{PKS 2221-02. (a) Unsharp-masked image. (b) Median filtered image.}
\label{pks2221_online} 
\end{figure*}

\begin{figure*}
\centering
\subfigure[]{\includegraphics[width=8.0cm]{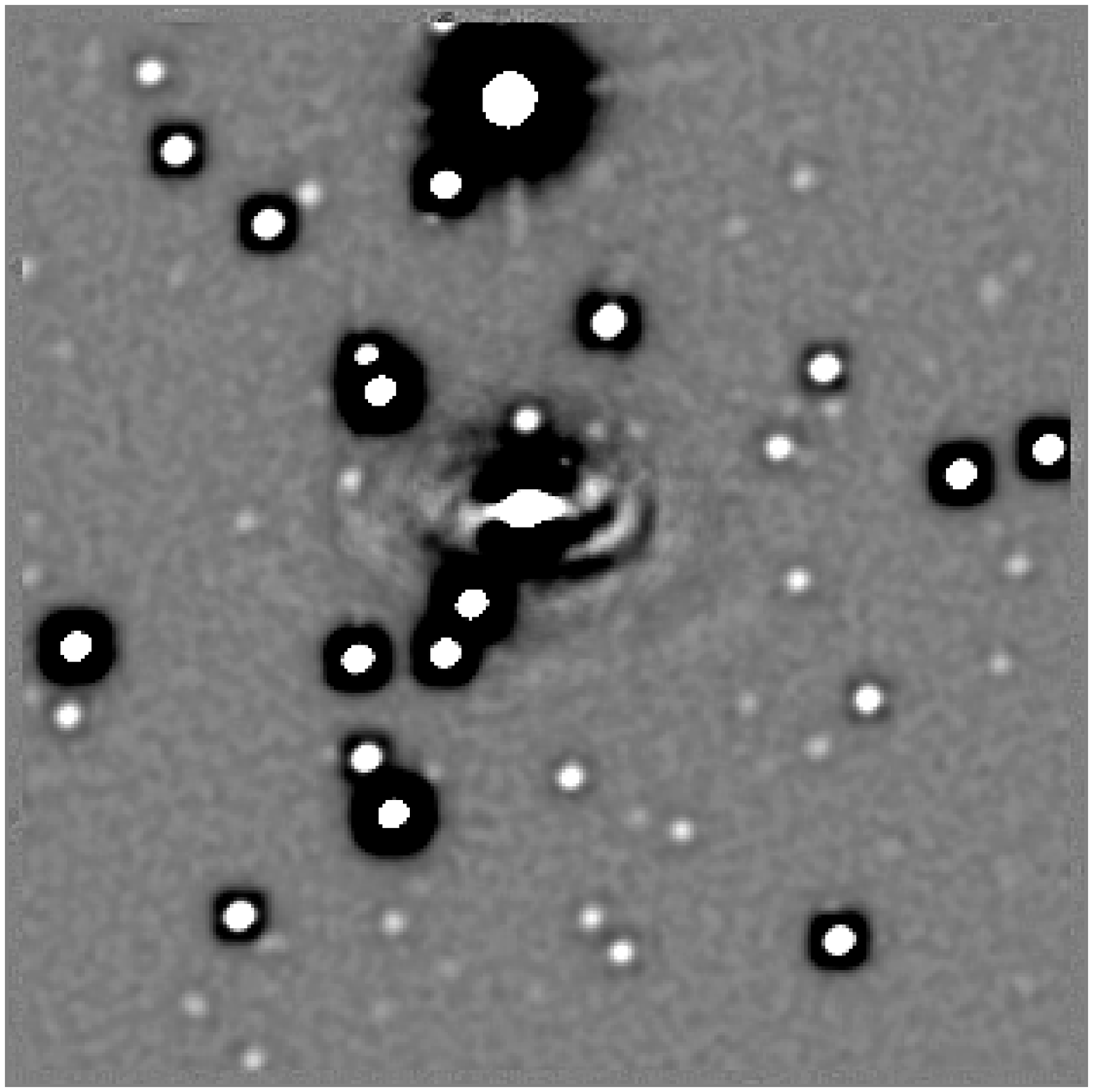}}
\subfigure[]{\includegraphics[width=8.0cm]{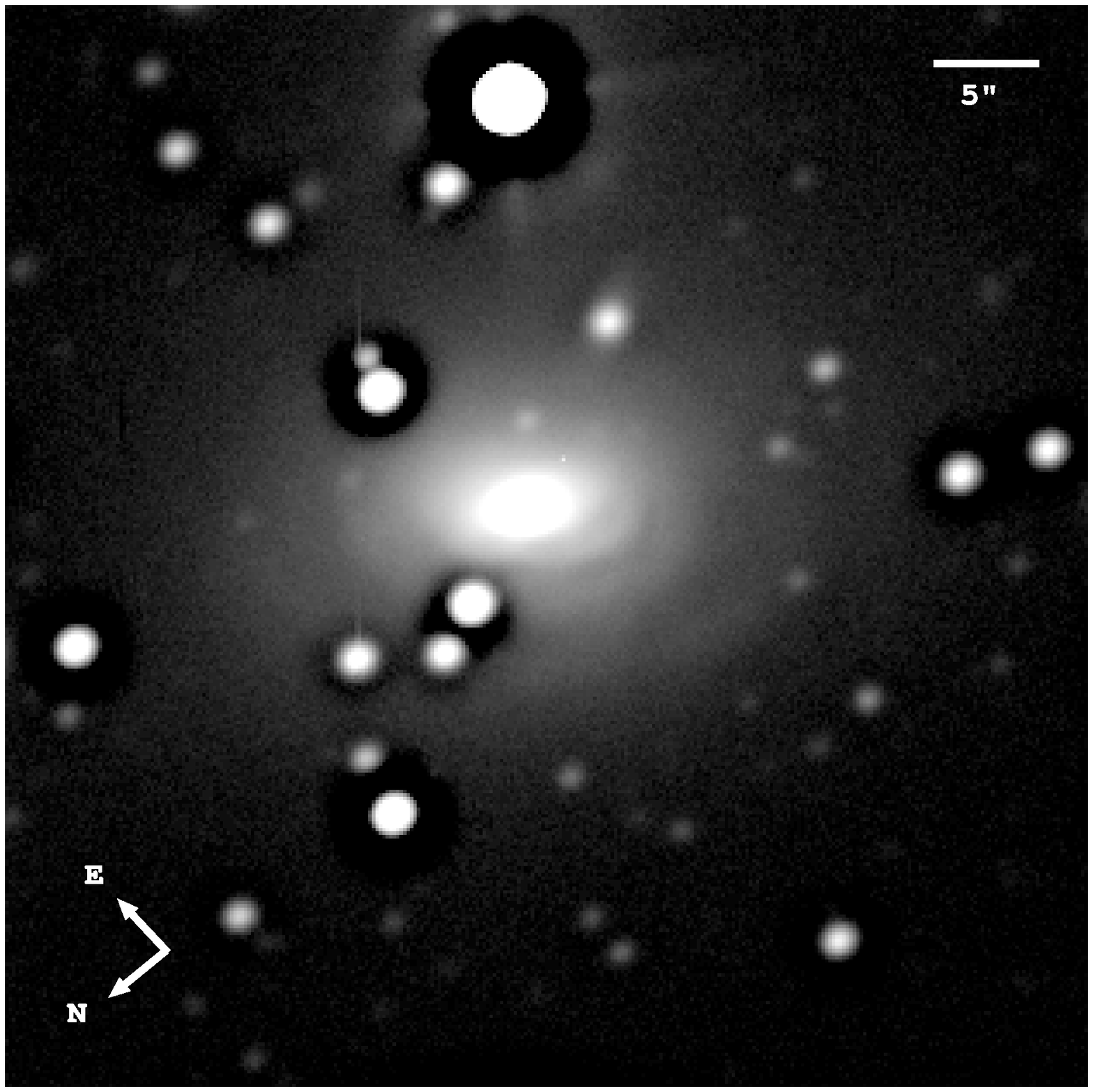}}
\caption{PKS 1949+02. (a) Smooth galaxy-subtracted image. (b) Unsharp-masked image.}
\label{pks1949_online} 
\end{figure*}

\begin{figure*}
\centering
\subfigure[]{\includegraphics[width=8.0cm]{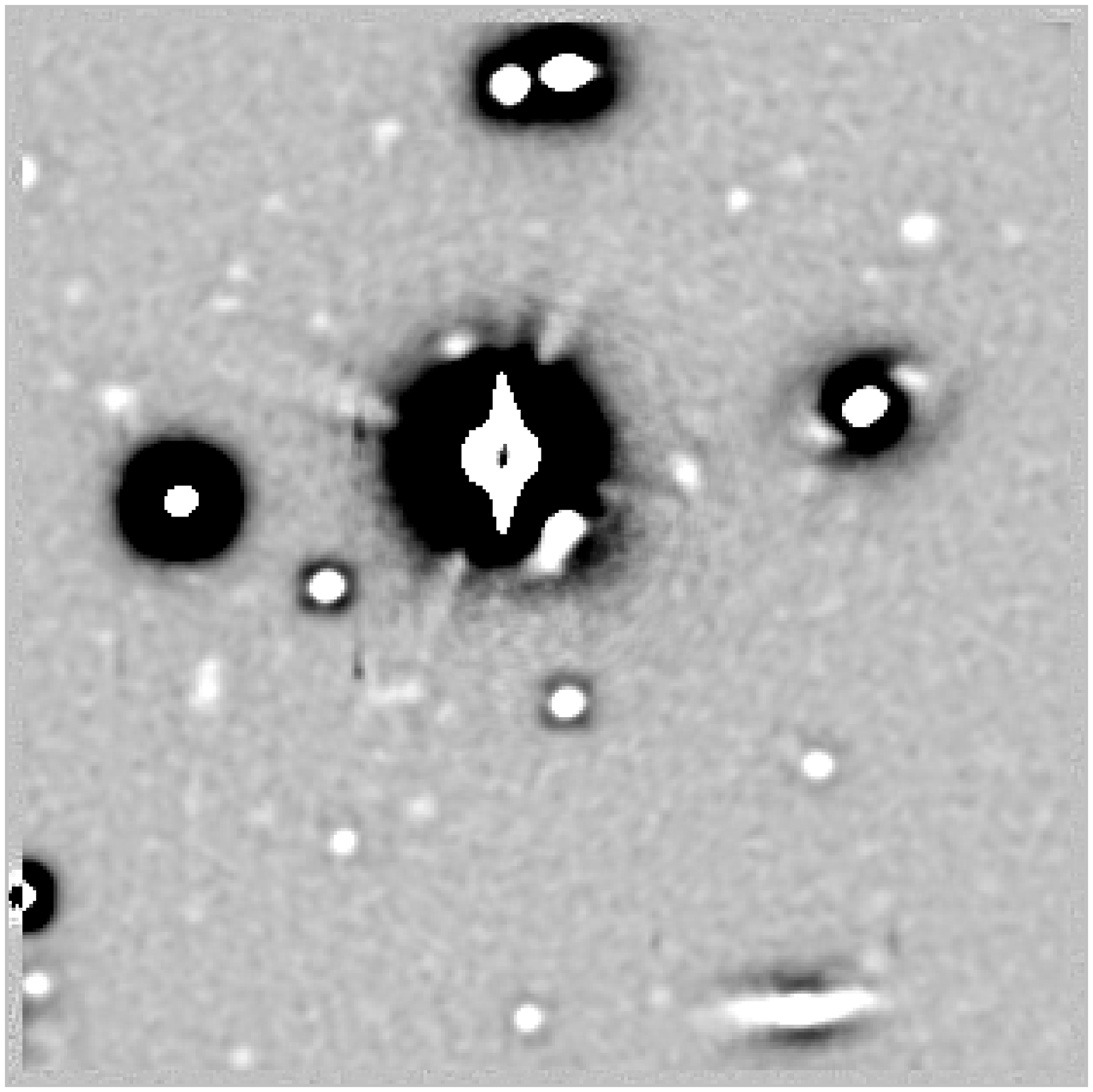}}
\subfigure[]{\includegraphics[width=8.0cm]{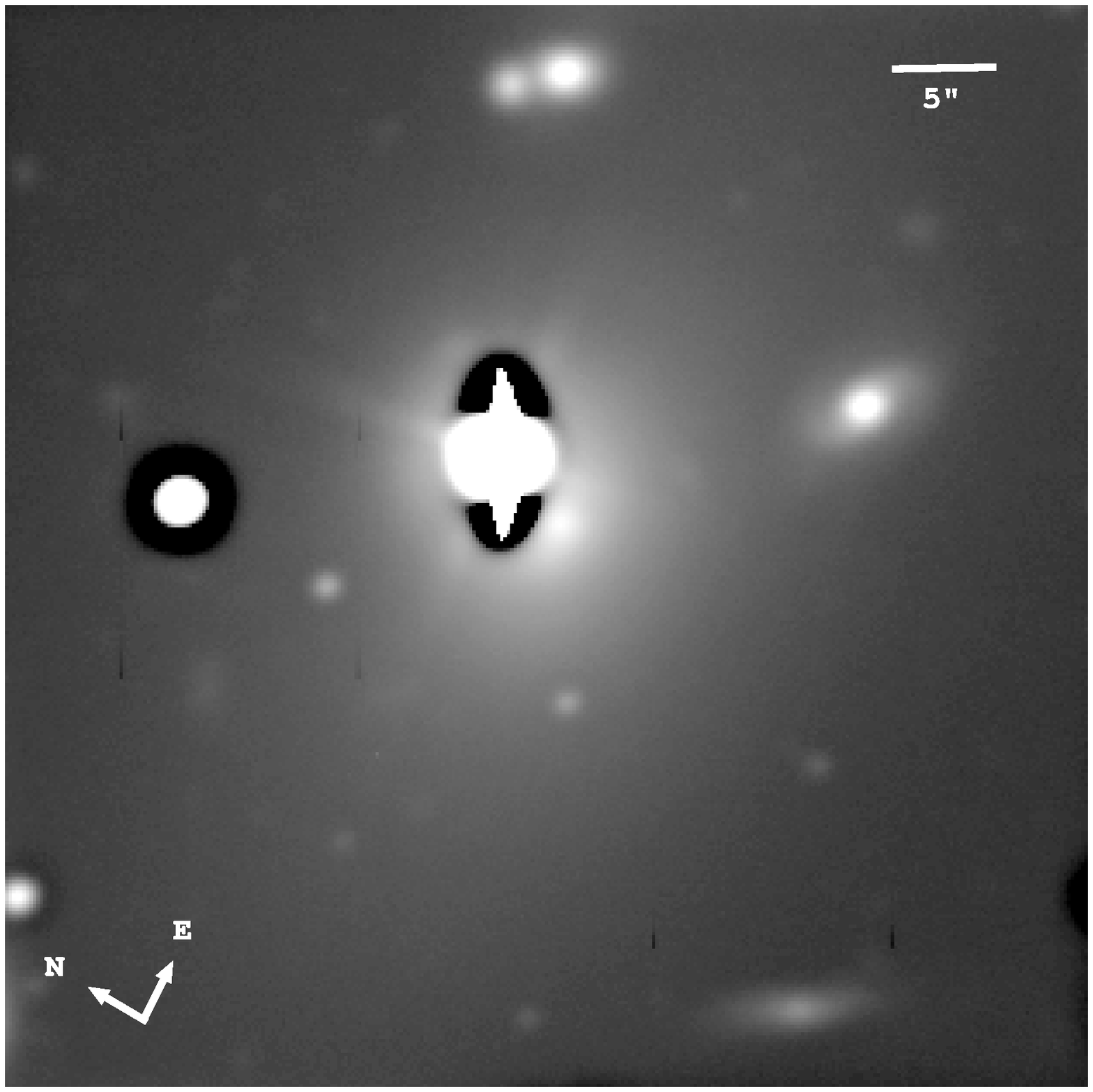}}
\caption{PKS 1954-55. (a) Smooth galaxy-subtracted image. (b) Unsharp-masked image.}
\label{pks1954_online} 
\end{figure*}

\begin{figure*}
\centering
\subfigure[]{\includegraphics[width=8.0cm,angle=-90]{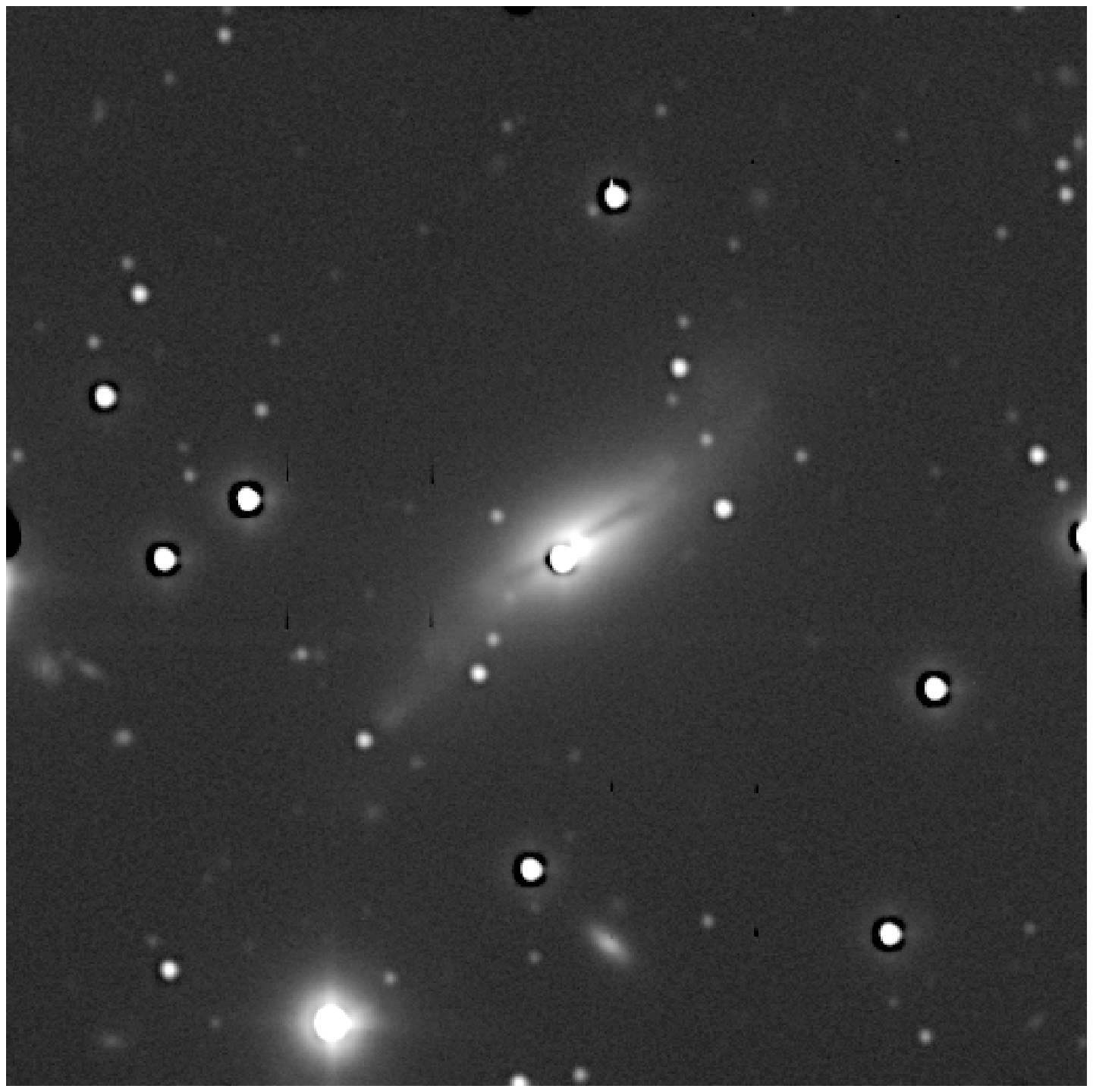}}
\subfigure[]{\includegraphics[width=8.0cm,angle=-90]{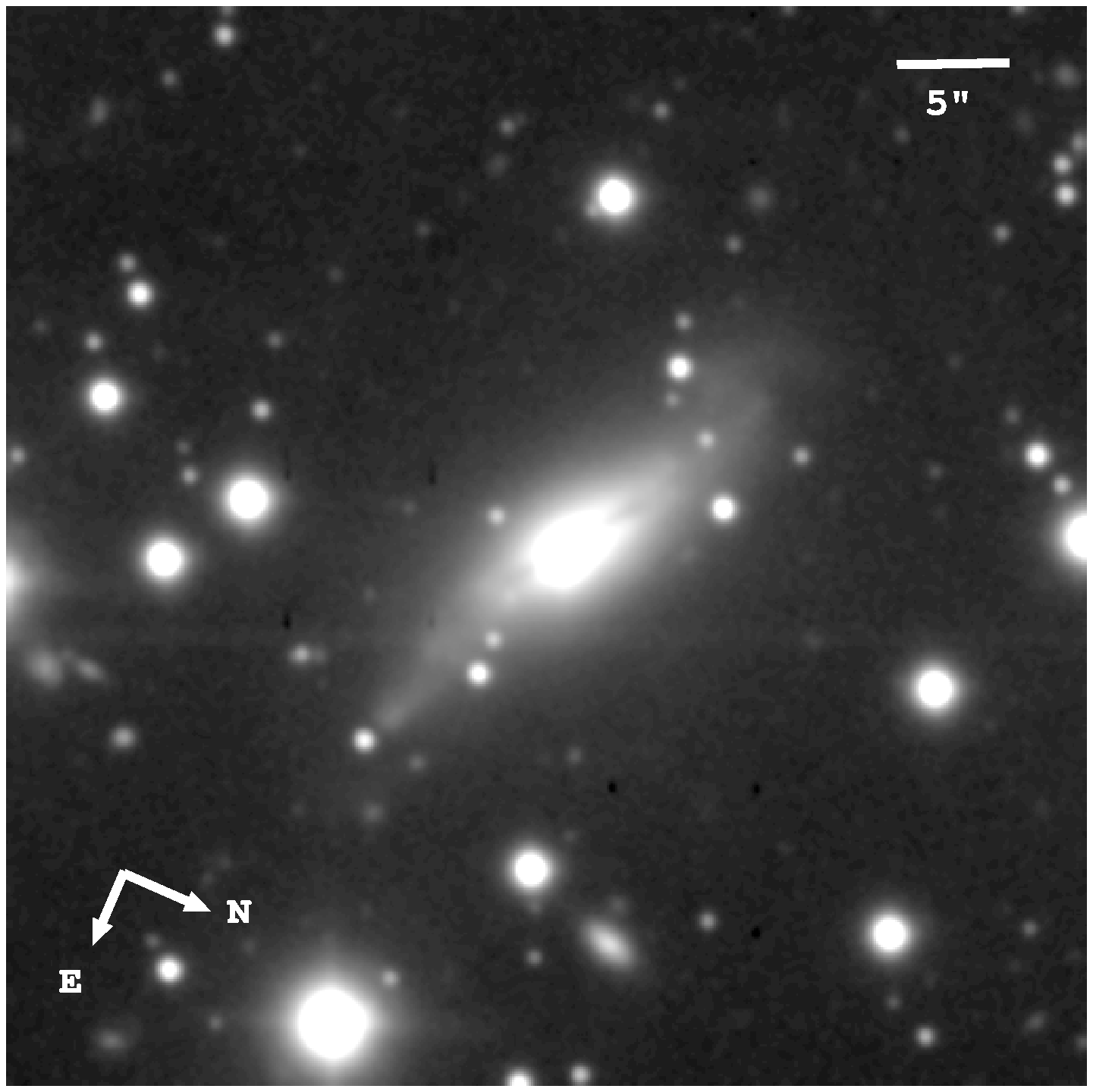}}
\caption{PKS 1814-63. (a) Unsharp-masked image. (b) Median filtered image.}
\label{pks1814_online} 
\end{figure*}

\begin{figure*}
\centering
\subfigure[]{\includegraphics[width=8.0cm]{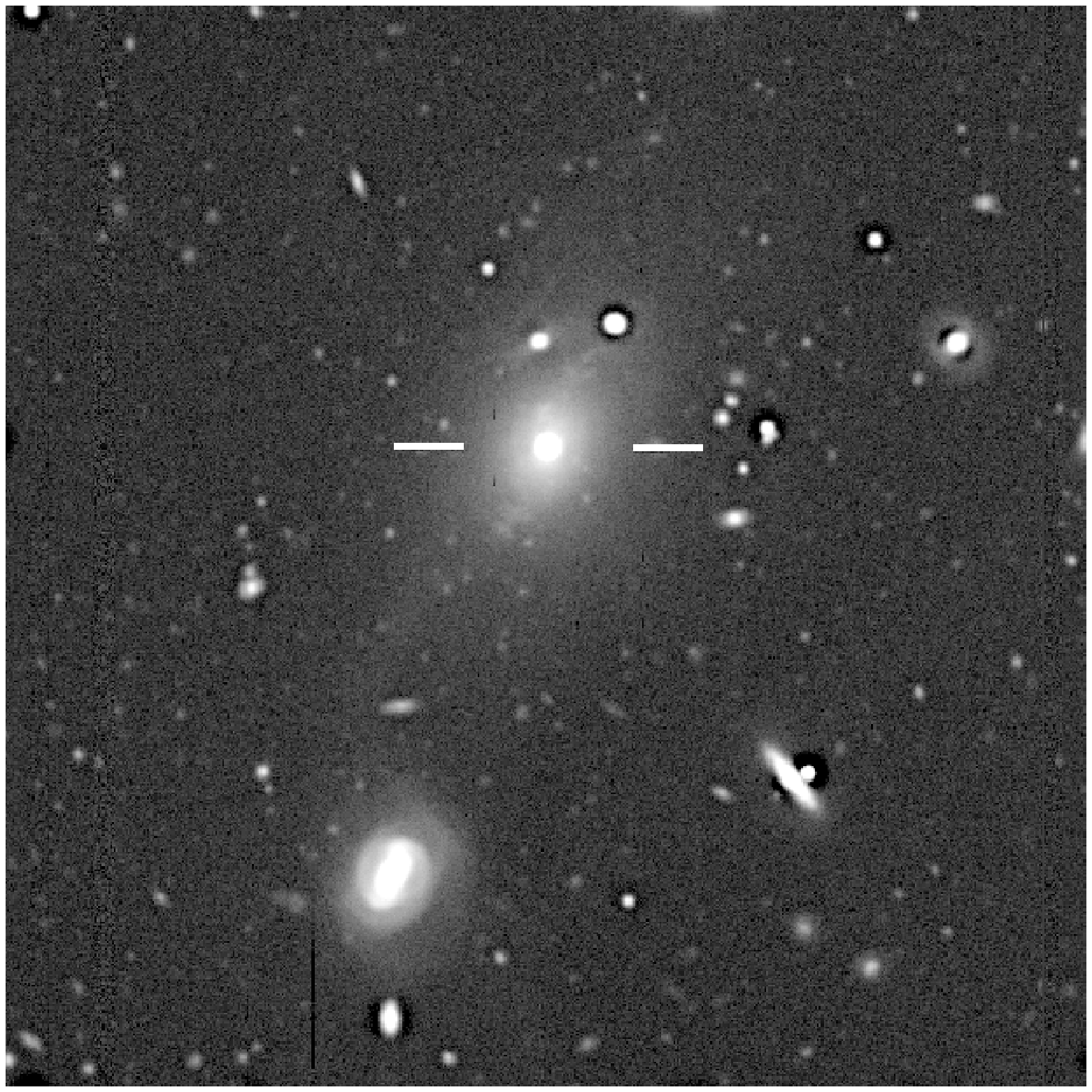}}
\subfigure[]{\includegraphics[width=8.0cm]{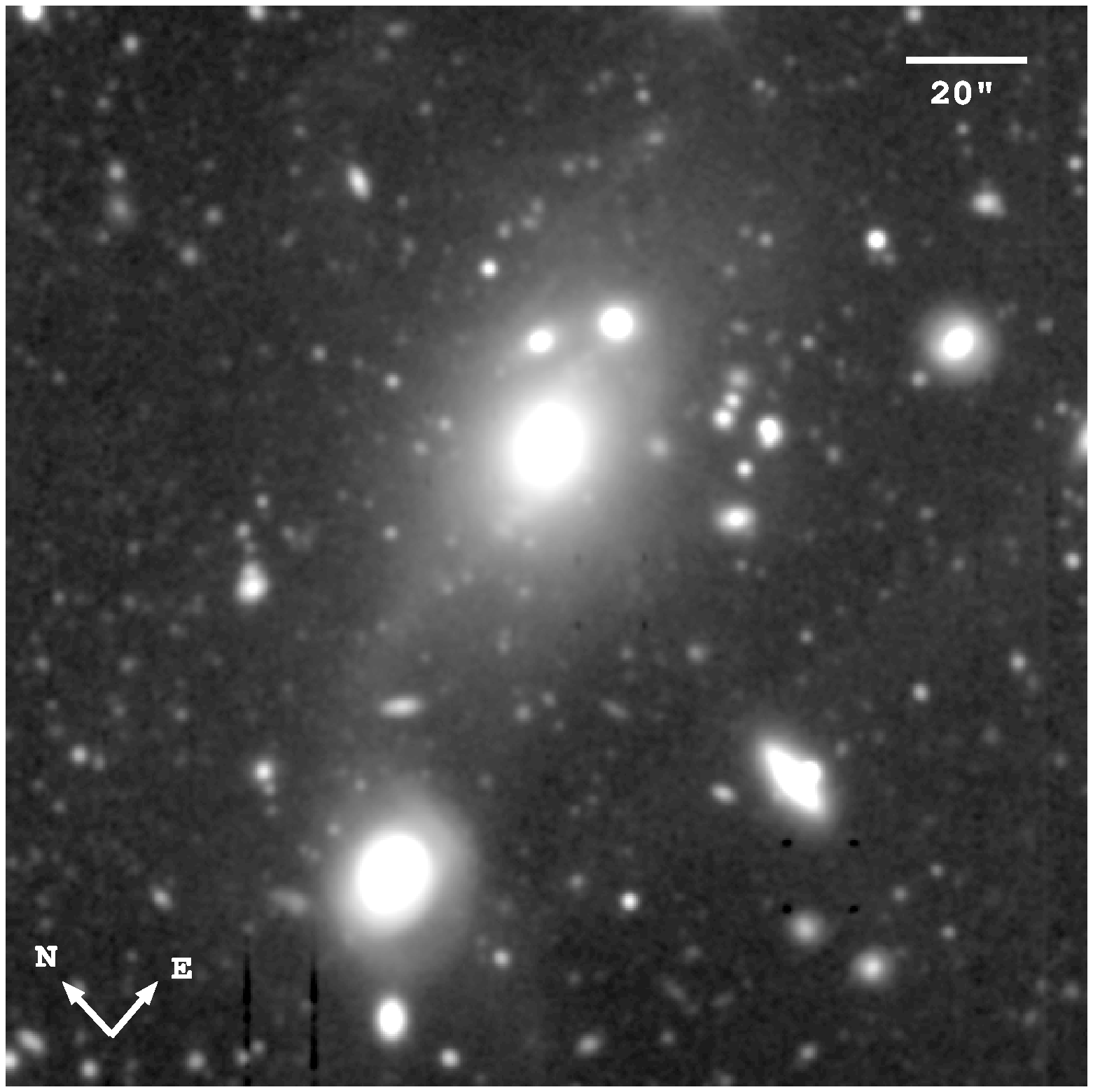}}
\caption{PKS 0349-27. (a) Unsharp-masked image. (b) Median filtered image.}
\label{pks0349_online} 
\end{figure*}

\begin{figure*}
\centering
\subfigure[]{\includegraphics[width=8.0cm]{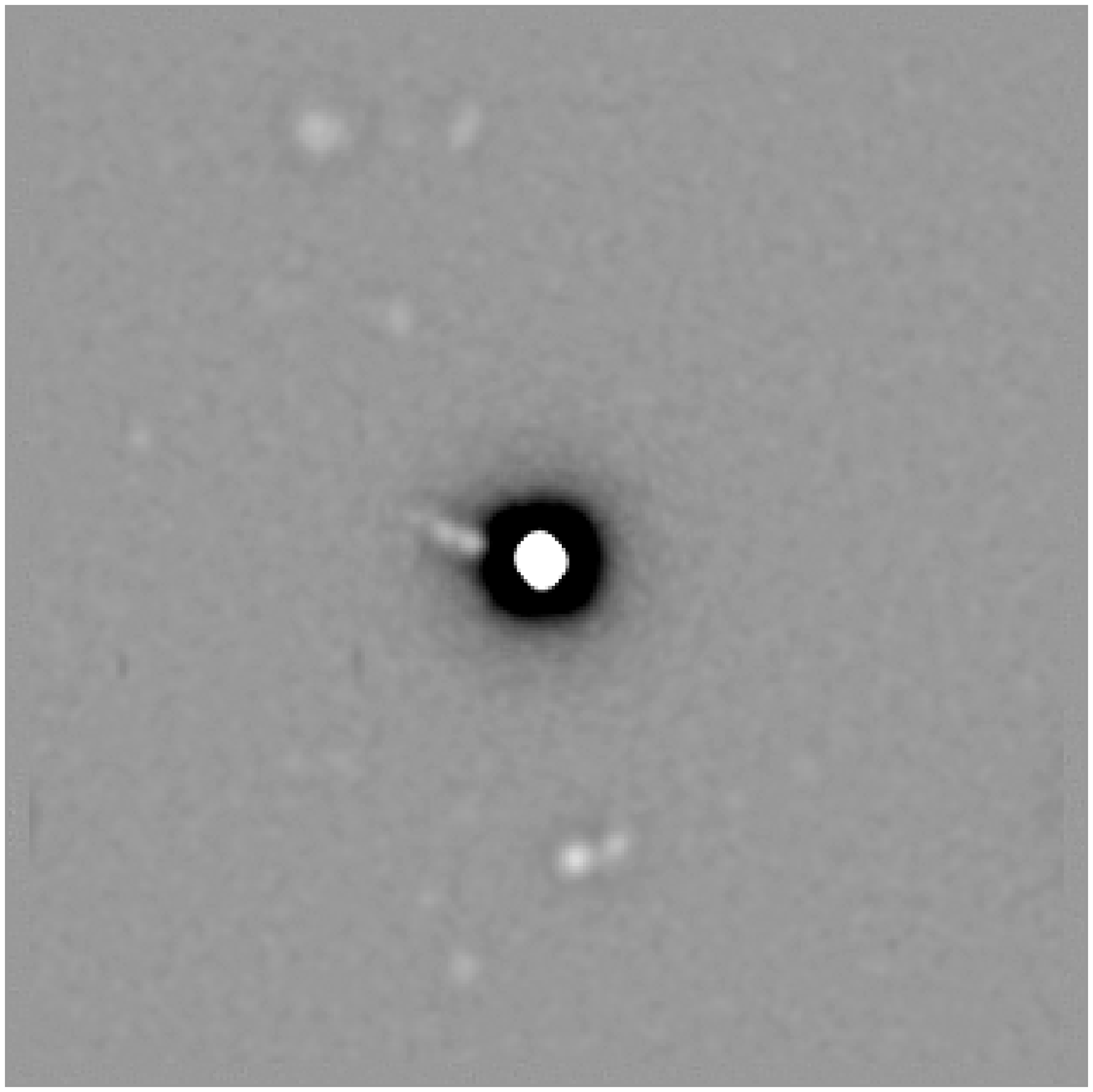}}
\subfigure[]{\includegraphics[width=8.0cm]{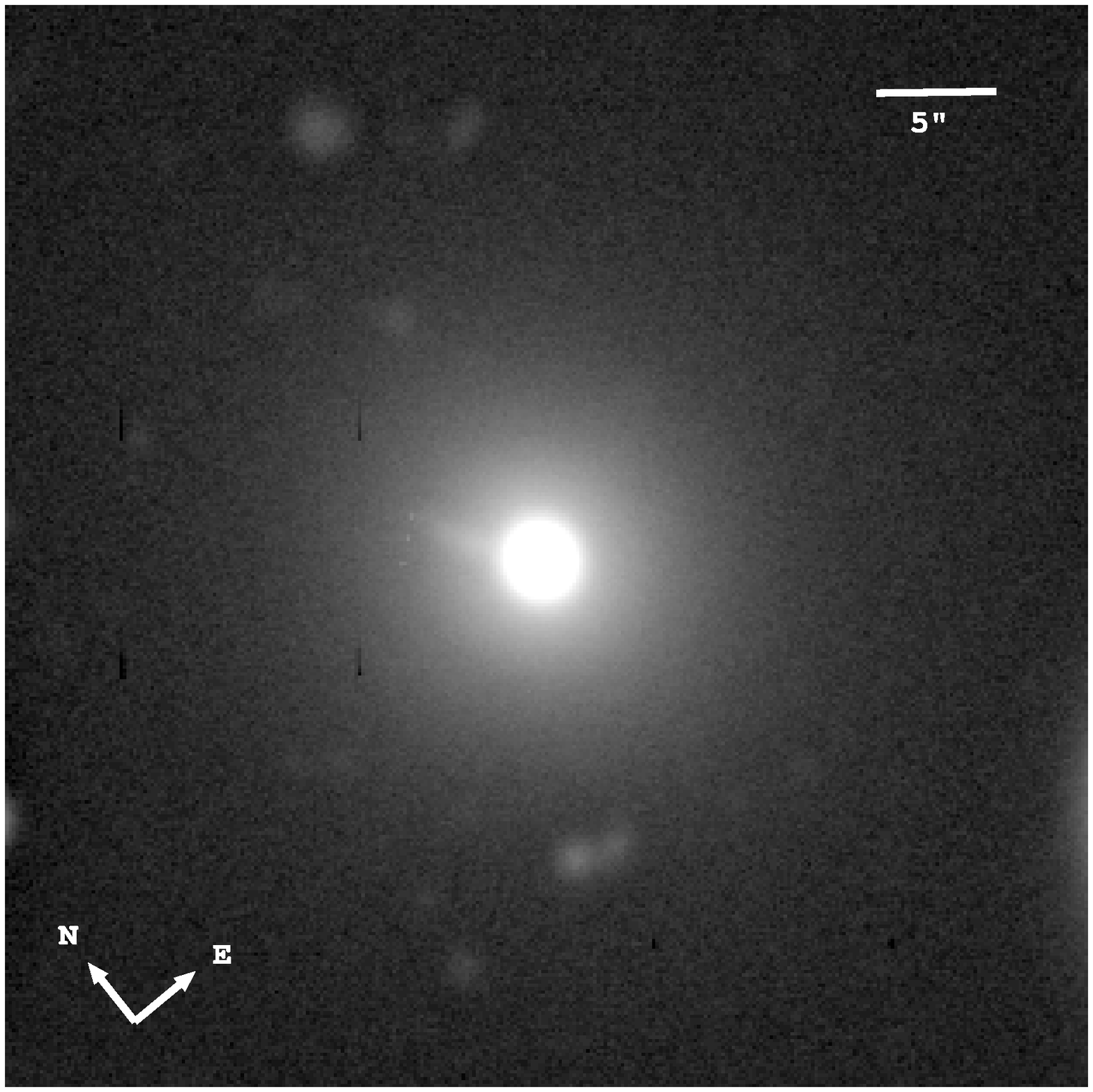}}
\caption{PKS 0034-01. (a) Smooth galaxy-subtracted image. (b) Unsharp-masked image.}
\label{pks0034_online} 
\end{figure*}

\begin{figure*}
\centering
\subfigure[]{\includegraphics[width=8.0cm]{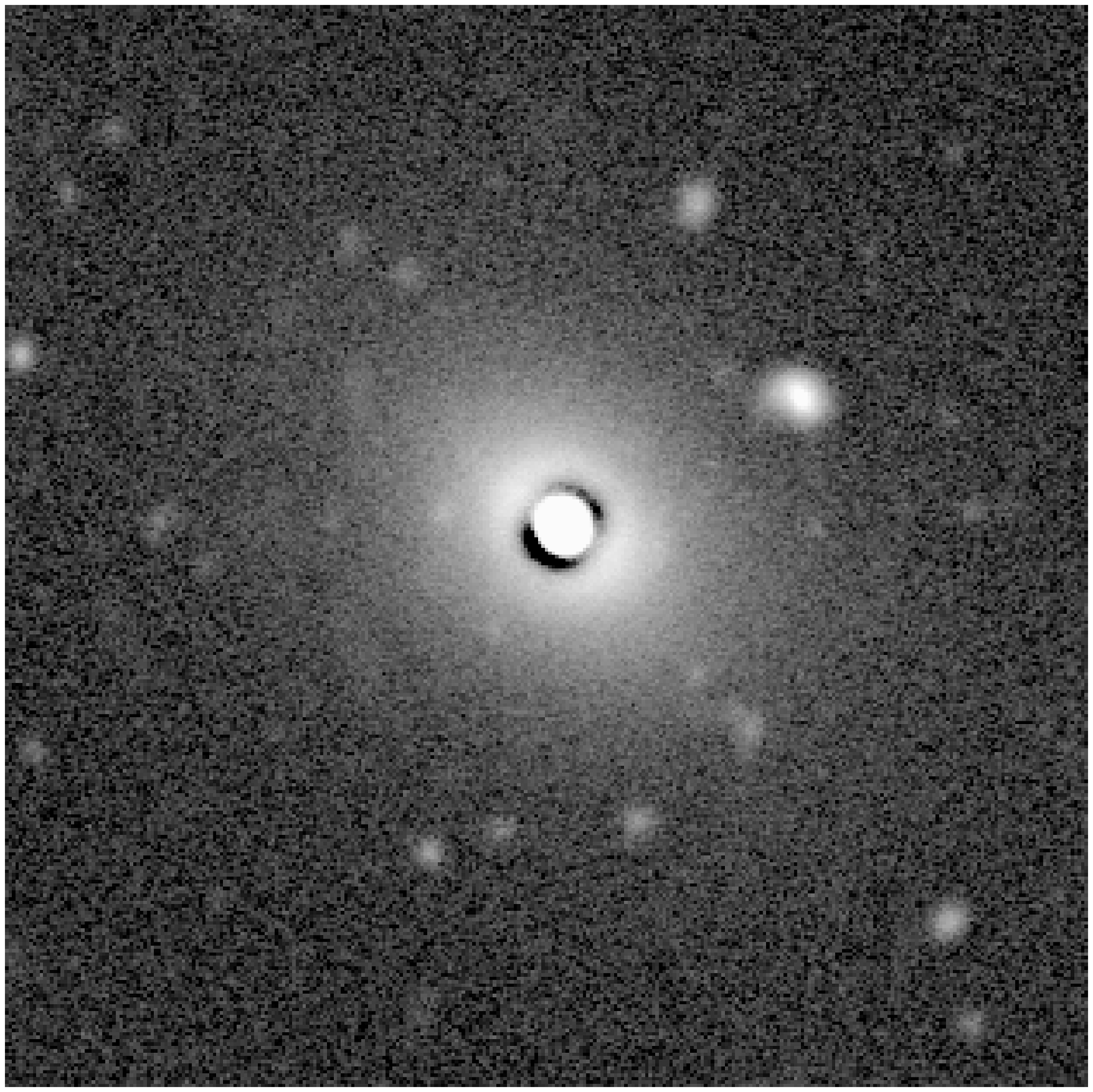}}
\subfigure[]{\includegraphics[width=8.0cm]{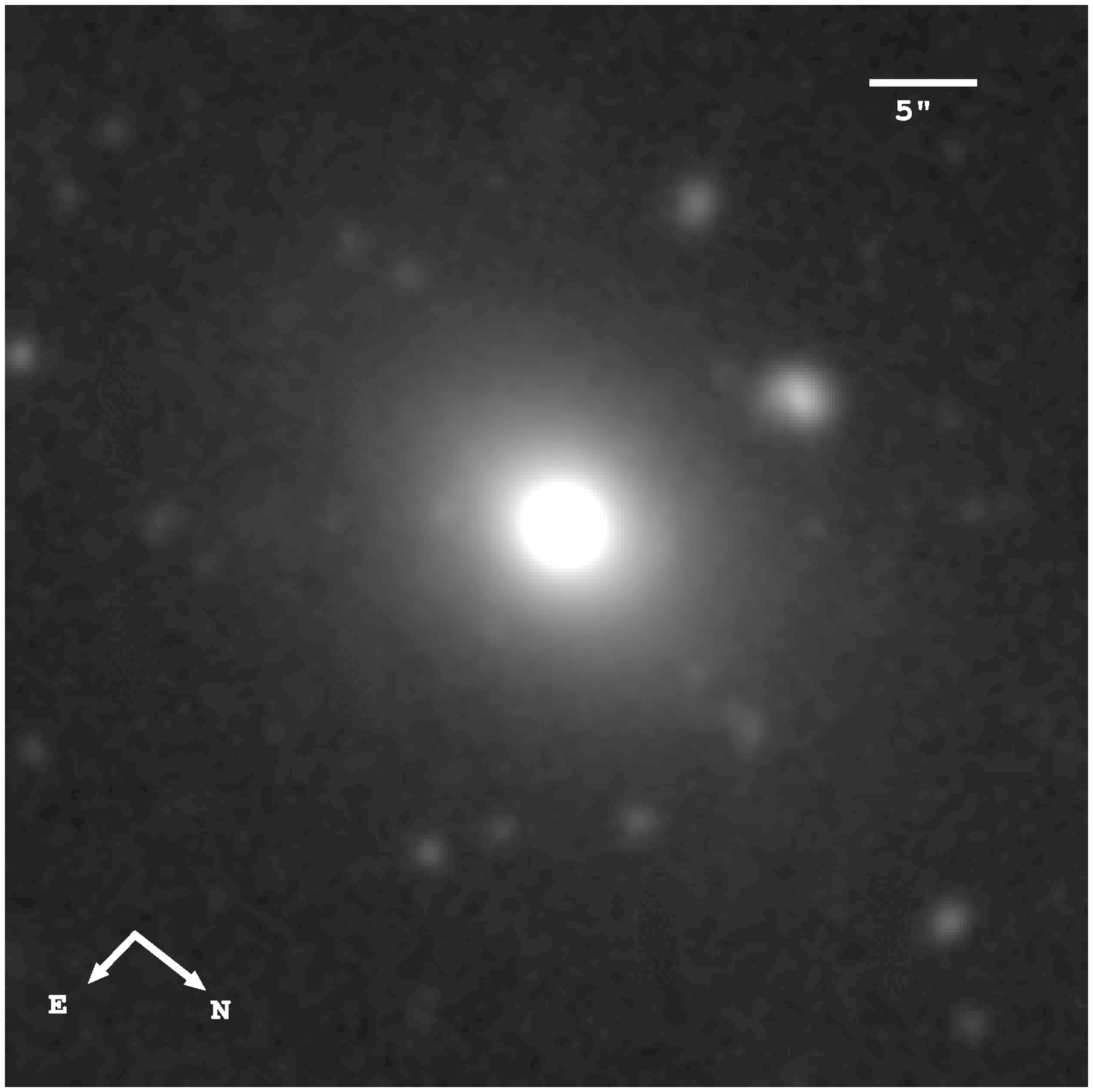}}
\caption{PKS 0945+07. (a) Unsharp-masked image. (b) Median filtered image.}
\label{pks0945_online} 
\end{figure*}

\begin{figure*}
\centering
\subfigure[]{\includegraphics[width=8.0cm]{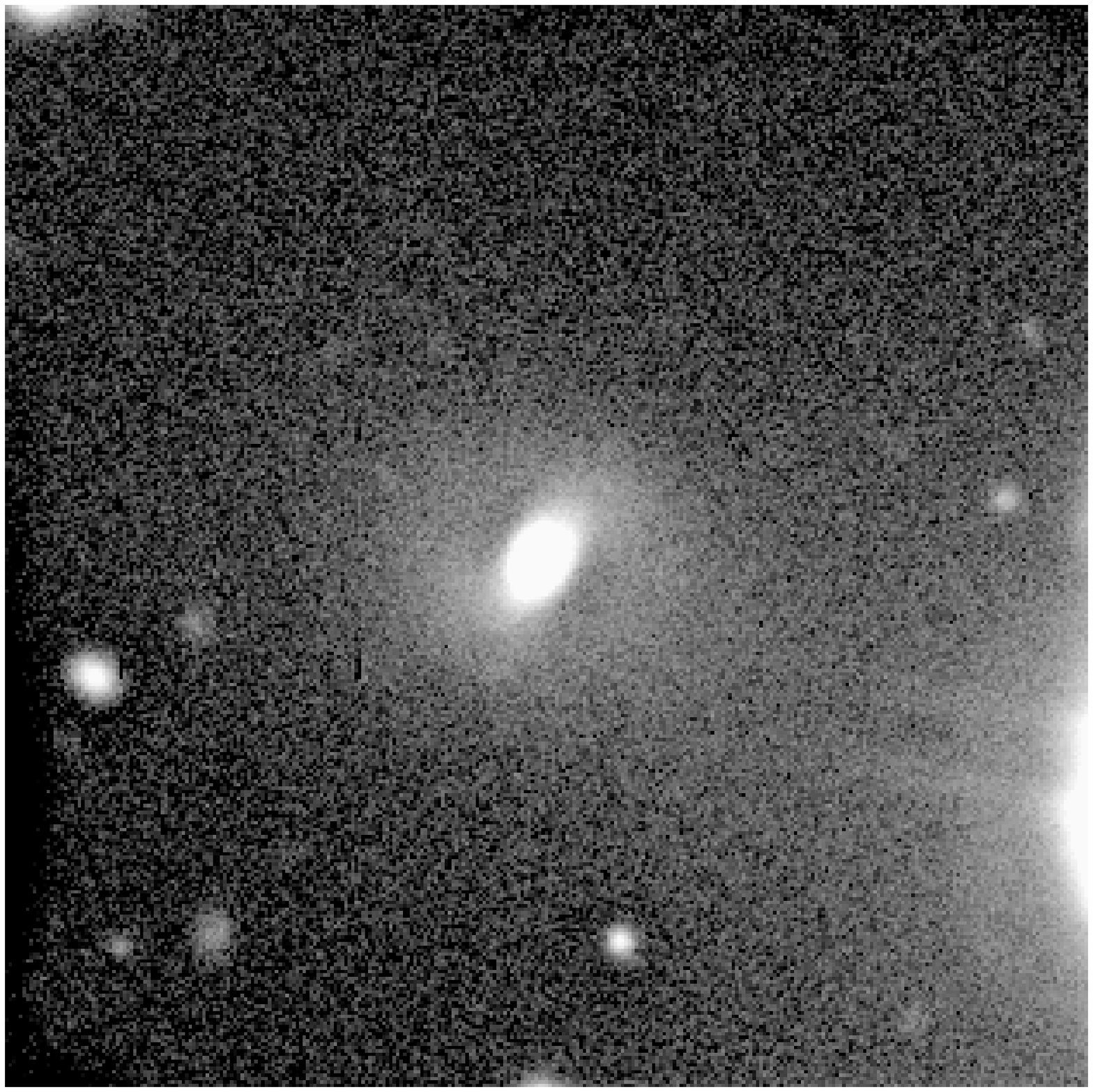}}
\subfigure[]{\includegraphics[width=8.0cm]{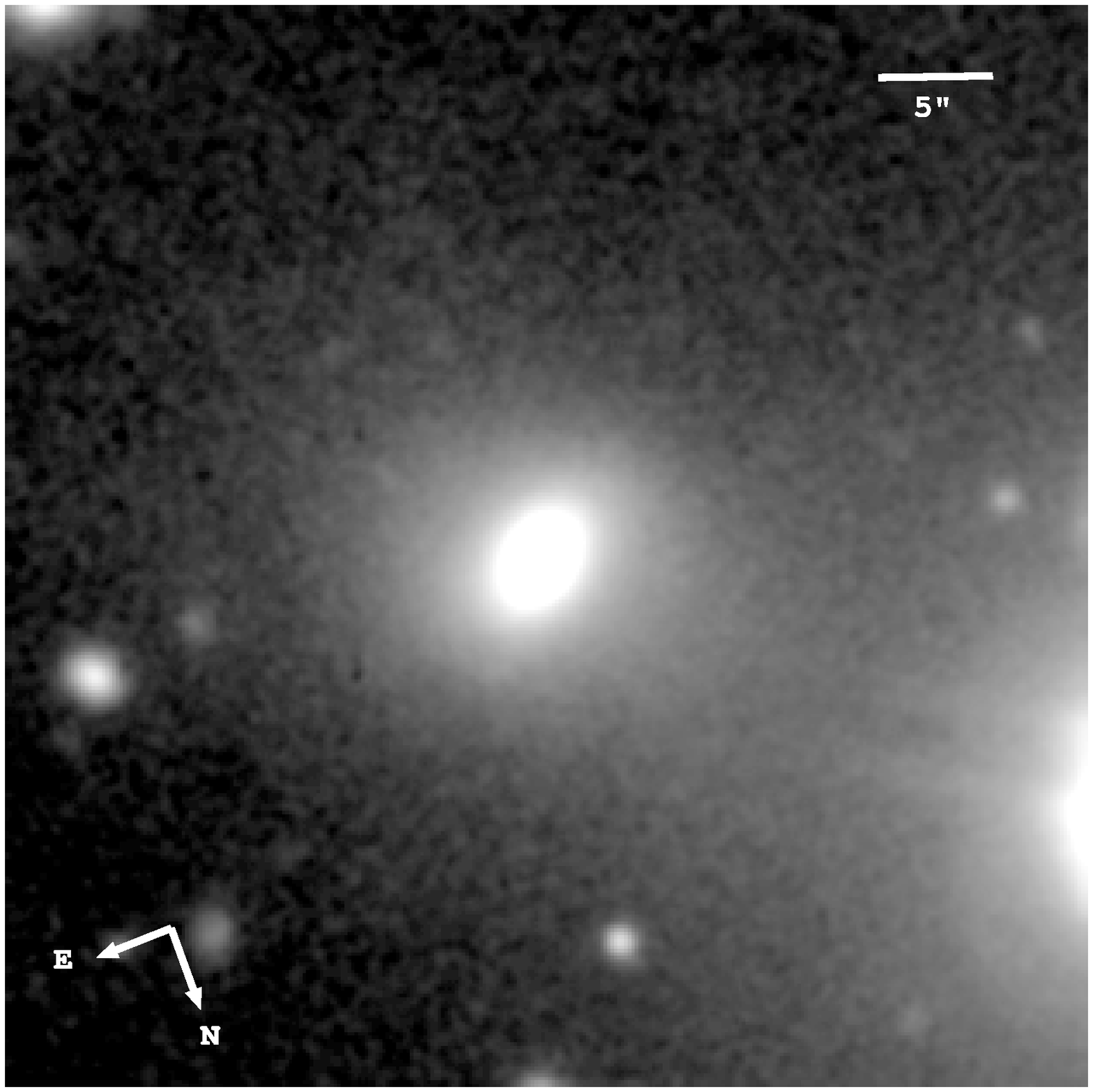}}
\caption{PKS 0404+03. (a) Unsharp-masked image. (b) Median filtered image.}
\label{pks0404_online} 
\end{figure*}

\begin{figure*}
\centering
\subfigure[]{\includegraphics[width=8.0cm]{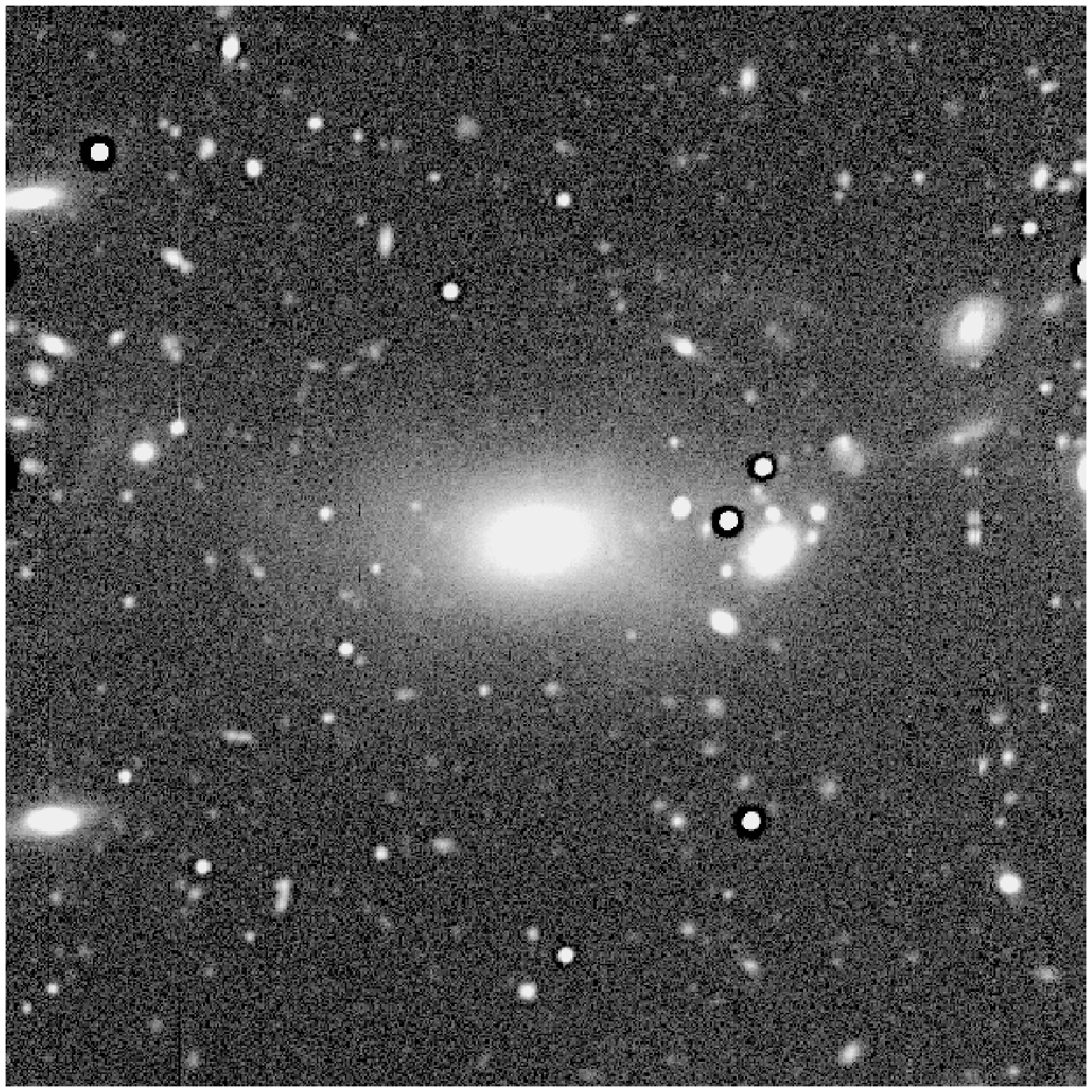}}
\subfigure[]{\includegraphics[width=8.0cm]{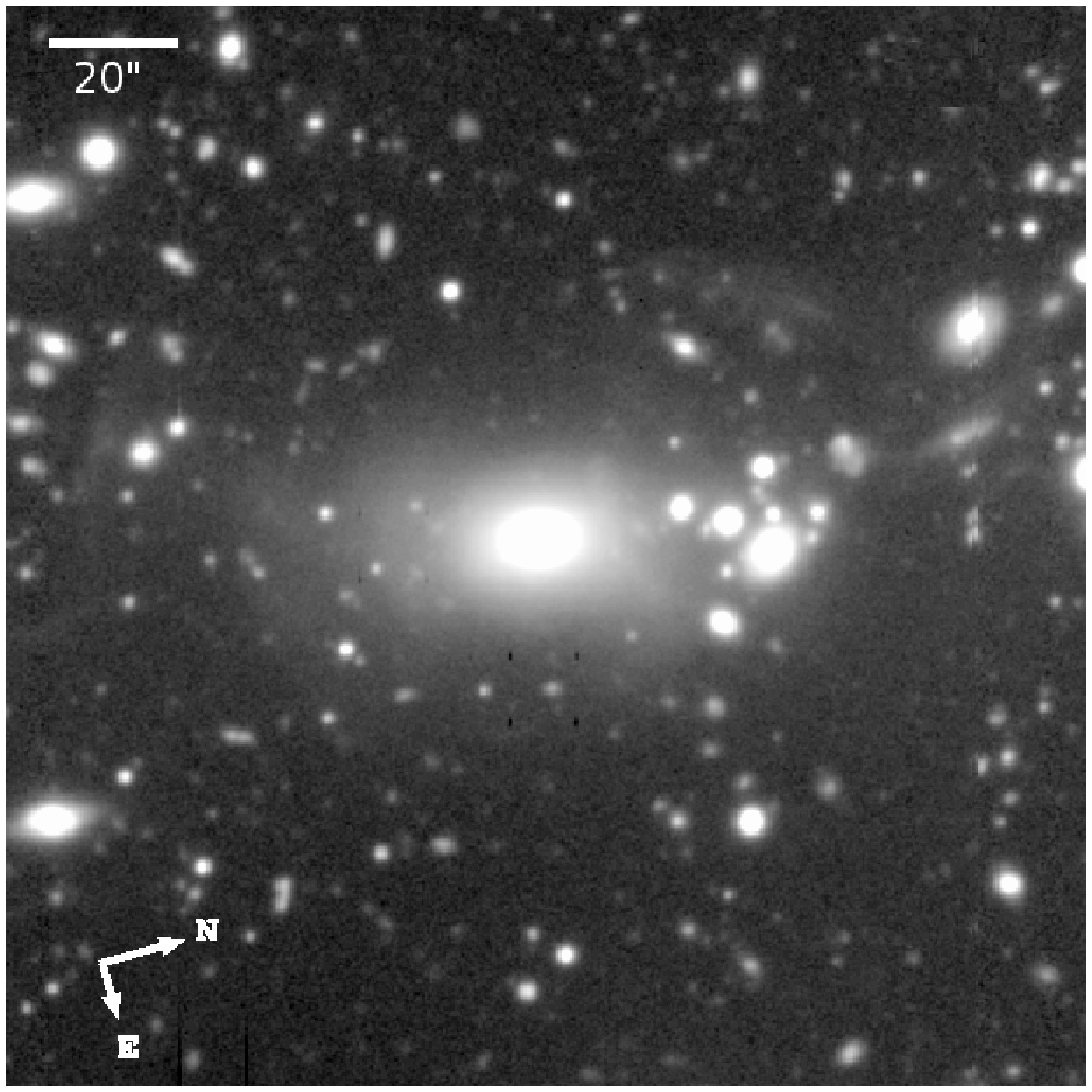}}
\caption{PKS 2356-61. (a) Unsharp-masked image. (b) Median filtered image.}
\label{pks2356_online} 
\end{figure*}

\begin{figure*}
\centering
\subfigure[]{\includegraphics[width=8.0cm]{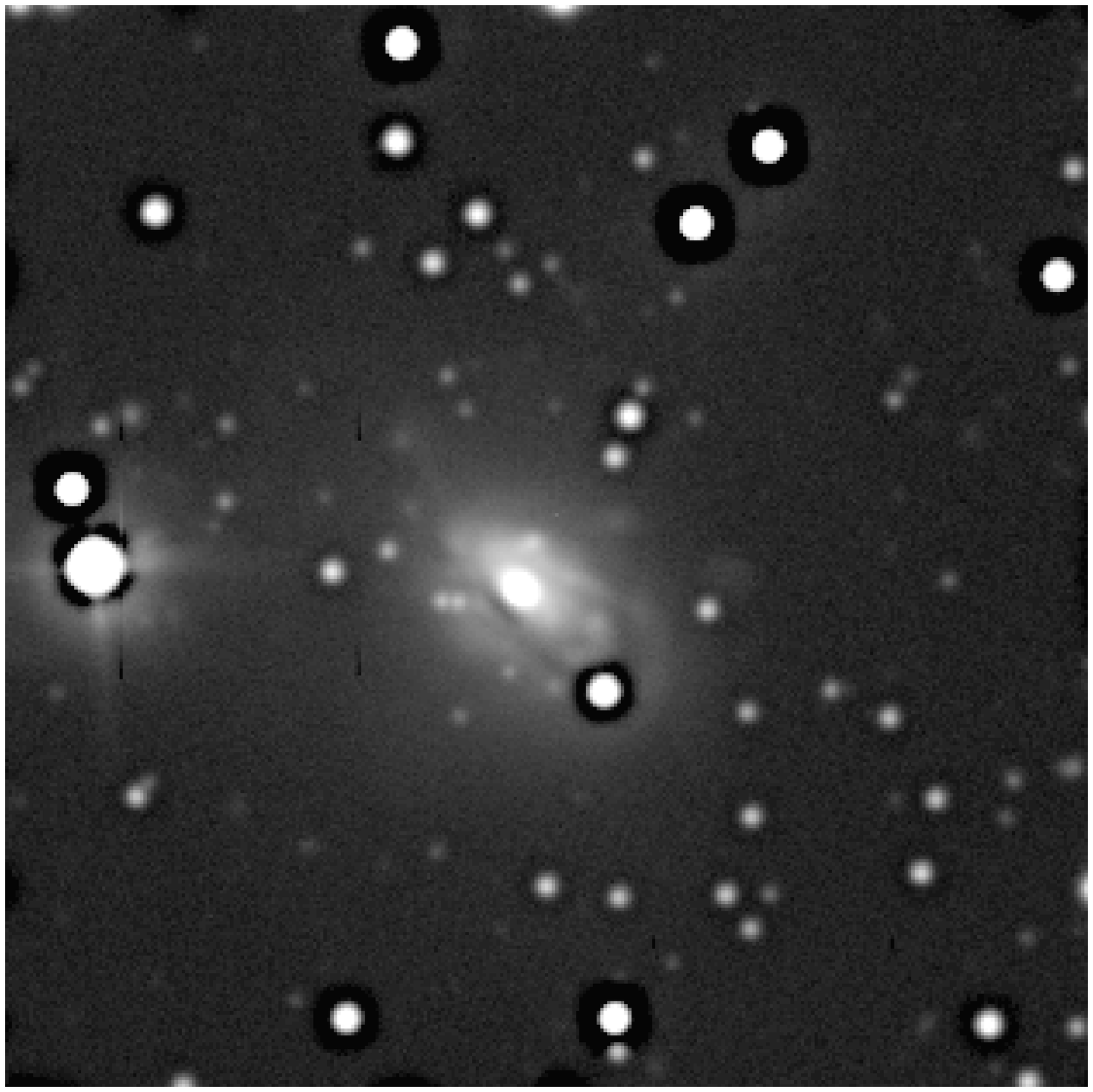}}
\subfigure[]{\includegraphics[width=8.0cm]{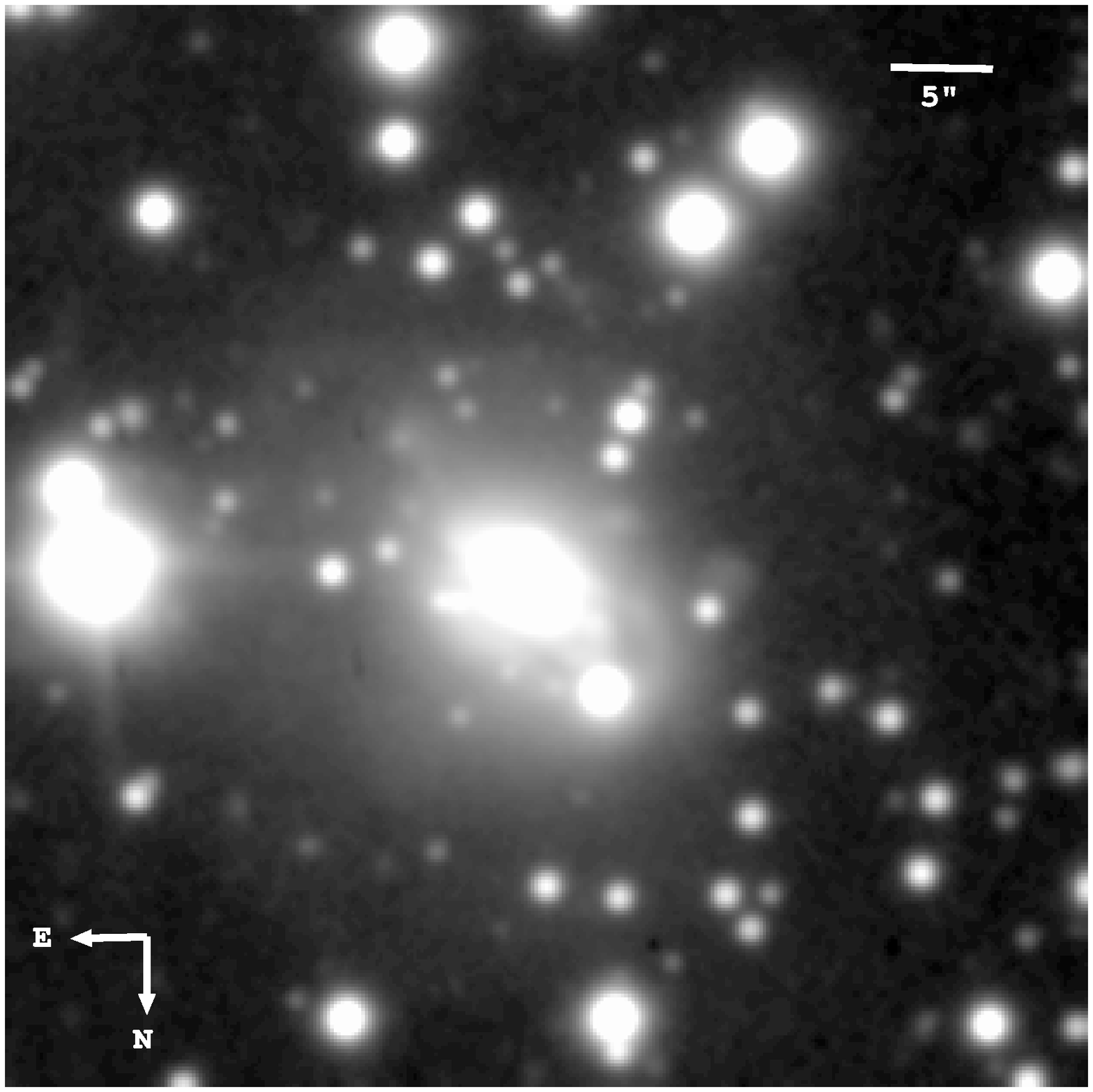}}
\caption{PKS 1733-56. (a) Unsharp-masked image. (b) Median filtered image.}
\label{pks1733_online} 
\end{figure*}

\begin{figure*}
\centering
\subfigure[]{\includegraphics[width=8.0cm]{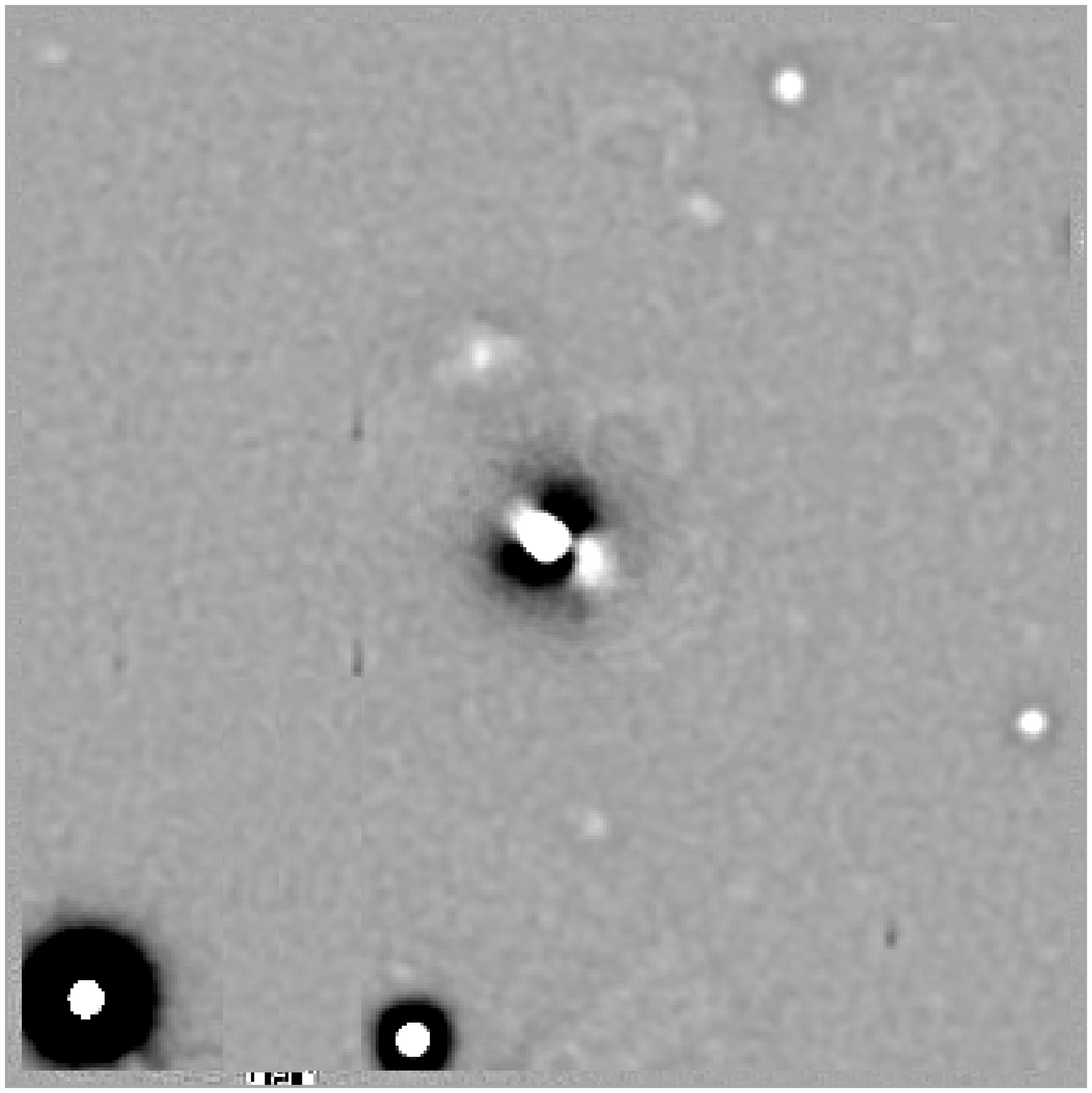}}
\subfigure[]{\includegraphics[width=8.0cm]{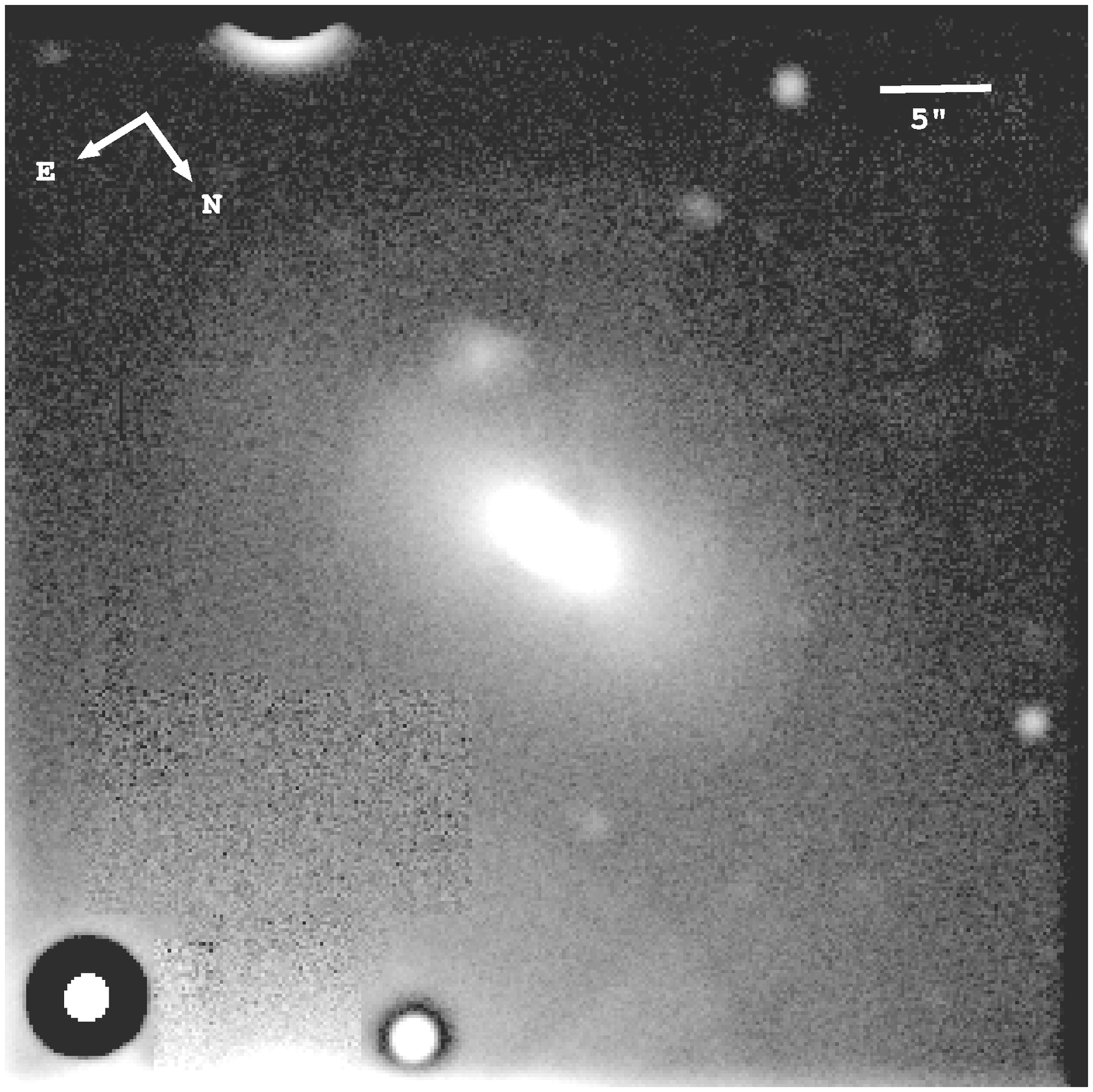}}
\caption{PKS 1559+02. (a) Smooth galaxy-subtracted image. (b) Unsharp-masked image.}
\label{pks1559_online} 
\end{figure*}

\begin{figure*}
\centering
\subfigure[]{\includegraphics[width=8.0cm]{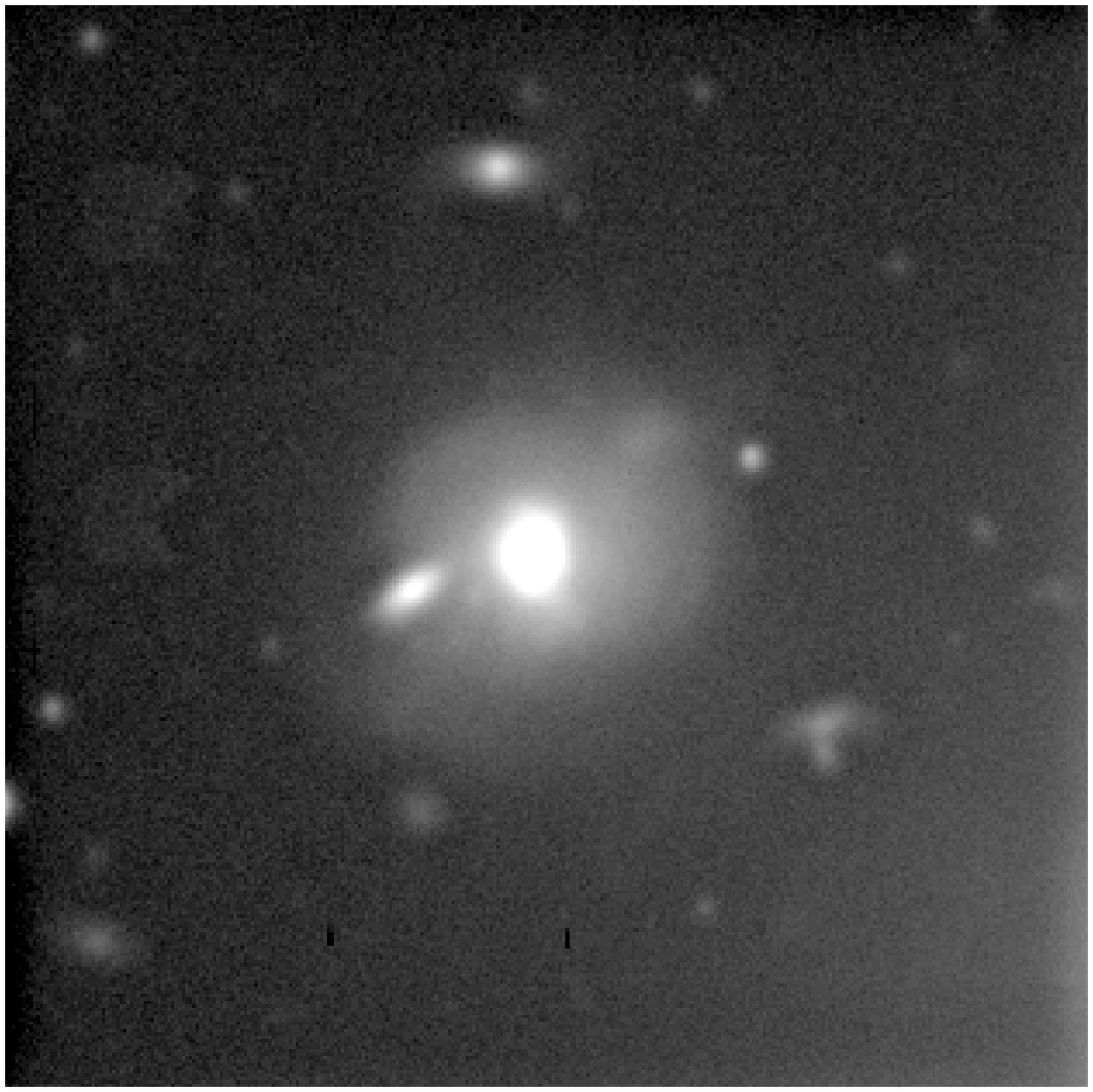}\label{pks0806_online_a}}
\subfigure[]{\includegraphics[width=8.0cm]{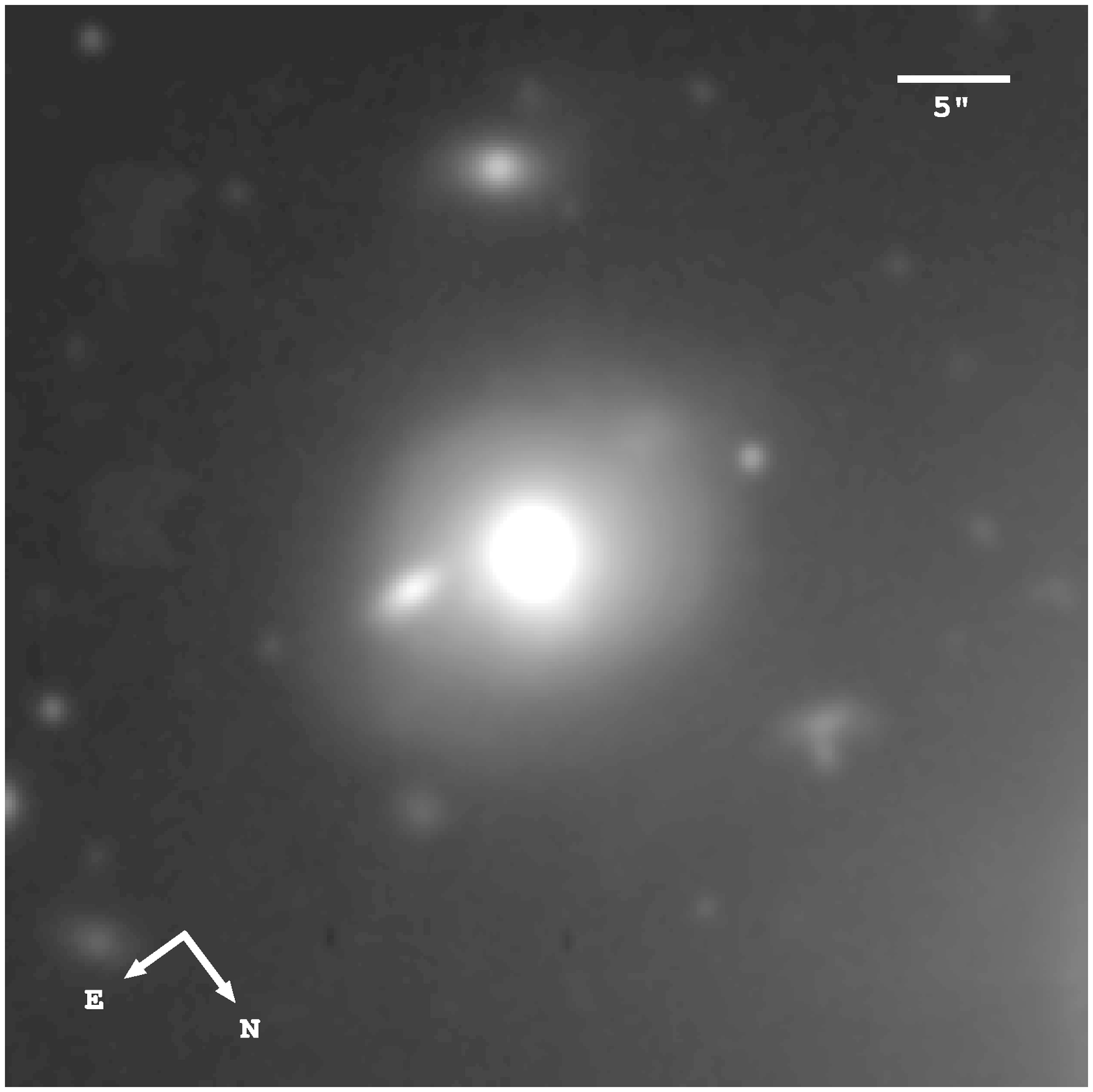}\label{pks0806_online_b}}
\caption{PKS 0806-10. (a) Unsharp-masked image. (b) Median filtered image.}
\label{pks0806_online} 
\end{figure*}

\begin{figure*}
\centering
\subfigure[]{\includegraphics[width=8.0cm]{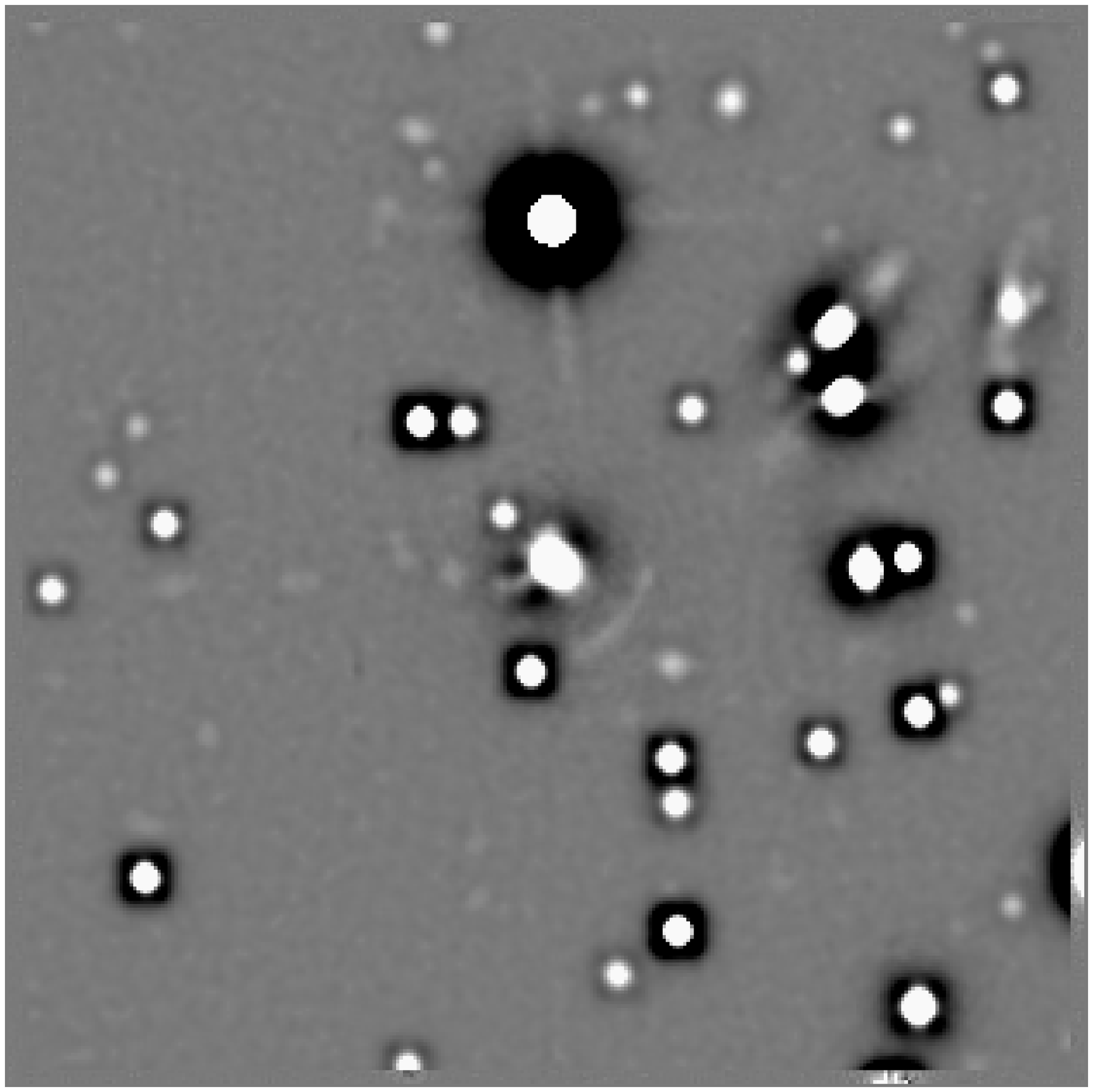}}
\subfigure[]{\includegraphics[width=8.0cm]{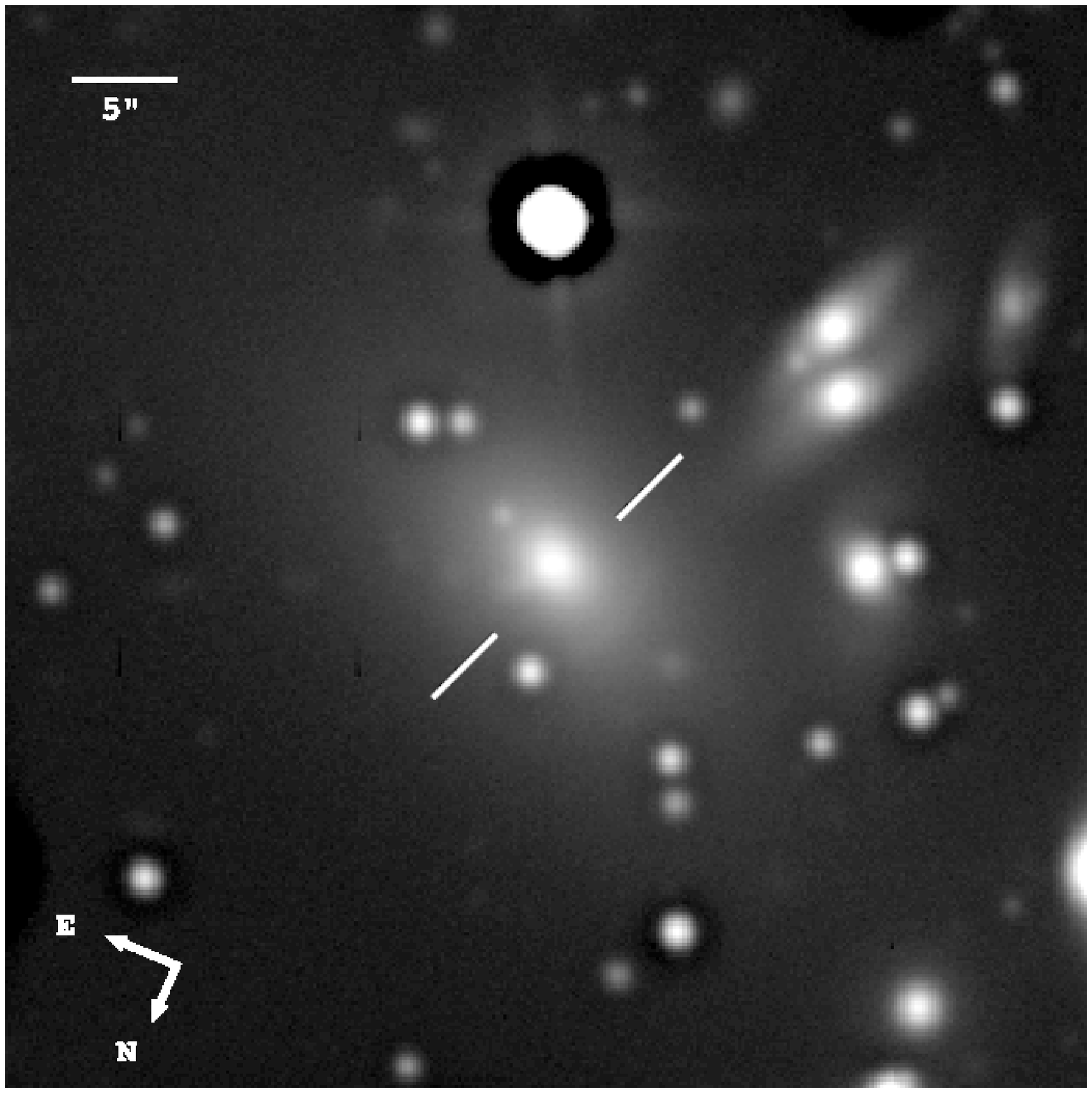}}
\caption{PKS 1839-48. (a) Smooth galaxy-subtracted image. (b) Unsharp-masked image.}
\label{pks1839_online} 
\end{figure*}

\begin{figure*}
\centering
\subfigure[]{\includegraphics[width=8.0cm]{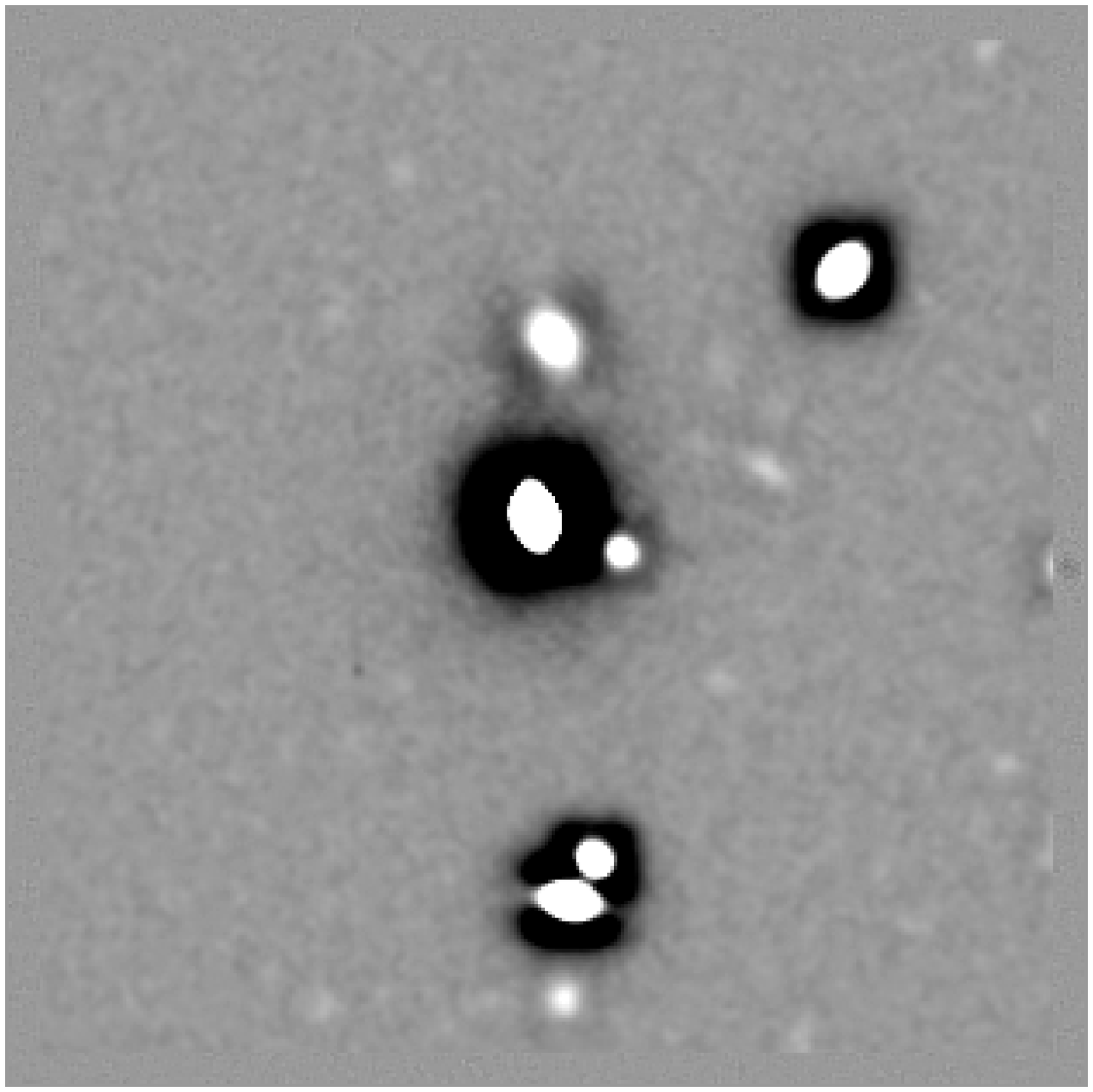}\label{pks0043_online_a}}
\subfigure[]{\includegraphics[width=8.0cm]{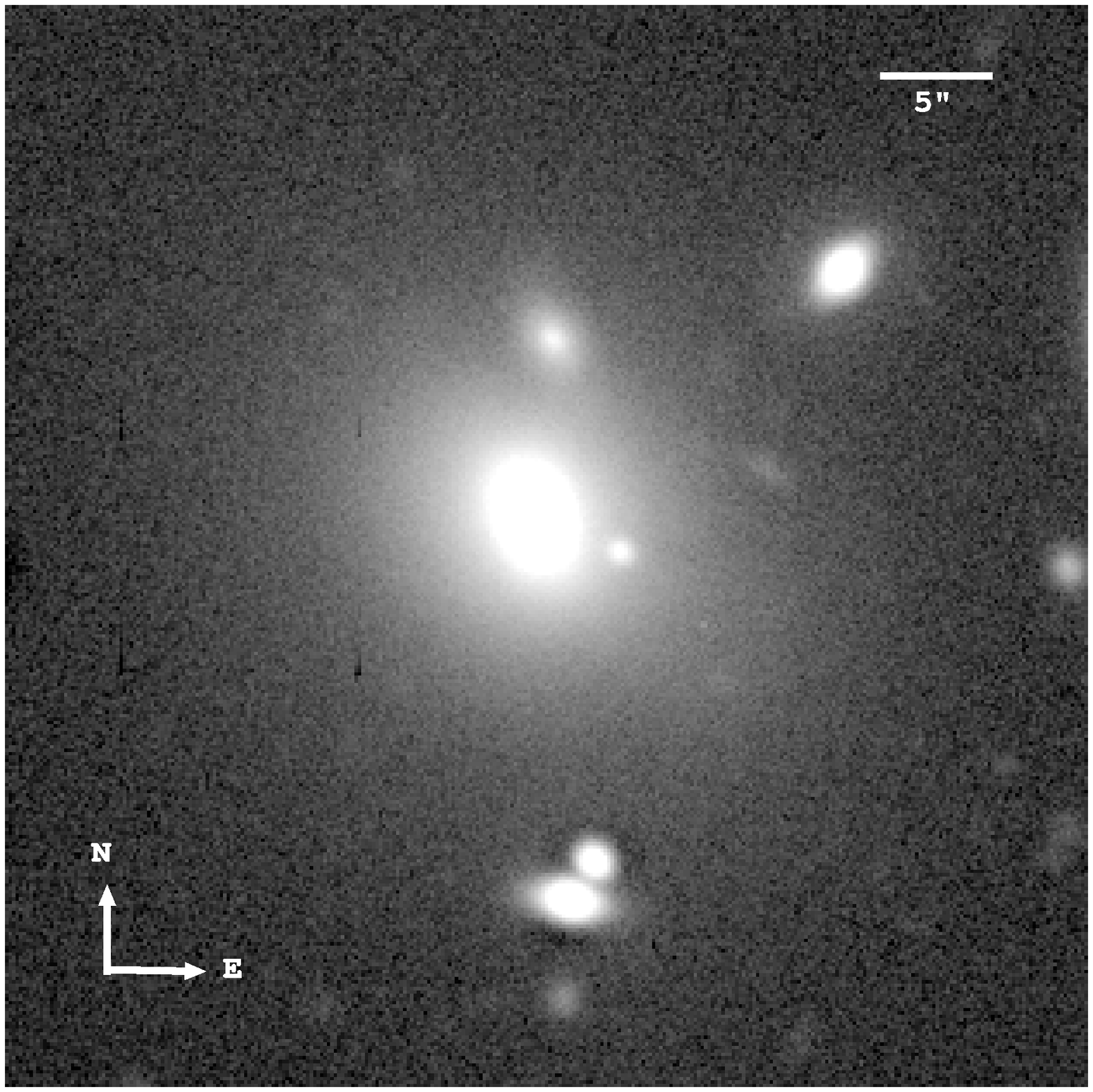}\label{pks0043_online_b}}
\caption{PKS 0043-42. (a) Smooth galaxy-subtracted image. (b) Unsharp-masked image.}
\label{pks0043_online} 
\end{figure*}

\begin{figure*}
\centering
\subfigure[]{\includegraphics[width=8.0cm]{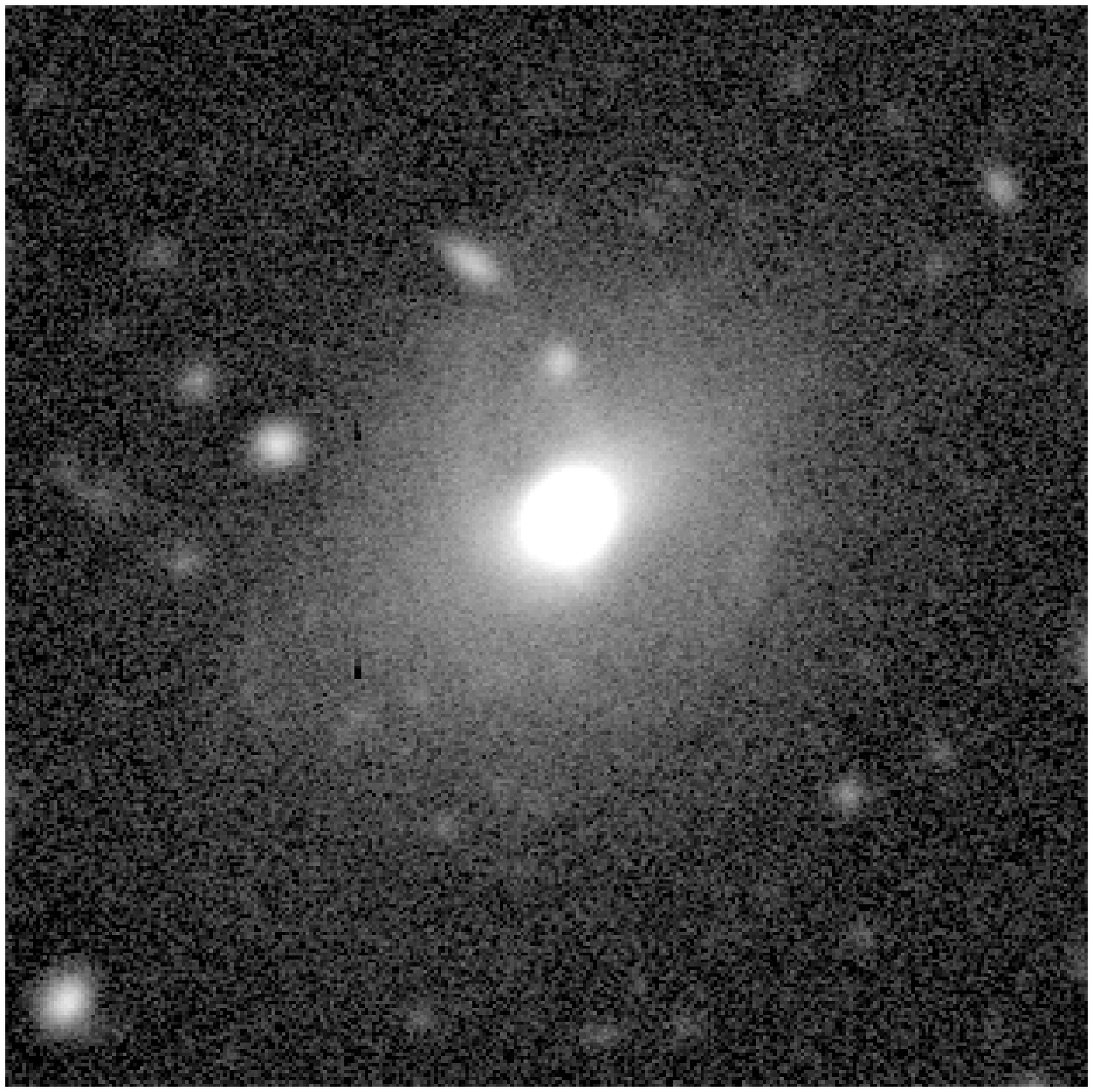}\label{pks0213_online_a}}
\subfigure[]{\includegraphics[width=8.0cm]{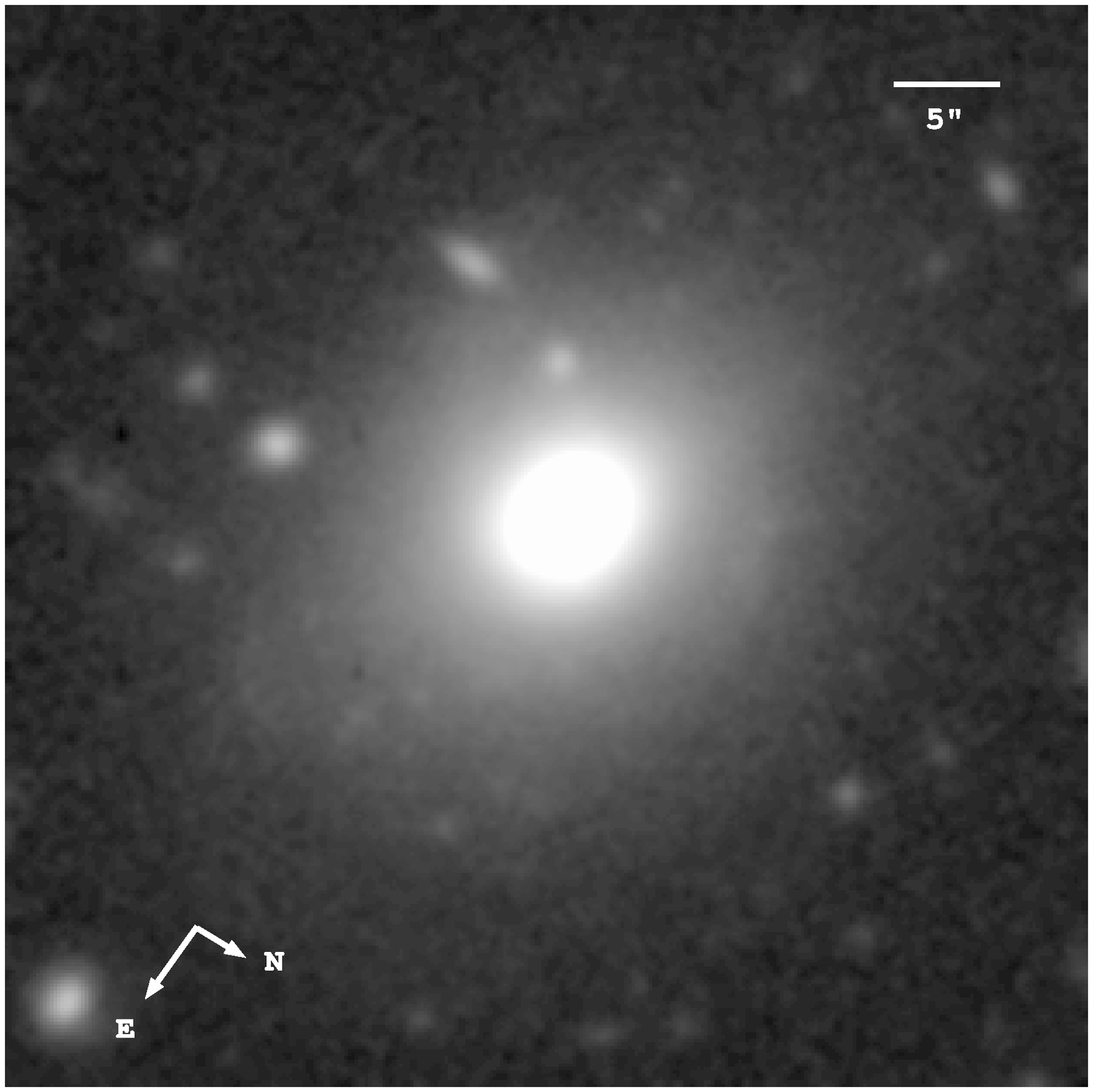}\label{pks0213_online_b}}
\caption{PKS 0213-13. (a) Unsharp-masked image. (b) Median filtered image.}
\label{pks0213_online} 
\end{figure*}

\begin{figure*}
\centering
\subfigure[]{\includegraphics[width=8.0cm]{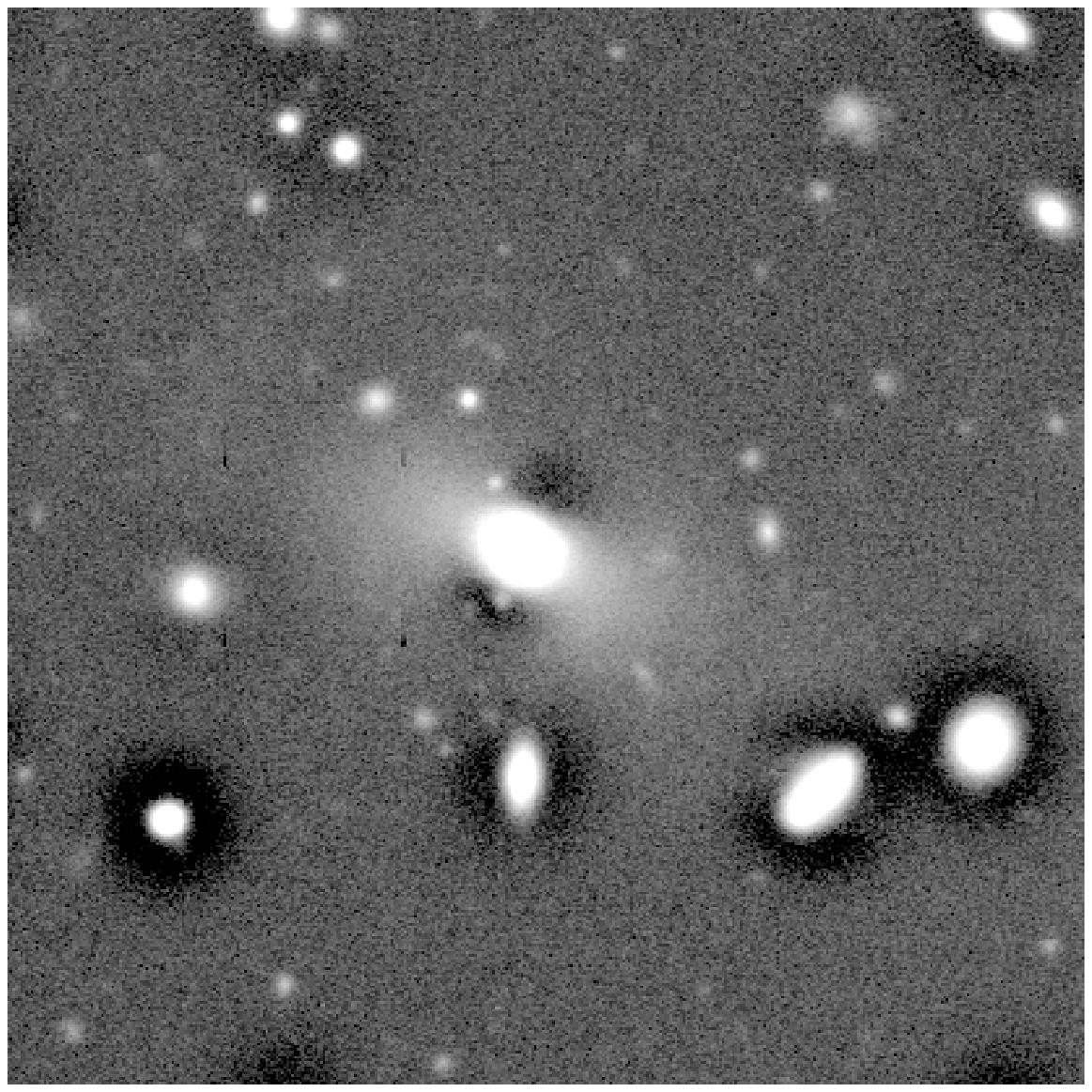}}
\subfigure[]{\includegraphics[width=8.0cm]{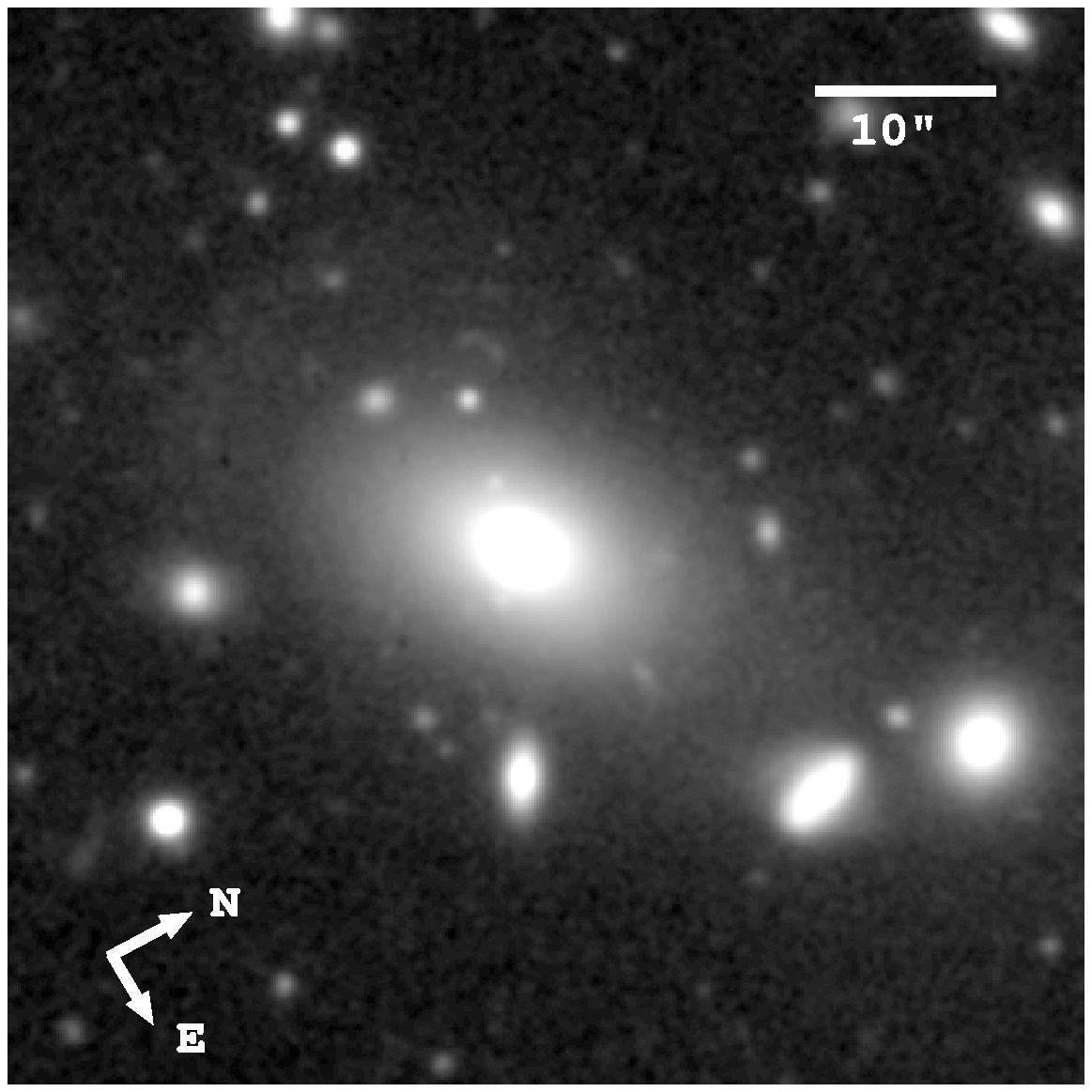}}
\caption{PKS 0442-28. (a) Unsharp-masked image. (b) Median filtered image.}
\label{pks0442_online} 
\end{figure*}

\begin{figure*}
\centering
\subfigure[]{\includegraphics[width=8.0cm]{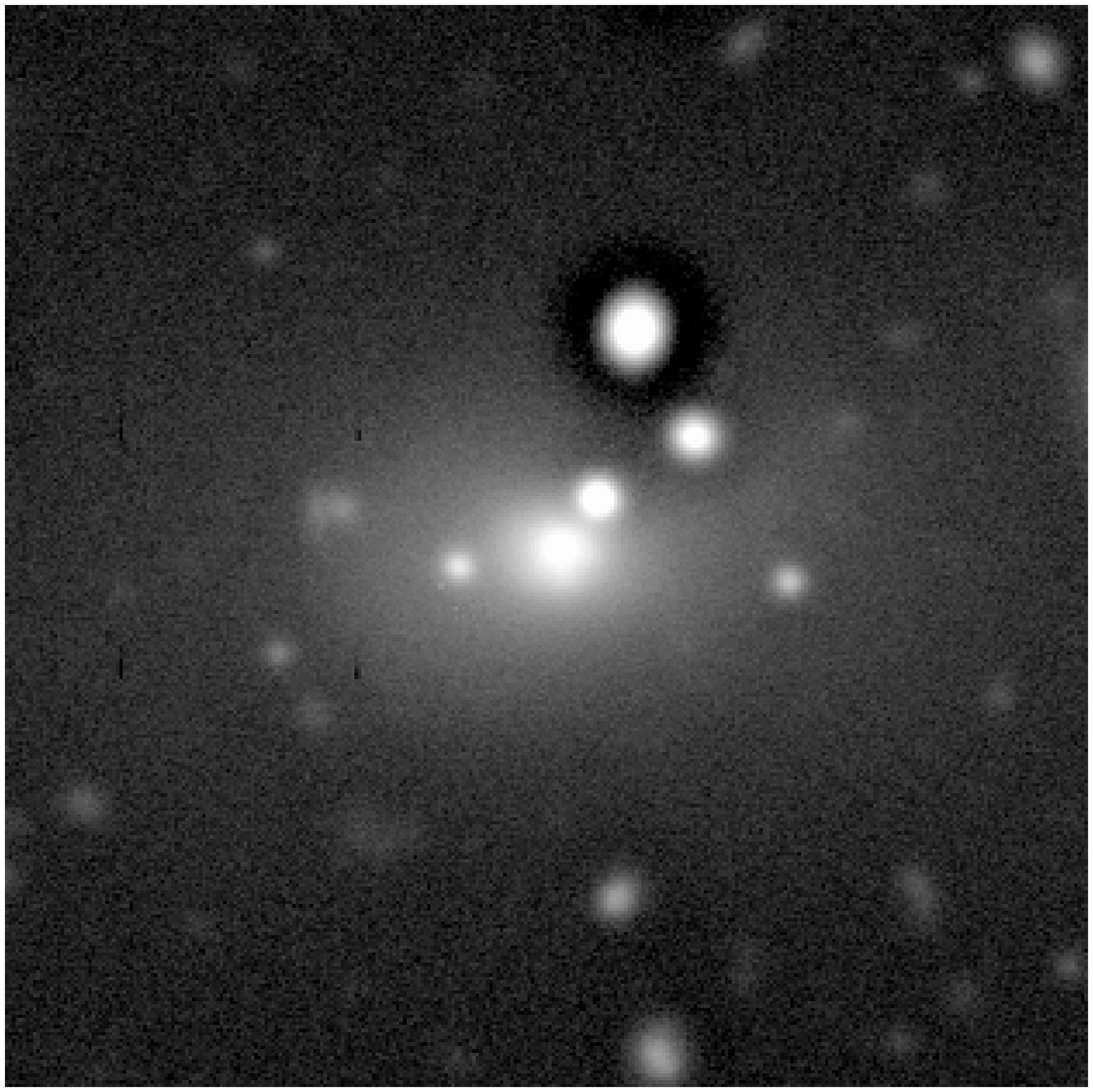}}
\subfigure[]{\includegraphics[width=8.0cm]{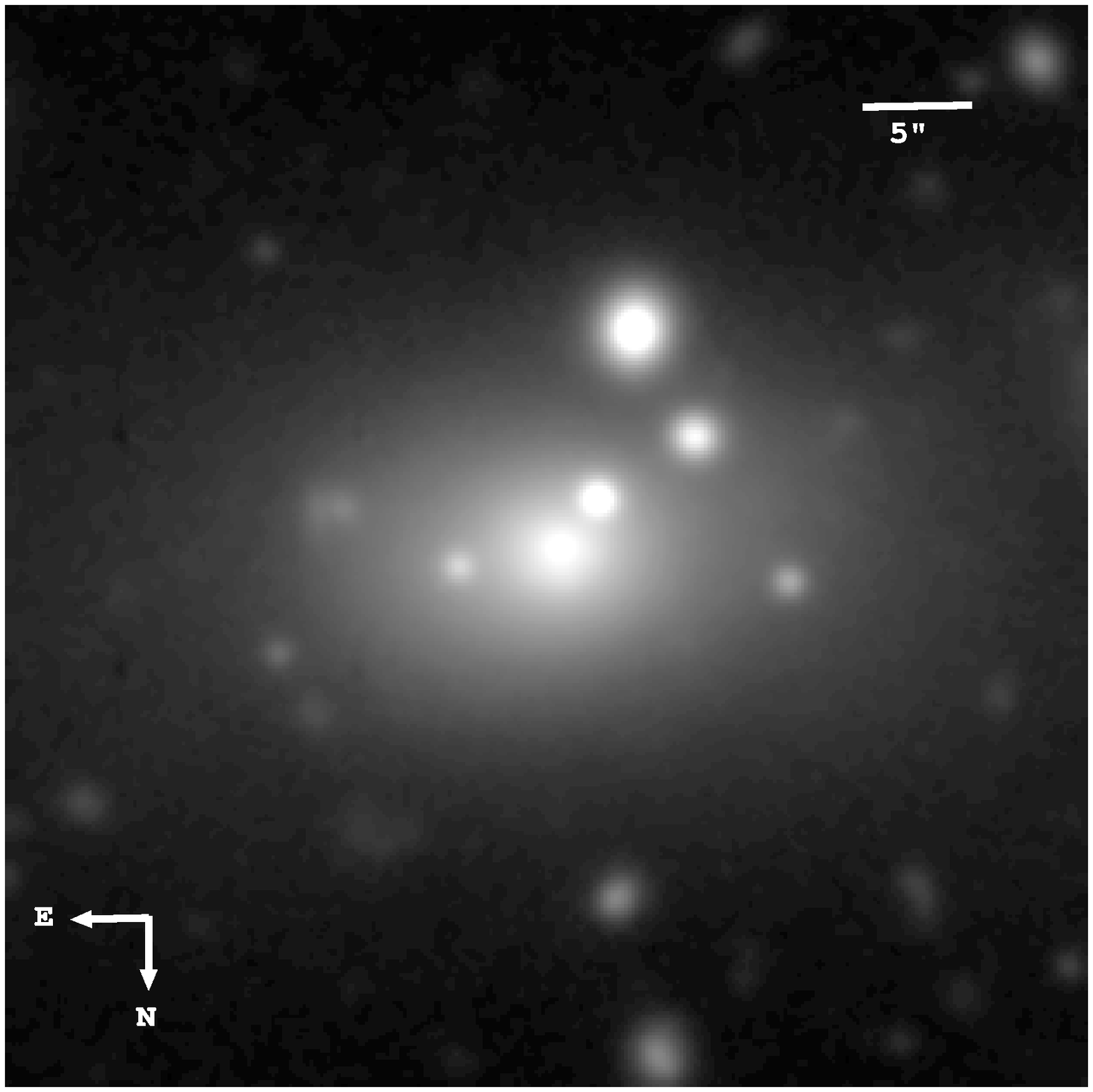}}
\caption{PKS 2211-17. (a) Unsharp-masked image. (b) Median filtered image.}
\label{pks2211_online} 
\end{figure*}

\begin{figure*}
\centering
\subfigure[]{\includegraphics[width=8.0cm]{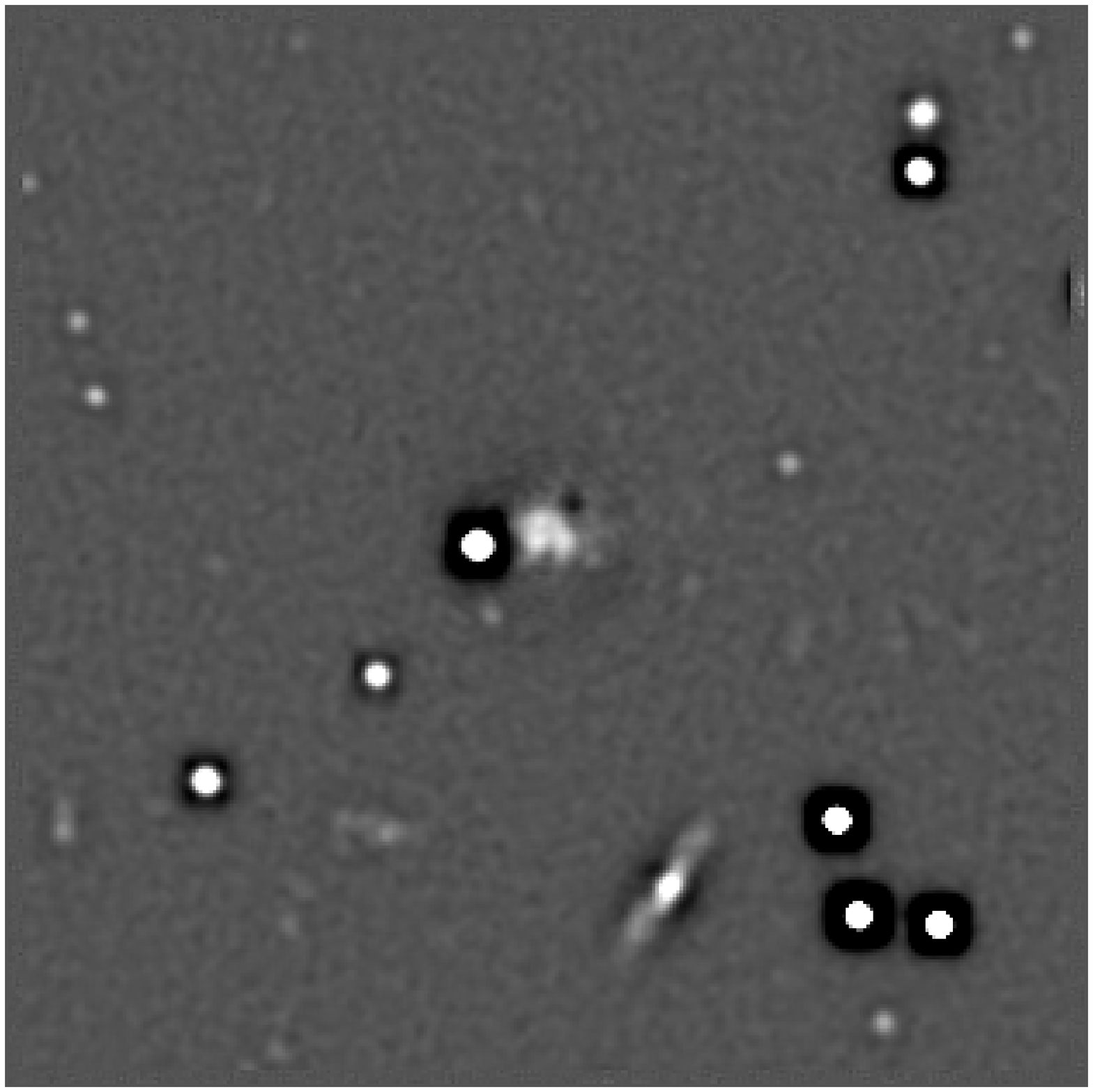}}
\subfigure[]{\includegraphics[width=8.0cm]{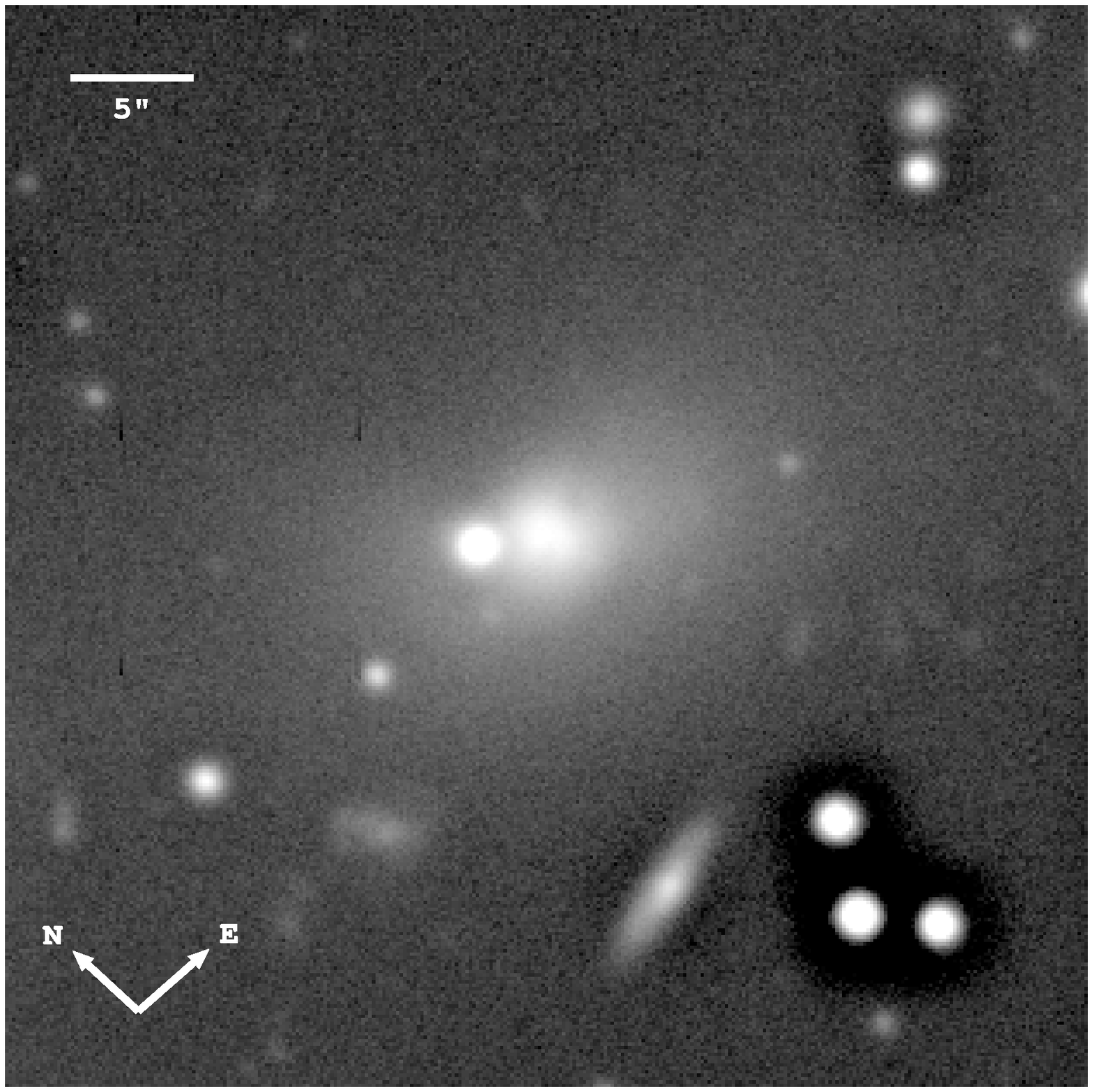}}
\caption{PKS 1648+05. (a) Smooth galaxy-subtracted image. (b) Unsharp-masked image.}
\label{pks1648_online} 
\end{figure*}

\begin{figure*}
\centering
\subfigure[]{\includegraphics[width=8.0cm]{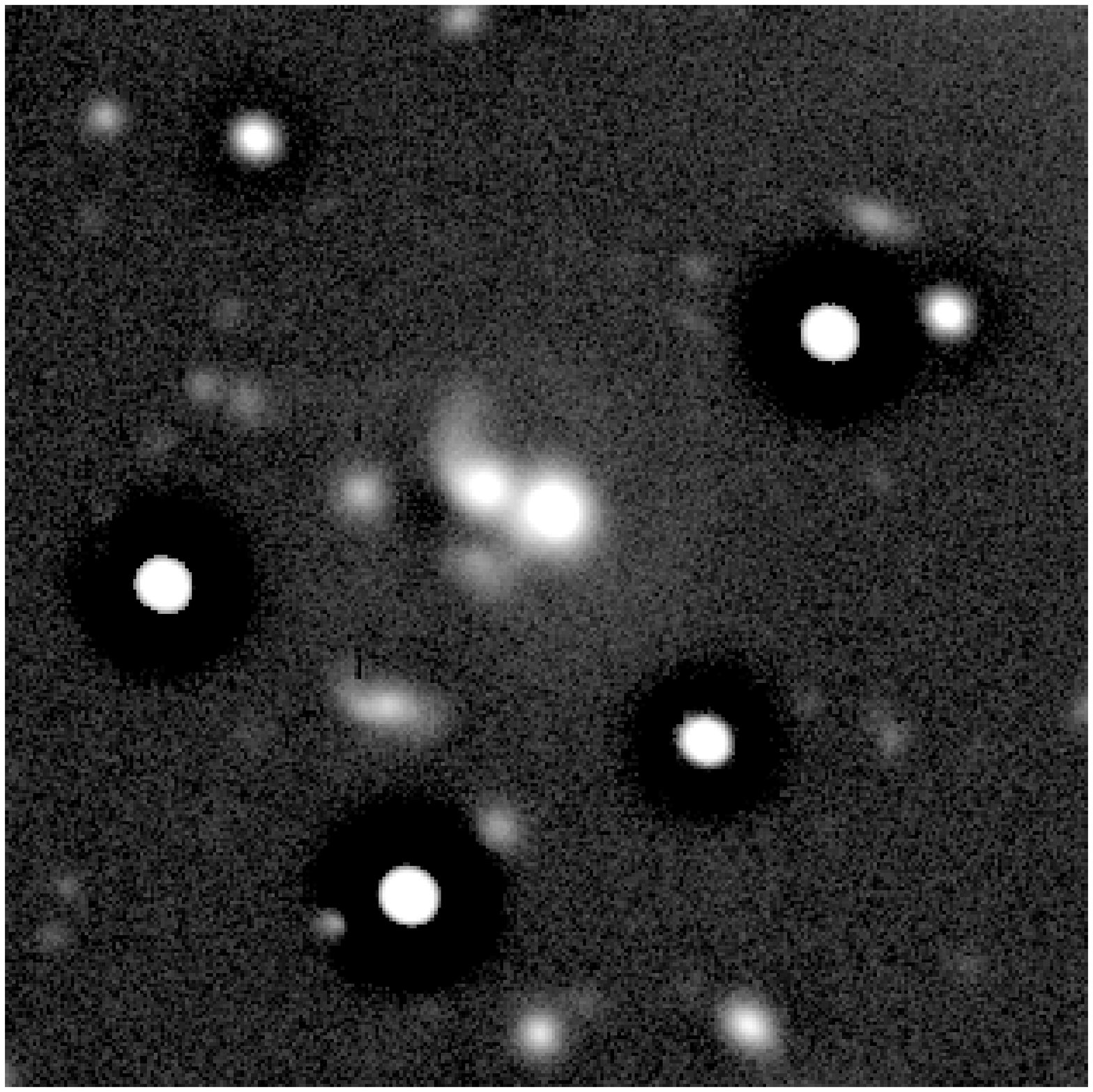}}
\subfigure[]{\includegraphics[width=8.0cm]{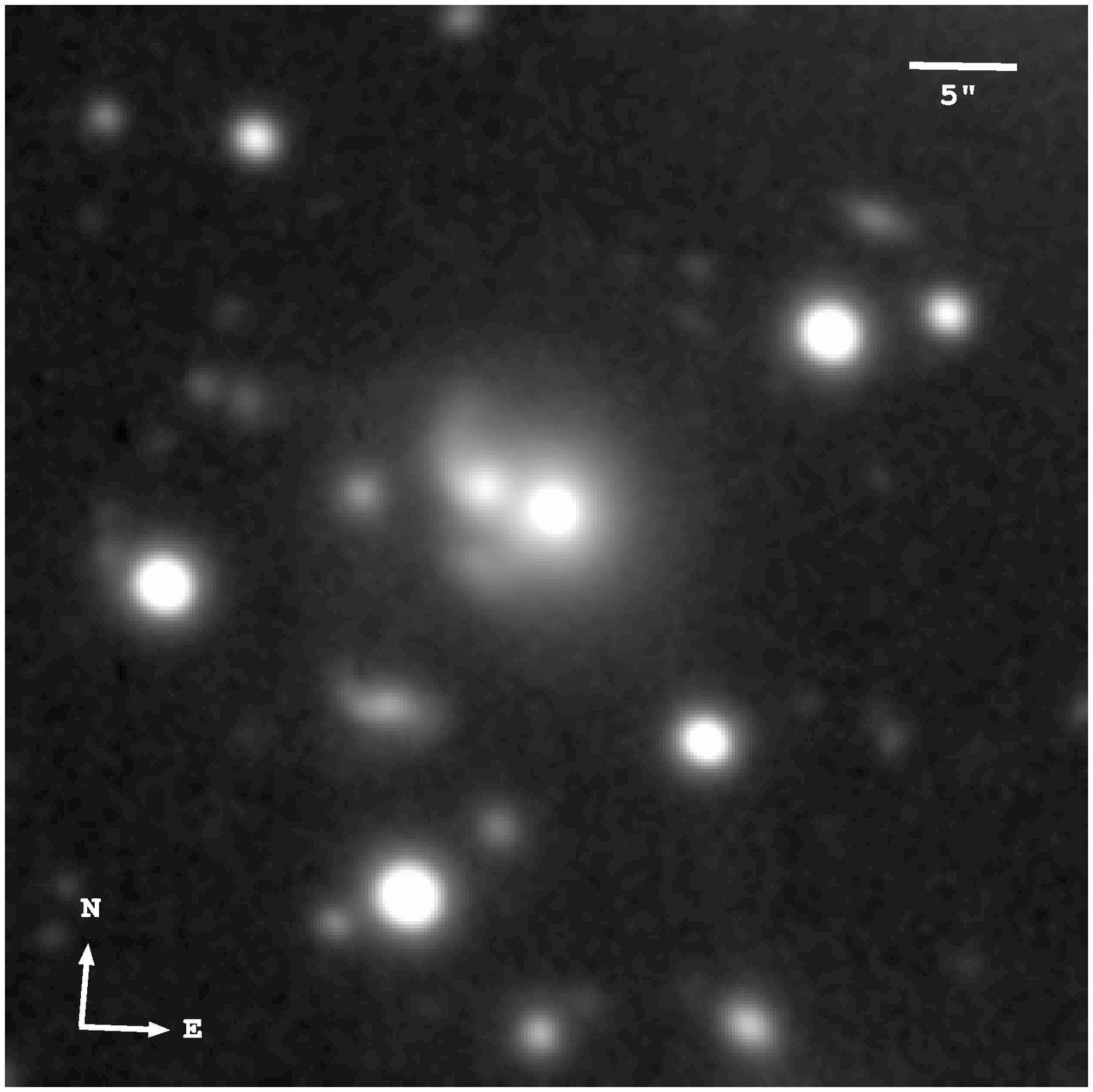}}
\caption{PKS 1934-63. (a) Unsharp-masked image. (b) Median filtered image.}
\label{pks1934_online} 
\end{figure*}

\begin{figure*}
\centering
\subfigure[]{\includegraphics[width=8.0cm]{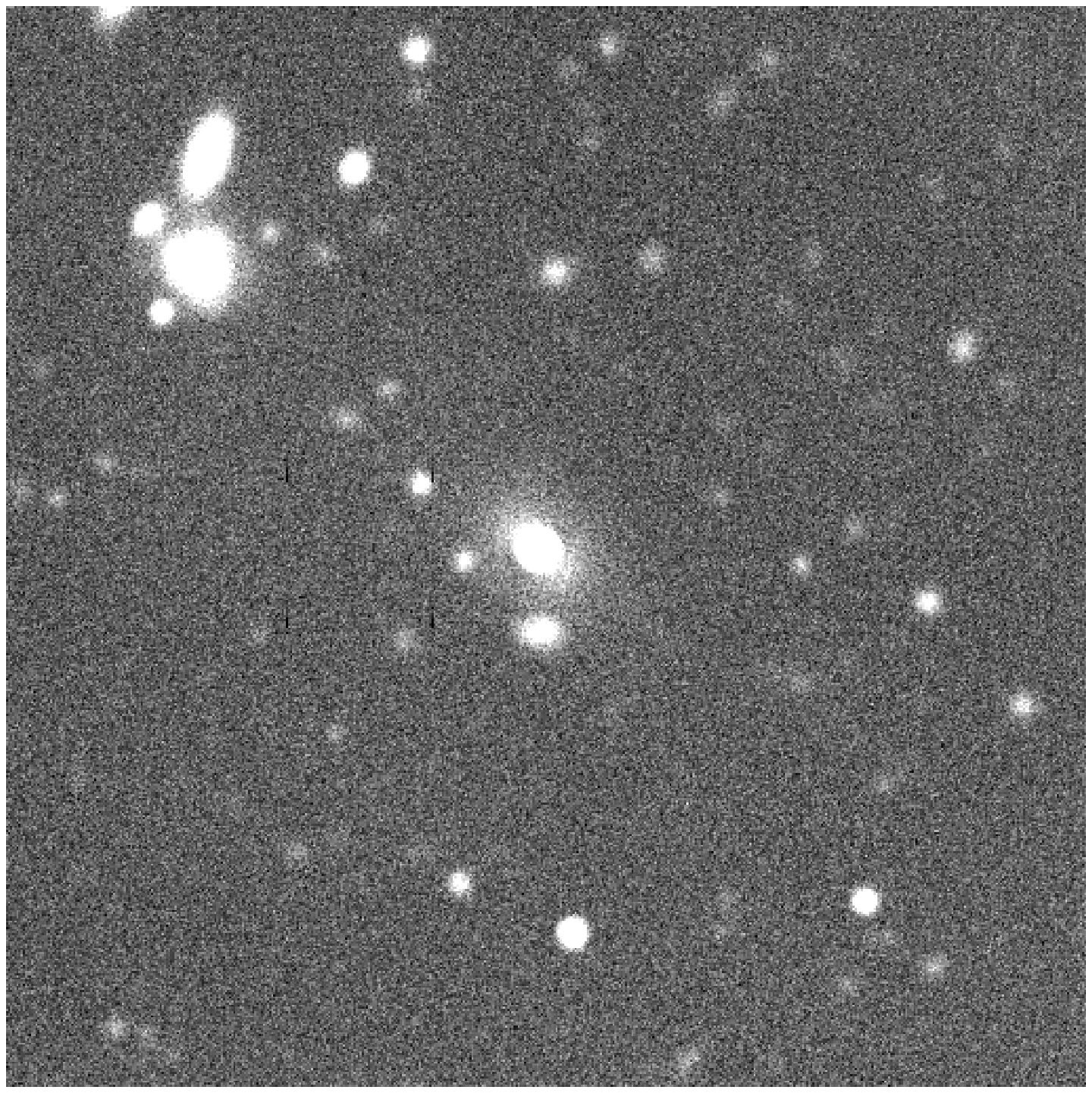}}
\subfigure[]{\includegraphics[width=8.0cm]{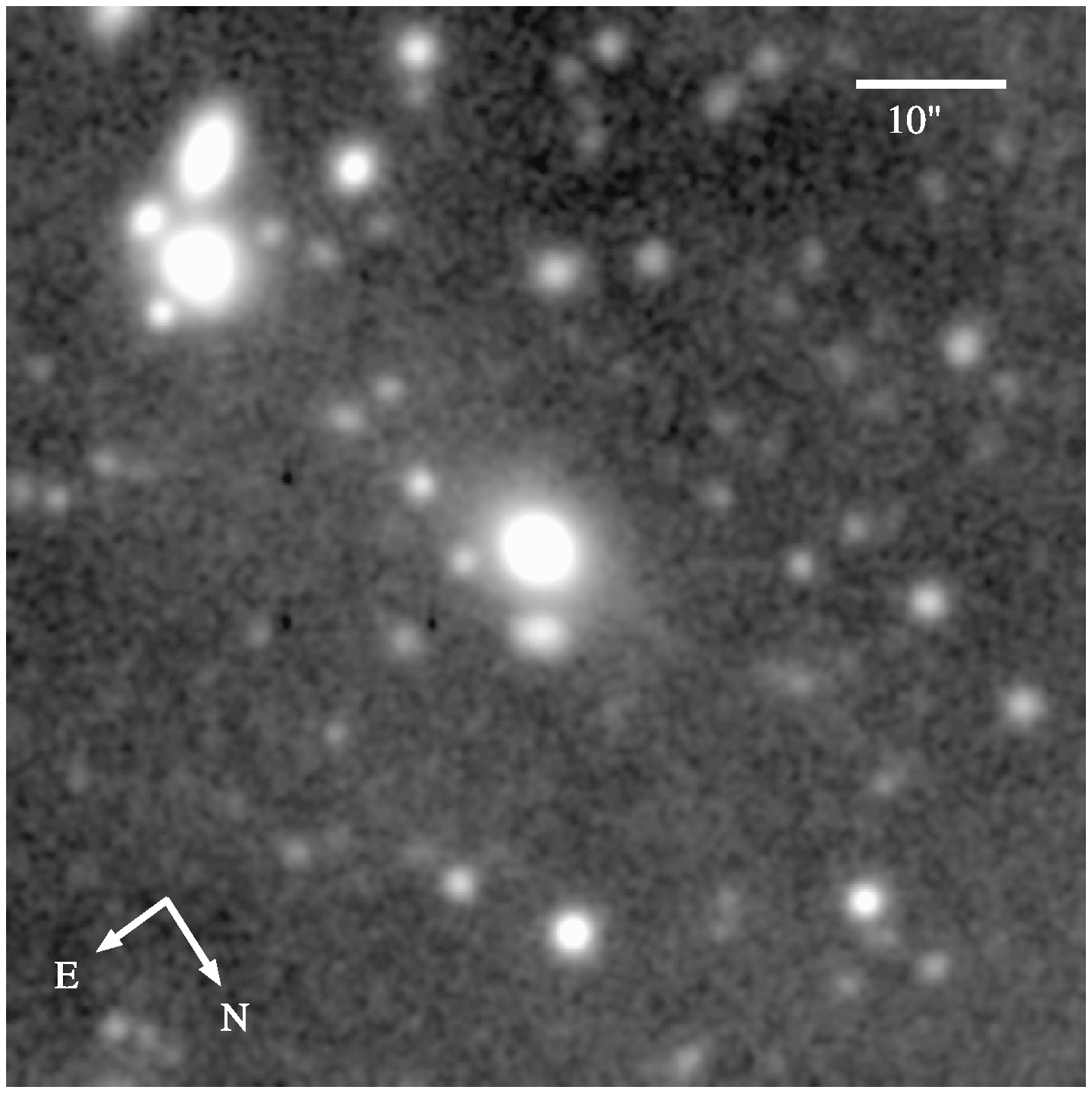}}
\caption{PKS 0038+09. (a) Unsharp-masked image. (b) Median filtered image.}
\label{pks0038_online} 
\end{figure*}

\begin{figure*}
\centering
\subfigure[]{\includegraphics[width=8.0cm]{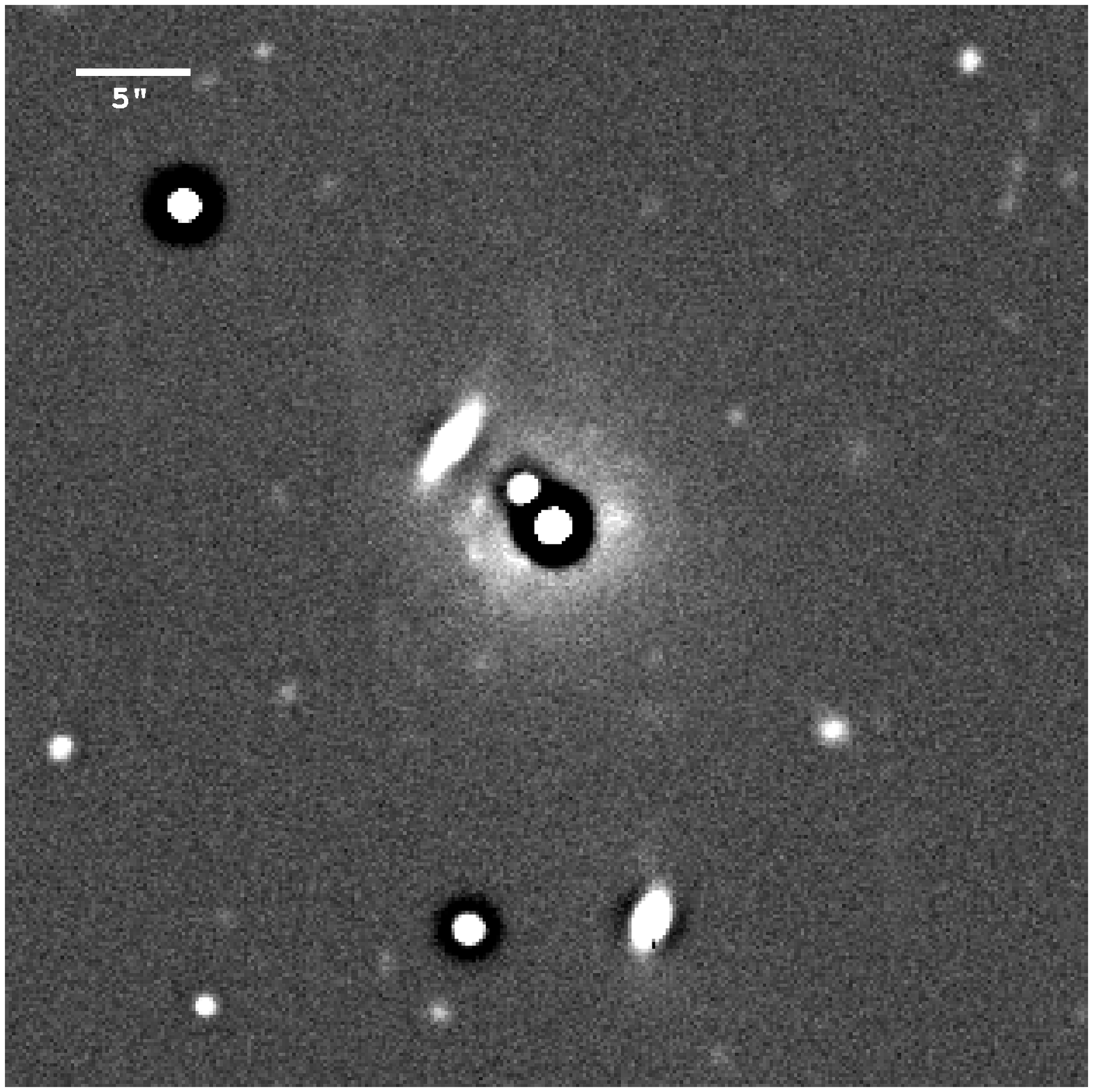}\label{pks2135_14_online_a}}
\subfigure[]{\includegraphics[width=8.0cm]{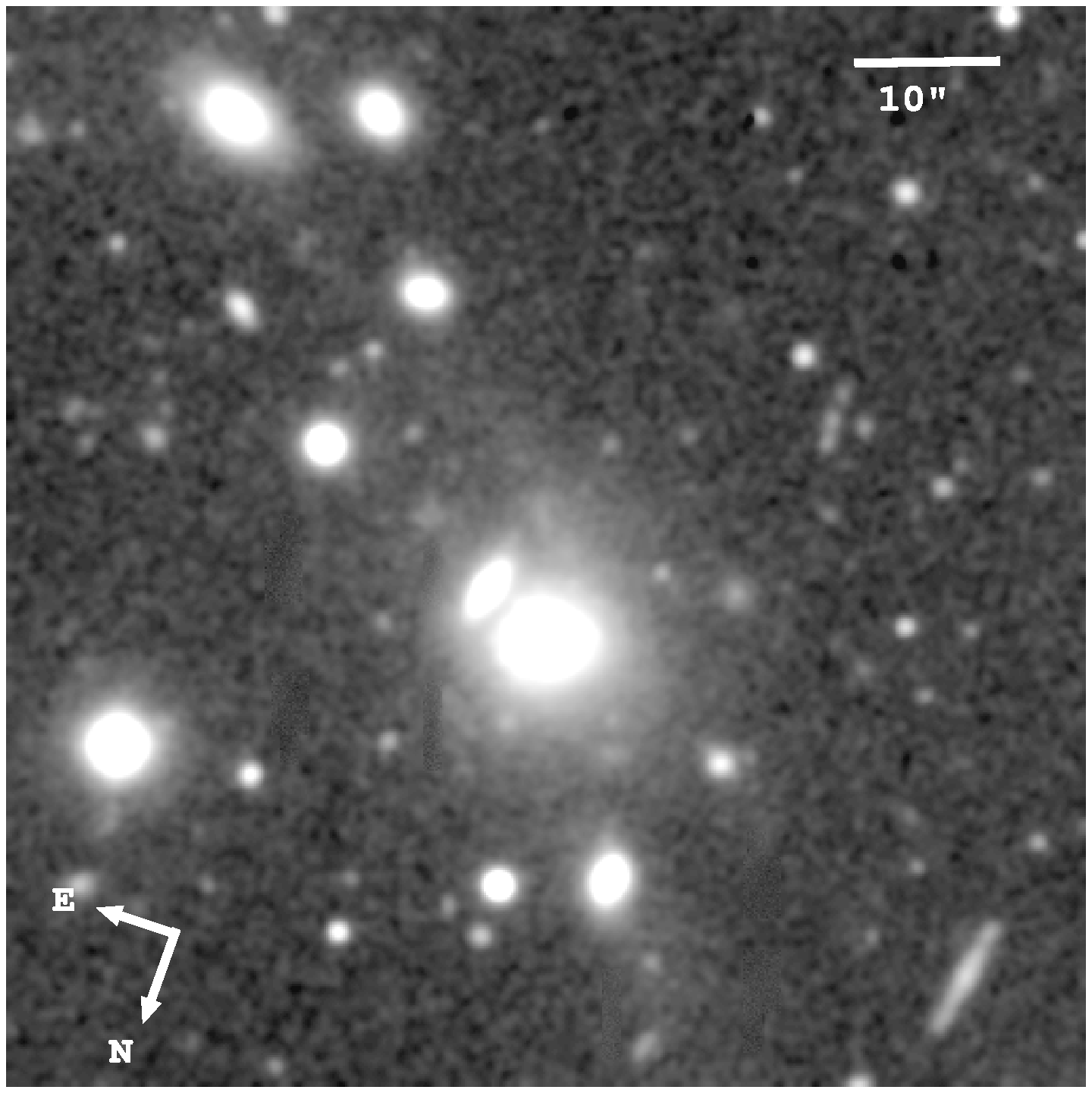}\label{pks2135_14_online_b}}
\caption{PKS 2135-14. (a) Unsharp-masked image. (b) Median filtered image.}
\label{pks2135_14_online} 
\end{figure*}

\begin{figure*}
\centering
\subfigure[]{\includegraphics[width=8.0cm]{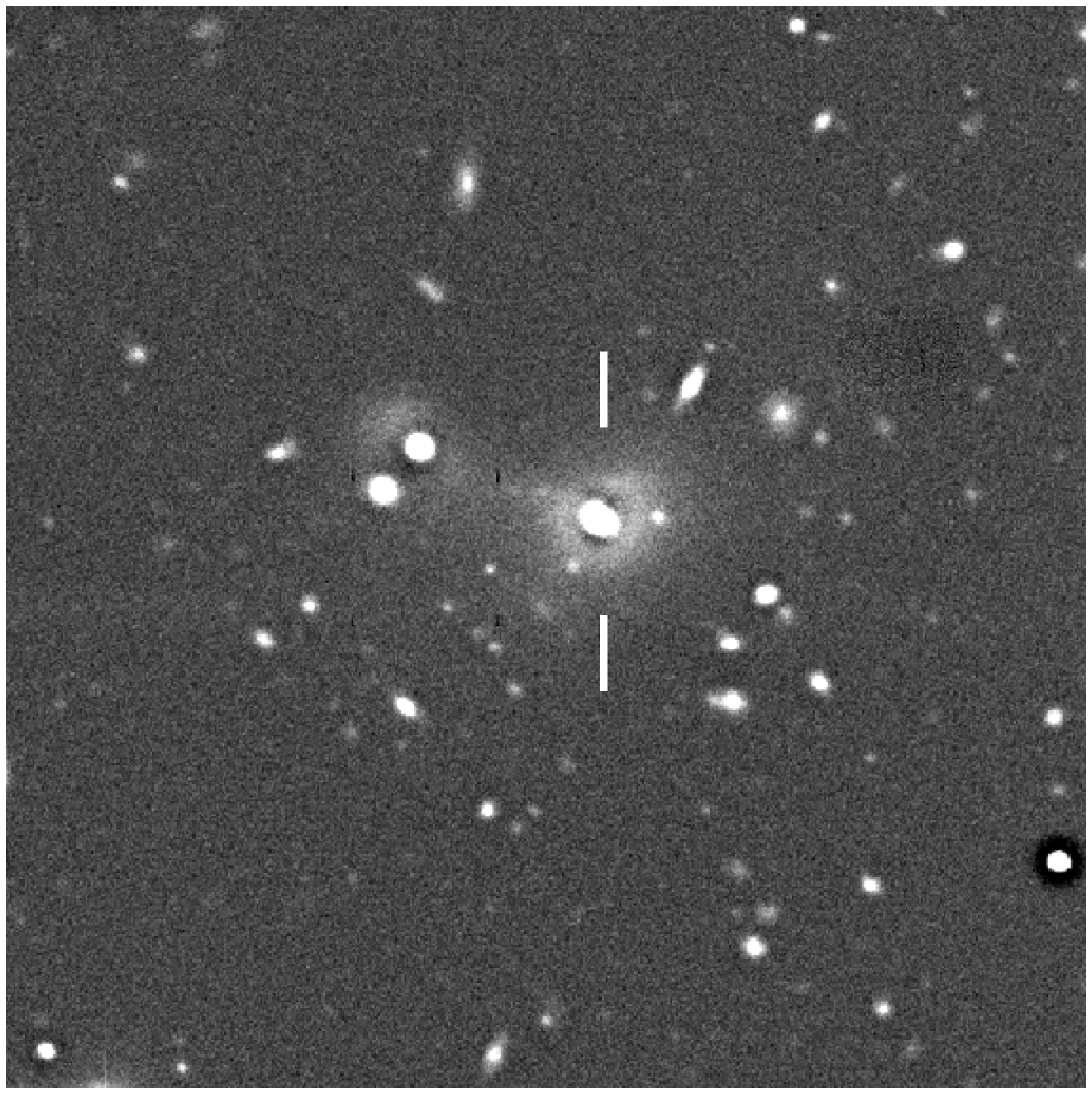}}
\subfigure[]{\includegraphics[width=8.0cm]{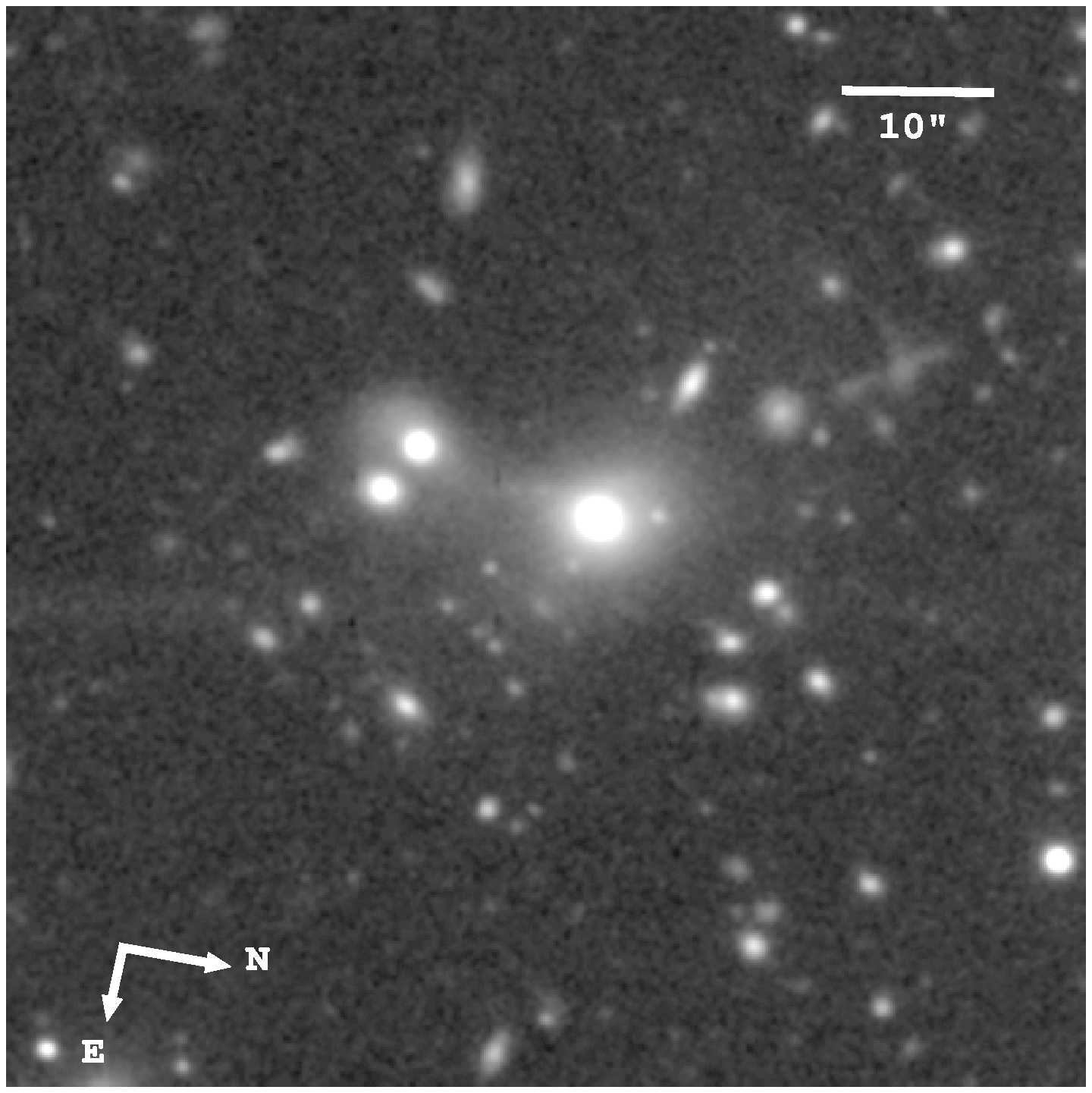}}
\caption{PKS 0035-02. (a) Unsharp-masked image. (b) Median filtered image.}
\label{pks0035_online} 
\end{figure*}

\begin{figure*}
\centering
\subfigure[]{\includegraphics[width=8.0cm]{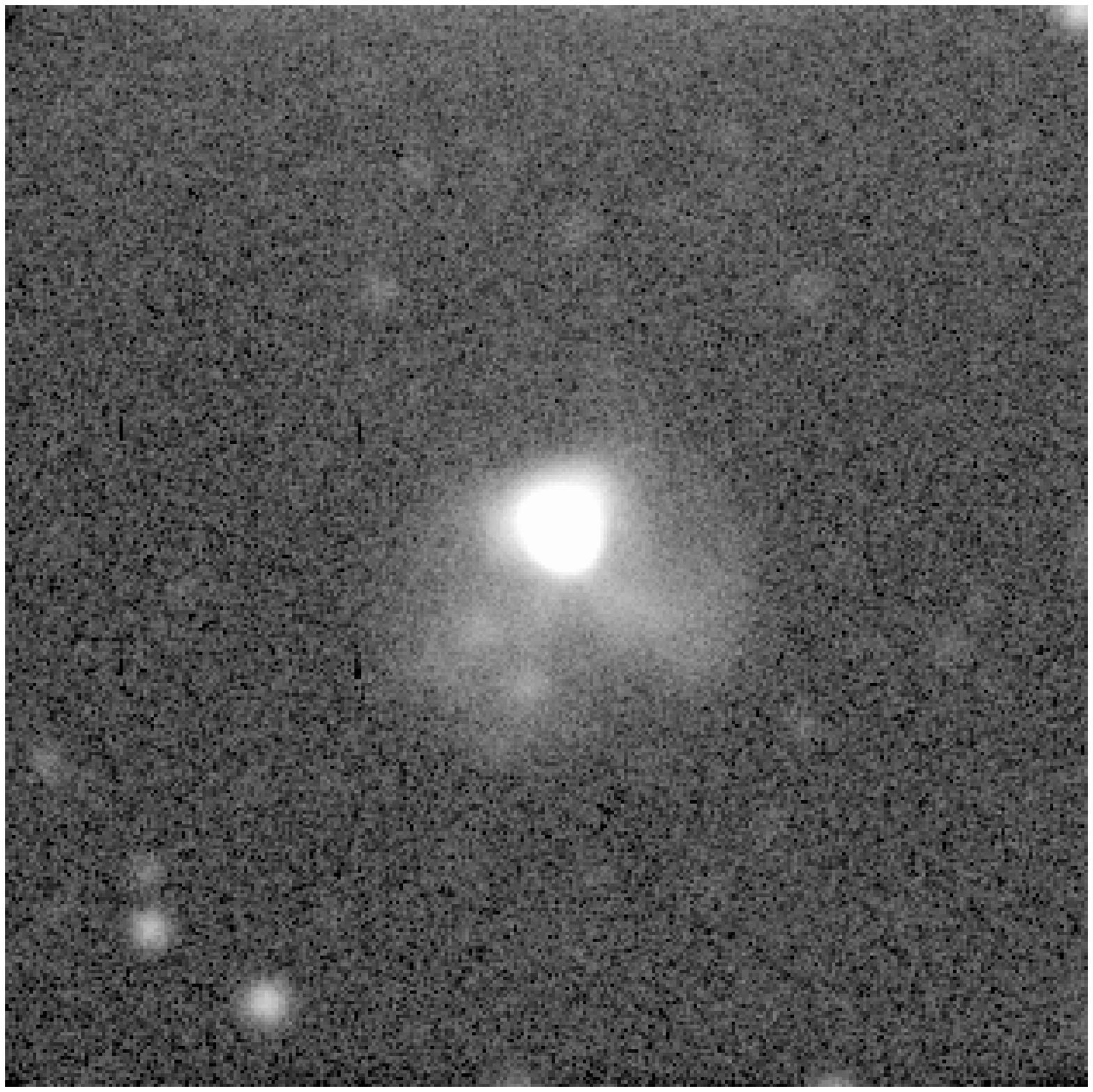}}
\subfigure[]{\includegraphics[width=8.0cm]{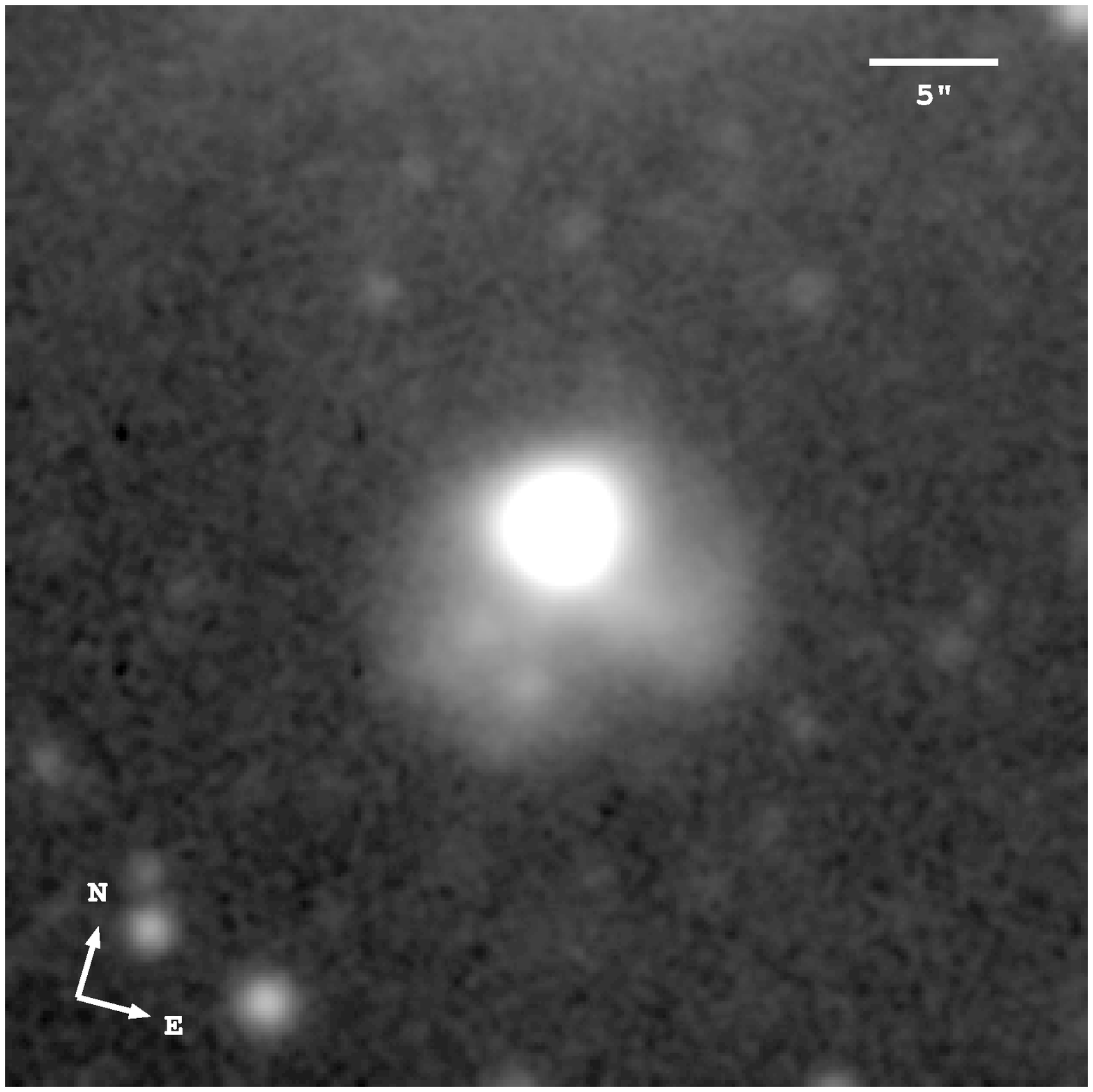}}
\caption{PKS 2314+03. (a) Unsharp-masked image. (b) Median filtered image.}
\label{pks2314_online} 
\end{figure*}

\begin{figure*}
\centering
\subfigure[]{\includegraphics[width=8.0cm]{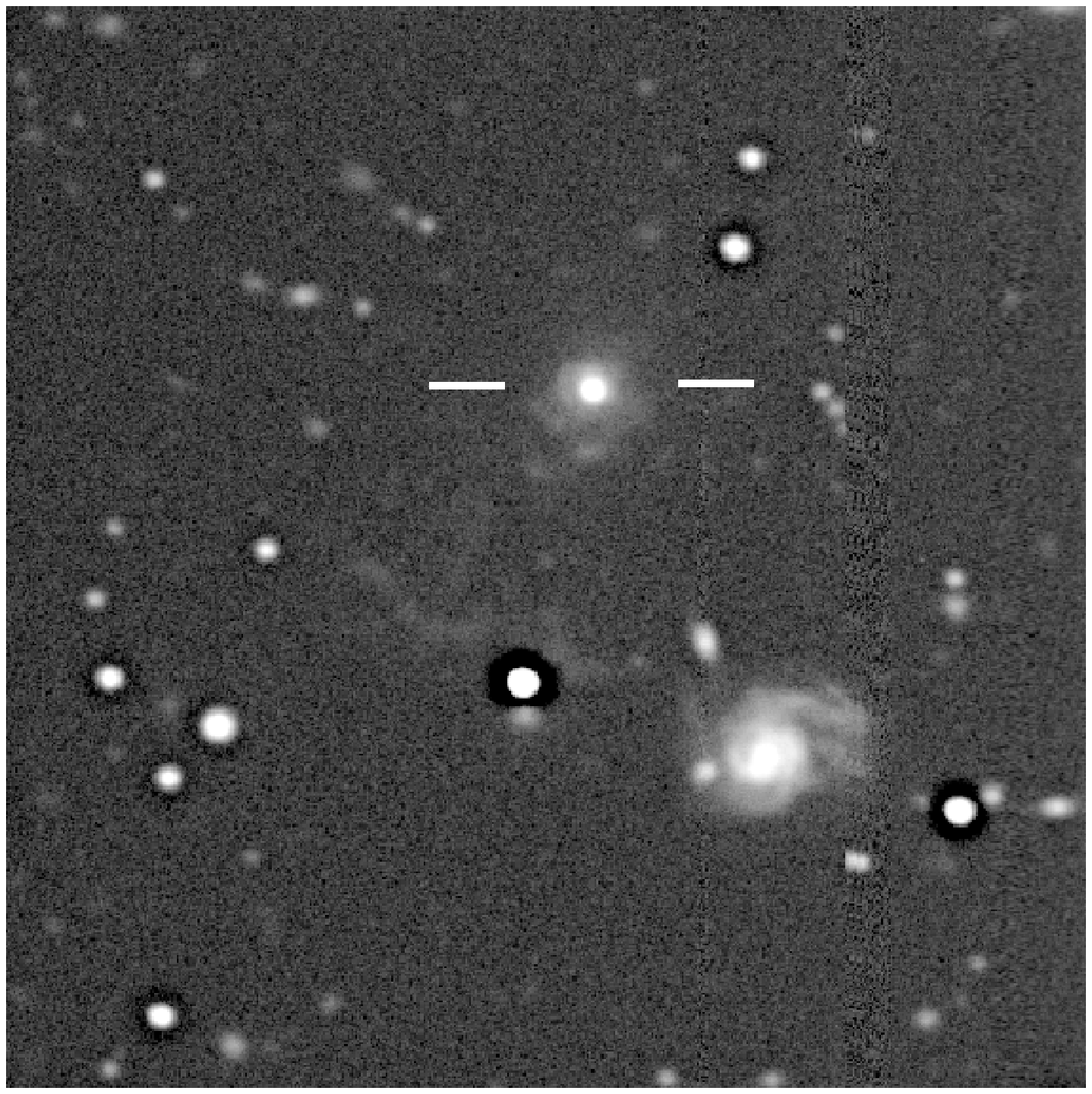}}
\subfigure[]{\includegraphics[width=8.0cm]{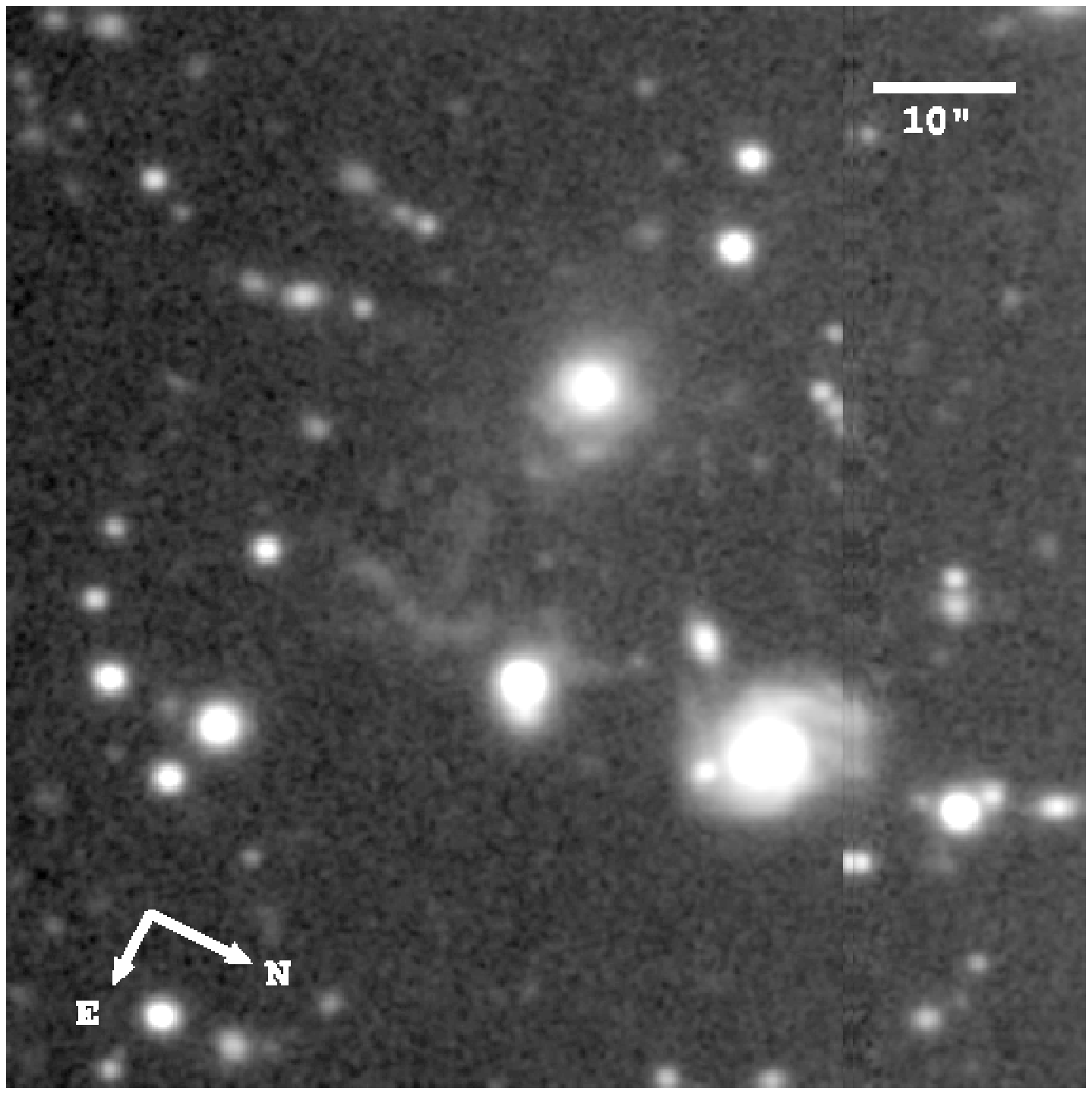}}
\caption{PKS 1932-46. (a) Unsharp-masked image. (b) Median filtered image.}
\label{pks1932_online} 
\end{figure*}

\begin{figure*}
\centering
\subfigure[]{\includegraphics[width=8.0cm]{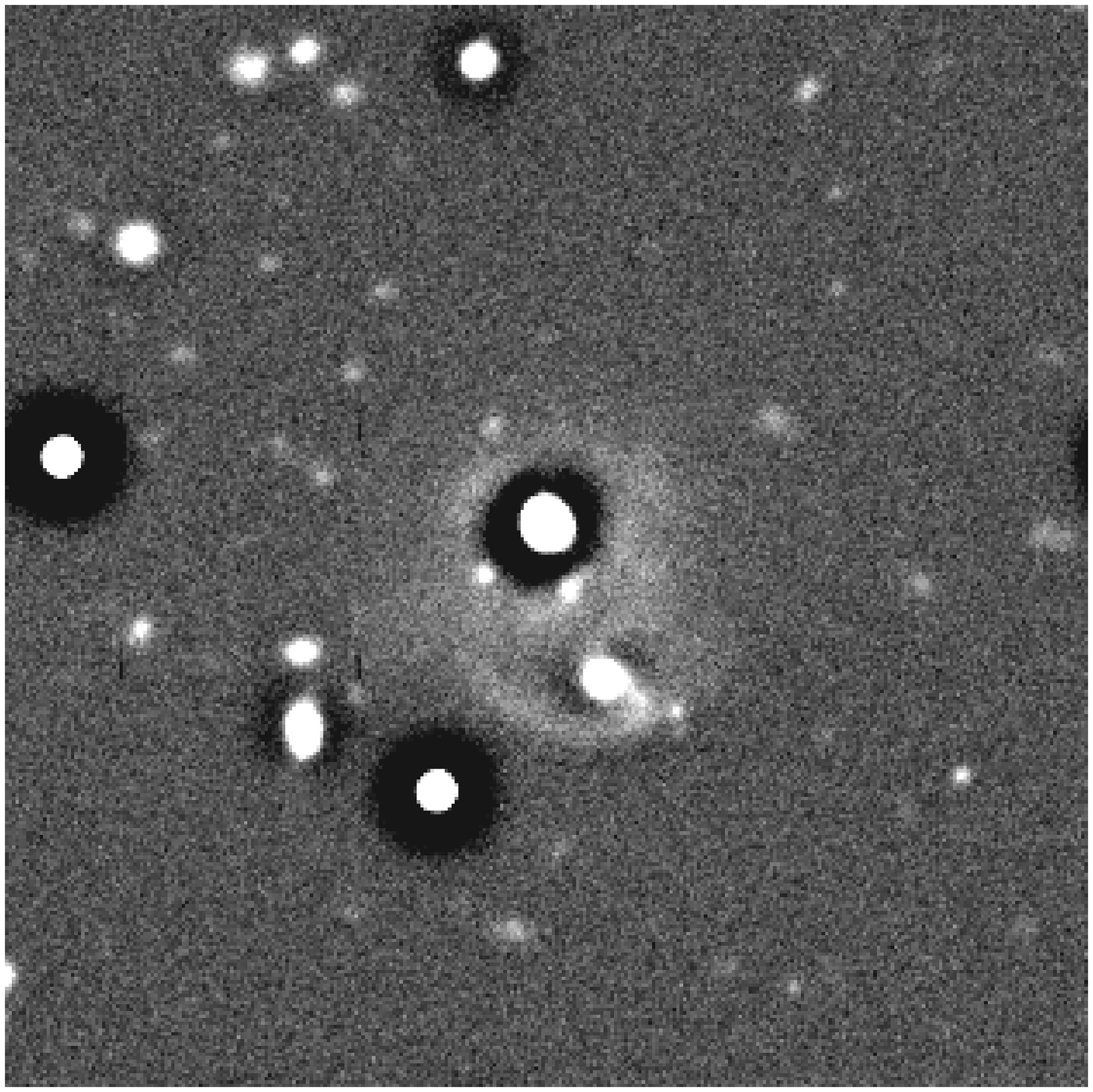}}
\subfigure[]{\includegraphics[width=8.0cm]{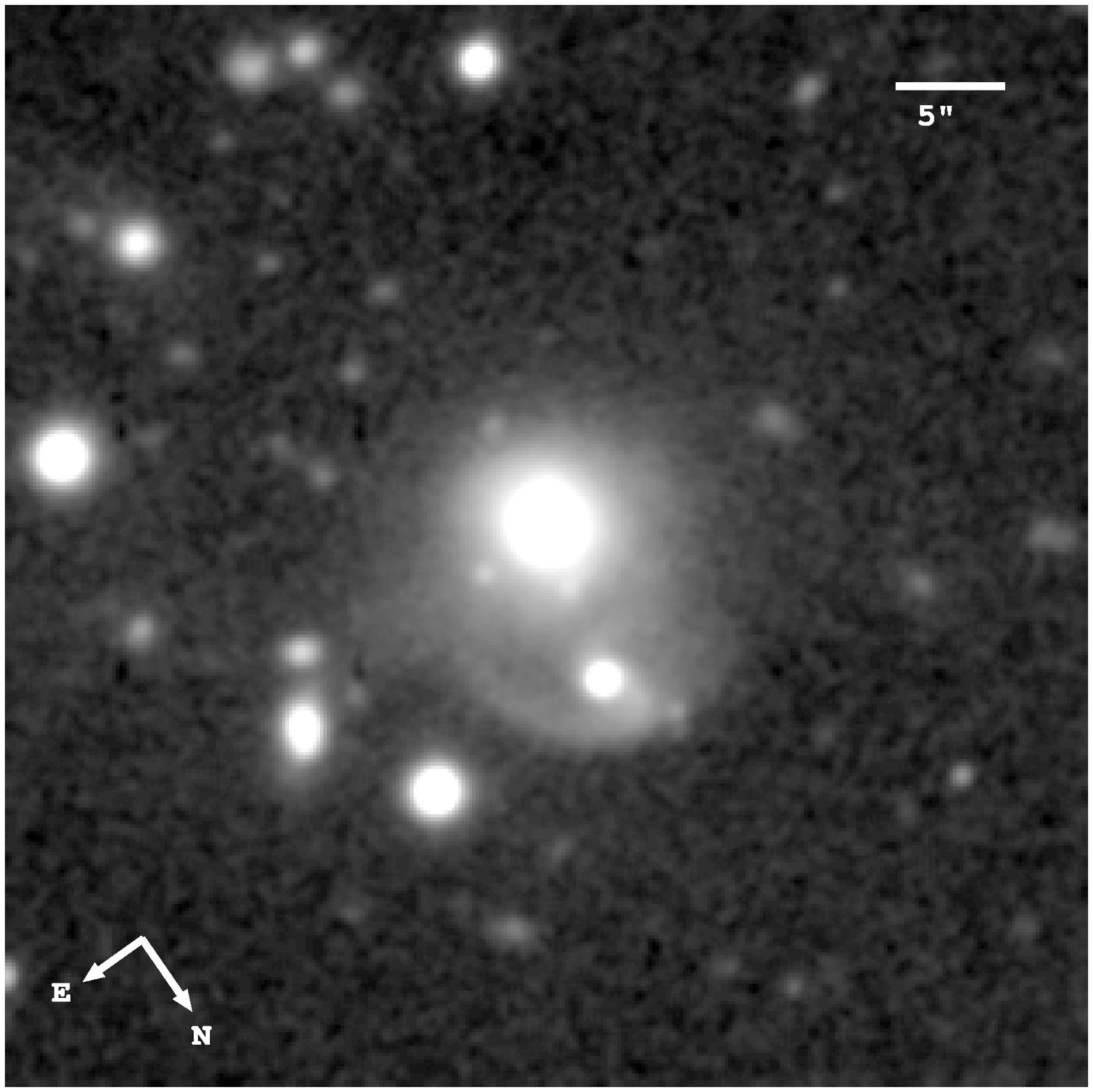}}
\caption{PKS 1151-34. (a) Unsharp-masked image. (b) Median filtered image.}
\label{pks1151_online} 
\end{figure*}

\begin{figure*}
\centering
\subfigure[]{\includegraphics[width=8.0cm]{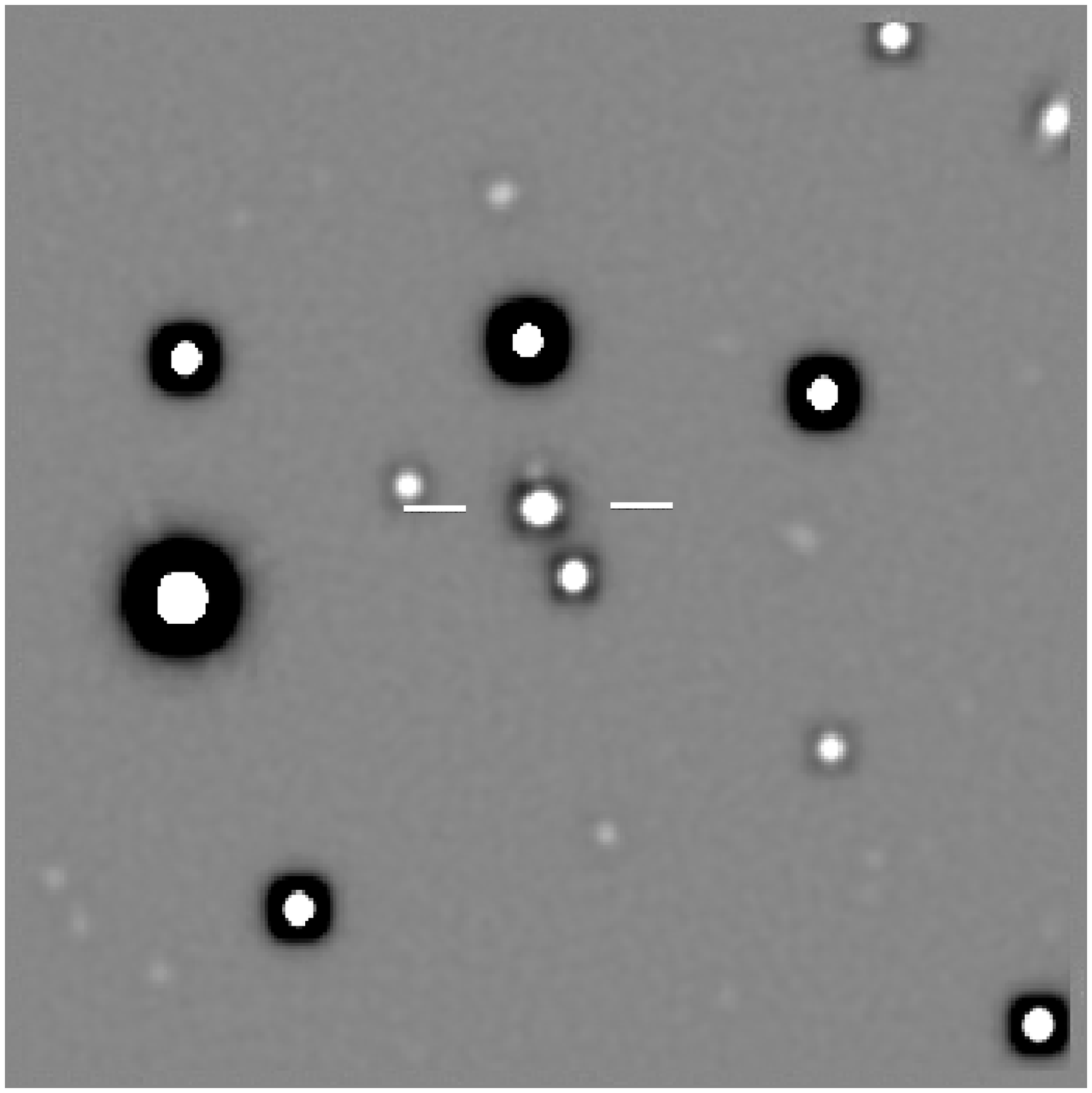}}
\subfigure[]{\includegraphics[width=8.0cm]{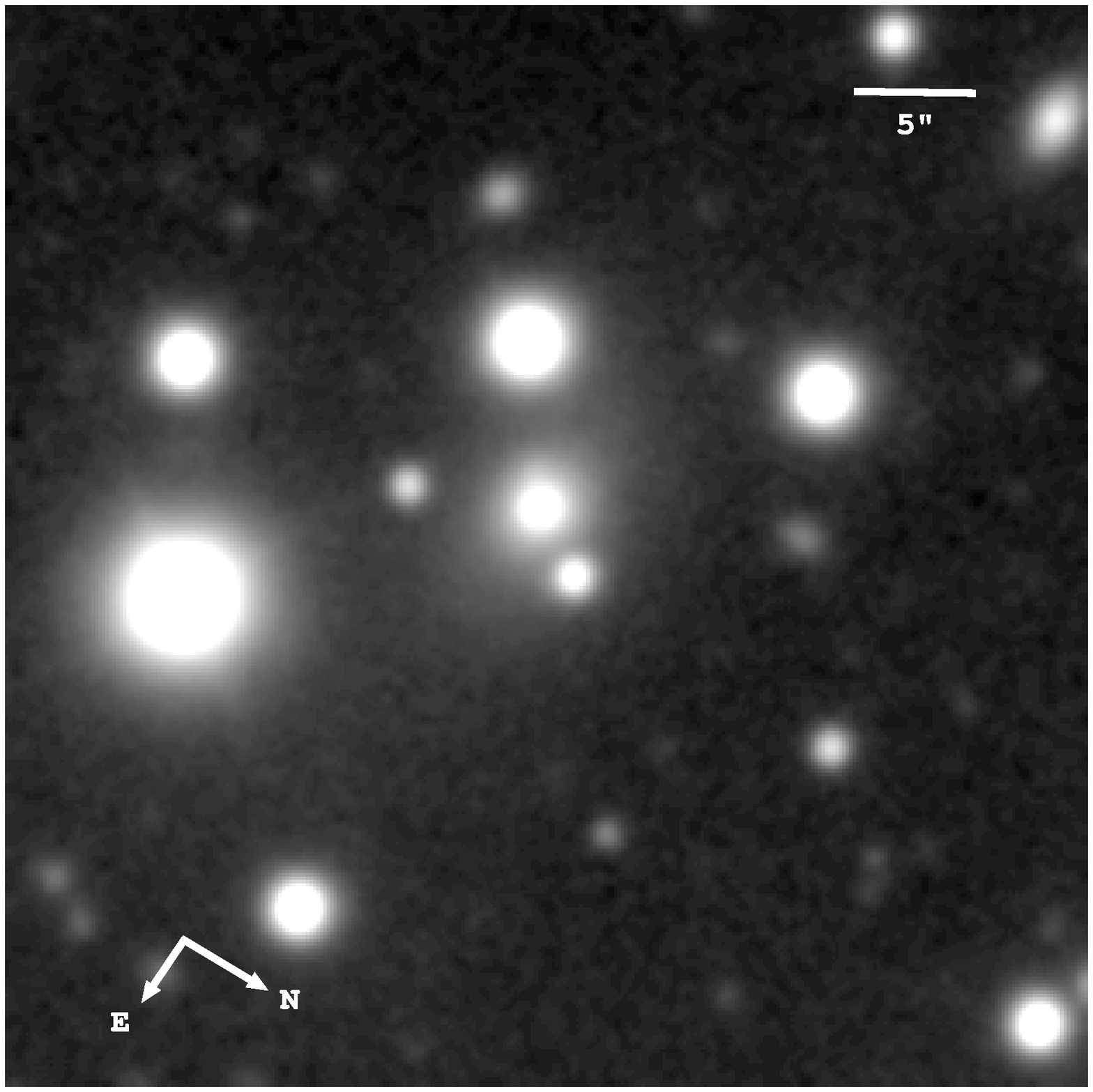}}
\caption{PKS 0859-25. (a) Smooth galaxy-subtracted image. (b) Median filtered image.}
\label{pks0859_online} 
\end{figure*}

\begin{figure*}
\centering
\subfigure[]{\includegraphics[width=8.0cm]{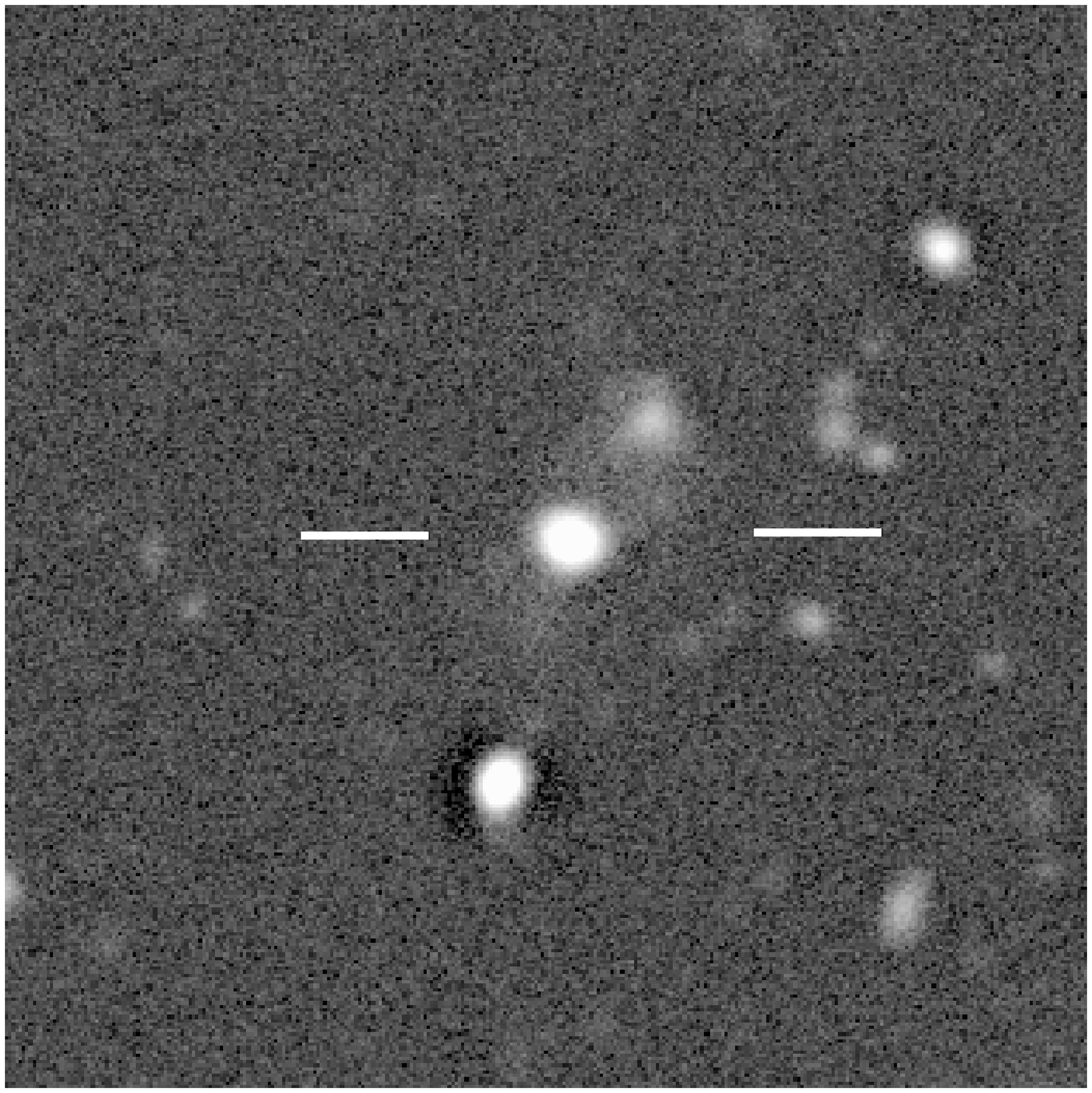}}
\subfigure[]{\includegraphics[width=8.0cm]{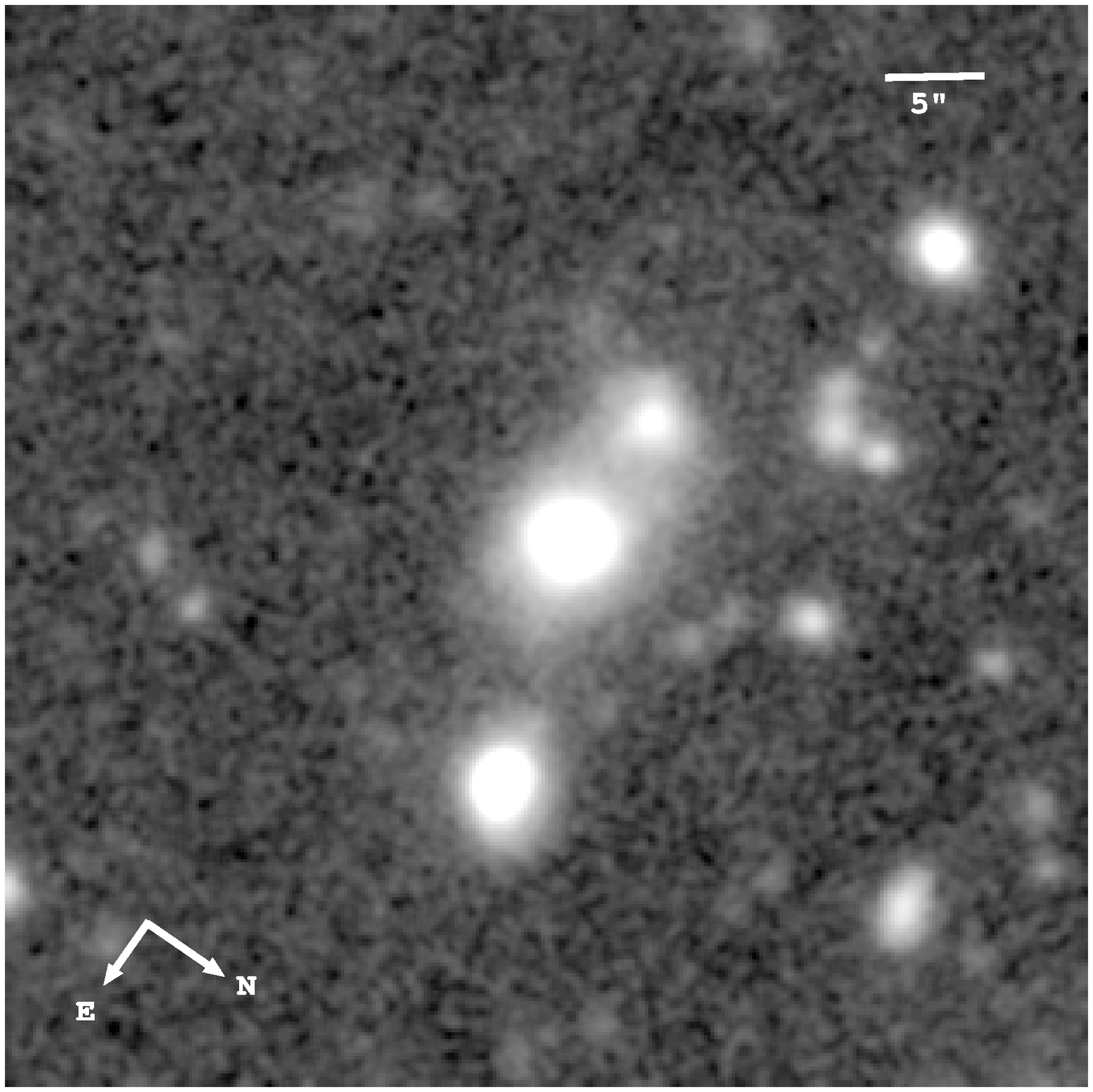}}
\caption{PKS 2250-41. (a) Unsharp-masked image. (b) Median filtered image.}
\label{pks2250_online} 
\end{figure*}

\begin{figure*}
\centering
\subfigure[]{\includegraphics[width=8.0cm]{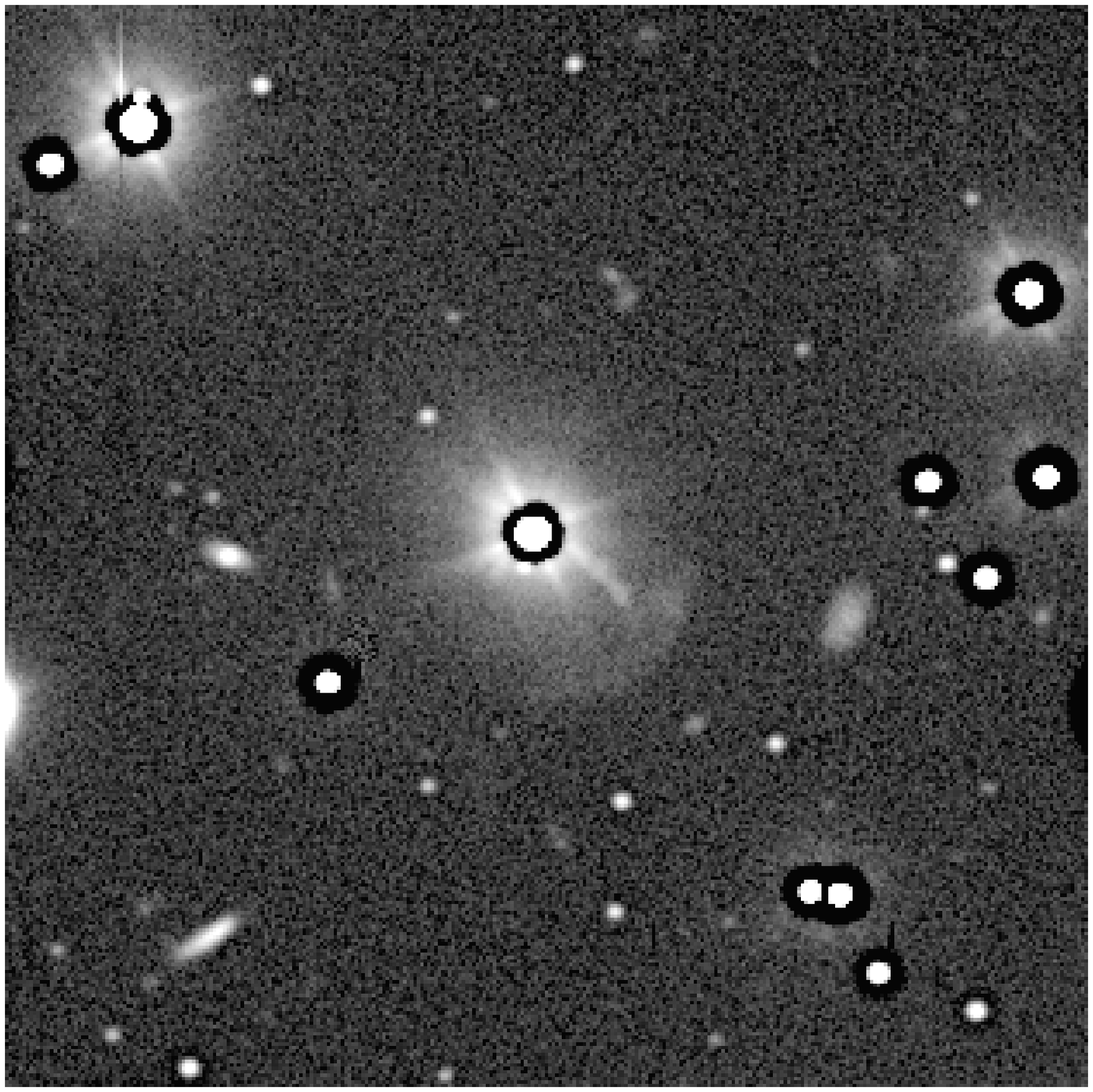}}
\subfigure[]{\includegraphics[width=8.0cm]{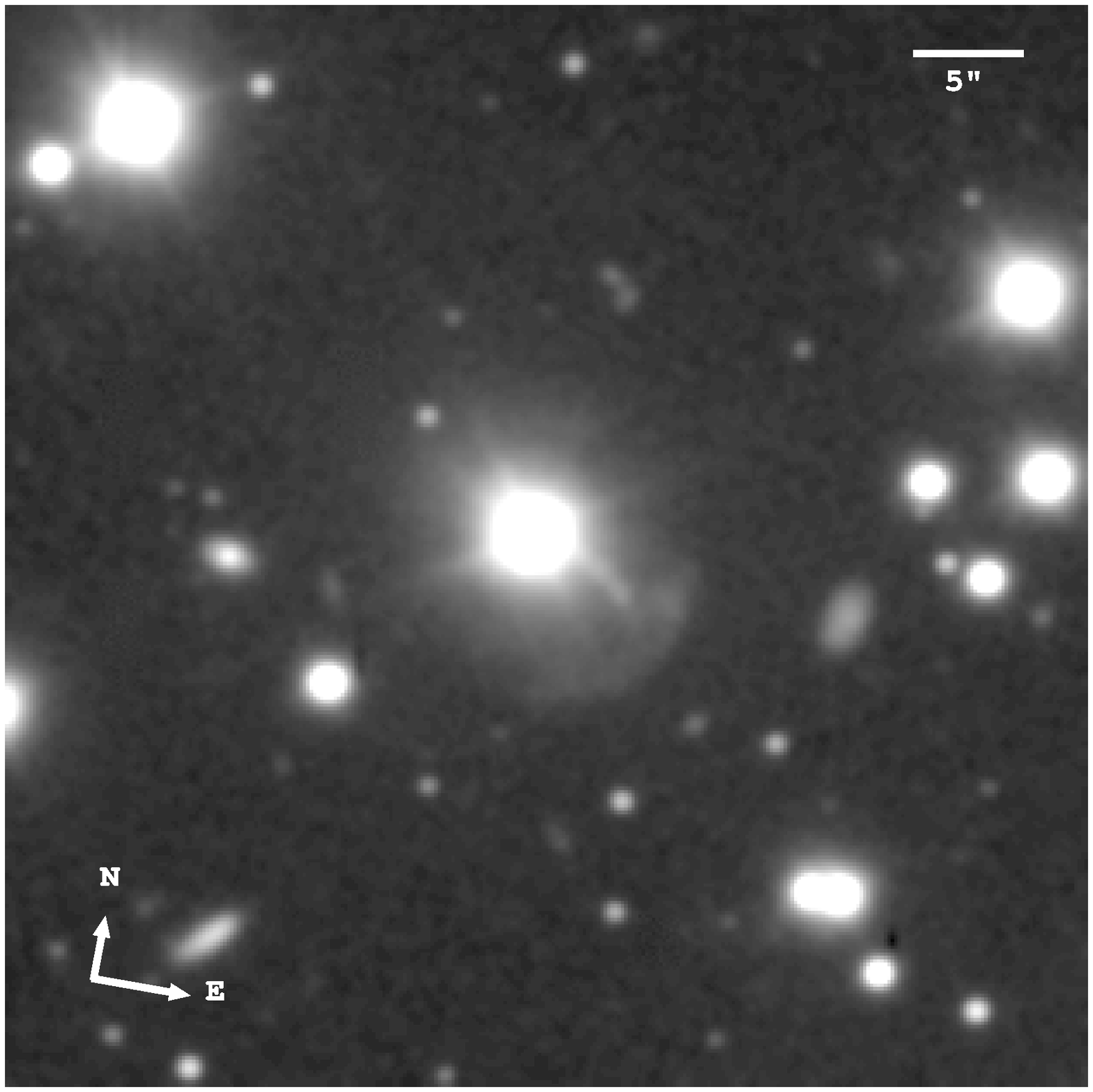}}
\caption{PKS 1355-41. (a) Unsharp-masked image. (b) Median filtered image.}
\label{pks1355_online} 
\end{figure*}

\begin{figure*}
\centering
\subfigure[]{\includegraphics[width=8.0cm]{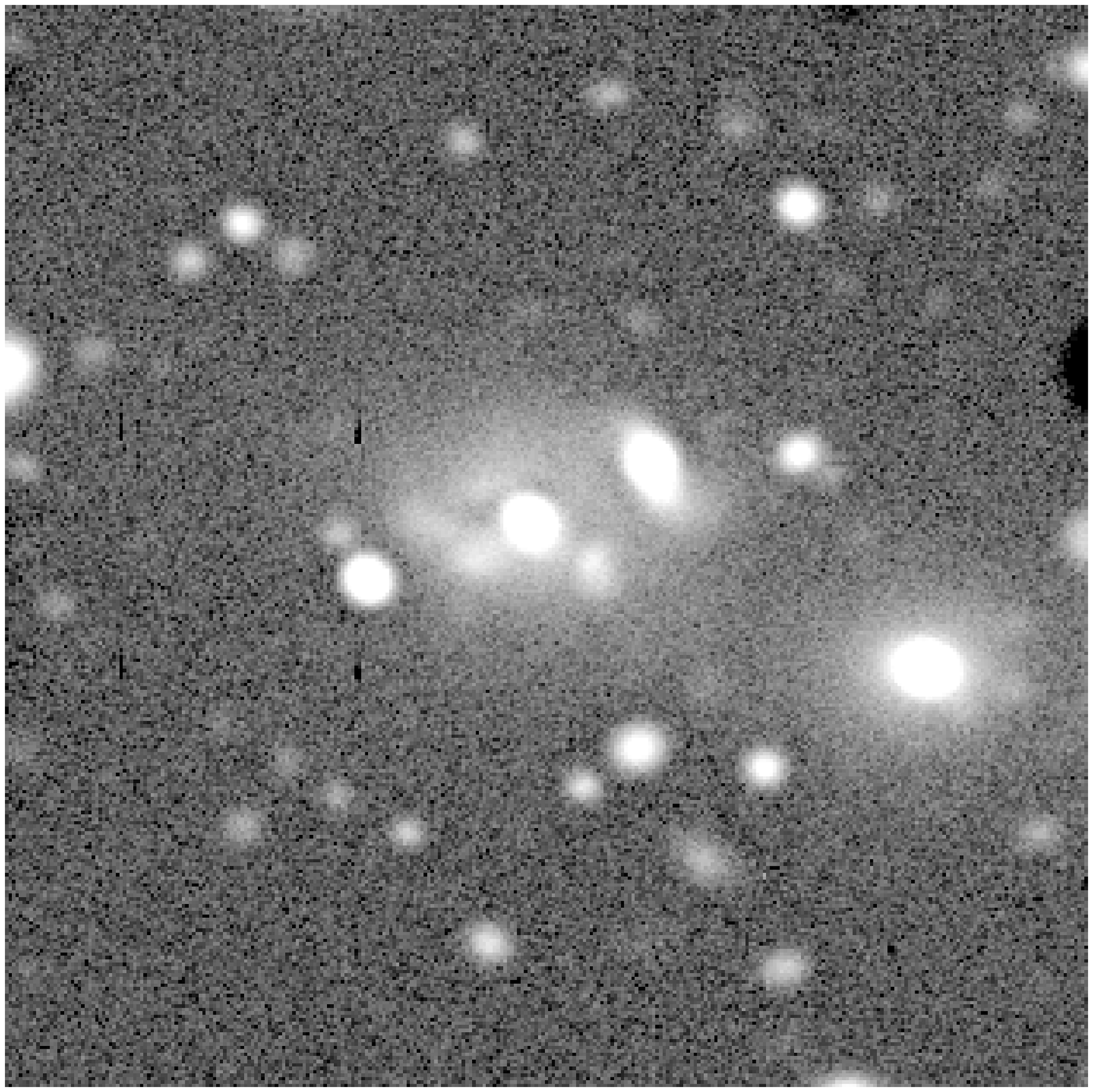}}
\subfigure[]{\includegraphics[width=8.0cm]{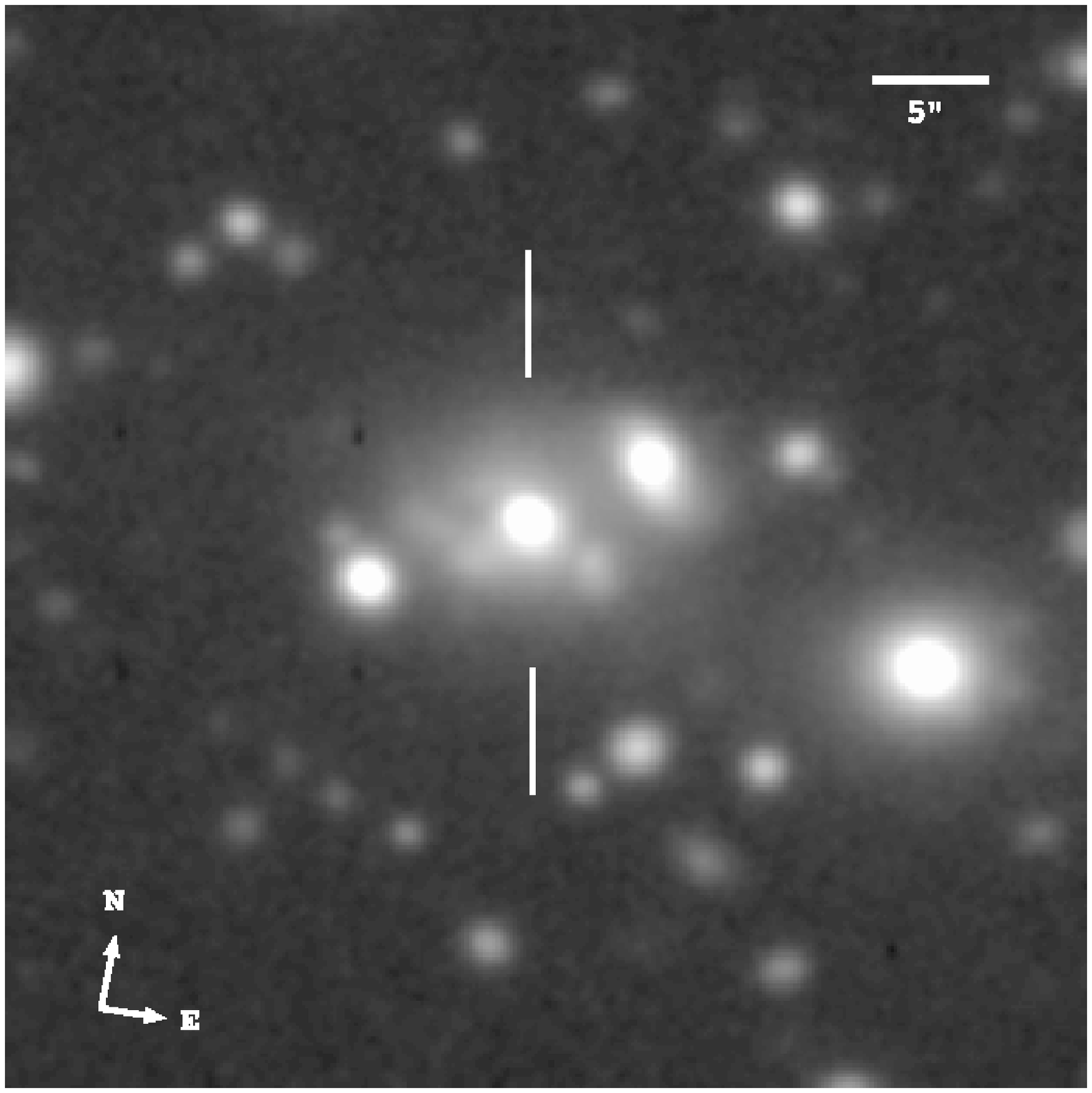}}
\caption{PKS 0023-26. (a) Unsharp-masked image. (b) Median filtered image.}
\label{pks0023_online} 
\end{figure*}

\begin{figure*}
\centering
\subfigure[]{\includegraphics[width=8.0cm]{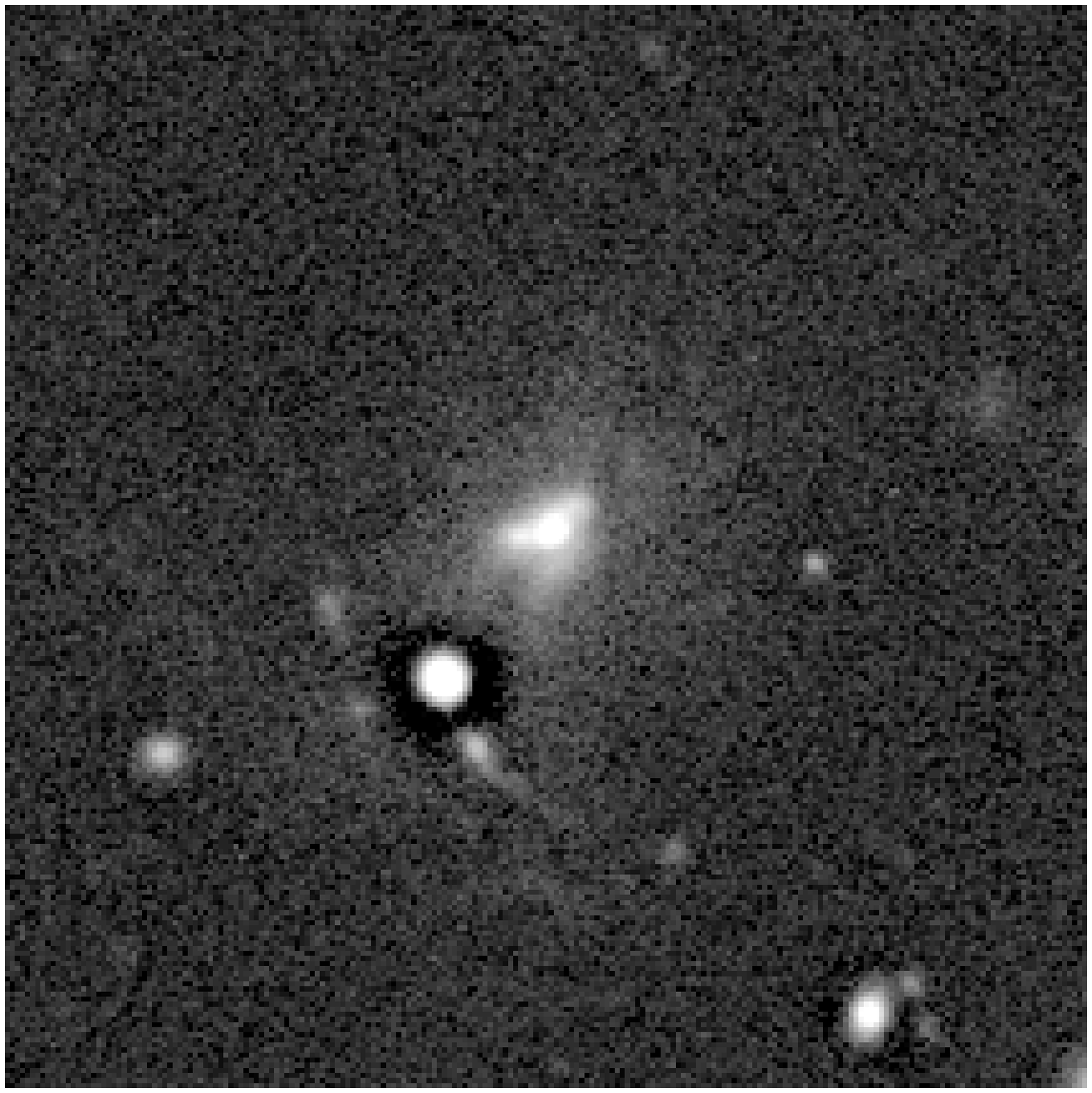}}
\subfigure[]{\includegraphics[width=8.0cm]{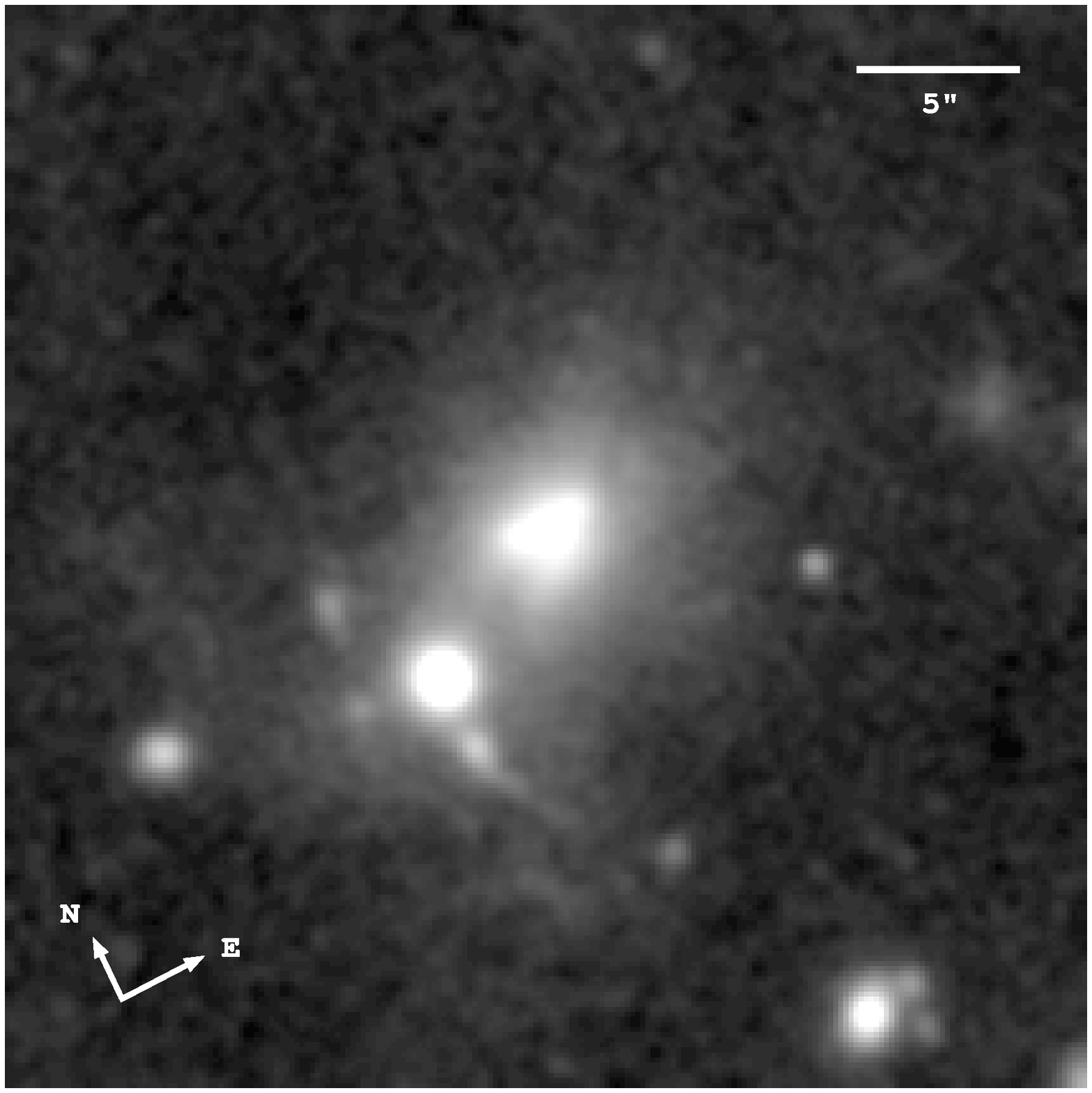}}
\caption{PKS 0347+05. (a) Unsharp-masked image. (b) Median filtered image.}
\label{pks0347_online} 
\end{figure*}

\begin{figure*}
\centering
\subfigure[]{\includegraphics[width=8.0cm]{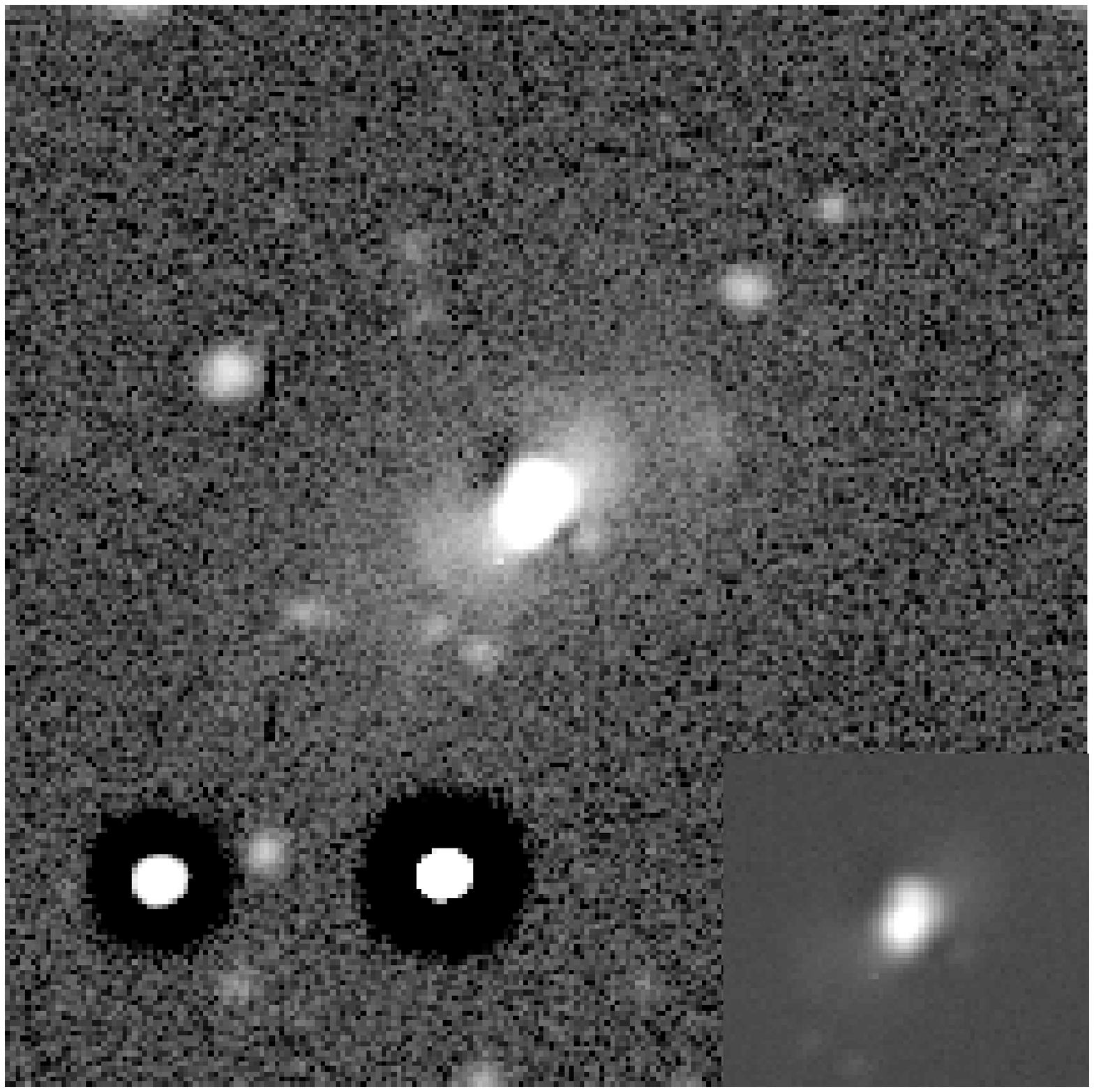}}
\subfigure[]{\includegraphics[width=8.0cm]{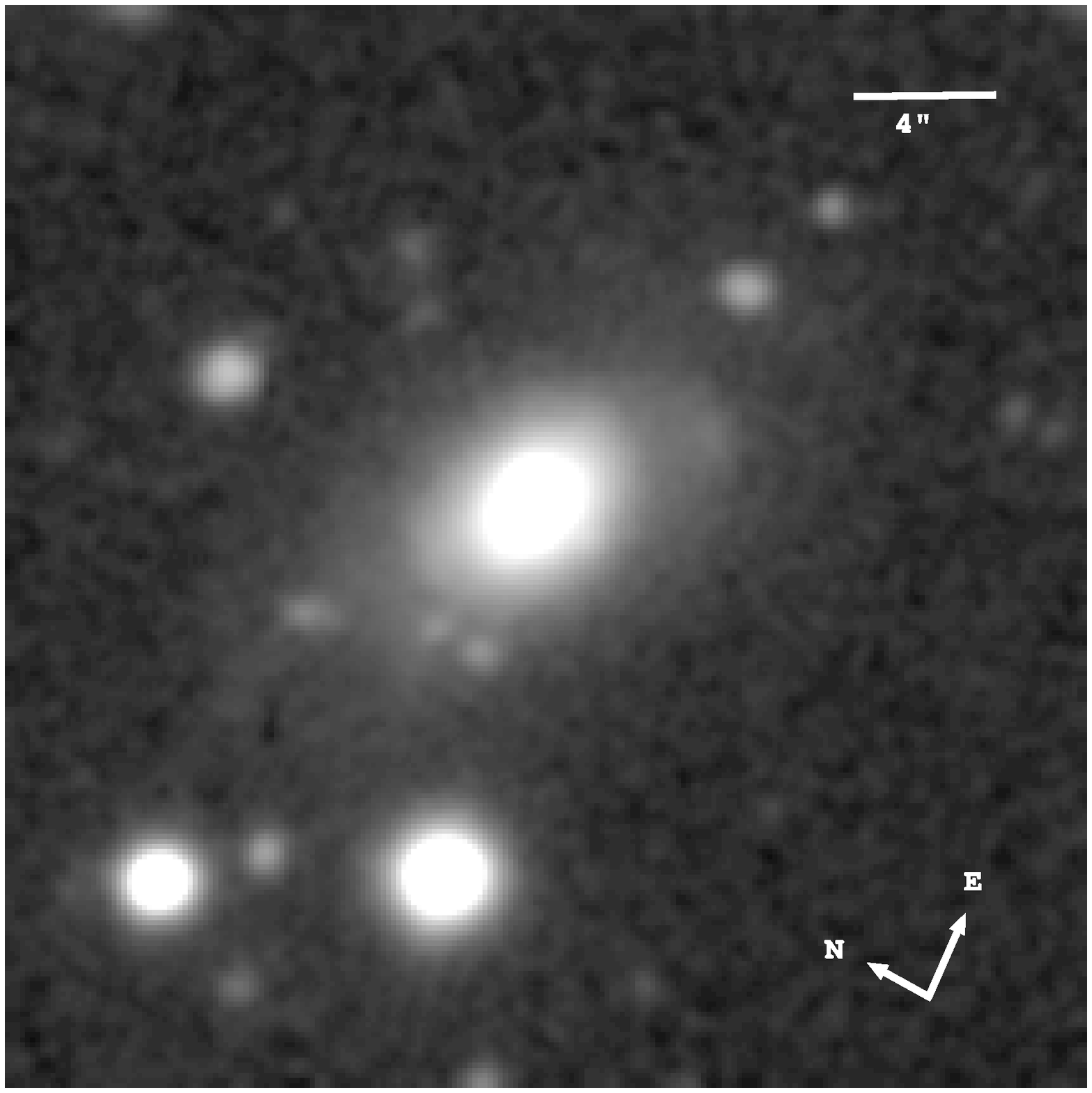}}
\caption{PKS 0039-44. (a) Unsharp-masked image. (b) Median filtered image.}
\label{pks0039_online} 
\end{figure*}

\begin{figure*}
\centering
\subfigure[]{\includegraphics[width=8.0cm]{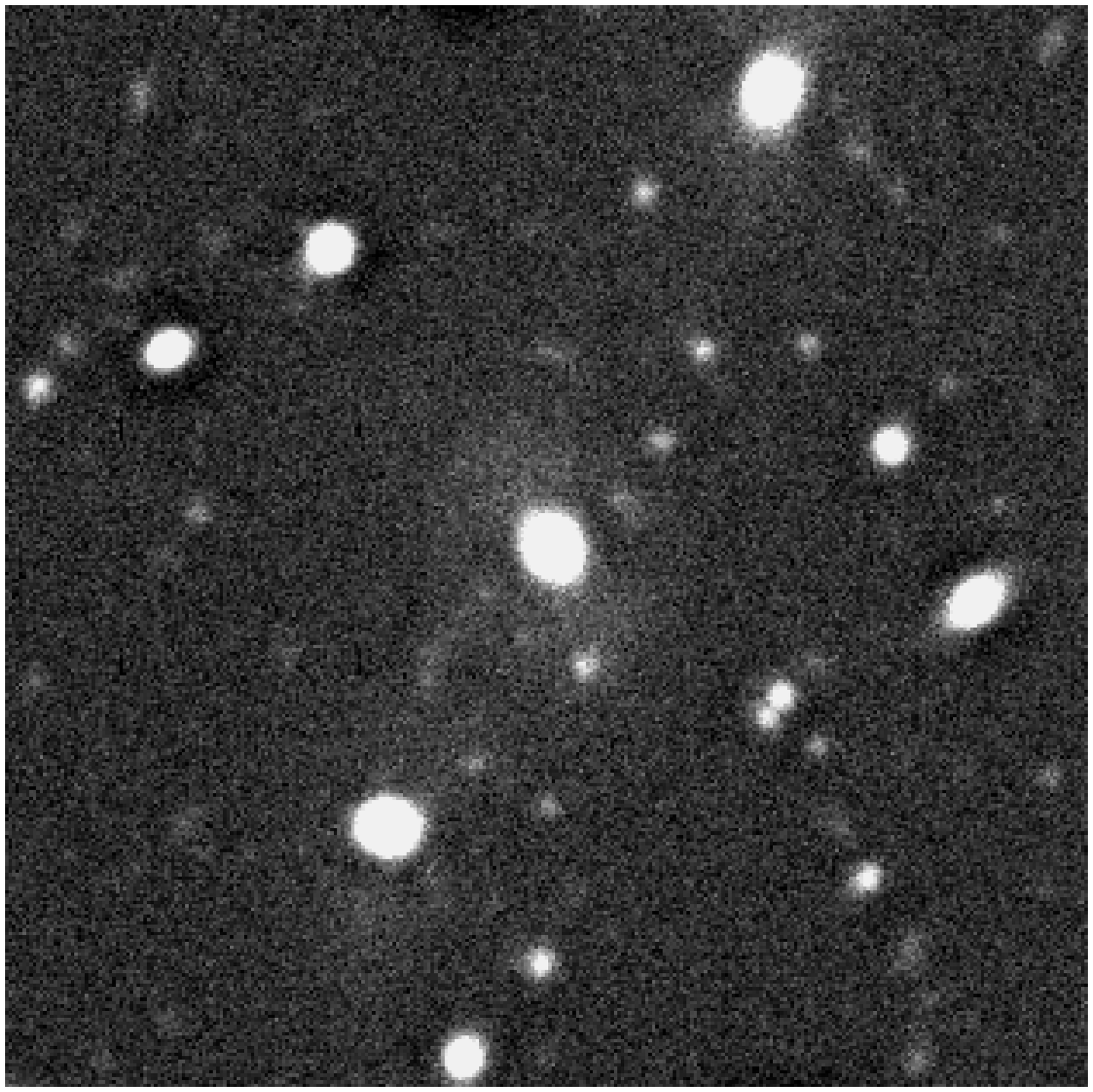}}
\subfigure[]{\includegraphics[width=8.0cm]{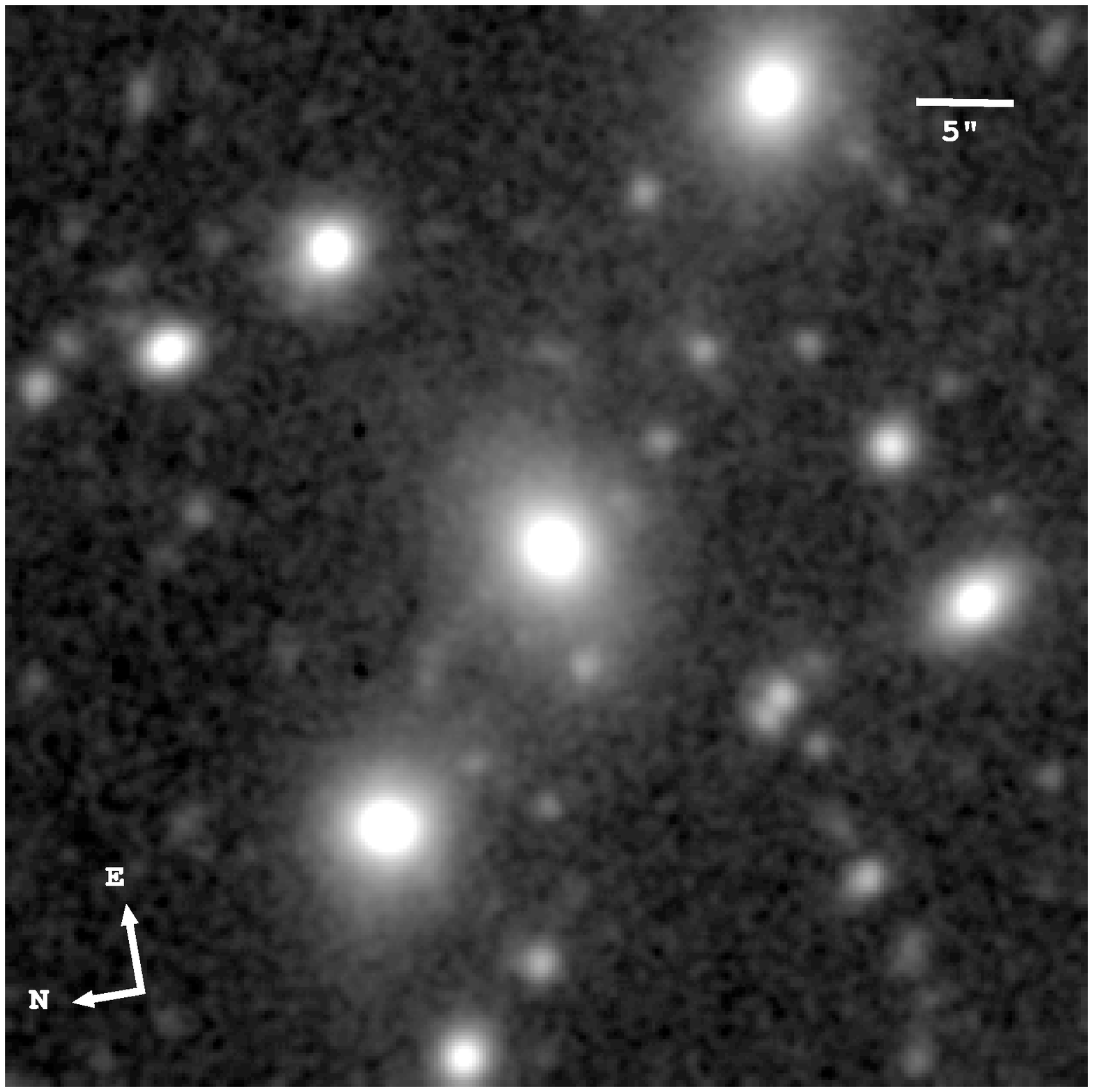}}
\caption{PKS 0105-16. (a) Unsharp-masked image. (b) Median filtered image.}
\label{pks0105_online} 
\end{figure*}

\begin{figure*}
\centering
\subfigure[]{\includegraphics[width=8.0cm]{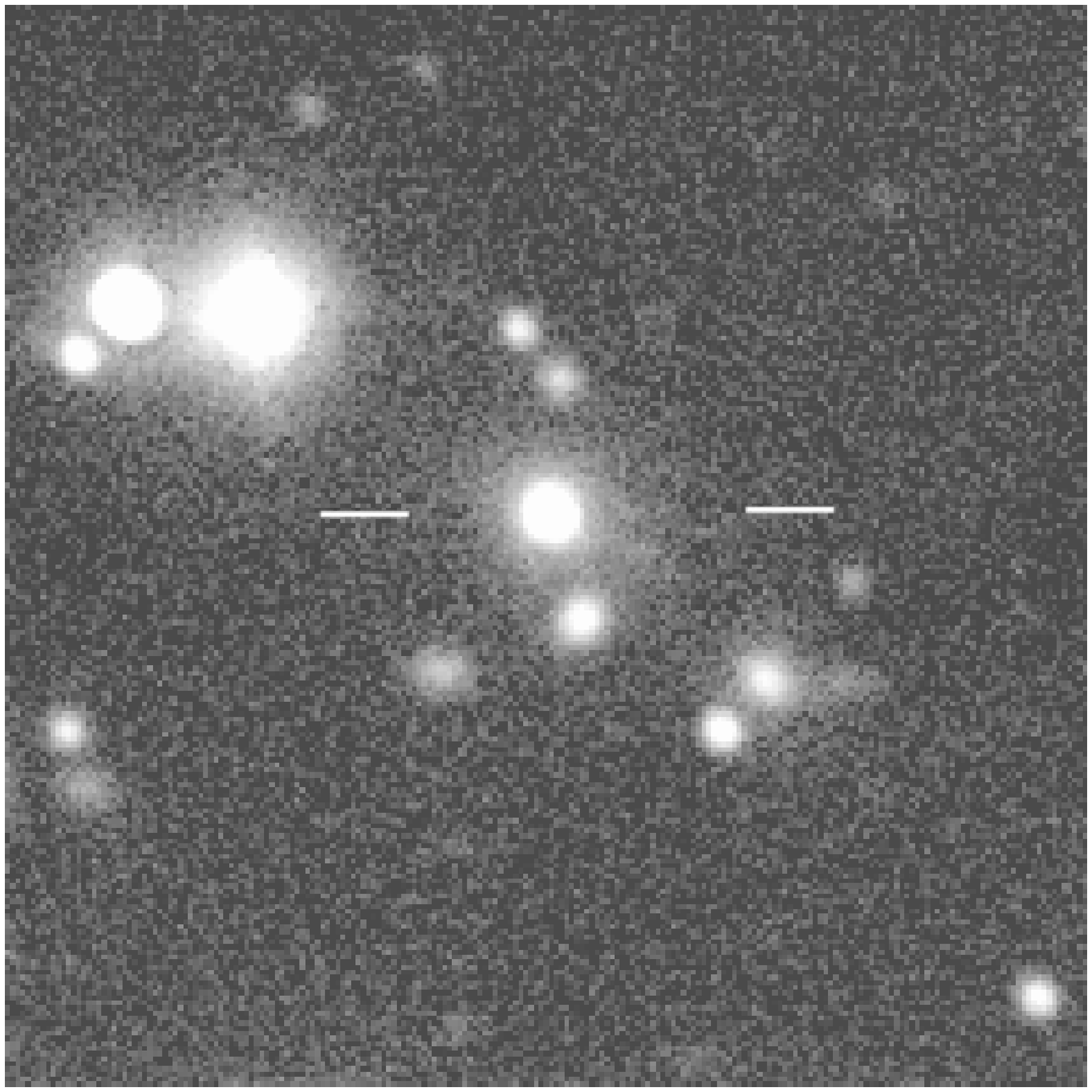}}
\subfigure[]{\includegraphics[width=8.0cm]{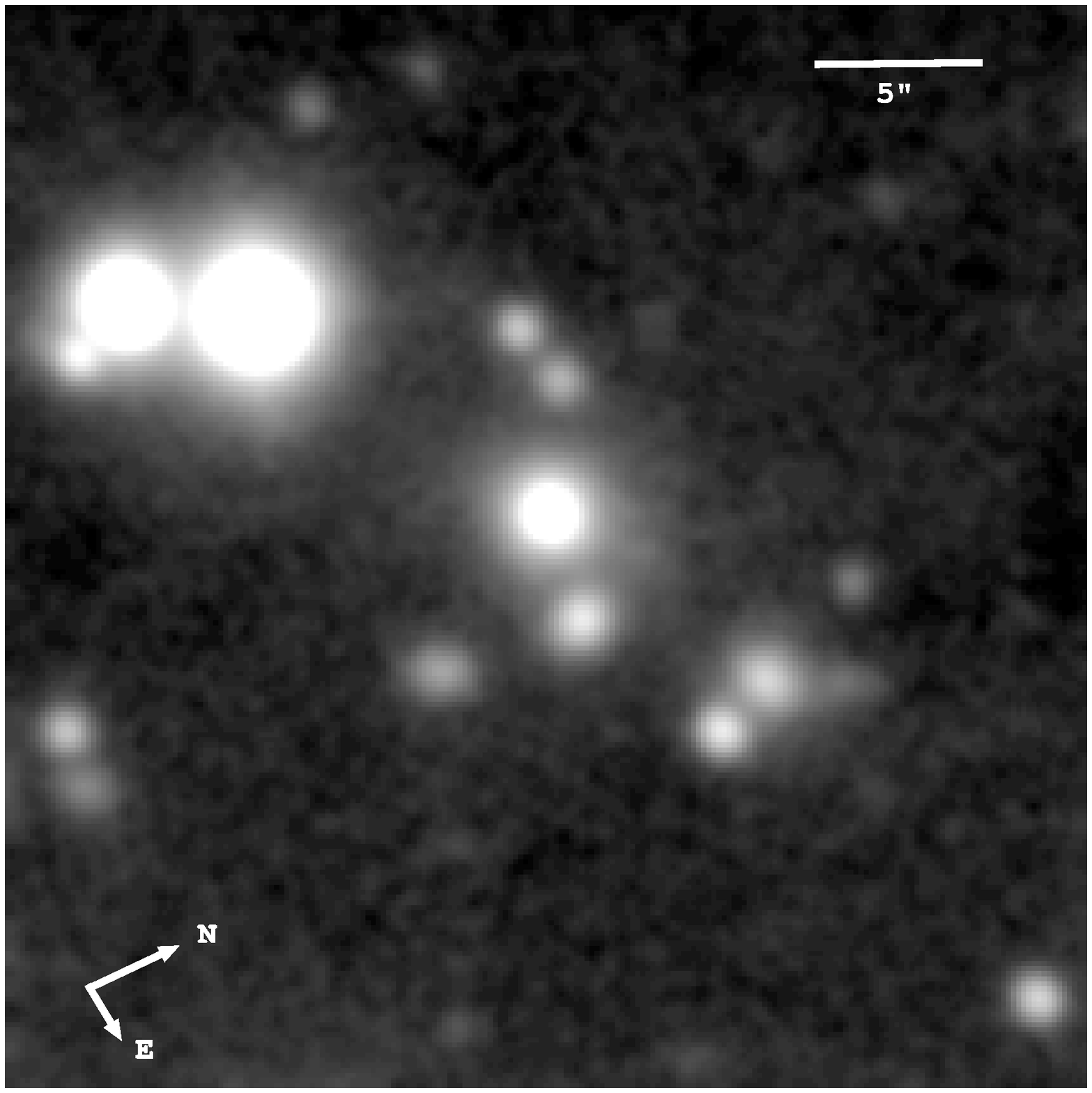}}
\caption{PKS 1938-15. (a) Unsharp-masked image. (b) Median filtered image.}
\label{pks1938_online} 
\end{figure*}

\begin{figure*}
\centering
\subfigure[]{\includegraphics[width=8.0cm]{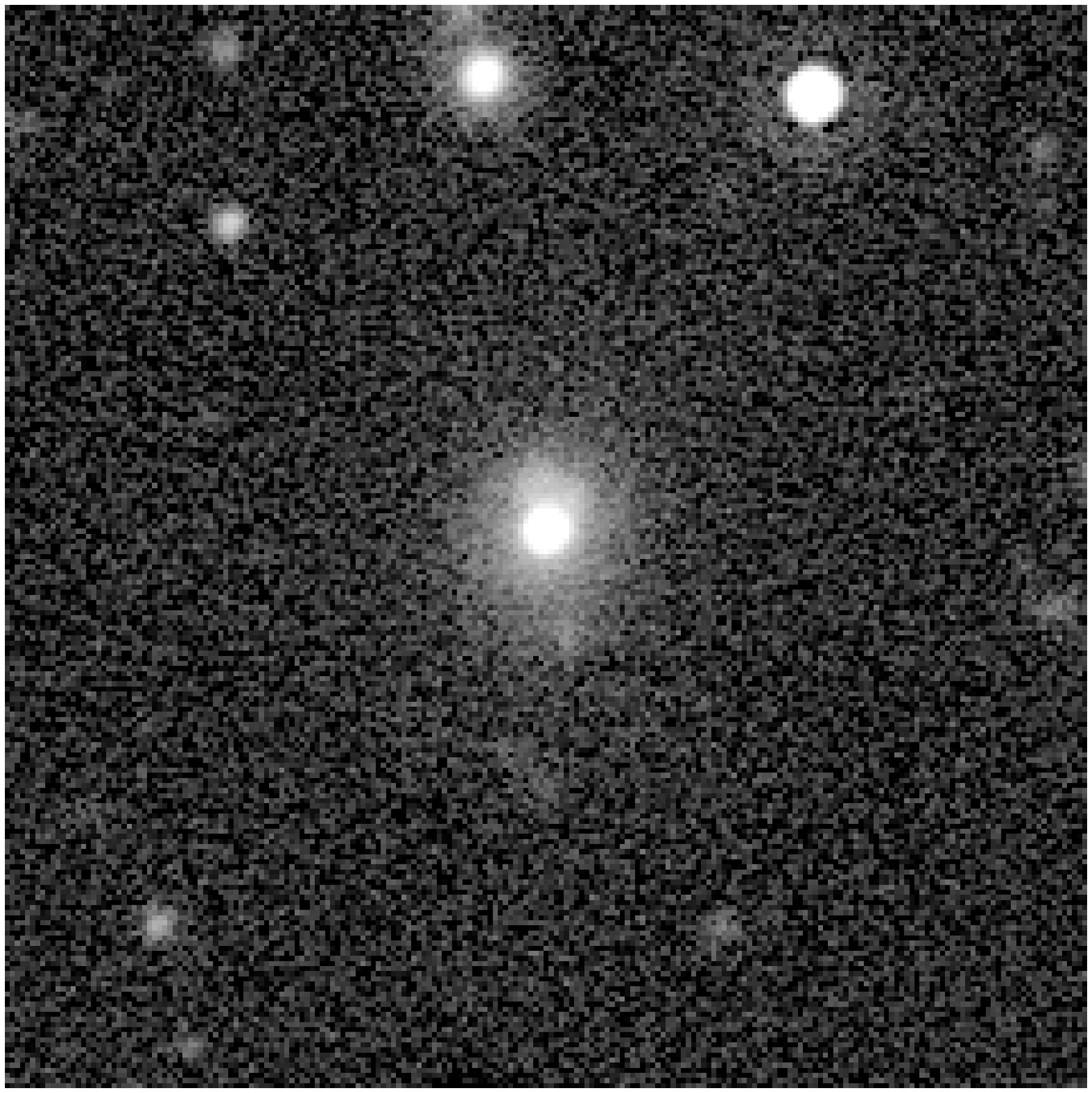}}
\subfigure[]{\includegraphics[width=8.0cm]{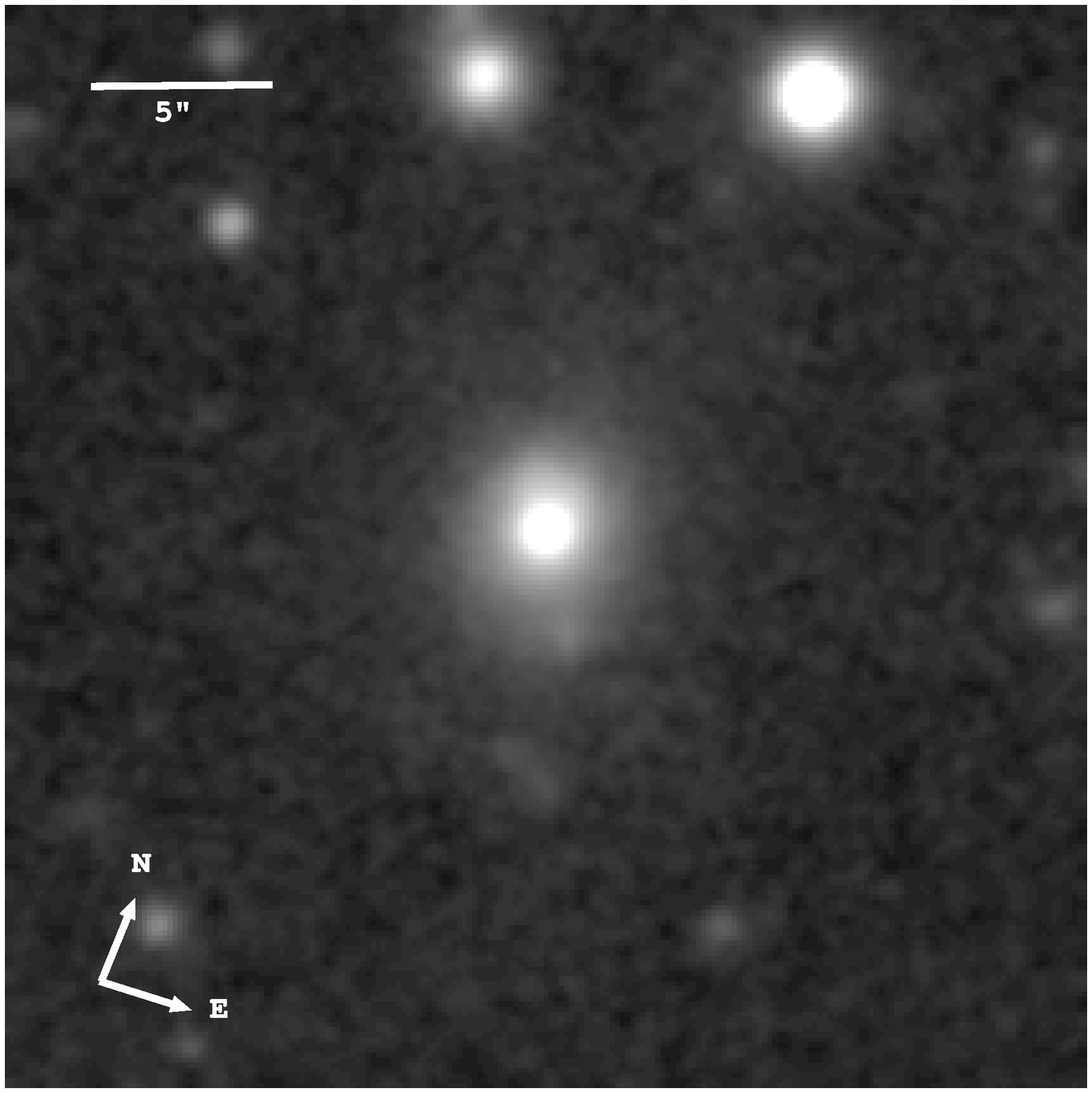}}
\caption{PKS 1602+01. (a) Unsharp-masked image. (b) Median filtered image.}
\label{pks1602_online} 
\end{figure*}

\begin{figure*}
\centering
\subfigure[]{\includegraphics[width=8.0cm]{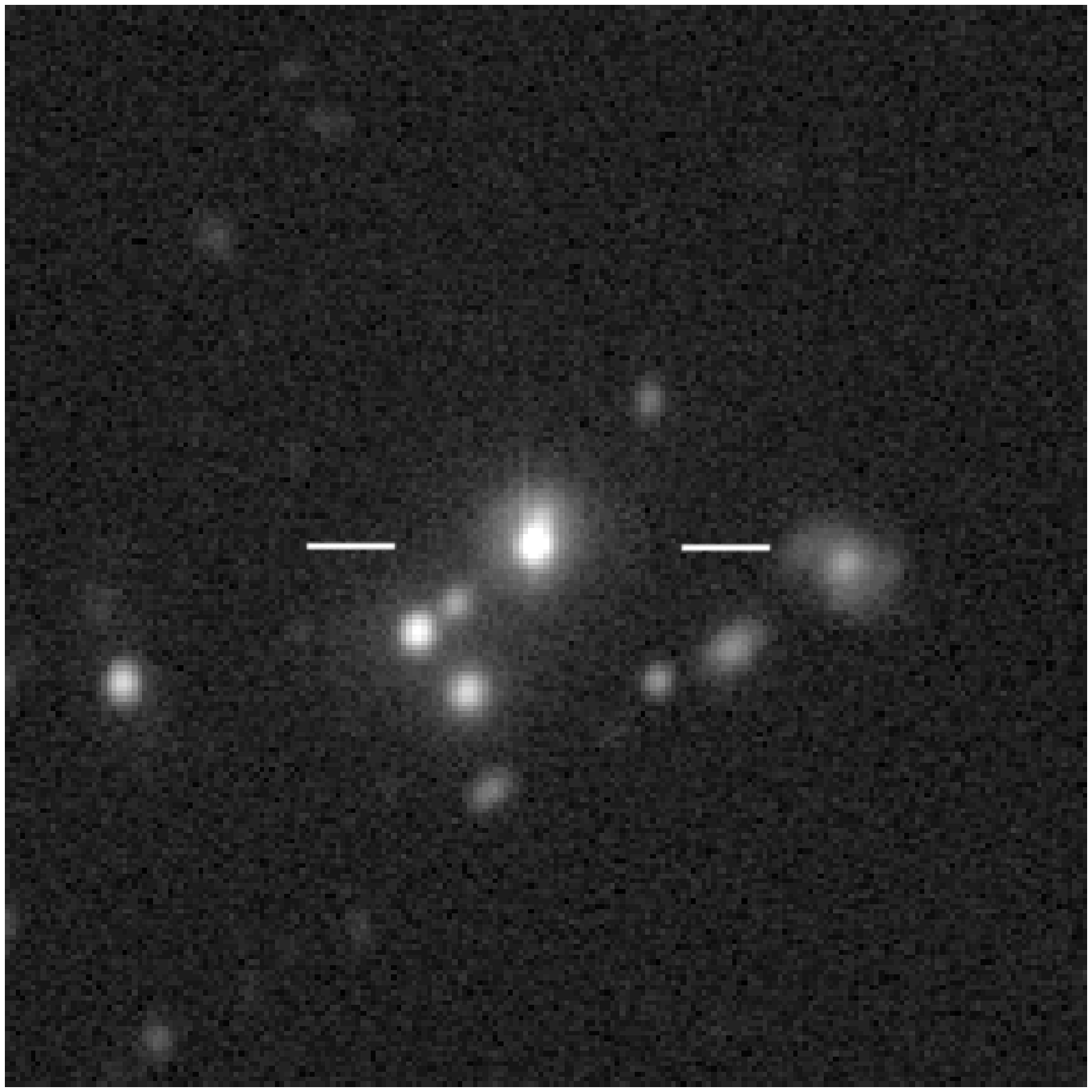}}
\subfigure[]{\includegraphics[width=8.0cm]{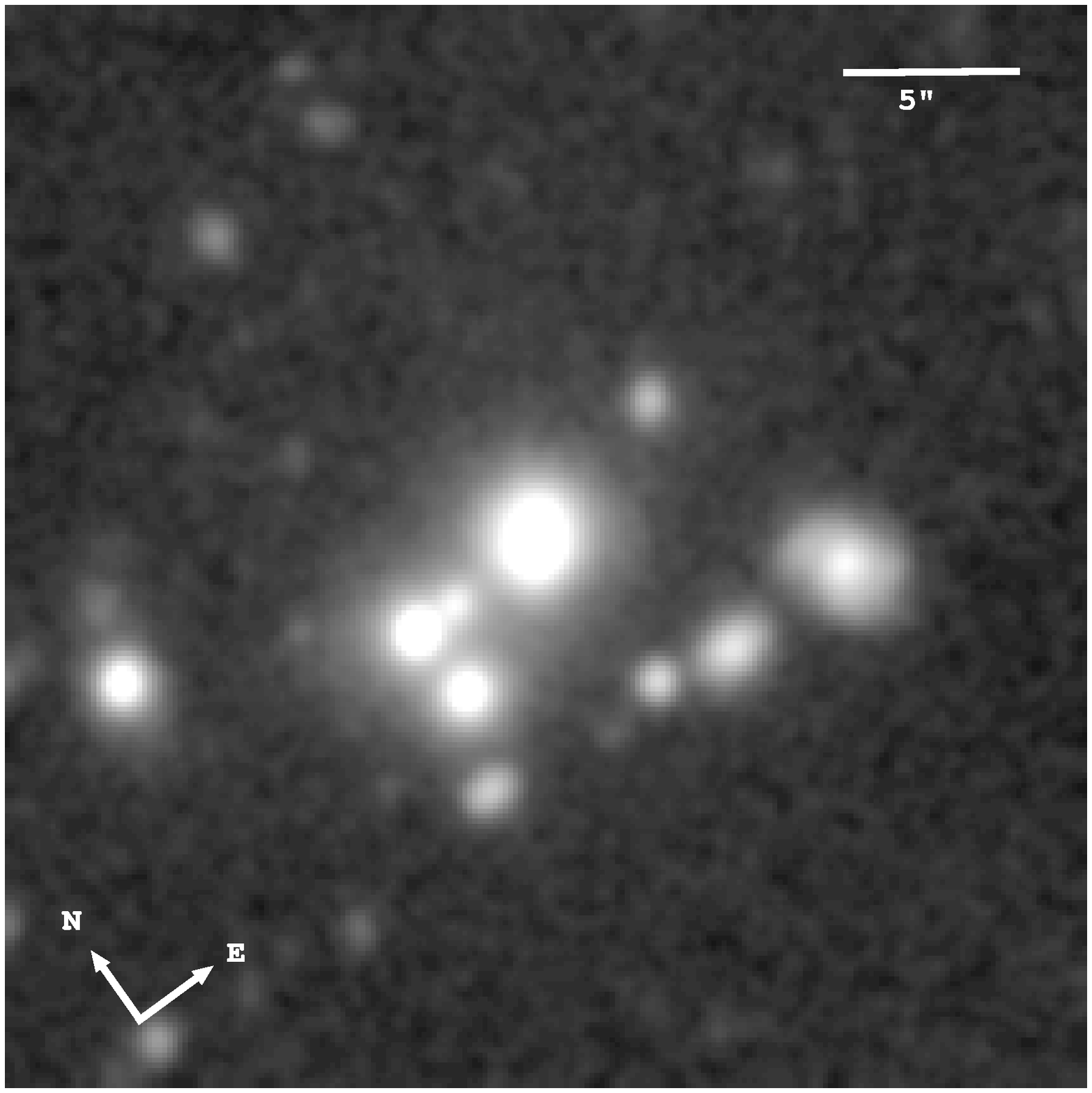}}
\caption{PKS 1306-09. (a) Unsharp-masked image. (b) Median filtered image.}
\label{pks1306_online} 
\end{figure*}

\begin{figure*}
\centering
\subfigure[]{\includegraphics[width=8.0cm]{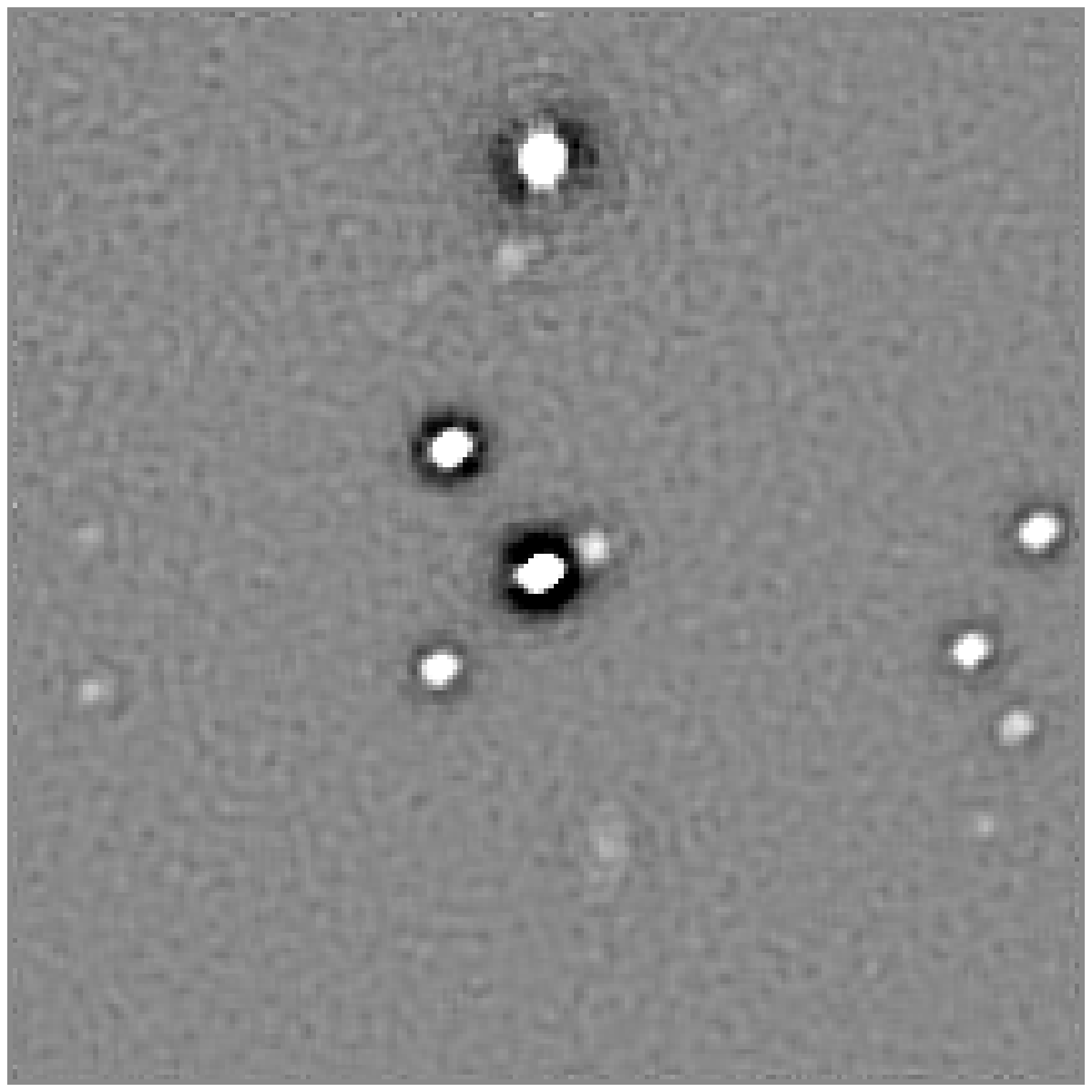}}
\subfigure[]{\includegraphics[width=8.0cm]{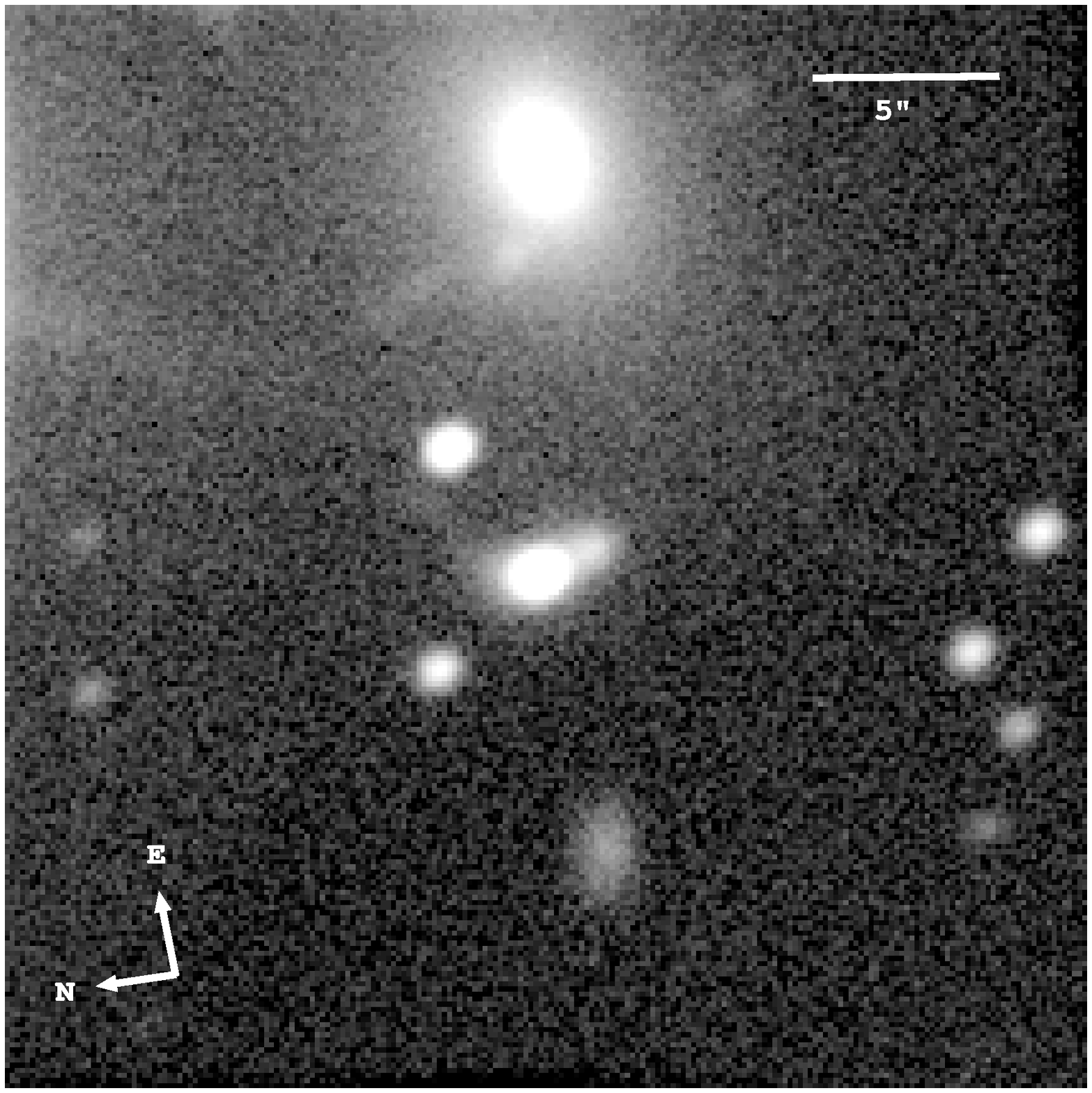}}
\caption{PKS 1547-79. (a) Smooth galaxy-subtracted image. (b) Unsharp-masked image.}
\label{pks1547_online} 
\end{figure*}

\begin{figure*}
\centering
\subfigure[]{\includegraphics[width=8.0cm]{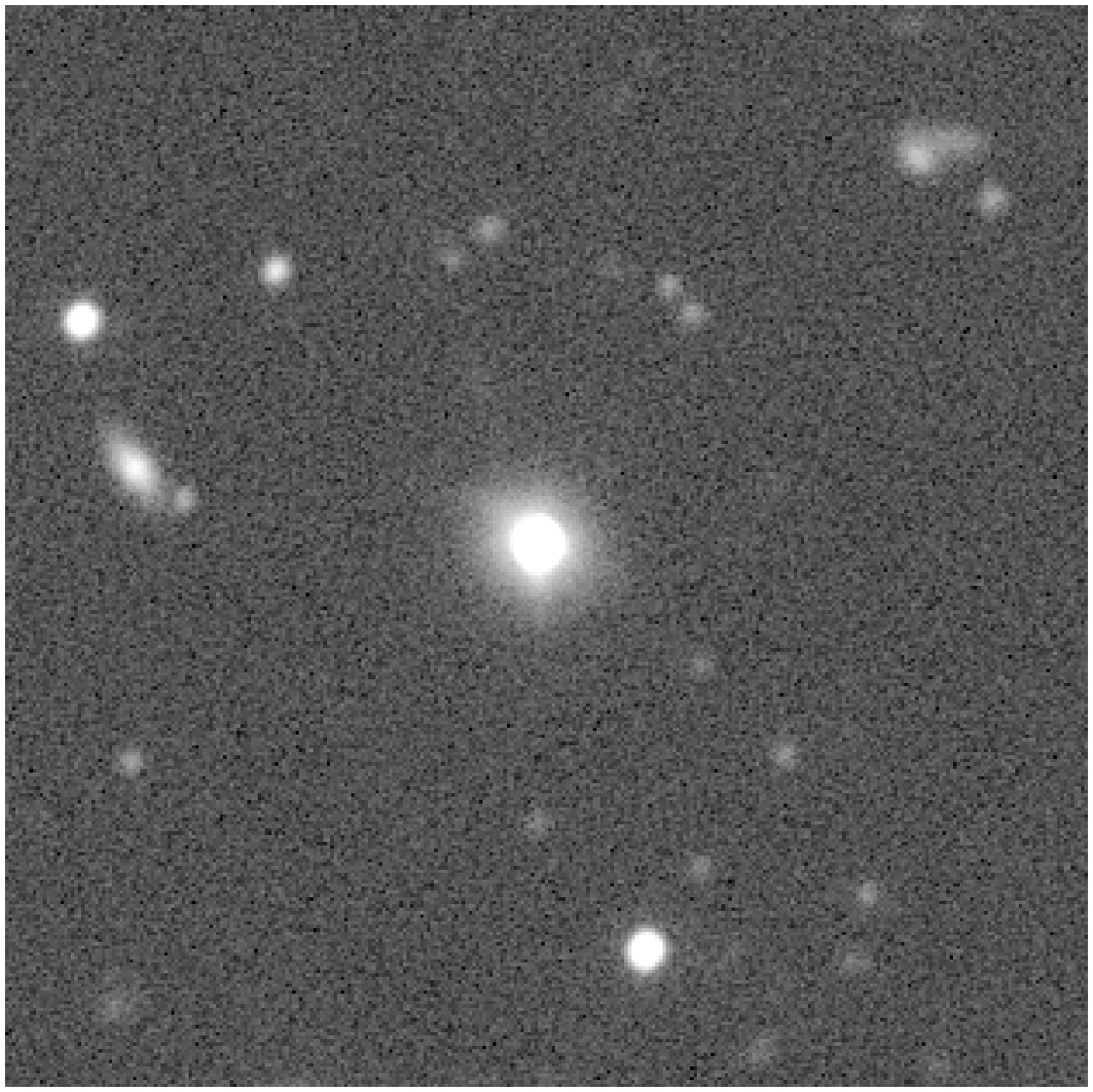}}
\subfigure[]{\includegraphics[width=8.0cm]{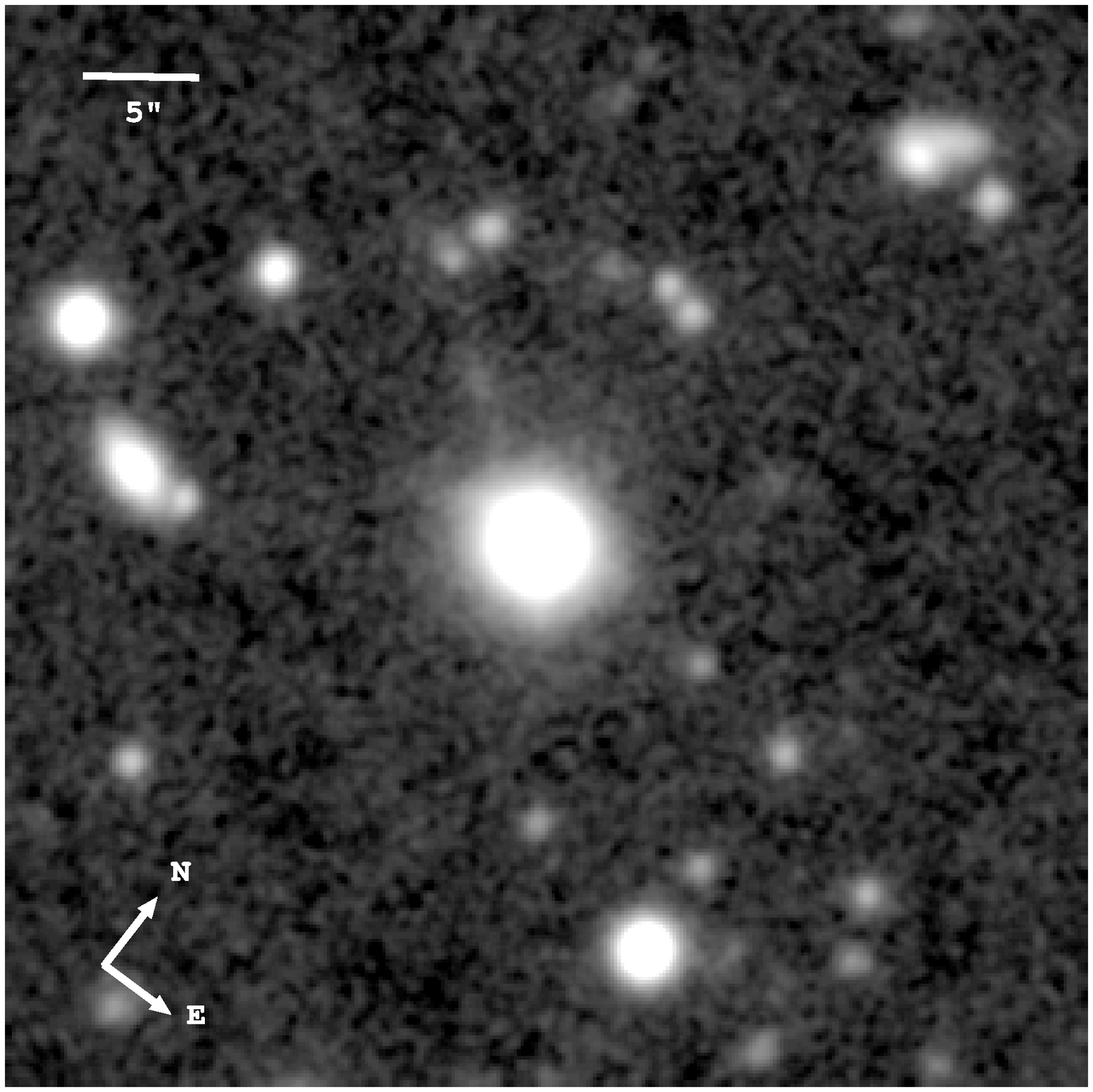}}
\caption{PKS 1136-13. (a) Unsharp-masked image. (b) Median filtered image.}
\label{pks1136_online} 
\end{figure*}

\begin{figure*}
\centering
\subfigure[]{\includegraphics[width=8.0cm]{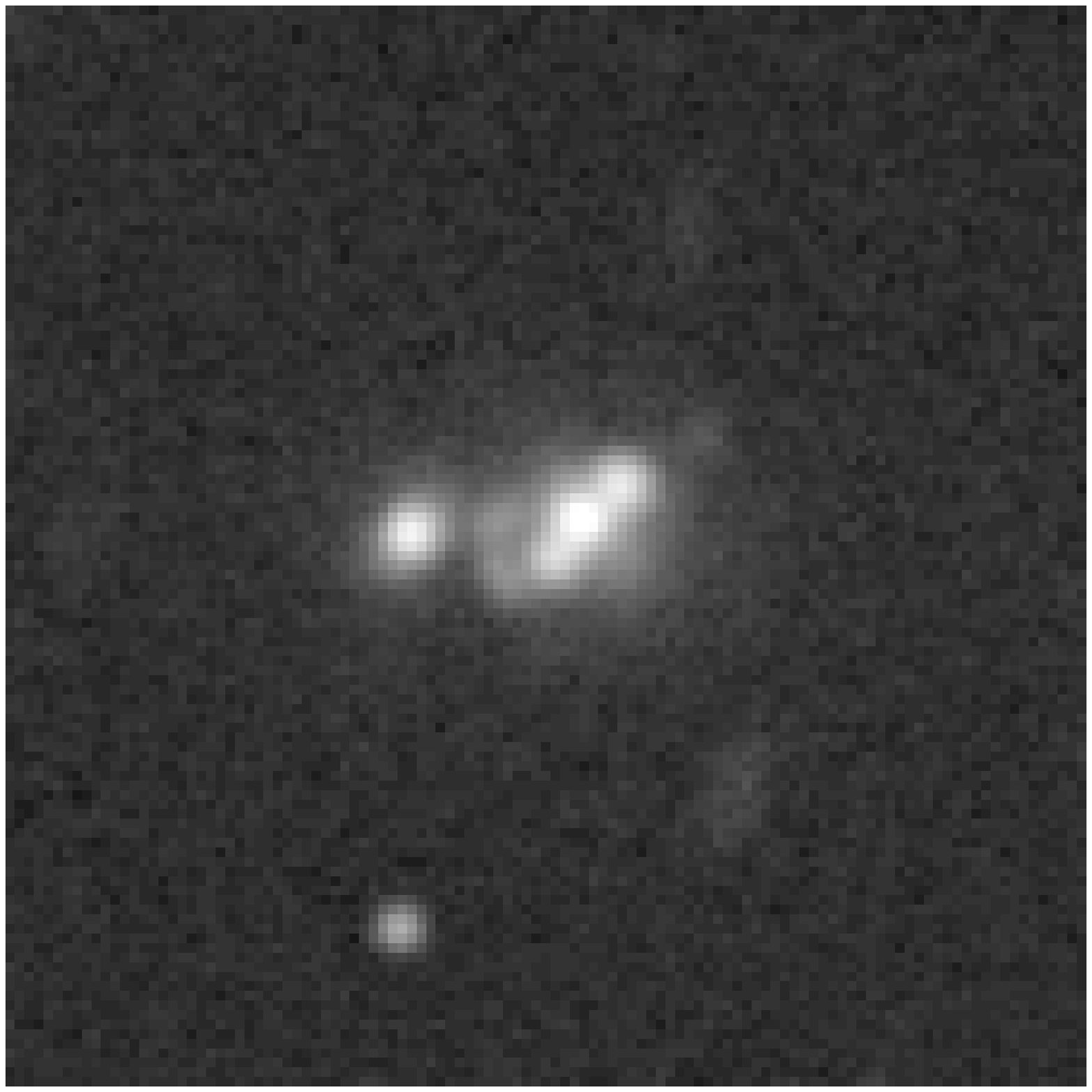}}
\subfigure[]{\includegraphics[width=8.0cm]{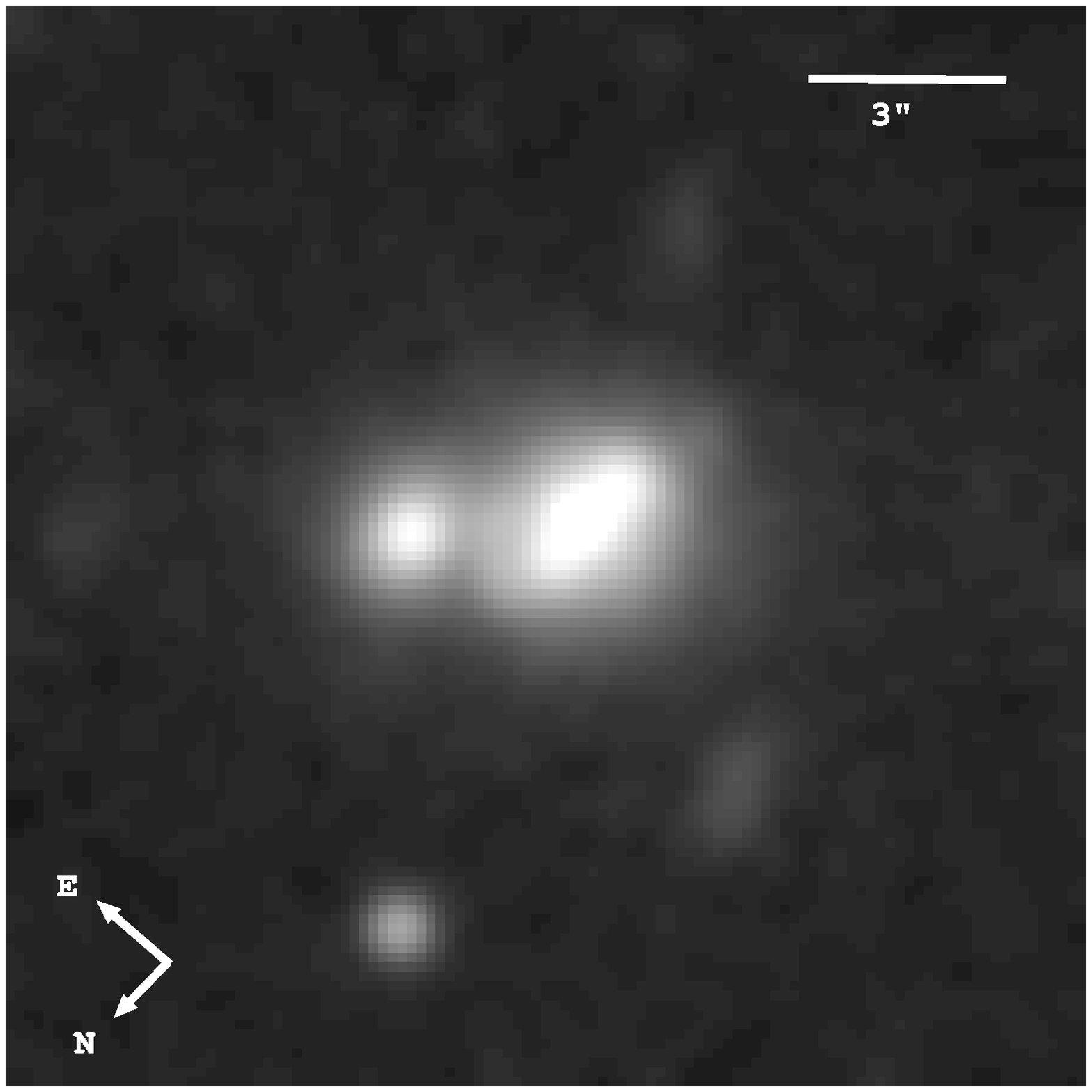}}
\caption{PKS 0117-15. (a) Unsharp-masked image. (b) Median filtered image.}
\label{pks0117_online} 
\end{figure*}

\begin{figure*}
\centering
\subfigure[]{\includegraphics[width=8.0cm]{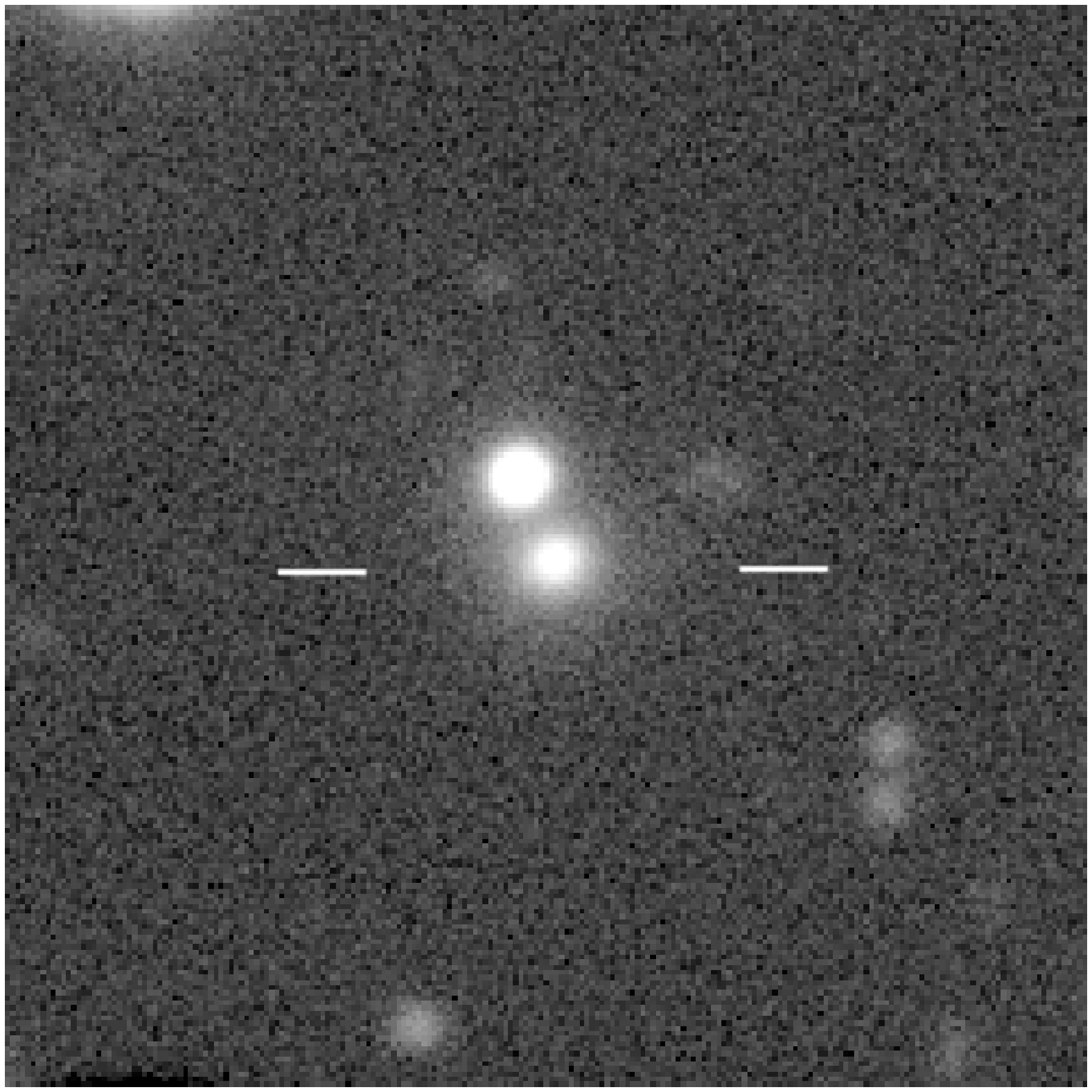}}
\subfigure[]{\includegraphics[width=8.0cm]{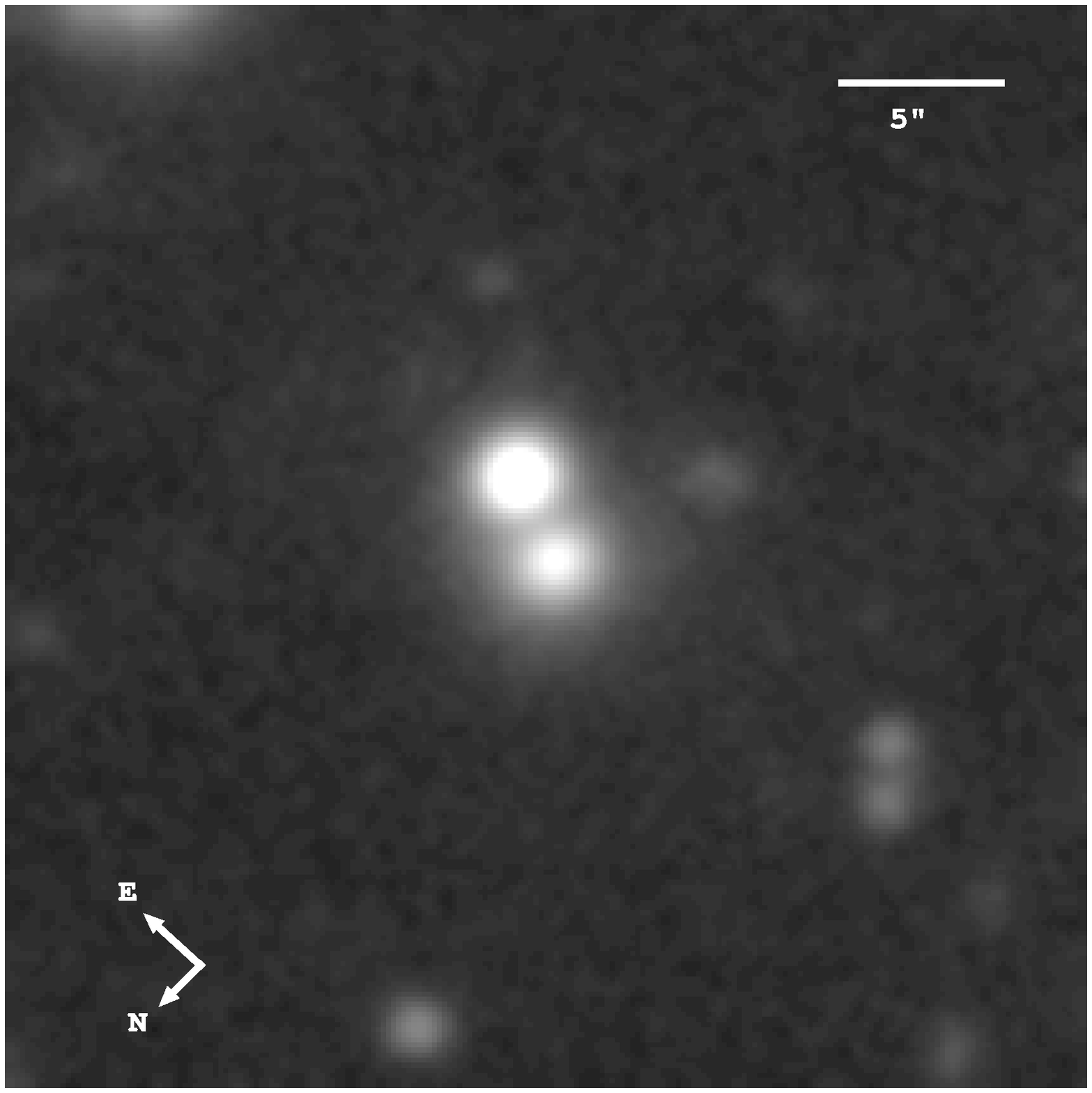}}
\caption{PKS 0252-71. (a) Unsharp-masked image. (b) Median filtered image.}
\label{pks0252_online} 
\end{figure*}

\begin{figure*}
\centering
\subfigure[]{\includegraphics[width=8.0cm]{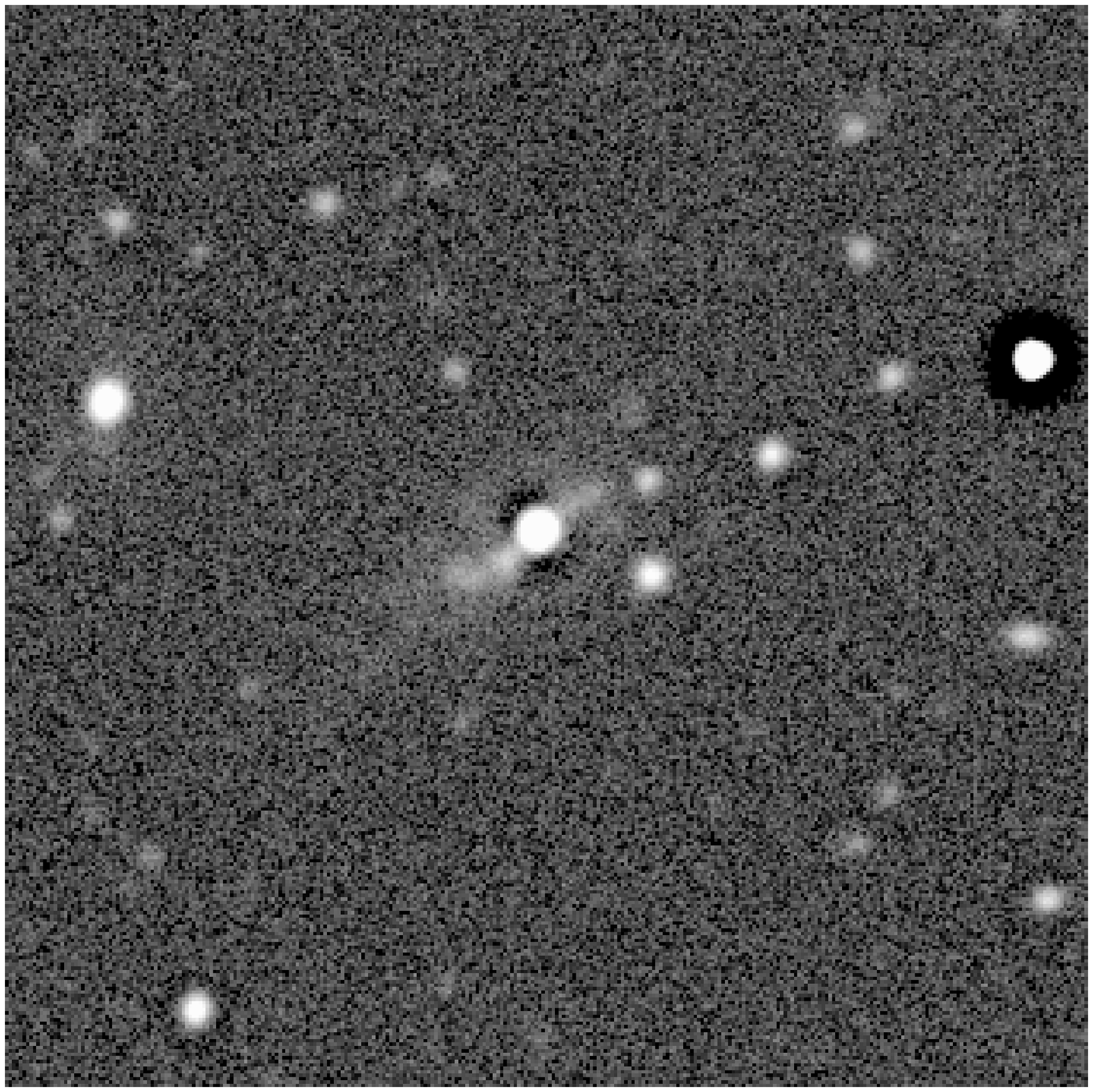}}
\subfigure[]{\includegraphics[width=8.0cm]{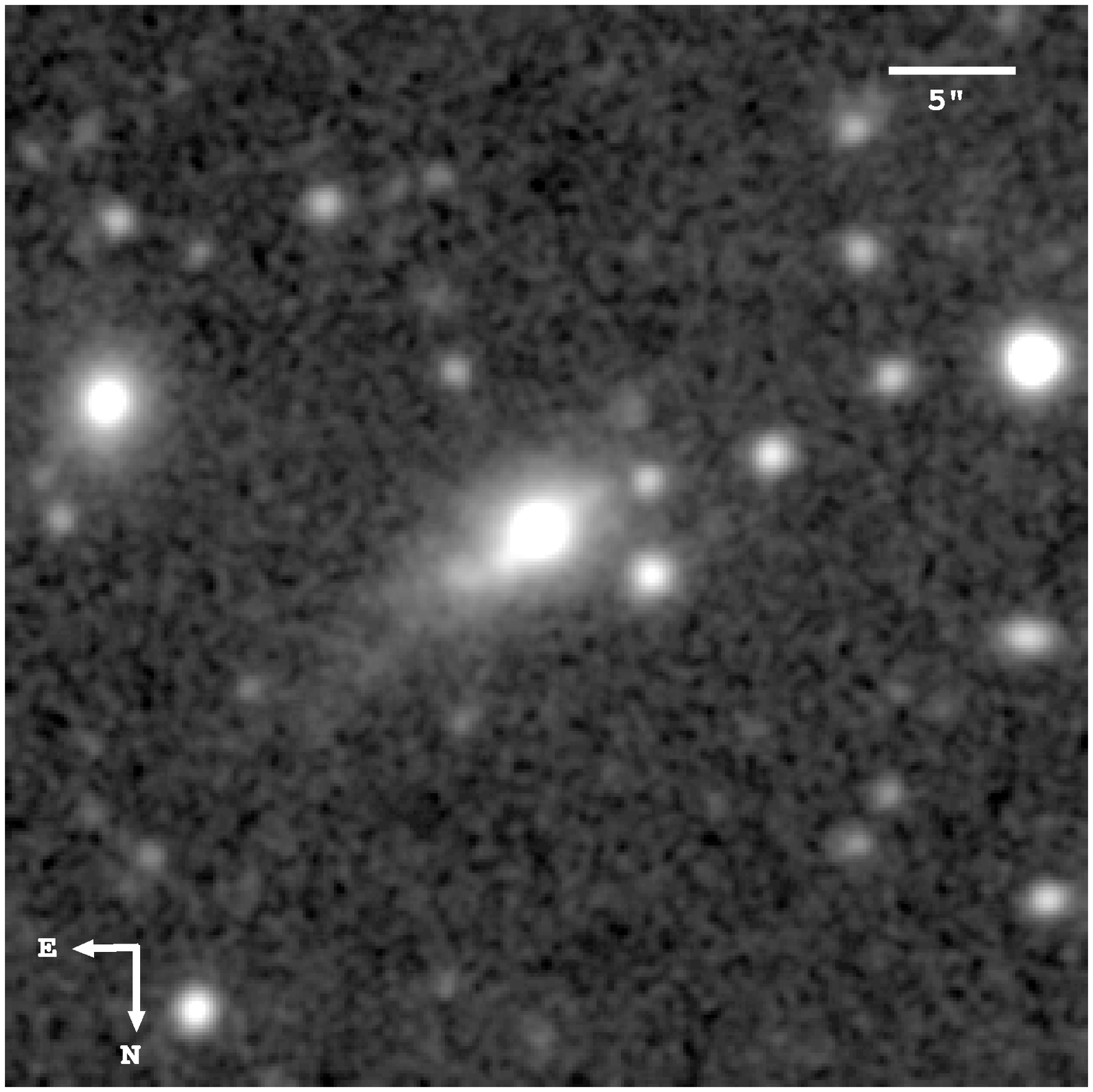}}
\caption{PKS 0235-19. (a) Unsharp-masked image. (b) Median filtered image.}
\label{pks0235_online} 
\end{figure*}

\begin{figure*}
\centering
\subfigure[]{\includegraphics[width=8.0cm]{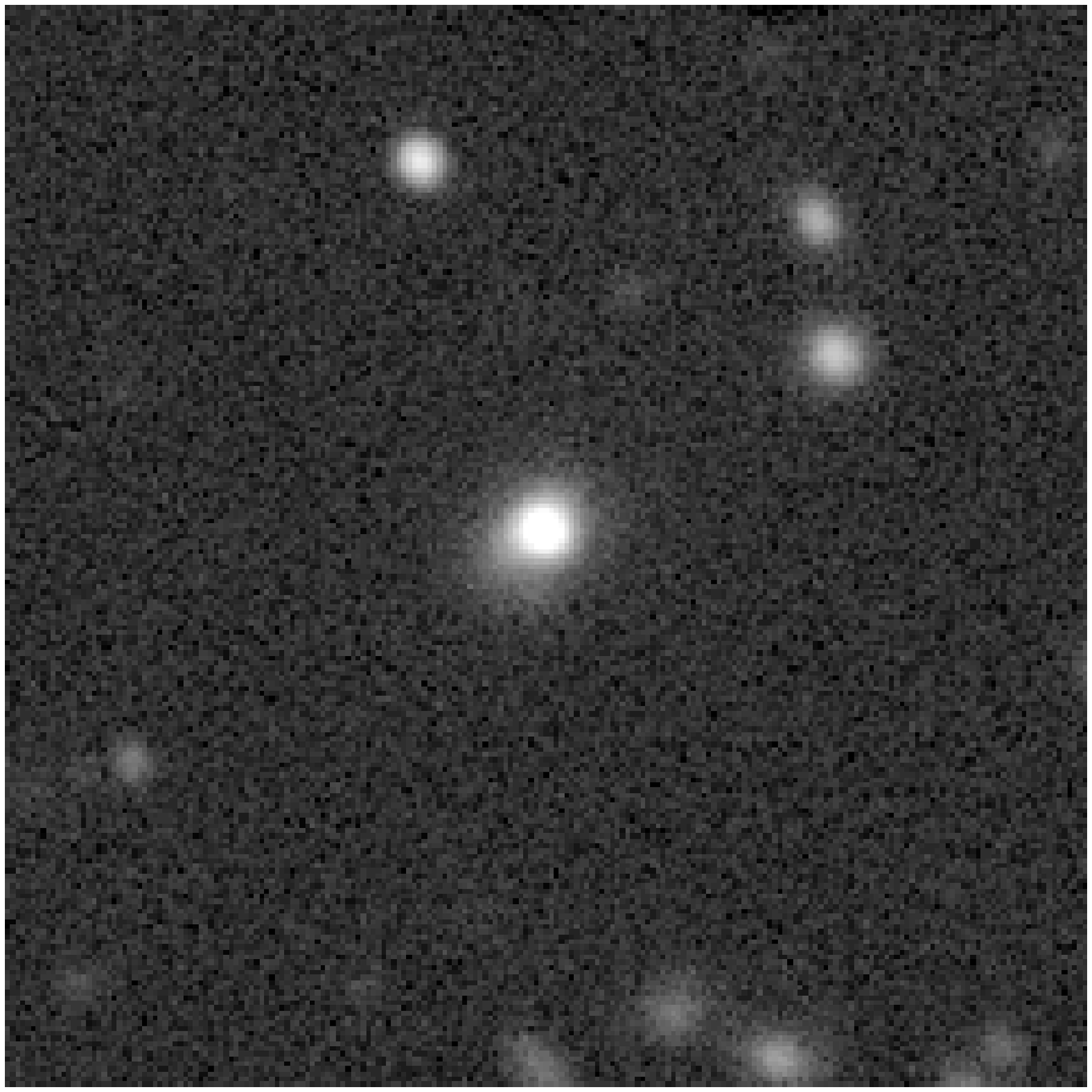}}
\subfigure[]{\includegraphics[width=8.0cm]{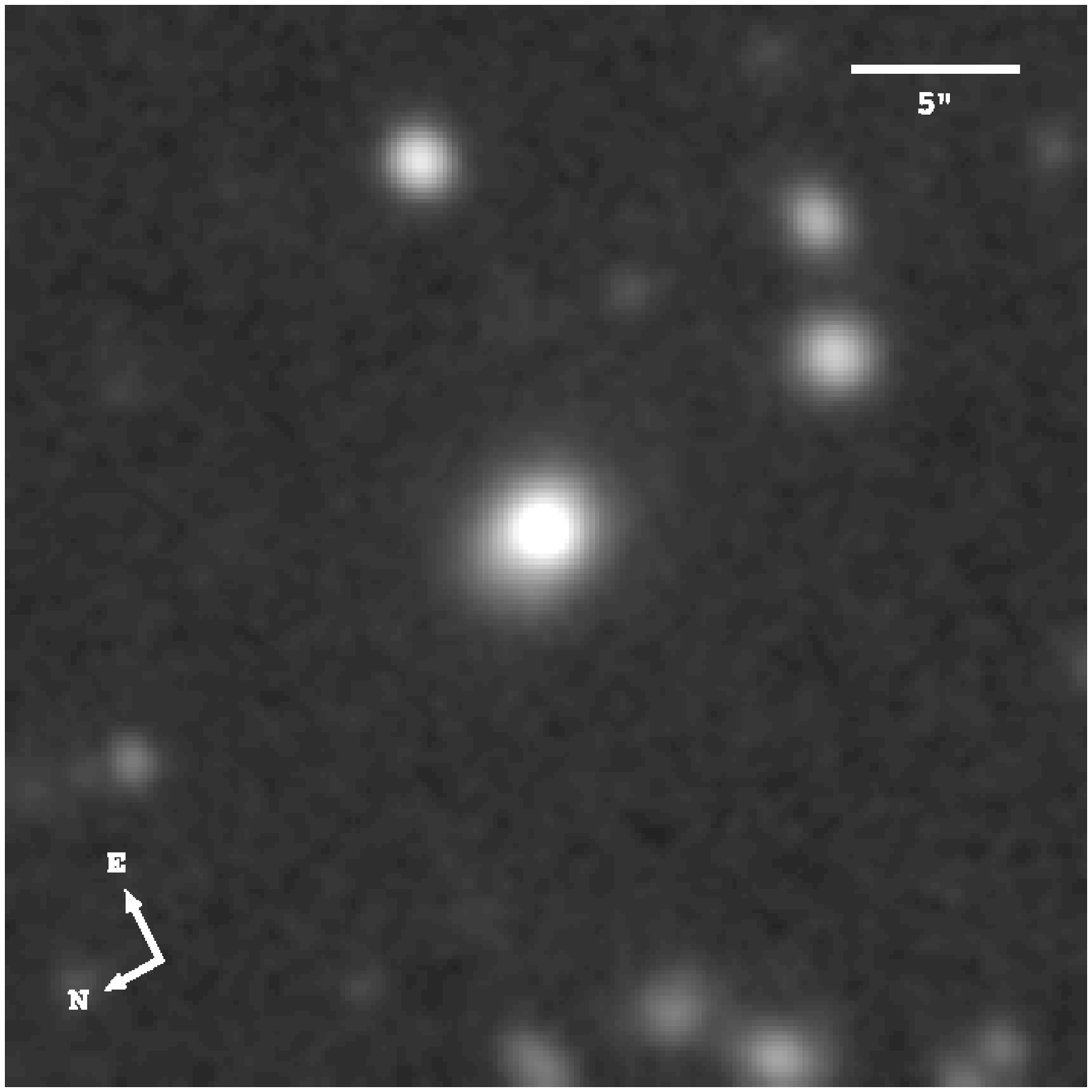}}
\caption{PKS 2135-20. (a) Unsharp-masked image. (b) Median filtered image.}
\label{pks2135_20_online} 
\end{figure*}

\begin{figure*}
\centering
\subfigure[]{\includegraphics[width=8.0cm]{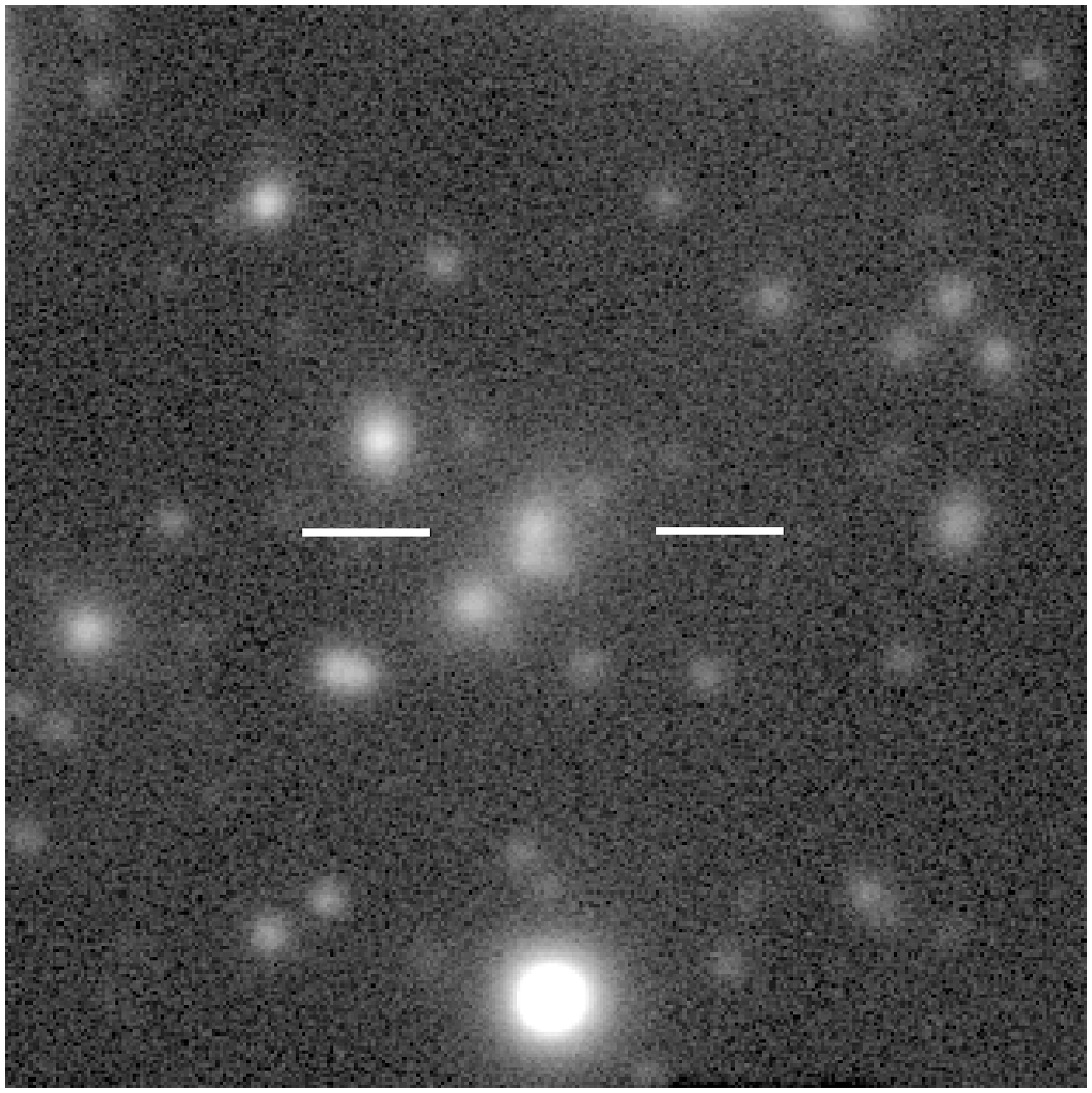}}
\subfigure[]{\includegraphics[width=8.0cm]{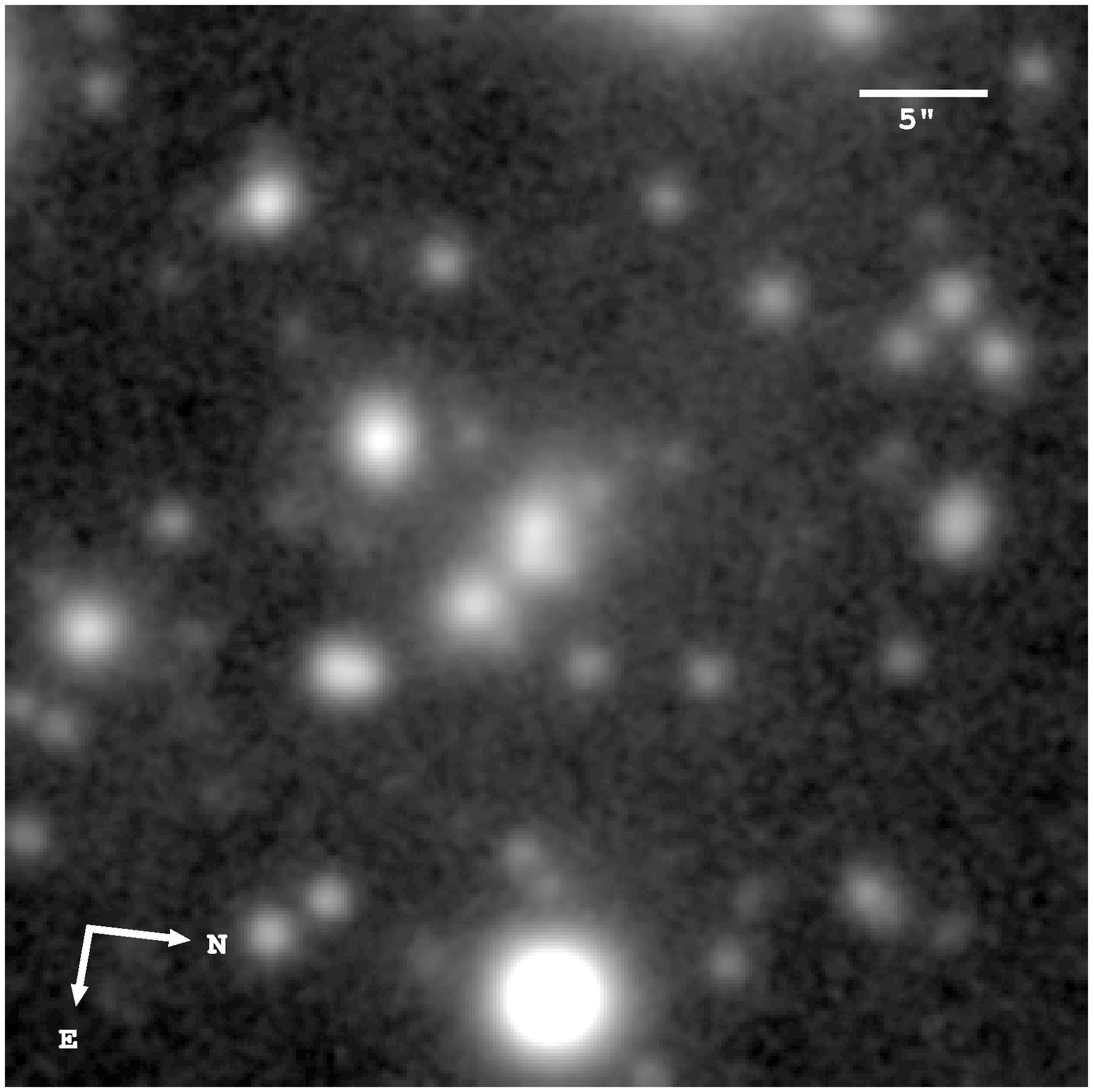}}
\caption{PKS 0409-75. (a) Unsharp-masked image. (b) Median filtered image.}
\label{pks0409_online} 
\end{figure*}

Here we present our processed Gemini GMOS-S images for all the galaxies in the sample, ordered by redshift, as
in Tables \ref{data}, \ref{mag}, and \ref{features} and in Appendix \ref{individual}. We tried the three techniques
described in Section \ref{analysis} with all the galaxies (namely, {\it Image filtering}, {\it Unsharp-masking}, 
and {\it} Smoothed galaxy subtraction), and present here the results for the two techniques in which the features
appear most clearly. The PRGs are placed at the centre of each field (Figures \ref{pks0620_online} to \ref{pks0409_online}),
unless otherwise indicated.

\clearpage

\clearpage

\clearpage

\clearpage

\label{lastpage}

\end{document}